\documentclass[a4paper,12pt,twoside,english]{report} 

\usepackage{babel}
\usepackage[T1]{fontenc}

\usepackage{ae}
\usepackage{url,xspace,boxedminipage}
\usepackage{amsfonts}
\usepackage{amssymb}
\usepackage{oldlfont} 
\input{tcilatex}
\renewcommand{\text}{\mbox}

\newtheorem{theorem}{Theorem}
\newtheorem{lemma}[theorem]{Lemma}
\newtheorem{proposition}[theorem]{Proposition}

\newtheorem{definition}[theorem]{Definition}
\newtheorem{corollary}[theorem]{Corollary}

\newcommand{\D}{\displaystyle}
\newcommand{\T}{\textstyle}
\newcounter{example}
\newcommand{\example}{\noindent{\bf Example \nopagebreak \stepcounter{example}%
                                     \arabic{example} }}
\renewcommand{\endproof}{\hspace*{\fill}$\Box$}

\newcommand{\C}{{\mathbb  C}}
\newcommand{\Ckl}{{\mathbb  C}} 
\newcommand{\Ckkl}{{\mathbb  C}}

\newcommand{\Z}{{\mathbb  Z}}
\newcommand{\R}{{\mathbb  R}}
\newcommand{\p}{{\mathbb  P}}
\newcommand{\Q}{{\mathbb  Q}}
\newcommand{\A}{{\mathbb  A}}
\newcommand{\N}{{\mathbb  N}}
\newcommand{\1}{{1\!\! 1}}
\newcommand{\E}{{\mathbb  E}}

\newcommand{\lnorm}{ \left| \! \left| \! \left| }
\newcommand{\rnorm}{ \right| \! \right| \! \right| }
\newcommand{\m}{{\rm m}}
\newcommand{\De}{{\rm D}}

\newcommand{\assumption}{\noindent{\bf Assumption }} 
\renewcommand{\note}{\noindent{\bf Note. \quad }} 
\newcommand{\remark}{\noindent{\bf Remark. \quad }} 
\newcommand{\remarks}{\noindent{\bf Remarks. \quad }}

\usepackage[colorlinks,
						bookmarks,
						pdftitle={Recent Results in
                 		Infinite Dimensional Analysis and 
                 		Applications to Feynman Integrals},
            pdfsubject={Feynman Integrals with White Noise Analysis},
            pdfkeywords={White Noise Analysis, Gaussian Analysis,
            Non-Gaussian Analysis, infinite dimensional Analysis,
            Feynman Integrals},
            pdfauthor={Werner Westerkamp},
            pagebackref,
            plainpages=false,
            linkcolor={blue},
            citecolor={blue},
            urlcolor={blue}  		
						]{hyperref}

\begin{document}

\pagenumbering{roman} 
\author{         vorgelegt von \smallskip \\
                 Werner Westerkamp \smallskip \\
                 aus Pr.~Oldendorf \\
                 \\
                 \href{mailto:Werner.Westerkamp@web.de}
                       {Werner.Westerkamp@web.de}
                 }
\title{          Recent Results in\\
                 Infinite Dimensional Analysis and\\ 
                 Applications to Feynman Integrals\\[15mm]
         {\large Dissertation zur Erlangung \\
                 des Doktorgrades der \\
                 Fakult\"at f\"ur Physik der \\
                 Universit\"at Bielefeld\\}}
\date{\vspace*{3cm}
\begin{tabular}{ll}
1.~Gutachter: & Prof.~Dr.~L.~Streit \\
2.~Gutachter: & Prof.~Dr.~Ph.~Blanchard \\[3mm]
Tag der Disputation: & 19.~Oktober 1995
\end{tabular}                            }
\maketitle
\tableofcontents

\LaTeXparent{dis3.tex}

\chapter{Introduction}

\setcounter{page}{1} \pagenumbering{arabic} \noindent
In recent years Gaussian analysis and in particular white noise analysis
have developed to a useful tool in applied mathematics and mathematical
physics. White noise analysis is a mathematical framework which offers
various generalizations of concepts known from finite dimensional analysis,
among them are differential operators and Fourier transform. For a detailed
exposition of the theory and for many examples of applications we refer the
reader to the recent monographs \cite{BeKo88,HKPS93,Ob94,Hi80} and the
introductory articles \cite{Kuo92,Po91,S94,W93}. \medskip

\noindent This work consists of three different main parts:

\begin{itemize}
\item  The generalization of the theory to an infinite dimensional analysis
with underlying {\it non-Gaussian} measure.

\item  Further development of Gaussian analysis.

\item  Applications to the theory of path-integrals.
\end{itemize}

Some of the results presented here have already been published as joint
works, see \cite{KLPSW94,LLSW94a,LLSW94b,CDLSW95,KoSW94,KSWY95}. We present
here a systematic exposition of this circle of ideas. \bigskip

\noindent {\large {\bf Non-Gaussian infinite dimensional analysis}}\\[2mm]
An approach to such a theory was recently proposed by \cite{AKS93}. For
smooth probability measures on infinite dimensional linear spaces a
biorthogonal decomposition is a natural extension of the orthogonal one that
is well known in Gaussian analysis. This biorthogonal ``Appell'' system has
been constructed for smooth measures by Yu.L.~Daletskii \cite{Da91}. For a
detailed description of its use in infinite dimensional analysis we refer to 
\cite{ADKS94}.\bigskip\ 

\noindent {\it Aim of the present work (Chapter 3). }We consider the case of
non--degenerate measures on co-nuclear spaces with analytic characteristic
functionals. It is worth emphasizing that no further condition such as
quasi--invariance of the measure or smoothness of logarithmic derivatives
are required. The point here is that the important example of Poisson noise
is now accessible.

\noindent For any such measure $\mu $ we construct an Appell system $\A^\mu $
as a pair $(\p ^\mu ,\Q ^\mu )$ of Appell polynomials $\p^\mu $ and a
canonical system of generalized functions $\Q ^\mu $, properly associated to
the measure $\mu $.\bigskip\ 

\pagebreak[3] \noindent {\it Central results. }Within the above framework

\begin{itemize}
\item  we obtain an explicit description of the test function space
introduced in \cite{ADKS94} (Theorem \ref{N1E1min})

\item  in particular this space is in fact identical for all the measures
that we consider

\item  characterization theorems for generalized as well as test functions
are obtained analogously as in Gaussian analysis \cite{KLS94} for more
references see \cite{KLPSW94} (Theorems \ref{CmuChar} and \ref{CharTh})

\item  the well known Wick product and the corresponding Wick calculus \cite
{KLS94} extends rather directly (Section \ref{Wick})

\item  similarly, a full description of positive distributions (as measures)
will be given (Section \ref{PosDist}).
\end{itemize}

\noindent Finally we should like to underline here the important conceptual
role of holomorphy here as well as in earlier studies of Gaussian analysis
(see e.g., \cite{PS91a,Ou91,KLPSW94,KLS94} as well as the references cited
therein).

All these results are collected in Chapter 3 as well as \cite{KSWY95}. 
\bigskip

\noindent {\large {\bf Gaussian analysis}} \\[2mm] In recent years there was
an increasing interest in white noise analysis, due to its rapid
developments in mathematical structure and applications in various domains.
Especially, the circle of ideas going under the heading `characterization
theorems' has played quite an important role in the last few years. These
results \cite{Ko80a,Lee89,PS91a}, and their variations and refinements (see,
e.g., \cite{KPS91,MY90,Ob91,SW93,Yan90,Zh92}, and references quoted there),
provide a deep insight into the structure of spaces of smooth and
generalized random variables over the white noise space or -- more generally
-- Gaussian spaces. Also, they allow for rather straightforward applications
of these notions to a number of fields: for example, Feynman integration 
\cite{FPS91,HS83,KaS92,LLSW94a}, representation of quantum field theory \cite
{AHPS89,PS93}, stochastic equations \cite{CLP93,Po91,Po92,Po94},
intersection local times \cite{FHSW94,Wa91}, Dirichlet forms \cite
{AHPRS90a,AHPRS90b,HPS88}, infinite dimensional harmonic analysis \cite{Hi89}
and so forth. Moreover, characterization theorems have been at the basis of
new methods for the construction of smooth and generalized random variables 
\cite{KoS93,MY90} which seem to be useful in applications untractable by
existing methods (e.g., \cite{HLOUZ93a,HLOUZ93b}). \medskip

One of the basic technical ideas in the development of the theory is the use
of dual pairs of spaces of test and generalized functionals. Since the
usefulness of a particular test function space depends on the application
one has in mind various dual pairs appear in the literature. In this work we
are particulary interested in the following spaces:

\begin{itemize}
\item  {\bf The Hida spaces}\\We construct a nuclear rigging 
$$
({\cal N})\,\subset \,L^2(\mu )\subset \,({\cal N})^{\prime } 
$$
We give the construction of the second quantized space $({\cal N})$ solely
in terms of the topology of ${\cal N}$, independent of the particular
representation as a projective limit. The purpose of Section 4.1 is
four-fold: We wish 1.\ to clarify and generalize the structure of the
existing characterization theorems, and at the same time, 2.\ to review and
unify recent developments in this direction, 3.\ to establish the connection
to rich, related mathematical literature \cite{AR73,Co82,Di81,Si69,Za76},
which might be helpful in future developments, and -- last but not least --
4.\ to fill a gap in the article \cite{PS91a}. In the course of doing this,
we also establish some new results, for instance an analytic extension
property of $U$--functionals, and the topological invariance of certain
spaces of generalized random variables with respect to different
construction schemes. \\The material presented in Section 4.1 is the central
part of \cite{KLPSW94}

\item  {\bf The test function spaces }${\cal G}$ {\bf \ and }${\cal M}$\\In
section 4.3 we discuss the space ${\cal G}$ introduced in \cite{PT94}. This
space and is dual are interesting because all terms in the chaos expansion
are given by Hilbert space kernels. So also the distributions have an
expansion in a series of $n$-fold stochastic integrals. \\A second useful
property is that ${\cal G}$ is an algebra under pointwise multiplication 
\cite{PT94} which is larger than $({\cal N})$. Since we are interested in
more general pointwise products, we introduce a second test function space $%
{\cal M}$ which again is bigger than ${\cal G}$. One can not expect that $%
{\cal M}$ is closed under multiplication but we will show that pointwise
multiplication is a separately continuous bilinear map ${\cal G}\times {\cal %
M}\rightarrow {\cal M}$. (Corollary \ref{GMtoM}). \\We will see that the
shift operator $\tau _\eta :\varphi \mapsto \varphi (\cdot +\eta )\ ,\quad
\eta \in {\cal H}$ is well defined from ${\cal G}$ into ${\cal G}$ and $%
{\cal M}$ into ${\cal M}$ and that we can extend $\tau _\eta $ to complex $%
\eta \in {\cal H}_{\Ckl }$ (Theorem \ref{tauetaMG}) Using this it is easy to
see that ${\cal G}$ and ${\cal M}$ are closed under G\^ateaux
differentiation. \\In section 4.3.3.2 we consider the composition of test
functions with projection operations on ${\cal N}^{\prime }$. This is of
particular interest to understand the action of Donsker's delta on test
functions. Note that Donsker's delta is in ${\cal M}^{\prime }$ (Theorem \ref
{DeltaInMSt}). We will understand what it means to integrate out this delta
distribution (Proposition \ref{DeltaMode}).

\item  {\bf The Meyer--Yan triple }\\We sketch the well-known construction
of the triple \cite{MY90} and state a convenient form of the
characterization theorem for generalized functions \cite{KoS93}. For later
use we add a corollary controlling the convergence of a sequence of
generalized functions. Furthermore the integration of a family of
distributions is discussed and controlled in terms of $S$-transform.
\end{itemize}

Besides the discussion of various spaces of test and generalized functions
Chapter 4 also contains a discussion of the scaling operator $\sigma _z$
which suggests one of the possible approaches to path integrals in a white
noise framework. We will collect some properties of $\sigma _z$ and specify
its domain and range where it acts continuously (similar to \cite{HKPS93}).
But applications to path integrals require extended domains of $\sigma _z$.
This naturally leads to the study of traces. In Proposition \ref{JoKal} we
give sufficient conditions on $\varphi \in L^2(\mu )$ to ensure that $\sigma
_z\varphi $ exists in some useful sense.

We close Chapter 4 by a detailed discussion of Donsker's delta function. In
particular we study its behavior under $\sigma _z$. Most of this results
have already been published in \cite{LLSW94b}.\bigskip

\pagebreak[3] \noindent {\large {\bf Applications to Feynman integrals}} 
\\[2mm] Path integrals are a useful tool in many branches of theoretical
physics including quantum mechanics, quantum field theory and polymer
physics. We are interested in a rigorous treatment of such path integrals.
As our basic example we think of a quantum mechanical particle.

On one hand it is possible to represent solutions of the heat equation by a
path integral representation, based on the Wiener measure in a
mathematically rigorous way. This is stated by the famous Feynman Kac
formula. On the other hand there have been a lot of attempts to write
solutions of the Schr\"odinger equation as a Feynman (path) integral in a
useful mathematical sense.

Unfortunately, however there can be no hope of extending the theory of
invariant measure from finite to infinite dimensional spaces. For example,
one may easily prove that no reasonably well-behaved translation invariant
measure exists on any infinite-dimensional Hilbert space. More specifically,
for any translation invariant measure on a infinite dimensional Hilbert
space such that all balls are measurable sets there must be many balls whose
measure is either zero or $\infty $. This is the reason why the formal
expression ${\cal D}^\infty x$ used in some physical textbooks is
problematic and misleading.

One may have some hope that the ill defined term ${\cal D}^\infty x$
combines with the kinetic energy term to produce a well defined complex
measure with imaginary variance $\sigma ^2=i$, or that this combination is
the limit of Gaussian measures. But this causes problems if we assume that
cylinder functions are integrated in the obvious way, see 
\cite[p.217]{Ex85}.

\begin{theorem}
{\rm (Cameron \cite{C60})}\\Any finite (complex or real) measure with $N$%
-dimensional densities%
$$
\rho _{t_N>\cdots >t_0}(x_N,\ldots ,x_0)=\prod_{j=1}^N\frac 1{\sqrt{2\pi
i\gamma (t_j-t_{j-1})}}~\exp \left( i\frac{(x_j-x_{j-1})^2}{2\gamma \left(
t_j-t_{j-1}\right) }\right) 
$$
must have $i\gamma \in \R _{+}$.
\end{theorem}

So there is no hope for measure theory to solve the problems with path
integration.

The successful methods are always more involved and less direct than in the
Euclidean (i.e.,\ Feynman Kac) case. Among them are analytic continuation,
limits of finite dimensional approximations and Fourier transform. We are
not interested in giving full reference on various theories of Feynman
integrals (brief surveys can be found in \cite{Ex85,Ta75}) but we like to
mention the method in \cite{AHK76} using Fresnel integrals. Here we have
chosen a white noise approach.\medskip\ 

The idea of realizing Feynman integrals within the white noise framework
goes back to \cite{HS83}. The ``average over all paths'' is performed with a
Hida distribution as the weight (instead of a measure). The existence of
such Hida distributions corresponding to Feynman integrands has been
established in \cite{FPS91}.

In this white noise framework we define the Feynman {\it integrand} as a
white noise distribution. Its expectation reduces to the Feynman {\it %
integral}, which had to be defined. But the Feynman {\it integrand} defined
before may also be very useful. The point is that at least the pairing with
the corresponding test functions is well defined. In this sense the Feynman
integrand serves as an integrator. We will illustrate this by two examples.

\begin{enumerate}
\item  Since all of our test function spaces contain smooth polynomials, all
``moments'' of the Feynman integrand are well defined. Furthermore the
singularity of the Feynman integrand gives some information on the growth of
the moments (w.r.t. $n$) i.e., if we have 
$$
\left| \left\langle \!\!\!\left\langle {\rm I},\;\tprod_{j=1}^n\langle \cdot
,\xi _j\rangle \right\rangle \!\!\!\right\rangle \right| \leq
(n!)^xC^n\tprod_{j=1}^n|\xi _j|_p\ ,\qquad \xi _j\in {\cal S}(\R ) 
$$
for some $C>0$ and some continuous norm on ${\cal S}(\R )$, then \\ $%
x=1\Leftrightarrow {\rm I}\in ({\cal S})^{-1},$\\$x=1/2\Leftrightarrow {\rm I%
}\in ({\cal S})^{\prime }$.

\item  Exponential functions are test functions (at least of finite order).
So if {\rm I} is a generalized white noise functional we can study the
pairing 
$$
\left\langle \!\!\left\langle {\rm I},\,e^{i\langle \cdot ,\xi \rangle
}\right\rangle \!\!\right\rangle \ ,\qquad \xi \in {\cal S}(\R )\,, 
$$
which is the Fourier-Gauss transform of {\rm I}. From the singularity of the
distribution {\rm I} we get some additional information of the analyticity
of the above pairing with respect to $\xi $.
\end{enumerate}

A second advantage is that the general white noise mathematics allows some
manipulations with the Feynman integrand {\rm I,} since there are many
operators acting continuously on the corresponding distribution space. This
allows well defined calculations. We will illustrate this advantage in
Chapter \ref{look}. There we will {\it prove} Ehrenfest's theorem and derive a
functional form of the canonical commutation relations for one particular
class of potentials. The first argument is essentially been done by applying
the adjoint of a differentiation operator to {\rm I}. After calculating this
expression we take expectation and obtain Ehrenfest's theorem.

We also want to stress that the white noise setting gives a good
``conceptual background'' to discuss some of the numerous independent
definitions of path integrals in a ``common language''. In some sense this
has been done in Chapters \ref{Quantum}--\ref{AHKclass}:

\begin{itemize}
\item  We present an analytic continuation approach related to the work of
Doss \cite{D80} based on our discussion of the scaling operator in section
4.5. As a by-product we will also see the relation to the definition of Hu
and Meyer \cite{HM88}. We will also use our discussion in section 4.3.3 and
integrate out Donsker's delta (introduced to fix the endpoints of the
paths). This gives a convenient form of the propagator.

\item  In the white noise framework the first attempt to include interaction
with a potential was done in \cite{KaS92}. Khandekar and Streit constructed
the Feynman integrand for a large class of potentials including singular
ones. Basically they constructed a strong Dyson series converging in the
space of Hida distributions. This approach only works for one space
dimension. We will generalize this construction to (one dimensional)
time-dependent potentials of non-compact support (Theorem \ref{3.1}). In
Section \ref{Verifying} we will show that the expectation of the constructed Feynman
integrand is indeed the physical propagator, i.e., it solves the
Schr\"odinger equation. This results can also be found in \cite{LLSW94a}.

\item  The above construction is not restricted to perturbations of the free
Feynman integrand. For example we may expand around the Feynman integrand of
the harmonic oscillator. This construction works for small times for the
same large class of potentials (Section \ref{perHarmOsc} or \cite{CDLSW95}).

\item  Modifications and generalizations of the Khandekar Streit
construction as above suffer from the restriction to one dimensional quantum
systems. In the work \cite{AHK76} Feynman integrals for potentials which are
Fourier transforms of bounded complex measures are discussed (with
independent methods). This class of potentials can also be considered in the
white noise framework, without restriction to the space dimension $d$. We
need some integrability condition of the measure associated to the potential
to ensure that the expansion the Feynman integrand converges in $({\cal S}%
_d)^{-1}$ (Theorem \ref{Iin(Sd)-1}). \\A smaller distribution space to
control the convergence of the perturbative expansion may be obtained by
sharpening the integrability condition on the measure. \\We will also allow
time dependent potentials which surprisingly may be more singular than in
the previous construction (Theorem \ref{TIin(Sd)-1}). \\For this class of
Feynman integrands we will show that our mathematical background allows to
prove some relations which were based before on some heuristic arguments,
(see Feynman and Hibbs \cite[p.175]{FH65}). In Chapter \ref{look} we will do this for
Ehrenfest's theorem and for the functional form of the canonical commutation
relations.
\end{itemize}

\vfill 

\begin{quote}
{\large {\bf Acknowledgements.}} \\[2mm] {\it First of all I want to thank
Prof.~Dr.~Ludwig~Streit for proposing such an interesting research activity.
Without his guidance, technical and academical advice this thesis would
never have been possible. I have profited greatly from working with
Prof.~Dr.~Yuri~G.~Kondratiev and I am grateful for all he taught me about
(non-)~Gaussian analysis. I am indebted to Prof.~Dr.~J\"urgen~Potthoff for
interesting discussions and a fruitful collaboration. I thank my colleagues
Peter~Leukert, Angelika~Lascheck, Jos\'e Lu\'\i s da Silva, Cust\'odia
Drumond and M\'ario Cunha for successful joint work. It is a pleasure to
thank my parents who always supported me. Finally, I thank my wife Monika
Westerkamp for her patience and support. \\[2mm] Financial support of a
scholarship from `Graduiertenf\"orderung des Landes Nord\-rhein-Westfalen'
is gratefully acknowledged.}
\end{quote}


\chapter{Preliminaries\label{Preliminaries}}

\section{Some facts on nuclear triples}

\LaTeXparent{nga4.tex}

We start with a real separable Hilbert space ${\cal H}$ with inner product $%
(\cdot ,\cdot )$ and norm $\left| \cdot \right| $ . For a given separable
nuclear space ${\cal N}$ (in the sense of Grothendieck) densely
topologically embedded in ${\cal H}$ we can construct the nuclear triple{\tt %
\ } 
$$
{\cal N}\subset {\cal H\subset N^{\prime }}. 
$$
The dual pairing $\langle \cdot ,\cdot \rangle $ of ${\cal N}^{\prime }$ and 
${\cal N}$ then is realized as an extension of the inner product in ${\cal H}
$%
$$
\langle f,\,\xi \rangle =(f,\xi )\quad f\in {\cal H},\ \xi \in {\cal N} 
$$
Instead of reproducing the abstract definition of nuclear spaces (see e.g., 
\cite{Sch71}) we give a complete (and convenient) characterization in terms
of projective limits of Hilbert spaces.

\begin{theorem}
The nuclear Fr\'echet space ${\cal N}$ can be represented as 
$$
{\cal N=}\bigcap\limits_{p\in \N }{\cal H}_p\text{,} 
$$
where $\{{\cal H}_p$, $p\in {\N\}}$ is a family of Hilbert spaces such that
for all $p_1,p_2\in {\N}$ there exists $p\in {\N}$ such that the embeddings $%
{\cal H}_p\hookrightarrow {\cal H}_{p_1}$ and ${\cal H}_p\hookrightarrow 
{\cal H}_{p_2}$ are of Hilbert-Schmidt class. The topology of ${\cal N}$ is
given by the projective limit topology, i.e., the coarsest topology on ${\cal %
N}$ such that the canonical embeddings ${\cal N\hookrightarrow H}_p$ are
continuous for all $p\in \N $.
\end{theorem}

The Hilbertian norms on ${\cal H}_p$ are denoted by $\left| \cdot \right| _p$%
. Without loss of generality we always suppose that $\forall p\in \N %
,\forall \xi \in {\cal N}:\left| \xi \right| \leq \left| \xi \right| _p$ and
that the system of norms is ordered, i.e., $\left| \cdot \right| _p$ $\leq
\left| \cdot \right| _q$ if $p<q$. By general duality theory the dual space $%
{\cal N}^{\prime }\,$ can be written as 
$$
{\cal N}^{\prime }=\bigcup\limits_{p\in \N }{\cal H}_{-p}\text{.} 
$$
with inductive limit topology $\tau _{ind}$ by using the dual family of
spaces $\{{\cal H}_{-p}:={\cal H}_p^{\prime },\ p\in {\N\}}$. The inductive
limit topology (w.r.t. this family) is the finest topology on ${\cal N}%
^{\prime }$ such that the embeddings ${\cal H}_{-p}\hookrightarrow {\cal N}%
^{\prime }$ are continuous for all $p\in \N $. It is convenient to denote
the norm on ${\cal H}_{-p}$ by $\left| \cdot \right| _{-p}$. Let us mention
that in our setting the topology $\tau _{ind}$ coincides with the Mackey
topology $\tau ({\cal N}^{\prime },{\cal N})$ and the strong topology $\beta
({\cal N}^{\prime },{\cal N})$. Further note that the dual pair $\langle 
{\cal N}^{\prime },{\cal N}\rangle $ is reflexive if ${\cal N}^{\prime }$ is
equipped with $\beta ({\cal N}^{\prime },{\cal N})$. In addition we have
that convergence of sequences is equivalent in $\beta ({\cal N}^{\prime },%
{\cal N})$ and the weak topology $\sigma ({\cal N}^{\prime },{\cal N})$, see
e.g., \cite[Appendix 5]{HKPS93}.

Further we want to introduce the notion of tensor power of a nuclear space.
The simplest way to do this is to start from usual tensor powers ${\cal H}%
_p^{\otimes n}\ ,\ n\in \N $ of Hilbert spaces. Since there is no danger of
confusion we will preserve the notation $\left| \cdot \right| _p$ and $%
\left| \cdot \right| _{-p}$ for the norms on ${\cal H}_p^{\otimes n}$ and $%
{\cal H}_{-p}^{\otimes n}$ respectively. Using the definition%
$$
{\cal N}^{\otimes n}:=\ \stackunder{p\in \N }{\rm pr\ lim}{\cal H}_p^{\otimes
n} 
$$
one can prove \cite{Sch71} that ${\cal N}^{\otimes n}$ is a nuclear space
which is called the $n^{th}$ tensor power of ${\cal N.}$ The dual space of $%
{\cal N}^{\otimes n}$ can be written%
$$
\left( {\cal N}^{\otimes n}\right) ^{\prime }=\ \stackunder{p\in \N }{\rm ind\
lim}{\cal H}_{-p}^{\otimes n} 
$$

Most important for the applications we have in mind is the following 'kernel
theorem', see e.g., \cite{BeKo88}.

\begin{theorem}
\label{KernelTh}Let $\xi _1,...,\xi _n\mapsto F_n\left( \xi _1,...,\xi
_n\right) $ be an n-linear form on ${\cal N}^{\otimes n}$ which is ${\cal H}%
_p$-continuous , i.e.,%
$$
|F_n\left( \xi _1,...,\xi _n\right) |\leq C\prod_{k=1}^n|\xi _k|_p 
$$
for some $p\in {\N}$ and $C>0.$\\Then for all $p^{\prime }>p$ such that the
embedding $i_{p^{\prime },p}:{\cal H}_{p^{\prime }}\hookrightarrow {\cal H}_p
$ is Hilbert-Schmidt there exist a unique $\Phi ^{(n)}\in {\cal H}%
_{-p^{\prime }}^{\otimes n}$ such that 
$$
F_n\left( \xi _1,\ldots ,\xi _n\right) =\langle \Phi ^{(n)},\xi _1\otimes
\cdots \otimes \xi _n\rangle \ ,\quad \text{ }\xi _1,...,\xi _n\in {\cal N} 
$$
and the following norm estimate holds%
$$
\left| \Phi ^{(n)}\right| _{-p^{\prime }}\leq C\;\left\| i_{p^{\prime
},p}\right\| _{HS}^n 
$$
using the Hilbert-Schmidt norm of $i_{p^{\prime },p}$.
\end{theorem}

\begin{corollary}
\label{KernelCor}Let $\xi _1,...,\xi _n\mapsto F\left( \xi _1,...,\xi
_n\right) $ be an $n$-linear form on ${\cal N}^{\otimes n}$ which is ${\cal H%
}_{-p}$-continuous, i.e.,%
$$
|F_n\left( \xi _1,\ldots ,\xi _n\right) |\leq C\prod_{k=1}^n|\xi _k|_{-p} 
$$
for some $p\in {\N}$ and $C>0$.\\Then for all $p^{\prime }<p$ such that the
embedding $i_{p,p^{\prime }}:{\cal H}_p\hookrightarrow {\cal H}_{p^{\prime }}
$ is Hilbert-Schmidt there exist a unique $\Phi ^{(n)}\in {\cal H}%
_{p^{\prime }}^{\otimes n}$ such that 
$$
F_n\left( \xi _1,...,\xi _n\right) =\langle \Phi ^{(n)},\xi _1\otimes \cdots
\otimes \xi _n\rangle ,\quad \text{ }\xi _1,...,\xi _n\in {\cal N} 
$$
and the following norm estimate holds%
$$
\left| \Phi ^{(n)}\right| _{p^{\prime }}\leq C\;\left\| i_{p,p^{\prime
}}\right\| _{HS}^n\text{ .} 
$$
\end{corollary}

If in Theorem \ref{KernelTh} (and in Corollary \ref{KernelCor} respectively
) we start from a symmetric $n$-linear form $F_n$ on ${\cal N}^{\otimes n}$
i.e., $F_n(\xi _{\pi _1},\ldots ,\xi _{\pi _n})=F_n\left( \xi _1,\ldots ,\xi
_n\right) $ for any permutation $\pi $, then the corresponding kernel $\Phi
^{(n)}$ can be localized in ${\cal H}_{p^{\prime }}^{\hat \otimes n}\subset 
{\cal H}_{p^{\prime }}^{\otimes n}$ (the n$^{th}$ symmetric tensor power of
the Hilbert space ${\cal H}_{p^{\prime }}$). For $f_1,\ldots ,f_n\in {\cal H}
$ let $\hat \otimes $ also denote the symmetrization of the tensor product%
$$
f_1\hat \otimes \cdots \hat \otimes f_n:=\frac 1{n!}\sum_\pi f_{\pi
_1}\otimes \cdots \otimes f_{\pi _n}\ , 
$$
where the sum extends over all permutations of $n$ letters. All the above
quoted theorems also hold for complex spaces, in particular the complexified
space ${\cal N}_{\Ckl }$. By definition an element $\theta \in $ ${\cal N}_{%
\Ckl }$ decomposes into $\theta =\xi +i\eta \ ,\ \xi ,\eta \in {\cal N}$. If
we also introduce the corresponding complexified Hilbert spaces ${\cal H}_{p,%
\Ckl }$ the inner product becomes%
$$
(\theta _1,\theta _2)_{{\cal H}_{p,\Ckkl }}=(\theta _1,\bar \theta _2)_{%
{\cal H}_p}=(\xi _1,\xi _2)_{{\cal H}_p}+(\eta _1,\eta _2)_{{\cal H}%
_p}+i(\eta _1,\xi _2)_{{\cal H}_p}-i(\xi _1,\eta _2)_{{\cal H}_p} 
$$
for $\theta _1,\theta _2\in {\cal H}_{p,\Ckl },\ \theta _1=\xi _1+i\eta _1\
,\theta _2=\xi _2+i\eta _2\ ,\ \xi _1,\xi _2,\eta _1,\eta _2\in {\cal H}_p$.
Thus we have introduced a nuclear triple%
$$
{\cal N}_{\Ckl }^{\hat \otimes n}\subset {\cal H}_{\Ckl }^{\hat \otimes
n}\subset \left( {\cal N}_{\Ckl }^{\hat \otimes n}\right) ^{\prime } 
$$
We also want to introduce the (Boson or symmetric) Fock space $\Gamma ({\cal %
H})$ of ${\cal H}$ by%
$$
\Gamma ({\cal H})=\bigoplus_{n=0}^\infty {\cal H}_{\Ckl }^{\hat \otimes n} 
$$
with the convention ${\cal H}_{\Ckl }^{\hat \otimes 0}:=\C $ and the
Hilbertian norm%
$$
\left\| \vec \varphi \right\| _{\Gamma ({\cal H})}^2=\sum_{n=0}^\infty
n!\;\left| \varphi ^{(n)}\right| ^2\ ,\quad \vec \varphi =\left\{ \varphi
^{(n)}\;\Big|\;n\in \N _0\right\} \in \Gamma ({\cal H})\;. 
$$

\section{Holomorphy on locally convex spaces\label{Holomorphy}}

\LaTeXparent{nga4.tex}

We shall collect some facts from the theory of holomorphic functions in
locally convex topological vector spaces ${\cal E}$ (over the complex field $%
{\C}$), see e.g., \cite{Di81}. Let ${\cal L}({\cal E}^n)$ be the space of
n-linear mappings from ${\cal E}^n$ into ${\C}$ and ${\cal L}_s({\cal E}^n)$
the subspace of symmetric n-linear forms. Also let ${\sl P}^n({\cal E})$
denote the n-homogeneous polynomials on ${\cal E}$. There is a linear
bijection ${\cal L}_s({\cal E}^n)\ni A\longleftrightarrow \widehat{A}\in 
{\sl P}^n({\cal E})$. Now let ${\cal U}\subset {\cal E}$ be open and
consider a function $G:{\cal U}\rightarrow {\C}$.

$G$ is said to be {\bf G-holomorphic} if for all $\theta _0\in {\cal U}$ and
for all $\theta \in {\cal E}$ the mapping from ${\C}$ to ${\C :}$ $\lambda
\rightarrow G(\theta _0+\lambda \theta )$ is holomorphic in some
neighborhood of zero in ${\C}$. If $G$ is G-holomorphic then there exists
for every $\eta \in {\cal U}$ a sequence of homogeneous polynomials $\frac
1{n!}\widehat{{\rm d}^n G(\eta )}$ such that 
$$
G(\theta +\eta )=\sum\limits_{n=0}^\infty \frac 1{n!}\widehat{{\rm d}%
^nG(\eta )}(\theta ) 
$$
for all $\theta $ from some open set ${\cal V}\subset {\cal U}$. $G$ is said
to be {\bf holomorphic}, if for all $\eta $ in ${\cal U}$ there exists an
open neighborhood ${\cal V}$ of zero such that $\sum\limits_{n=0}^\infty
\frac 1{n!}\widehat{{\rm d}^nG(\eta )}(\theta )$ converges uniformly on $%
{\cal V}$ to a continuous function. We say that $G$ is holomorphic at $%
\theta _0$ if there is an open set ${\cal U}$ containing $\theta _0$ such
that $G$ is holomorphic on ${\cal U}$. The following proposition can be
found e.g., in \cite{Di81}.

\begin{proposition}
\label{GHolLocB} $G$ is holomorphic if and only if it is G-holomorphic and
locally bounded.
\end{proposition}

\noindent Let us explicitly consider a function holomorphic at the point $%
0\in {\cal E}={\cal N}_{\Ckl }$, then

1) there exist $p$ and $\varepsilon >0$ such that for all $\xi _0\in {\cal N}%
_{\Ckl }$ with $\left| \xi _0\right| _p\leq \varepsilon $ and for all $\xi
\in {\cal N}_{\Ckl }$ the function of one complex variable $\lambda
\rightarrow G(\xi _0+\lambda \xi )$ is analytic at $0\in {\C}$, and

2) there exists $c>0$ such that for all $\xi \in {\cal N}_{\Ckl }$ with $%
\left| \xi \right| _p\leq \varepsilon $ : $\left| G(\xi )\right| \leq c$.

\noindent As we do not want to discern between different restrictions of one
function, we consider germs of holomorphic functions, i.e., we identify $F$
and $G$ if there exists an open neighborhood ${\cal U}:0\in {\cal U}\subset 
{\cal N}_{\Ckl }$ such that $F(\xi )=G(\xi )$ for all $\xi \in {\cal U}$.
Thus we define ${\rm Hol}_0({\cal N}_{\Ckl })$ as the algebra of germs of
functions holomorphic at zero equipped with the inductive topology given by
the following family of norms%
$$
{\rm n}_{p,l,\infty }(G)=\sup _{\left| \theta \right| _p\leq 2^{-l}}\left|
G(\theta )\right| ,\quad p,l\in \N . 
$$
\bigskip

\begin{sloppypar}
Let use now introduce spaces of entire functions which will be useful later.
Let ${\cal E}_{2^{-l}}^k({\cal H}_{-p,\Ckl })$ denote the set of all entire
functions on ${\cal H}_{-p,\Ckl }$ of growth $k\in [1,2]$ and type $2^{-l},\
p,l\in \Z $. This is a linear space with norm 
$$
{\rm n}_{p,l,k}(\varphi )=\sup _{z\in {\cal H}_{-p,\Ckkl }}\left| \varphi
(z)\right| \exp \left( -2^{-l}|z|_{-p}^k\right) ,\qquad \varphi \in {\cal E}%
_{2^{-l}}^k({\cal H}_{-p,\Ckl }) 
$$
The space of entire functions on ${\cal N}_{\Ckl }^{\prime }$ of growth $k$
and minimal type is naturally introduced by 
$$
{\cal E}_{\min }^k({\cal N}_{\Ckl }^{\prime }):=\ \stackunder{p,l\in \N }{\rm pr\ lim\,}{\cal E}_{2^{-l}}^k({\cal H}_{-p,\Ckl })\ , 
$$
see e.g., \cite{Ou91}. We will also need the space of entire functions on $%
{\cal N}_{\Ckl }$ of growth $k$ and finite type:%
$$
{\cal E}_{\max }^k({\cal N}_{\Ckl }):=\ \stackunder{p,l\in \N }{\rm ind\ lim}%
{\cal E}_{2^l}^k({\cal H}_{p,\Ckl })\ . 
$$
In the following we will give an equivalent description of ${\cal E}_{\min
}^k({\cal N}_{\Ckl }^{\prime })$ and ${\cal E}_{\max }^k({\cal N}_{\Ckl })$.
Cauchy's inequality and Corollary \ref{KernelCor} allow to write the Taylor
coefficients in a convenient form. Let $\varphi \in {\cal E}_{\min }^k({\cal %
N}_{\Ckl }^{\prime })$ and $z\in {\cal N}_{\Ckl }^{\prime }$, then there
exist kernels $\varphi ^{(n)}\in {\cal N}_{\Ckl }^{\hat \otimes n}$ such
that 
$$
\langle z^{\otimes n},\varphi ^{(n)}\rangle =\frac 1{n!}\widehat{{\rm d}^n\varphi (0)}(z) 
$$
i.e., 
\begin{equation}
\label{phi(z)}\varphi (z)=\sum_{n=0}^\infty \langle z^{\otimes n},\varphi
^{(n)}\ \rangle . 
\end{equation}
This representation allows to introduce a nuclear topology on ${\cal E}%
_{\min }^k({\cal N}_{\Ckl }^{\prime })$, see \cite{Ou91} for details. Let 
{\rm E}$_{p,q}^\beta $ denote the space of all functions of the form (\ref
{phi(z)}) such that the following Hilbertian norm 
\begin{equation}
\label{3StrichNorm}\lnorm  \varphi \rnorm  _{p,q,\beta
}^2:=\sum_{n=0}^\infty (n!)^{1+\beta }2^{nq}\left| \varphi ^{(n)}\right|
_p^2\;,\quad p,q\in \N  
\end{equation}
is finite for $\beta \in [0,1]$. (By $\left| \varphi ^{(0)}\right| _p$ we 
simply mean the complex modulus for
all $p$.)  The space {\rm E}$_{-p-,q}^{-\beta }$ with
the norm $\lnorm  \varphi \rnorm  _{-p,-q,-\beta }$ is defined analogously.
\end{sloppypar}

\begin{theorem}
\label{Ekminprlim}The following topological identity holds: 
$$
\stackunder{p,q\in \N }{\rm pr\ lim}\;{\rm E}_{p,q}^\beta ={\cal E}_{\min
}^{\frac 2{1+\beta }}({\cal N}_{\Ckl }^{\prime })\quad . 
$$
\end{theorem}

The proof is an immediate consequence of the following two lemmata which
show that the two systems of norms are in fact equivalent.

\begin{lemma}
\label{nplk3Strich}Let $\varphi \in ${\rm E}$_{p,q}^\beta $ then $\varphi
\in {\cal E}_{2^{-l}}^{\frac 2{1+\beta }}({\cal H}_{-p,\Ckl })$ for $l=\frac
q{1+\beta }$. Moreover 
\begin{equation}
\label{nplk3StrichNorm}{\rm n}_{p,l,k}(\varphi )\leq \lnorm  \varphi \rnorm  %
_{p,q,\beta }\ ,\ \ k=\tfrac 2{1+\beta }\ .
\end{equation}
\end{lemma}

\TeXButton{Proof}{\proof} We look at the convergence of the series $\varphi
(z)=\sum_{n=0}^\infty \langle z^{\otimes n},\varphi ^{(n)}\ \rangle \ $, $%
z\in {\cal H}_{-p,\Ckl }\ ,\ \varphi ^{(n)}\in {\cal H}_{p,\Ckl }$ if $%
\sum_{n=0}^\infty (n!)^{1+\beta }2^{nq}|\varphi ^{(n)}|_p^2=\lnorm 
\varphi \rnorm  _{p,q,\beta }^2$ is finite. The following estimate holds:%
\begin{eqnarray*}
\sum_{n=0}^\infty |\langle z^{\otimes n},\varphi ^{(n)}\ \rangle | %
& \leq & \left( \sum_{n=0}^\infty (n!)^{1+\beta }2^{nq}|\varphi %
^{(n)}|_p^2\right) ^{1/2}\left( \sum_{n=0}^\infty %
\frac 1{(n!)^{1+\beta }}2^{-nq}|z|_{-p}^{2n}\right) ^{1/2} %
\\& \leq & \lnorm  \varphi \rnorm  _{p,q,\beta }\cdot \left( %
\sum_{n=0}^\infty \left\{ \frac 1{n!}2^{-\frac{nq}{1+\beta }}%
|z|_{-p}^{\frac{2n}{1+\beta }}\right\} ^{1+\beta }\right) ^{1/2} %
\\& \leq & \lnorm  \varphi \rnorm  _{p,q,\beta }\left( %
\sum_{n=0}^\infty \frac 1{n!}2^{-\frac{nq}{1+\beta }}|z|_{-p}^{\frac{2n}{1+\beta }}\right) ^{(1+\beta )/2} \\& \leq & \lnorm  \varphi \rnorm %
_{p,q,\beta }\exp \left( %
2^{-\frac q{1+\beta }}|z|_{-p}^{\frac 2{1+\beta }}\right) . 
\end{eqnarray*}
\TeXButton{End Proof}{\endproof}

\begin{lemma}
\label{3Strichnplk}For any $p^{\prime },q\in \N $ there exist $p,l\in \N $
such that 
$$
{\cal E}_{2^{-l}}^{\frac 2{1+\beta }}({\cal H}_{-p,\Ckl })\subset {\rm E}%
_{p^{\prime },q}^\beta  
$$
i.e., there exists a constant $C>0$ such that 
$$
\lnorm  \varphi \rnorm  _{p^{\prime },q,\beta }\leq C\;{\rm n}%
_{p,l,k}(\varphi ),\quad \varphi \in {\cal E}_{2^{-l}}^k({\cal H}_{-p,\Ckl %
}),\quad k=\tfrac 2{1+\beta }. 
$$
\end{lemma}

\TeXButton{Remark}{\remark } More precisely we will prove the following:\ If $%
\varphi \in {\cal E}_{2^{-l}}^k({\cal H}_{-p,\Ckl })$ then $\varphi \in $%
{\rm E}$_{p^{\prime },q}^\beta $ for $k=\frac 2{1+\beta }$ and $\rho
:=2^{q-2l/k}k^{2/k}e^2\left\| i_{p^{\prime },p}\right\| _{HS}^2<1$ (in
particular this requires $p^{\prime }>p$ to be such that the embedding $%
i_{p^{\prime },p}:{\cal H}_{p^{\prime }}\hookrightarrow {\cal H}_p$ is
Hilbert-Schmidt). \\Moreover the following bound holds 
\begin{equation}
\label{3Normnplk}\lnorm  \varphi \rnorm  _{p^{\prime },q,\beta }\leq {\rm n}%
_{p,l,k}(\varphi )\cdot \left( 1-\rho \right) ^{-1/2}\ . 
\end{equation}

\TeXButton{Proof}{\proof} The assumption $\varphi \in {\cal E}_{2^{-l}}^k(%
{\cal H}_{-p,\Ckl })$ implies a bound of the growth of $\varphi :$%
$$
|\varphi (z)|\leq {\rm n}_{p,l,k}(\varphi )\exp (2^{-l}|z|_{-p}^k)\ . 
$$
For each $\rho >0\ ,\ z\in {\cal H}_{-p,\Ckl }$ the Cauchy inequality from
complex analysis \cite{Di81} gives%
$$
\left| \frac 1{n!}\widehat{{\rm d}^n\varphi (0)}(z)\right| \leq {\rm n}%
_{p,l,k}(\varphi )\rho ^{-n}\exp (\rho ^k2^{-l})\;|z|_{-p}^n\ . 
$$
By polarization \cite{Di81} it follows for $z_1,\ldots ,z_n\in {\cal H}_{-p,\Ckl }$%
$$
\left| \frac 1{n!}{\rm d}^n\varphi (0)(z_1,\ldots ,z_n)\right| \leq {\rm n}%
_{p,l,k}(\varphi )\frac 1{n!}\left( \frac n\rho \right) ^n\exp (\rho
^k2^{-l})\prod_{k=1}^n|z_k|_{-p}\ . 
$$
For $p^{\prime }>p$ such that $\left\| i_{p^{\prime },p}\right\| _{HS}$ is
finite, an application of the kernel theorem guarantees the existence of
kernels $\varphi ^{(n)}\in {\cal H}_{p^{\prime },\Ckl }^{\hat \otimes n}$
such that 
$$
\varphi (z)=\sum_{n=0}^\infty \langle z^{\hat \otimes n},\varphi ^{(n)}\
\rangle 
$$
with the bound%
$$
\left| \varphi ^{(n)}\right| _{p^{\prime }}\leq {\rm n}_{p,l,k}(\varphi
)\frac 1{n!}\left( \frac n\rho \left\| i_{p^{\prime },p}\right\|
_{HS}\right) ^n\exp (\rho ^k\cdot 2^{-l})\ . 
$$
We can optimize the bound with the choice of an $n$-dependent $\rho $.
Setting $\rho ^k=2^ln/k$ we obtain 
\begin{eqnarray*}
\left| \varphi ^{(n)}\right| _{p^{\prime }} & \leq & {\rm n}_{p,l,k}(\varphi %
)\frac 1{n!}n^{n(1-1/k)}\left( \tfrac 1k2^l\right) ^{-n/k}\left\| %
i_{p^{\prime },p}\right\| _{HS}^ne^{n/k} %
\\& \leq & {\rm n}_{p,l,k}(\varphi )\;(n!)^{-1/k}\left\{ (k2^{-l})^{1/k}e %
\left\| i_{p^{\prime },p}\right\| _{HS}\right\} ^n\ , 
\end{eqnarray*}
where we used $n^n\leq n!\,e^n$ in the last estimate. Now choose $\beta \in
[0,1]$ such that $k=\frac 2{1+\beta }$ to estimate the following norm:%
\begin{eqnarray*}
\lnorm  \varphi \rnorm  _{p^{\prime },q,\beta }^2 %
& \leq & {\rm n}%
_{p,l,k}^2(\varphi )\sum_{n=0}^\infty (n!)^{1+\beta -\frac 2k}2^{qn}\left\{ %
(k2^{-l})^{1/k}e\left\| i_{p^{\prime },p}\right\| _{HS}\right\} ^{2n} %
\\& \leq & {\rm n}_{p,l,k}^2(\varphi )\left( 1-2^q\left\{ (k2^{-l})^{1/k}e %
\left\| i_{p^{\prime },p}\right\| _{HS}\right\} ^2\right) ^{-1} 
\end{eqnarray*}
for sufficiently large $l$. This completes the proof.\TeXButton{End Proof}
{\endproof}\bigskip\ 

Analogous estimates for these systems of norms also hold if $\beta ,p,q,l$
become negative. This implies the following theorem. For related results see
e.g., \cite[Prop.8.6]{Ou91}.

\begin{theorem}
\label{indlimEkmax} \hfill \\If $\beta \in [0,1)$ then the following
topological identity holds:%
$$
\stackunder{p,q\in \N }{\rm ind\ lim}\ {\rm E}_{-p,-q}^{-\beta }={\cal E}%
_{\max }^{2/(1-\beta )}({\cal N}_{\Ckl }). 
$$
If $\beta =1$ we have%
$$
\stackunder{p,q\in \N }{\rm ind\ lim}\ {\rm E}_{-p,-q}^{-1}={\rm Hol}_0(%
{\cal N}_{\Ckl })\ . 
$$
\end{theorem}

\noindent This theorem and its proof will appear in the context of section 
\ref{Characterization}. The characterization of distributions in infinite
dimensional analysis is strongly related to this theorem. From this point of
view it is natural to postpone its proof to section \ref{Characterization}.

\chapter{Generalized functions in infinite dimensional analysis}

\LaTeXparent{nga4.tex}

\section{Measures on linear topological spaces}

To introduce probability measures on the vector space ${\cal N}^{\prime }$,
we consider ${\cal C}_\sigma ({\cal N^{\prime }})$ the $\sigma $-algebra
generated by cylinder sets on ${\cal N}^{\prime }$, which coincides with the
Borel $\sigma $-algebras ${\cal B}_\sigma ({\cal N}^{\prime })$ and ${\cal B}%
_\beta ({\cal N}^{\prime })$ generated by the weak and strong topology on $%
{\cal N}^{\prime }$ respectively. Thus we will consider this $\sigma $%
-algebra as the {\em natural} $\sigma $-algebra on ${\cal N}^{\prime }$.
Detailed definitions of the above notions and proofs of the mentioned
relations can be found in e.g., \cite{BeKo88}.

We will restrict our investigations to a special class of measures $\mu $ on 
${\cal C}_\sigma ({\cal N^{\prime }})$, which satisfy two additional
assumptions. The first one concerns some analyticity of the Laplace
transformation%
$$
l_\mu (\theta )=\int_{{\cal N^{\prime }}}\exp \left\langle x,\theta
\right\rangle \ {\rm d}\mu (x)=:\E _\mu (\exp \left\langle \cdot ,\theta
\right\rangle )\text{ , }\theta \in {\cal N}_{\Ckl }\text{.} 
$$
Here we also have introduced the convenient notion of expectation $\E _\mu $
of a $\mu $-integrable function.

\TeXButton{Assumption}{\assumption} 1 \quad The measure $\mu $ has an analytic
Laplace transform in a neighborhood of zero. That means there exists an open
neighborhood ${\cal U}\subset {\cal N}_{\Ckl }$ of zero, such that $l_\mu $
is holomorphic on ${\cal U}$, i.e., $l_\mu \in {\rm Hol}_0({\cal N}_{\Ckl })$
. This class of {\em analytic measures} is denoted by ${\cal M}_a({\cal N}%
^{\prime }).$\bigskip\ 

\noindent An equivalent description of analytic measures is given by the
following lemma.

\begin{lemma}
\label{equiLemma}The following statements are equivalent\medskip\ \\{\bf 1)}%
\quad $\mu \in {\cal M}_a({\cal N}^{\prime })$\smallskip\\{\bf 2)}$\quad \D%
\exists p_\mu \in {\N},\quad \exists C>0:\qquad \left| \int_{{\cal N}%
^{\prime }}\ \langle x,\theta \rangle ^n\;{\rm d}\mu (x)\right| \leq
n!\,C^n\left| \theta \right| _{p_\mu }^n\;,\quad \theta \in {\cal H}_{p_\mu ,%
\Ckl }$\smallskip\\{\bf 3)}$\quad \D\exists p_\mu ^{\prime }\in {\N},\quad
\exists \varepsilon _\mu >0:\ \qquad \int_{{\cal N}^{\prime }}\ \exp
(\varepsilon _\mu \left| x\right| _{-p_\mu ^{\prime }})\,{\rm d}\mu
(x)<\infty $
\end{lemma}

\noindent \TeXButton{Proof}{\proof}The proof can be found in \cite{KoSW94}.
We give its outline in the following. The only non-trivial step is the proof
of 2)$\Rightarrow $3).

\noindent By polarization \cite{Di81} 2) implies 
\begin{equation}
\label{PolBound}\left| \int_{{\cal N}^{\prime }}\langle x^{\otimes
n},\tbigotimes_{j=1}^n\xi _j\rangle \ {\rm d}\mu (x)\right| \leq
n!\;C^n\prod_{j=1}^n\left| \xi _j\right| _{p_\mu }\ ,\quad \xi _j\in {\cal H}%
_{p^{\prime }} 
\end{equation}
for a (new) constant $C>0$. Choose $p^{\prime }>p_\mu $ such that the
embedding $i_{p^{\prime },p_\mu }:{\cal H}_{p^{\prime }}\rightarrow {\cal H}%
_{p_\mu }$ is of Hilbert-Schmidt type. Let $\left\{ e_k,\ k\in {\N}\right\}
\subset {\cal N}$ be an orthonormal basis in ${\cal H}_{p^{\prime }}$. Then $%
\left| x\right| _{-p^{\prime }}^2=\sum\limits_{k=1}^\infty \left\langle
x,e_k\right\rangle ^2$, $x\in {\cal H}_{-p^{\prime }}$. We will first
estimate the moments of even order%
$$
\int_{{\cal N}^{\prime }}\left| x\right| _{-p^{\prime }}^{2n}\ {\rm d}\mu
(x)=\sum\limits_{k_1=1}^\infty \cdots \sum\limits_{k_n=1}^\infty \int_{{\cal %
N}^{\prime }}\left\langle x,e_{k_1}\right\rangle ^2\ \cdots \left\langle
x,e_{k_n}\right\rangle ^2\ {\rm d}\mu (x)\ \text{,} 
$$
where we changed the order of summation and integration by a monotone
convergence argument. Using the bound (\ref{PolBound}) we have%
\begin{eqnarray*}
\int_{{\cal N}^{\prime }}\left| x\right| _{-p^{\prime }}^{2n}\ {\rm d}\mu(x)%
&\leq & \ C^{2n}\ (2n)!\sum\limits_{k_1=1}^\infty \cdots%
\sum\limits_{k_n=1}^\infty \left| e_{k_1}\right| _{p_\mu}^2\cdots \left|%
e_{k_n}\right| _{p_\mu}^2 %
\\&=& \ C^{2n}\ (2n)!\left( \sum\limits_{k=1}^\infty \left| e_k\right|%
_{p_\mu}^2\right) ^n %
\\&=& \ \left( C\cdot \left\| i_{p^{\prime },p_\mu}\right\| _{HS}%
\right) ^{2n}(2n)!%
\end{eqnarray*}
because%
$$
\sum\limits_{k=1}^\infty \left| e_k\right| _{p_\mu}^2=\left\| i_{p^{\prime
},p_\mu }\right\| _{HS}^2\text{ .} 
$$
The moments of arbitrary order can now be estimated by the Schwarz inequality%
\begin{eqnarray*}
\int \left| x\right| _{-p^{\prime }}^n\ {\rm d}\mu (x) %
& \leq & \sqrt{\mu ({\cal N}^{\prime })}\left( \int \left| x\right| _{-p}^{2n}%
\ {\rm d}\mu (x)\right)^{\frac 12} %
\\&\leq & \sqrt{\mu ({\cal N}^{\prime })}\left( C\left\| %
i_{p^{\prime },p_\mu}\right\| _{HS}\right) ^n\sqrt{(2n)!} %
\\&\leq & \sqrt{\mu ({\cal N}^{\prime })}\left( 2 C\left\| %
i_{p^{\prime },p_\mu}\right\| _{HS}\right) ^nn! 
\end{eqnarray*}
since $(2n)!\leq 4^n(n!)^2$ . \\Choose $\varepsilon <$ $\left( \ 2C\left\|
i_{p^{\prime },p_\mu }\right\| _{HS}\right) ^{-1}$ then%
\begin{eqnarray} \label{exponent}
\int e^{\varepsilon \left| x\right| _{-p^{\prime }}}{\rm d}\mu%
(x) &=& \sum_{n=0}^\infty \frac{\varepsilon ^n}{n!}\int \left| x\right|%
_{-p^{\prime }}^n\ {\rm d}\mu (x) \nonumber %
\\&\leq & \sqrt{\mu ({\cal N}^{\prime })}%
\ \sum_{n=0}^\infty \left( \varepsilon \ 2 C%
\left\| i_{p^{\prime },p_\mu }\right\| _{HS}\right) ^n<\infty 
\end{eqnarray}
Hence the lemma is proven.\TeXButton{End Proof}{\endproof}\bigskip\ \ 

For $\mu \in {\cal M}_a({\cal N}^{\prime })$ the estimate in statement 2 of
the above lemma allows to define the moment kernels ${\rm M}_n^\mu \in (%
{\cal N}^{\hat \otimes n})^{\prime }.$ This is done by extending the above
estimate by a simple polarization argument and applying the kernel theorem.
The kernels are determined by%
$$
l_\mu (\theta )=\sum_{n=0}^\infty \frac 1{n!}\langle {\rm M}_n^\mu ,\theta
^{\otimes n}\rangle 
$$
or equivalently%
$$
\langle {\rm M}_n^\mu ,\theta _1\hat \otimes \cdots \hat \otimes \theta
_n\rangle =\left. \frac{\partial ^n}{\partial t_1\cdots \partial t_n}l_\mu
(t_1\theta _1+\cdots +t_n\theta _n)\right| _{t_1=\cdots =t_n=0}\ . 
$$
Moreover, if $p>p_\mu $ is such that embedding $i_{p,p_\mu }:{\cal H}%
_p\hookrightarrow {\cal H}_{p_\mu }$ is Hilbert-Schmidt then 
\begin{equation}
\label{MnmuNorm}\left| {\rm M}_n^\mu \right| _{-p}\leq \left( nC\left\|
i_{p,p_\mu }\right\| _{HS}\right) ^n\leq {n!}\ \left( eC\left\| i_{p,p_\mu
}\right\| _{HS}\right) ^n\ . 
\end{equation}

\begin{definition}
\label{P(N)}A function $\varphi :{\cal N}^{\prime }\rightarrow {\C}$ of the
form $\varphi (x)=\sum_{n=0}^N\langle x^{\otimes n},\varphi ^{(n)}\rangle $, 
$x\in {\cal N}^{\prime }$, $N\in {\N,}$ is called a continuous polynomial
(short $\varphi \in {\cal P}({\cal N}^{\prime })$ ) iff $\varphi ^{(n)}\in 
{\cal N}_{\Ckl }^{\hat \otimes n}$, $\forall n\in {\N}_0=\N \cup \{0\}$.
\end{definition}

Now we are ready to formulate the second assumption: \bigskip\ 

\TeXButton{Assumption}{\assumption} 2 \quad For all $\varphi \in {\cal P}({\cal N}%
^{\prime })$ with $\varphi =0$ $\mu $-almost everywhere we have $\varphi
\equiv 0$. In the following a measure with this property will be called {\em %
non-degenerate}. \bigskip\ 

\TeXButton{Note}{\note } Assumption 2 is equivalent to:\\Let $\varphi \in {\cal P}({\cal N}%
^{\prime })$ with $\int_A\varphi \,{\rm d}\mu =0$ for all $A\in {\cal C}%
_\sigma ({\cal N}^{\prime })$ then $\varphi \equiv 0$.\\A sufficient
condition can be obtained by regarding admissible shifts of the measure $\mu 
$. If $\mu (\cdot +\xi )$ is absolutely continuous with respect to $\mu $
for all $\xi \in {\cal N,}$ i.e., there exists the Radon-Nikodym derivative$%
{\cal \quad }$%
$$
\rho _\mu (\xi ,x)=\frac{{\rm d}\mu (x+\xi )}{{\rm d}\mu (x)}\ ,\quad x\in 
{\cal N}^{\prime }{\rm \;,} 
$$
Then we say that $\mu $ is ${\cal N}${\em --quasi-invariant} see e.g., \cite
{GV68,Sk74}. This is sufficient to ensure Assumption 2, see e.g., \cite
{KoTs91,BeKo88}.\bigskip\ 

\example In Gaussian Analysis (especially White Noise Analysis) the Gaussian
measure $\gamma _{{\cal H}}$ corresponding to the Hilbert space ${\cal H}$
is considered. Its Laplace transform is given by 
$$
l_{\gamma _{{\cal H}}}(\theta )=e^{\frac 12\langle \theta ,\theta \rangle }\
,\qquad \theta \in {\cal N}_{\Ckl }\ , 
$$
hence $\gamma _{{\cal H}}\in {\cal M}_a({\cal N}^{\prime })$. It is well
known that $\gamma _{{\cal H}}$ is ${\cal N}$--quasi-invariant (moreover $%
{\cal H}$--quasi-invariant) see e.g., \cite{Sk74,BeKo88}. Due to the
previous note $\gamma _{{\cal H}}$ satisfies also Assumption 2.\bigskip\ 

\example  {\it (Poisson measures)} \smallskip \\ Let use consider the
classical (real) Schwartz triple 
$$
{\cal S}(\R )\subset L^2(\R )\subset {\cal S}^{\prime }(\R )\,. 
$$
The Poisson white noise measure $\mu _p$ is defined as a probability measure
on ${\cal C}_\sigma ({\cal S}^{\prime }(\R ))$ with the Laplace transform%
$$
l_{\mu _p}(\theta )=\exp \left\{ \int_{\R }(e^{\theta (t)}-1)\;{\rm d}%
t\right\} =\exp \left\{ \langle e^\theta -1,1\rangle \right\} ,\qquad \theta
\in {\cal S}_{\Ckl }(\R )\,, 
$$
see e.g., \cite{GV68}. It is not hard to see that $l_{\mu _p}$ is a
holomorphic function on ${\cal S}_{\Ckl }(\R )$, so Assumption 1 is
satisfied. But to check Assumption 2, we need additional considerations.

First of all we remark that for any $\xi \in {\cal S}(\R )\,,\ \xi \neq 0$
the measures $\mu _p$ and $\mu _p(\cdot +\xi )$ are orthogonal (see \cite
{VGG75} for a detailed analysis). It means that $\mu _p$ is not ${\cal S}(\R %
)$-quasi-invariant and the note after Assumption 2 is not applicable now.

Let some $\varphi \in {\cal P}({\cal S}^{\prime }(\R ))\,,\,\varphi =0~~\mu
_p$-a.s. be given. We need to show that then $\varphi \equiv 0$. To this end
we will introduce a system of orthogonal polynomials in the space $L^2(\mu
_p)$ which can be constructed in the following way. The mapping%
$$
\theta (\cdot )\mapsto \alpha (\theta )(\cdot )=\log (1+\theta (\cdot ))\in 
{\cal S}_{\Ckl }(\R )\,,\quad \theta \in {\cal S}_{\Ckl }(\R ) 
$$
is holomorphic on a neighborhood ${\cal U}\subset {\cal S}_{\Ckl }(\R %
)\,,\,0\in {\cal U}$. Then%
$$
e_{\mu _p}^\alpha (\theta ;x)=\frac{e^{\langle \alpha (\theta ),x\rangle }}{%
l_{\mu _p}(\alpha (\theta ))}=\exp \{\langle \alpha (\theta ),x\rangle
-\langle \theta ,1\rangle \}\,,\quad \theta \in {\cal U}\,,\ x\in {\cal S}%
^{\prime }(\R ) 
$$
is a holomorphic function on ${\cal U}$ for any $\,x\in {\cal S}^{\prime }(%
\R )$. The Taylor decomposition and the kernel theorem (just as in
subsection \ref{AppellSec} below) give 
$$
e_{\mu _p}^\alpha (\theta ;x)=\sum_{n=0}^\infty \frac 1{n!}\langle \theta
^{\otimes n},C_n(x)\rangle \,, 
$$
where $C_n:{\cal S}^{\prime }(\R )\rightarrow {\cal S}^{\prime }(\R )^{\hat
\otimes n}$ are polynomial mappings. For $\varphi ^{(n)}\in {\cal S}_{\Ckl }(%
\R )^{\hat \otimes n}\,,\,n\in \N _0$, we define Charlier polynomials%
$$
x\mapsto C_n(\varphi ^{(n)};x)=\langle \varphi ^{(n)},C_n(x)\rangle \in \C %
\,,\ \,x\in {\cal S}^{\prime }(\R )\,. 
$$
Due to \cite{Ito88,IK88} we have the following orthogonality property:%
$$
\forall \varphi ^{(n)}\in {\cal S}_{\Ckl }(\R )^{\hat \otimes n}\,,\,\forall
\psi ^{(m)}\in {\cal S}_{\Ckl }(\R )^{\hat \otimes n} 
$$
$$
\int C_n(\varphi ^{(n)})C_m(\psi ^{(m)})\;{\rm d}\mu _p=\delta _{nm}
n!\langle \varphi ^{(n)},\psi ^{(n)}\rangle \,. 
$$
Now the rest is simple. Any continuous polynomial $\varphi $ has a uniquely
defined decomposition%
$$
\varphi (x)=\sum_{n=0}^N\langle \varphi ^{(n)},C_n(x)\rangle \,\,,\quad
\,x\in {\cal S}^{\prime }(\R )\,, 
$$
where $\varphi ^{(n)}\in {\cal S}_{\Ckl }(\R )^{\hat \otimes n}$. If $%
\varphi =0~\mu _p$-a.e. then 
$$
\left\| \varphi \right\| _{L^2(\mu _p)}^2=\sum_{n=0}^Nn!\,\langle \varphi
^{(n)},\overline{\varphi ^{(n)}}\rangle =0. 
$$
Hence $\varphi ^{(n)}=0\,,\,n=0\,,\,\ldots ,\,N$, i.e., $\varphi \equiv 0$.
So Assumption 2 is satisfied.

\section{Concept of distributions in infinite dimensional analysis\label
{Concept}}

In this section we will introduce a preliminary distribution theory in
infinite dimensional non-Gaussian analysis. We want to point out in advance
that the distribution space constructed here is in some sense too big for
practical purposes. In this sense section \ref{Concept} may be viewed as a
stepping stone to introduce the more useful structures in \S \ref
{testfunctions} and \S \ref{Distributions}.

We will choose ${\cal P}({\cal N}^{\prime })$ as our (minimal) test function
space. (The idea to use spaces of this type as appropriate spaces of test
functions is rather old see \cite{KMP65}. They also discussed in which sense
this space is ``minimal''.) First we have to ensure that ${\cal P}({\cal N}%
^{\prime })$ is densely embedded in $L^2(\mu )$. This is fulfilled because
of our assumption 1 \cite[Sec.§10 Th.1]{Sk74}. The space ${\cal P}({\cal N}%
^{\prime })$ may be equipped with various different topologies, but there
exists a natural one such that ${\cal P}({\cal N}^{\prime })$ becomes a
nuclear space \cite{BeKo88}. The topology on ${\cal P}({\cal N}^{\prime })$
is chosen such that is becomes isomorphic to the topological direct sum of
tensor powers ${\cal N}_{\Ckl }^{\hat \otimes n}$ see e.g., 
\cite[Ch II 6.1, Ch III 7.4]{Sch71}%
$$
{\cal P}({\cal N}^{\prime })\simeq \bigoplus_{n=0}^\infty {\cal N}_{\Ckl %
}^{\hat \otimes n}\text{ .} 
$$
via 
$$
\varphi (x)=\sum_{n=0}^\infty \left\langle x^{\otimes n},\varphi
^{(n)}\right\rangle \longleftrightarrow \vec \varphi =\left\{ \varphi
^{(n)}\;\Big|\;n\in \N _0\right\} . 
$$
Note that only a finite number of $\varphi ^{(n)}$ is non-zero. We will not
reproduce the full construction here, but we will describe the notion of
convergence of sequences this topology on ${\cal P}({\cal N}^{\prime })$.
For $\varphi \in {\cal P}({\cal N}^{\prime })$, $\varphi
(x)=\sum_{n=0}^{N(\varphi )}\left\langle x^{\otimes n},\varphi
^{(n)}\right\rangle $ let $p_n:{\cal P}({\cal N}^{\prime })\rightarrow {\cal %
N}_{\Ckl }^{\hat \otimes n}$ denote the mapping $p_n$ is defined by $%
p_n\varphi :=\varphi ^{(n)}.$ A sequence $\left\{ \varphi _j,\ j\in {\N}%
\right\} $ of smooth polynomials converges to $\varphi \in {\cal P}({\cal N}%
^{\prime })$ iff the $N(\varphi _j)\ $are bounded and $p_n\varphi _j%
\stackunder{n\rightarrow \infty }{\longrightarrow }p_n\varphi $ in ${\cal N}%
_{\Ckl }^{\hat \otimes n}$ for all $n\in {\N}$.

Now we can introduce the dual space ${\cal P}_\mu ^{\prime }({\cal N}%
^{\prime })$ of ${\cal P}({\cal N}^{\prime })$ with respect to $L^2(\mu )$.
As a result we have constructed the triple 
$$
{\cal P}({\cal N}^{\prime })\subset L^2(\mu )\subset {\cal P}_\mu ^{\prime }(%
{\cal N}^{\prime }) 
$$
The (bilinear) dual pairing $\langle \!\langle \cdot ,\cdot \rangle
\!\rangle _\mu $ between ${\cal P}_\mu ^{\prime }({\cal N}^{\prime })$ and $%
{\cal P}({\cal N}^{\prime })$ is connected to the (sesqui\-linear) inner
product on $L^2(\mu )$ by%
$$
\langle \!\langle \varphi ,\;\psi \rangle \!\rangle _\mu =(\varphi ,\;%
\overline{\psi })_{L^2(\mu )}\ ,\quad \varphi \in L^2(\mu ),\ \psi \in {\cal %
P}({\cal N}^{\prime })\text{ .} 
$$
Since the constant function 1 is in ${\cal P}({\cal N}^{\prime })$ we may
extend the concept of expectation from random variables to distributions;
for $\Phi \in {\cal P}_\mu ^{\prime }({\cal N}^{\prime })$ 
$$
\E _\mu (\Phi ):=\left\langle \!\left\langle \Phi ,1\right\rangle
\!\right\rangle _\mu \;. 
$$
The main goal of this section is to provide a description of ${\cal P}_\mu
^{\prime }({\cal N}^{\prime })$ , see Theorem \ref{PStrichRep} below. The
simplest approach to this problem seems to be the use of so called $\mu $%
-Appell polynomials.

\subsection{Appell polynomials associated to the measure
\texorpdfstring{$\mu $ }{ \textmu }
\label{AppellSec}}

Because of the holomorphy of $l_\mu $ and $l_\mu (0)=1$ there exists a
neighborhood of zero%
$$
{\cal U}_0=\left\{ \theta \in {\cal N}_{\Ckl }\ \Big|\ 2^{q_0}\left| \theta
\right| _{p_0}<1\right\} 
$$
$p_0,q_0\in {\N,}$ $p_0\geq p_\mu ^{\prime }$ , $2^{-q_0}\leq \varepsilon
_\mu $ ($p_\mu ^{\prime },\varepsilon _\mu $ from Lemma \ref{equiLemma})
such that $l_\mu (\theta )\neq 0$ for $\theta \in {\cal U}_0$ and the
normalized exponential 
\begin{equation}
\label{emy}e_\mu (\theta ;z)=\frac{e^{\left\langle z,\theta \right\rangle }}{%
l_\mu (\theta )}\text{ \quad for }\theta \in {\cal U}_0,\quad z\in {\cal N}_{%
\Ckl }^{\prime }\;, 
\end{equation}
is well defined. We use the holomorphy of $\theta \mapsto e_\mu (\theta ;z)$
to expand it in a power series in $\theta $ similar to the case
corresponding to the construction of one dimensional Appell polynomials \cite
{Bo76}. We have in analogy to \cite{AKS93,ADKS94}%
$$
e_\mu (\theta ;z)=\sum_{n=0}^\infty \frac 1{n!}\widehat{{\rm d}^ne_\mu (0,z)}%
(\theta ) 
$$
where $\widehat{{\rm d}^ne_\mu (0;z)}$ is an n-homogeneous continuous
polynomial. But since $e_\mu (\theta ;z)$ is not only G-holomorphic but
holomorphic we know that $\theta \rightarrow $ $e_\mu (\theta ;z)$ is also
locally bounded. Thus Cauchy's inequality for Taylor series \cite{Di81} may
be applied, $\rho \leq 2^{-q_0}$ , $p\geq p_0$%
\begin{equation}
\label{Cauchyemy}\left| \frac 1{n!}\widehat{{\rm d}^ne_\mu (0;z)}(\theta
)\right| \leq \frac 1{\rho ^n}\sup \limits_{\left| \theta \right| _p=\rho
}\left| e_\mu (\theta ;z)\right| \left| \theta \right| _p^n\leq \frac 1{\rho
^n}\sup \limits_{\left| \theta \right| _p=\rho }\frac 1{l_\mu (\theta
)}e^{\rho \left| z\right| _{-p}}\left| \theta \right| _p^n 
\end{equation}
if $z\in {\cal H}_{-p,\Ckl }$. This inequality extends by polarization \cite
{Di81} to an estimate sufficient for the kernel theorem. Thus we have a
representation $\widehat{{\rm d}^ne_\mu (0;z)}(\theta )=\left\langle P_n^\mu
(z),\theta ^{\otimes n}\right\rangle $ where $P_n^\mu (z)\in \left( {\cal N}%
^{\hat \otimes n}_{\Ckl }\right) ^{\prime }$. The kernel theorem really
gives a little more: $P_n^\mu (z)\in {\cal H}_{-p^{\prime }}^{\hat \otimes n}
$ for any $p^{\prime }(>p\geq p_0)$ such that the embedding operator $%
i_{p^{\prime },p}:{\cal H}_{p^{\prime }}\hookrightarrow {\cal H}_{p\text{ }}$%
is Hilbert-Schmidt. Thus we have 
\begin{equation}
\label{Pgenerator}e_\mu (\theta ;z)=\sum_{n=0}^\infty \frac
1{n!}\left\langle P_n^\mu (z),\theta ^{\otimes n}\right\rangle \quad \text{%
for }\theta \in {\cal U}_0,\ z\in {\cal N}_{\Ckl }^{\prime }\text{ .} 
\end{equation}
We will also use the notation%
$$
P_n^\mu (\varphi ^{(n)})(z):=\left\langle P_n^\mu (z),\varphi
^{(n)}\right\rangle ,\qquad \varphi ^{(n)}\in {\cal N}_{\Ckl }^{\hat \otimes
n},\quad n\in {\N}. 
$$
Thus for any measure satisfying Assumption 1 we have defined the ${\p}^\mu $%
-system

$$
{\p}^\mu =\left\{ \left\langle P_n^\mu (\cdot ),\varphi ^{(n)}\right\rangle
\ \bigg|\ \varphi ^{(n)}\in {\cal N}_{\Ckl }^{\hat \otimes n},\ n\in {\N}%
\right\} . 
$$
\bigskip\ 

Let us collect some properties of the polynomials $P_n^\mu (z).$

\begin{proposition}
For $x,y\in {\cal N}^{\prime }\ ,\ n\in \N $ the following holds\\[6mm]
(P1) \nopagebreak \hfill \\[-11mm] 
\begin{equation}
\label{(P1)}P_n^\mu (x)=\sum_{k=0}^n\binom nkx^{\otimes k}\hat \otimes
P_{n-k}^\mu (0),
\end{equation}
\\[5mm]
(P2)\nopagebreak \hfill \\[-11mm] 
\begin{equation}
\label{(P2)}x^{\otimes n}=\sum_{k=0}^n\binom nkP_k^\mu (x)\hat \otimes {\rm M%
}_{n-k}^\mu 
\end{equation}
\\[5mm]
(P3)\nopagebreak \hfill \\[-11mm] 
$$
P_n^\mu (x+y)=\sum_{k+l+m=n}\frac{n!}{k!\,l!\,m!}P_k^\mu (x)\hat \otimes
P_l^\mu (y)\hat \otimes {\rm M}_m^\mu  
$$
\begin{equation}
\label{(P3)}=\sum_{k=0}^n\binom nkP_k^\mu (x)\hat \otimes y^{\otimes (n-k)}
\end{equation}
(P4)\ Further we observe 
\begin{equation}
\label{(P4)}\E _\mu (\langle P_m^\mu (\cdot ),\varphi ^{(m)}\rangle
)=0\qquad \text{for }m\neq 0\ ,\varphi ^{(m)}\in {\cal N}_{\Ckl }^{\hat
\otimes m}\ .
\end{equation}
(P5) For all $p>p_0$ such that the embedding ${\cal H}_p\hookrightarrow 
{\cal H}_{p_0}$ is Hilbert--Schmidt and for all $\varepsilon >0$ small
enough $\left( \varepsilon \leq \frac{2^{-q_0}}{e\left\| i_{p,p_0}\right\|
_{HS}}\right) $ there exists a constant $C_{p,\varepsilon }>0$ with 
\begin{equation}
\label{(P5)}\left| P_n^\mu (z)\right| _{-p}\leq C_{p,\varepsilon
}\,n!\,\varepsilon ^{-n}\,e^{\varepsilon |z|_{-p}},\quad z\in {\cal H}_{-p,%
\Ckl }
\end{equation}
\end{proposition}

\TeXButton{Proof}{\proof}We restrict ourselves to a sketch of proof, details
can be found in \cite{ADKS94}.

\noindent (P1) This formula can be obtained simply by substituting 
\begin{equation}
\label{1/L}\frac 1{l_\mu (\theta )}=\sum\limits_{n=0}^\infty \frac
1{n!}\left\langle P_n^\mu (0),\theta ^{\otimes n}\right\rangle ,\quad \theta
\in {\cal N}_{\Ckl },\left| \theta \right| _q<\delta 
\end{equation}
and 
$$
e^{\left\langle x,\theta \right\rangle }=\sum\limits_{n=0}^\infty \frac
1{n!}\left\langle x^{\otimes n},\theta ^{\otimes n}\right\rangle ,\quad
\theta \in {\cal N}_{\Ckl },x\in {\cal N}^{\prime } 
$$
in the equality $e_\mu (\theta ;x)=e^{\left\langle x,\theta \right\rangle
}l_\mu ^{-1}(\theta )$. A comparison with (\ref{Pgenerator}) proves (P1).
The proof of (P2) is completely analogous to the proof of (P1).

\noindent (P3) We start from the following obvious equation of the
generating functions%
$$
e_\mu (\theta ;x+y)=e_\mu (\theta ;x)\,e_\mu (\theta ;y)\,l_\mu (\theta ) 
$$
This implies%
$$
\sum_{n=0}^\infty \frac 1{n!}\langle P_n^\mu (x+y),\theta ^{\otimes
n}\rangle =\sum_{k,l,m=0}^\infty \frac 1{k!\,l!\,m!}\,\langle P_k(x)\hat
\otimes P_l(y)\hat \otimes {\rm M}_m,\;\theta ^{\otimes (k+l+m)}\rangle 
$$
from this (P3) follows immediately.

\noindent (P4) To see this we use, $\theta \in {\cal N}_{\Ckl }$,%
$$
\sum_{n=0}^\infty \frac 1{n!}\E _\mu (\langle P_m^\mu (\cdot ),\theta
^{\otimes n}\rangle )=\E _\mu (e_\mu (\theta ;\cdot ))=\frac{\E _\mu
(e^{\langle \cdot ,\theta \rangle })}{l_\mu (\theta )}=1\ . 
$$
Then a comparison of coefficients and the polarization identity gives the
above result.

\noindent (P5) We can use 
\begin{equation}
\label{Pnxnorm}|P_n^\mu (z)|_{-p^{\prime }}\leq n!\left( \sup _{|\theta
|_p=\rho }\frac 1{l_\mu (\theta )}\right) e^{\rho |z|_{-p}}\left( \frac
e\rho \,\left\| i_{p^{\prime },p}\right\| _{HS}\right) ^n\;,\quad z\in {\cal %
H}_{-p,\Ckl } 
\end{equation}
$p>p_0,p^{\prime },\rho $ defined above. (\ref{Pnxnorm}) is a simple
consequence of the kernel theorem by (\ref{Cauchyemy}). In particular we
have 
$$
\left| P_n^\mu (0)\right| _{-p}\leq n!\left( \sup _{|\theta |_{p_0}=\rho
}\frac 1{l_\mu (\theta )}\right) \left( \frac e\rho \left\|
i_{p,p_0}\right\| _{HS}\right) ^n 
$$
If $p>p_0$ such that $\left\| i_{p,p_0}\right\| _{HS}$ is finite. For $%
0<\varepsilon \leq 2^{-q_0}/e\left\| i_{p,p_0}\right\| _{HS}$ we can fix $%
\rho =\varepsilon \,e\,\left\| i_{p,p_0}\right\| _{HS}\leq 2^{-q_0}$. With 
$$
C_{p,\varepsilon }:=\sup _{|\theta |_{p_0}=\rho }\frac 1{l_\mu (\theta )} 
$$
we have%
$$
\left| P_n^\mu (0)\right| _{-p}\leq C_{p,\varepsilon }\,n!\,\varepsilon
^{-n}. 
$$
Using (\ref{(P1)}) the following estimates hold 
\begin{eqnarray*}
\left| P_n^\mu (z)\right| _{-p} & \leq & \sum_{k=0}^n\binom nk\left| P_k^\mu %
(0)\right| _{-p}\left| z\right| _{-p}^{n-k}\ ,\qquad z\in {\cal H}_{-p,\Ckl } %
\\& \leq & C_{p,\varepsilon }\sum_{k=0}^n\tbinom nkk!\,\varepsilon ^{-k}\left| %
z\right| _{-p}^{n-k} %
\\&=& C_{p,\varepsilon }\,n!\,\varepsilon ^{-n}\sum_{k=0}^n\tfrac 1{(n-k)!}%
(\varepsilon \,|z|_{-p})^{n-k} %
\\& \leq & C_{p,\varepsilon }\,n!\,%
\varepsilon ^{-n}\,e^{\varepsilon |z|_{-p}}\ . 
\end{eqnarray*}
This completes the proof.\TeXButton{End Proof}{\endproof} \bigskip

\TeXButton{Note }{\note } The formulae (\ref{(P1)}) and (\ref{1/L}) can also be used as an
alternative definition of the polynomials $P_n^\mu (x)$\bigskip\ .

\example 
\newcounter{GaussA} \setcounter{GaussA}{\value{example}} \label{GaussAP} Let
us compare to the case of Gaussian Analysis. Here one has 
$$
l_{\gamma _{{\cal H}}}(\theta )=e^{\frac 12\langle \theta ,\theta \rangle }\
,\qquad \theta \in {\cal N}_{\Ckl } 
$$
Then it follows%
$$
{\rm M}_{2n}^\mu =(-1)^nP_{2n}^\mu (0)=\frac{(2n)!}{n!\,2^n}{\rm Tr}^{\hat
\otimes n}\ ,\qquad n\in \N  
$$
and ${\rm M}_n^\mu =P_n^\mu (0)=0$ if $n$ is odd. Here ${\rm Tr}\in {\cal N}%
^{\prime \otimes 2}$ denotes the trace kernel defined by 
\begin{equation}
\label{Trace}\langle {\rm Tr},\eta \otimes \xi \rangle =(\eta ,\xi )\
,\qquad \eta ,\xi \in {\cal N} 
\end{equation}
A simple comparison shows that 
$$
P_n^\mu (x)=:x^{\otimes n}:\text{ } 
$$
and%
$$
e_\mu (\theta ;x)=:e^{\langle x,\theta \rangle }: 
$$
where the r.h.s. denotes usual Wick ordering see e.g., \cite{BeKo88,HKPS93}.
This procedure is uniquely defined by%
$$
\langle :x^{\otimes n}:,\xi ^{\otimes n}\rangle =2^{-\frac n2}|\xi
|^n\,H_n\left( \tfrac 1{\sqrt{2}|\xi |}\langle x,\xi \rangle \right) \
,\qquad \xi \in {\cal N} 
$$
where $H_n$ denotes the Hermite polynomial of order $n$ (see e.g., \cite
{HKPS93} for the normalization we use).\bigskip\ 

Now we are ready to give the announced description of ${\cal P}({\cal N}%
^{\prime })$.

\begin{lemma}
\label{PrepLemma}For any $\varphi \in {\cal P}({\cal N}^{\prime })$ there
exists a unique representation 
\begin{equation}
\label{Prep}\varphi (x)=\sum\limits_{n=0}^N\left\langle P_n^\mu (x),\varphi
^{(n)}\right\rangle \ ,\quad \text{ }\varphi ^{(n)}\in {\cal N}_{\Ckl %
}^{\hat \otimes n}
\end{equation}
and vice versa, any functional of the form (\ref{Prep}) is a smooth
polynomial.
\end{lemma}

\TeXButton{Proof}{\proof}The representations from Definition \ref{P(N)} and
equation (\ref{Prep}) can be transformed into one another using (\ref{(P1)})
and (\ref{(P2)}). \TeXButton{End Proof}{\endproof}

\subsection{The dual Appell system and the representation theorem for 
\texorpdfstring{${\cal P}_\mu ^{\prime }({\cal N}^{\prime })$} {P\textacute (N\textacute )  }
}

To give an internal description of the type (\ref{Prep}) for ${\cal P}_\mu
^{\prime }({\cal N}^{\prime })$ we have to construct an appropriate system
of generalized functions, the ${\Q}^\mu $-system. The construction we
propose here is different from that of \cite{ADKS94} where smoothness of the
logarithmic derivative of $\mu $ was demanded and used for the construction
of the ${\Q}^\mu $-system. To avoid this additional assumption (which
excludes e.g., Poisson measures) we propose to construct the ${\Q}^\mu $%
-system using differential operators.

Define a differential operator of order $n$ with constant coefficient $\Phi
^{(n)}\in \left( {\cal N}_{\Ckl }^{\hat \otimes n}\right) ^{\prime }$ 
$$
D(\Phi ^{(n)})\langle x^{\otimes m},\varphi ^{(m)}\rangle =\QDATOPD\{ . {%
\dfrac{m!}{(m-n)!}\langle x^{\otimes (m-n)}\hat \otimes \Phi ^{(n)},\varphi
^{(m)}\rangle \text{ \quad for }m\geq n}{\text{\hspace{1.5cm}}0\text{%
\hspace{4cm}for }m<n} 
$$
($\varphi ^{(m)}\in {\cal N}_{\Ckl }^{\hat \otimes m},m\in {\N}$) and extend
by linearity from the monomials to ${\cal P}({\cal N}^{\prime }).$

\begin{lemma}
\label{Dcontinuity}$D(\Phi ^{(n)})$is a continuous linear operator from $%
{\cal P}({\cal N}^{\prime })$ to ${\cal P}({\cal N}^{\prime })$ .
\end{lemma}

\TeXButton{Remark }{\remark } For $\Phi ^{(1)}\in {\cal N}^{\prime }$ we
have the usual G\^ateaux derivative as e.g., in white noise analysis \cite
{HKPS93}%
$$
D(\Phi ^{(1)})\varphi =D_{\Phi ^{(1)}}\varphi :=\frac{{\rm d}}{{\rm d}t}%
\varphi (\cdot +t\Phi ^{(1)})|_{t=0} 
$$
for $\varphi \in {\cal P}({\cal N})$ and we have $D(\left( \Phi
^{(1)}\right) ^{\otimes n})=(D_{\Phi ^{(1)}})^n$ thus $D(\left( \Phi
^{(1)}\right) ^{\otimes n})$ is in fact a differential operator of order $n$.

\TeXButton{Proof}{\proof}By definition ${\cal P}({\cal N}^{\prime })$ is
isomorphic to the topological direct sum of tensor powers ${\cal N}_{\Ckl %
}^{\hat \otimes n}$%
$$
{\cal P}({\cal N}^{\prime })\simeq \bigoplus_{n=0}^\infty {\cal N}_{\Ckl %
}^{\hat \otimes n}\text{ .} 
$$
Via this isomorphism $D(\Phi ^{(n)})$ transforms each component ${\cal N}_{%
\Ckl }^{\hat \otimes m}$, $m\geq n$ by 
$$
\varphi ^{(m)}\mapsto \frac{n!}{(m-n)!}(\Phi ^{(n)},\;\varphi ^{(m)})_{{\cal %
H}^{\hat \otimes n}} 
$$
where the contraction $(\Phi ^{(n)},\;\varphi ^{(m)})_{{\cal H}^{\hat
\otimes n}}$ $\in {\cal N}_{\Ckl }^{\otimes (m-n)}$ is defined by 
\begin{equation}
\label{contraction}\langle x^{\otimes (m-n)},\;(\Phi ^{(n)},\;\varphi
^{(m)})_{{\cal H}^{\hat \otimes n}}\rangle :=\langle x^{\otimes (m-n)}\hat
\otimes \Phi ^{(n)},\varphi ^{(m)}\rangle 
\end{equation}
for all $x\in {\cal N}^{\prime }$. It is easy to verify that%
$$
|(\Phi ^{(n)},\;\varphi ^{(m)})_{{\cal H}^{\hat \otimes n}}|_q\leq |\Phi
^{(n)}|_{-q}|\varphi ^{(m)}|_q\text{ ,\qquad }q\in \N  
$$
which guarantees that $(\Phi ^{(n)},\;\varphi ^{(m)})_{{\cal H}^{\hat
\otimes n}}\in {\cal N}_{\Ckl }^{\otimes (m-n)}$ and shows at the same time
that $D(\Phi ^{(n)})$ is continuous on each component. This is sufficient to
ensure the stated continuity of $D(\Phi ^{(n)})$ on ${\cal P}({\cal N}%
^{\prime }).$\TeXButton{End Proof}{\endproof}\bigskip\ 

\begin{lemma}
For $\Phi ^{(n)}\in {\cal N}_{\Ckl }^{\prime \hat \otimes n}$ , $\varphi
^{(m)}\in {\cal N}_{\Ckl }^{\hat \otimes m}$ we have\\[9mm](P6)\\[-14mm] 
\begin{equation}
\label{(P6)}D(\Phi ^{(n)})\langle P_m^\mu (x),\varphi ^{(m)}\rangle
=\QATOPD\{ . {\dfrac{m!}{(m-n)!}\left\langle P_{m-n}^\mu (x)\hat \otimes
\Phi ^{(n)},\;\varphi ^{(m)}\right\rangle \text{ for }m\geq n}{\text{%
\hspace{1.5cm}}0\text{\hspace{4.1cm} for }m<n}
\end{equation}
\end{lemma}

\TeXButton{Proof}{\proof}This follows from the general property of Appell
polynomials which behave like ordinary powers under differentiation. More
precisely, by using 
$$
\langle P_m^\mu ,\theta ^{\otimes m}\rangle =\left. \left( \frac{{\rm d}}{%
{\rm d}t}\right) ^me_\mu (t\theta ;\cdot )\right| _{t=0}\ ,\qquad \theta \in 
{\cal N}_{\Ckl } 
$$
we have%
\begin{eqnarray*}
D(\Phi ^{(1)})\langle P_m^\mu (x),\theta ^{\otimes m}\rangle %
&=& \left. \frac{{\rm d}}{{\rm d}\lambda }\langle P_m^\mu (x+\lambda %
\Phi ^{(1)}),\theta ^{\otimes m}\rangle \right| _{\lambda =0} %
\\&=& \left. \left( \frac \partial {\partial t}\right) ^m\frac \partial {\partial \lambda }e_\mu (t\theta ;x+\lambda \Phi ^{(1)})%
\right| _{\T \QATOP{t=0}{\lambda =0}} %
\\&=& \langle \Phi ^{(1)},\theta \rangle \left. \left( \tfrac \partial {\partial t}\right) ^mt\;e_\mu (t\theta ;x)\right| _{t=0} %
\\&=& \left. \langle \Phi ^{(1)},\theta \rangle \sum_{k=0}^m\tbinom mk\left( %
\left( \tfrac{{\rm d}}{{\rm d}t}\right) ^kt\right) \left( \tfrac{{\rm d}}{{\rm d}t}\right) ^{m-k}e_\mu (t\theta ;x)\right| _{t=0} %
\\&=& \left. m\,\langle \Phi ^{(1)},\theta \rangle \left( \tfrac{{\rm d}}{{\rm d}t}\right) ^{m-1}e_\mu (t\theta ;x)\right| _{t=0} %
\\&=& m\,\langle \Phi ^{(1)},\theta \rangle \left\langle P_{m-1}^\mu (x),\theta %
^{\otimes (m-1)}\right\rangle \text{ .} 
\end{eqnarray*}
This proves 
$$
D(\Phi ^{(1)})\langle P_m^\mu ,\varphi ^{(m)}\rangle =m\left\langle
P_{m-1}^\mu \hat \otimes \Phi ^{(1)},\;\varphi ^{(m)}\right\rangle . 
$$
The property (\ref{(P6)}), then follows by induction.\TeXButton{End Proof}
{\endproof}\bigskip\ 

In view of Lemma \pageref{Dcontinuity} it is possible to define the adjoint
operator $D(\Phi ^{(n)})^{*}:{\cal P}_\mu ^{\prime }({\cal N}^{\prime
})\rightarrow {\cal P}_\mu ^{\prime }({\cal N}^{\prime })$ for $\Phi
^{(n)}\in {\cal N}_{\Ckl }^{\prime \hat \otimes n}$ . Further we can
introduce the constant function $\1 \in {\cal P}_\mu ^{\prime }({\cal N}%
^{\prime })$ such that $\1 (x)\equiv 1$ for all $x\in {\cal N}^{\prime }$ ,
so 
$$
\langle \!\langle \1 ,\;\varphi \rangle \!\rangle _\mu =\int_{{\cal N}%
^{\prime }}\varphi (x)\,{\rm d\mu }(x)=\E _\mu (\varphi ). 
$$
Now we are ready to define our $\Q$-system.

\begin{definition}
For any $\Phi ^{(n)}\in \left( {\cal N}_{\Ckl }^{\hat \otimes n}\right)
^{\prime }$ we define $Q_n^\mu (\Phi ^{(n)})\in {\cal P}_\mu ^{\prime }(%
{\cal N}^{\prime })$ by%
$$
Q_n^\mu (\Phi ^{(n)})=D(\Phi ^{(n)})^{*}{\1}\ . 
$$
\end{definition}

We want to introduce an additional formal notation $Q_n^\mu (x)$ which
stresses the linearity of $\Phi ^{(n)}\mapsto Q_n^\mu (\Phi ^{(n)})\in P_\mu
^{\prime }({\cal N}^{\prime }):$%
$$
\langle Q_n^\mu ,\Phi ^{(n)}\rangle :=Q_n^\mu (\Phi ^{(n)})\ . 
$$

\example It is possible to put further assumptions on the measure $\mu $ to
ensure that the expression is more than formal. Let us assume a smooth
measure (i.e., the logarithmic derivative of $\mu $ is infinitely
differentiable, see \cite{ADKS94} for details) with the property%
$$
\exists q\in \N \ ,\ \exists \{C_n\geq 0,\;n\in \N \}:\forall \xi \in {\cal N%
} 
$$
$$
\left| \int D_\xi ^n\varphi \;{\rm d}\mu (x)\right| \leq C_n\left\| \varphi
\right\| _{L^2(\mu )}|\xi |_q^n 
$$
where $\varphi $ is any finitely based bounded ${\cal C}^\infty $-function
on ${\cal N}^{\prime }$. This obviously establishes a bound of the type 
$$
\left\| Q_n^\mu (\xi _1\otimes \cdots \otimes \xi _n)\right\| _{L^2(\mu
)}\leq C_n^{\prime }\prod_{j=1}^n|\xi _j|_q\ ,\qquad \xi _1,\ldots ,\xi
_n\in {\cal N\ },\ n\in \N  
$$
which is sufficient to show (by means of kernel theorem) that there exists $%
Q_n^\mu (x)\in \left( {\cal N}_{\Ckl }^{\hat \otimes n}\right) ^{\prime }$
for almost all $x\in {\cal N}^{\prime }$ such that we have the representation%
$$
Q_n^\mu (\varphi ^{(n)})(x)=\langle Q_n^\mu (x),\varphi ^{(n)}\rangle \
,\qquad \varphi ^{(n)}\in {\cal N}_{\Ckl }^{\hat \otimes n} 
$$
for almost all $x\in {\cal N}^{\prime }$. For any smooth kernel $\varphi
^{(n)}\in {\cal N}_{\Ckl }^{\hat \otimes n}$ we have then that the function%
$$
x\mapsto \langle Q_n^\mu (x),\varphi ^{(n)}\rangle \ =Q_n^\mu \left( \varphi
^{(n)}\right) (x)\ 
$$
belongs to $L^2(\mu ).$\bigskip\ 

\example The simplest non trivial case can be studied using finite
dimensional real analysis. We consider $\R $ as our basic Hilbert space and
as our nuclear space ${\cal N}$. Thus the nuclear ``triple'' is simply%
$$
\R \subseteq \R \subseteq \R  
$$
and the dual pairing between a ``test function'' and a ``distribution''
degenerates to multiplication. On $\R $ we consider a measure ${\rm d}\mu
(x)=\rho (x)\,{\rm d}x$ where $\rho $ is a positive ${\cal C}^\infty $%
--function on $\R $ such that Assumptions 1 and 2 are fulfilled. In this
setting the adjoint of the differentiation operator is given by%
$$
\left( \frac{{\rm d}}{{\rm d}x}\right) ^{*}f(x)=-\left( \tfrac{{\rm d}}{{\rm %
d}x}+\beta (x)\right) f(x)\ ,\qquad f\in {\cal C}^1(\R ) 
$$
where the logarithmic derivative $\beta $ of the measure $\mu $ is given by 
$$
\beta =\frac{\rho ^{\prime }}\rho 
$$
This enables us to calculate the $\Q ^\mu $-system. One has 
$$
Q_n^\mu (x)=\left( \left( \tfrac{{\rm d}}{{\rm d}x}\right) ^{*}\right) ^n\1 %
=(-1)^n\left( \tfrac{{\rm d}}{{\rm d}x}+\beta (x)\right) ^n\1  
$$
$$
=(-1)^n\frac{\rho ^{(n)}(x)}{\rho (x)}\ . 
$$
The last equality can be seen by simple induction. \\ If $\rho =\frac 1{%
\sqrt{2\pi }}e^{-\frac 12x^2}$ is the Gaussian density $Q_n^\mu $ is related
to the n$^{th}$ Hermite polynomial:%
$$
Q_n^\mu (x)=2^{-n/2}H_n\left( \tfrac x{\sqrt{2}}\right) \ . 
$$
\bigskip\ 

\begin{definition}
We define the $\Q^\mu $-system in ${\cal P}_\mu ^{\prime }({\cal N}^{\prime
})$ by 
$$
\Q ^\mu =\left\{ Q_n^\mu (\Phi ^{(n)})\ \Big|\ \qquad \Phi ^{(n)}\in \left( 
{\cal N}_{\Ckl }^{\hat \otimes n}\right) ^{\prime },\ n\in \N _0\ \right\} \
, 
$$
and the pair $(\p ^\mu ,\Q^\mu )$ will be called the Appell system $\A ^\mu $
generated by the measure $\mu $.
\end{definition}

Now we are going to discuss the central property of the Appell system $\A %
^\mu $.

\begin{theorem}
\label{BiorTh}{\rm (Biorthogonality w.r.t. }$\mu ${\rm )} 
\begin{equation}
\label{QnPnPair}\left\langle \!\!\left\langle \langle Q_n^\mu (\Phi
^{(n)}),\ \langle P_m^\mu ,\varphi ^{(m)}\rangle \right\rangle
\!\!\right\rangle _\mu =\delta _{m,n}\;n!\;\langle \Phi ^{(n)},\varphi
^{(n)}\rangle 
\end{equation}
for $\Phi ^{(n)}\in \left( {\cal N}_{\Ckl }^{\hat \otimes n}\right) ^{\prime
}$ and $\varphi ^{(m)}\in {\cal N}_{\Ckl }^{\hat \otimes m}$ .
\end{theorem}

\TeXButton{Proof}{\proof}It follows from (\ref{(P4)}) and (\ref{(P6)}) that%
\begin{eqnarray*}
\left\langle \!\!\left\langle Q_n^\mu (\Phi ^{(n)}),\ \langle P_m^\mu %
,\varphi ^{(m)}\rangle \right\rangle \!\!\right\rangle _\mu %
&=& \left\langle \!\!\left\langle \1 ,D(\Phi ^{(n)})\langle P_m^\mu ,\varphi %
^{(m)}\rangle \right\rangle \!\!\right\rangle _\mu %
\\&=& \frac{m!}{(m-n)!}\E _\mu \left( \langle P_{(m-n)}^\mu \hat \otimes \Phi %
^{(n)},\;\varphi ^{(m)}\rangle \right) %
\\&=& m!\;\delta _{m,n}\;\langle \Phi ^{(m)},\varphi ^{(m)}\rangle \ . 
\end{eqnarray*}
\TeXButton{End Proof}{\endproof}\bigskip\ 

Now we are going to characterize the space ${\cal P}_\mu ^{\prime }({\cal N}%
^{\prime })$

\begin{theorem}
\label{PStrichRep}For all $\Phi \in {\cal P}_\mu ^{\prime }({\cal N}^{\prime
})$ there exists a unique sequence $\{\Phi ^{(n)}\big|\ n\in \N _0\},\ \Phi
^{(n)}\in \left( {\cal N}_{\Ckl }^{\hat \otimes n}\right) ^{\prime }$ such
that 
\begin{equation}
\label{Qexpansion}\Phi =\sum_{n=0}^\infty Q_n^\mu (\Phi ^{(n)})\equiv
\sum_{n=0}^\infty \langle Q_n^\mu ,\Phi ^{(n)}\rangle 
\end{equation}
and vice versa, every series of the form (\ref{Qexpansion}) generates a
generalized function in ${\cal P}_\mu ^{\prime }({\cal N}^{\prime }).$
\end{theorem}

\TeXButton{Proof}{\proof}For $\Phi \in {\cal P}_\mu ^{\prime }({\cal N}%
^{\prime })$ we can uniquely define $\Phi ^{(n)}\in \left( {\cal N}_{\Ckl %
}^{\hat \otimes n}\right) ^{\prime }$ by 
$$
\langle \Phi ^{(n)},\varphi ^{(n)}\rangle =\frac 1{n!}\ \!\langle \!\langle
\Phi ,\;\langle P_n^\mu ,\varphi ^{(n)}\rangle \rangle \!\rangle _\mu \
,\qquad \varphi ^{(n)}\in {\cal N}_{\Ckl }^{\hat \otimes n} 
$$
This definition is possible because $\langle P_n^\mu ,\varphi ^{(n)}\rangle
\in {\cal P}({\cal N}^{\prime })$. The continuity of $\varphi ^{(n)}\mapsto
\langle \Phi ^{(n)},\varphi ^{(n)}\rangle $ follows from the continuity of $%
\varphi \mapsto \langle \!\langle \Phi ,\varphi \rangle \!\rangle \ ,\
\varphi \in $ ${\cal P}({\cal N}^{\prime })$. This implies that $\varphi
\mapsto \sum_{n=0}^\infty n!\;\langle \Phi ^{(n)},\varphi ^{(n)}\rangle $ is
continuous on ${\cal P}({\cal N}^{\prime })$. This defines a generalized
function in ${\cal P}_\mu ^{\prime }({\cal N}^{\prime })$, which we denote
by $\sum_{n=0}^\infty Q_n^\mu (\Phi ^{(n)})$. In view of Theorem \ref{BiorTh}
it is obvious that 
$$
\Phi =\sum_{n=0}^\infty Q_n^\mu (\Phi ^{(n)})\ . 
$$

To see the converse consider a series of the form (\ref{Qexpansion}) and $%
\varphi \in {\cal P}({\cal N}^{\prime })$. Then there exist $\varphi
^{(n)}\in {\cal N}_{\Ckl }^{\hat \otimes n}\ ,\ n\in \N $ and $N\in \N $
such that we have the representation%
$$
\varphi =\sum_{n=0}^NP_n^\mu (\varphi ^{(n)})\ . 
$$
So we have%
$$
\!\left\langle \!\!\!\left\langle \sum_{n=0}^\infty Q_n^\mu (\Phi
^{(n)}),\varphi \right\rangle \!\!\!\right\rangle _{\!\!\mu
}:=\sum_{n=0}^Nn!\;\langle \Phi ^{(n)},\varphi ^{(n)}\rangle 
$$
because of Theorem \ref{BiorTh}. The continuity of $\varphi \mapsto \langle
\!\langle \sum_{n=0}^\infty Q_n^\mu (\Phi ^{(n)}),\varphi \rangle \!\rangle
_\mu $ follows because $\varphi ^{(n)}\mapsto \langle \Phi ^{(n)},\varphi
^{(n)}\rangle $ is continuous for all $n\in \N $ .\TeXButton{End Proof}
{\endproof}
\LaTeXparent{nga4.tex}

\section{Test functions on a linear space with measure \label{testfunctions}}

In this section we will construct the test function space $({\cal N})^1$ and
study its properties. On the space ${\cal P}({\cal N}^{\prime })$ we can
define a system of norms using the representation from Lemma \ref{PrepLemma}%
. Let 
$$
\varphi =\sum_{n=0}^N\langle P_n^\mu ,\;\varphi ^{(n)}\rangle \in {\cal P}(%
{\cal N}^{\prime }) 
$$
be given, then $\varphi ^{(n)}\in {\cal H}_{p,\Ckl }^{\hat \otimes n}$ for
each $p\geq 0\ \ (n\in \N )$. Thus we may define for any $p,q\in \N $ a
Hilbertian norm on ${\cal P}({\cal N}^{\prime })$ by 
$$
\left\| \varphi \right\| _{p,q,\mu }^2=\sum_{n=0}^N(n!)^2\;2^{nq}\;|\varphi
^{(n)}|_p^2 
$$
The completion of ${\cal P}({\cal N}^{\prime })$ w.r.t. $\left\| \cdot
\right\| _{p,q,\mu }$ is called $({\cal H}_p)_{q,\mu }^1$ .

\begin{definition}
We define 
$$
({\cal N})_\mu ^1:=\ \stackunder{p,q\in \N }{\rm pr\ lim}({\cal H}_p)_{q,\mu
}^1\ . 
$$
\end{definition}

This space has the following properties

\begin{theorem}
$({\cal N})_\mu ^1$ is a nuclear space. The topology $({\cal N})_\mu ^1$ is
uniquely defined by the topology on ${\cal N}$: It does not depend on the
choice of the family of norms $\{|\cdot |_p\}$.
\end{theorem}

\TeXButton{Proof}{\proof}Nuclearity of $({\cal N})_\mu ^1$ follows
essentially from that of ${\cal N}.$ For fixed $p,q$ consider the embedding 
$$
I_{p^{\prime },q^{\prime },p,q}:\left( {\cal H}_{p^{\prime }}\right)
_{q^{\prime },\mu }^1\rightarrow \left( {\cal H}_p\right) _{q,\mu }^1 
$$
where $p^{\prime }$ is chosen such that the embedding%
$$
i_{p^{\prime },p}:\,{\cal H}_{p^{\prime }}\rightarrow {\cal H}_p 
$$
is Hilbert--Schmidt. Then $I_{p^{\prime },q^{\prime },p,q}$ is induced by%
$$
I_{p^{\prime },q^{\prime },p,q}\varphi =\sum_{n=0}^\infty \langle P_n^\mu
,i_{p^{\prime },p}^{\otimes n}\varphi ^{(n)}\rangle \quad \text{ for \quad }%
\varphi =\sum_{n=0}^\infty \langle P_n^\mu ,\varphi ^{(n)}\rangle \in \left( 
{\cal H}_{p^{\prime }}\right) _{q^{\prime },\mu }^1. 
$$
Its Hilbert--Schmidt norm is easily estimated by using an orthonormal basis
of $\left( {\cal H}_{p^{\prime }}\right) _{q^{\prime },\mu }^1$. The result
is the bound 
$$
\left\| I_{p^{\prime },q^{\prime },p,q}\right\| _{HS}^2\le \sum_{n=0}^\infty
2^{n(q-q^{\prime })}\left\| i_{p^{\prime },p}\right\| _{HS}^{2n} 
$$
which is finite for suitably chosen $q^{\prime }$.

Let us assume that we are given two different systems of Hilbertian norms $%
\left| \,\cdot \,\right| _p$ and $\left| \,\cdot \,\right| _k^{\prime }$,
such that they induce the same topology on ${\cal N}$ . For fixed $k$ and $l$
we have to estimate $\left\| \,\cdot \,\right\| _{k,l,\mu }^{\prime }$ by $%
\left\| \,\cdot \,\right\| _{p,q,\mu }$ for some $p,q$ (and vice versa which
is completely analogous). Since $\left| \,\cdot \,\right| _k^{\prime }$ has
to be continuous with respect to the projective limit topology on ${\cal N}$%
, there exists $p$ and a constant $C$ such that $\left| f\right| _k^{\prime
}\leq C\left| f\right| _p$, for all $f\in {\cal N}$, i.e., the injection $i$
from ${\cal H}_p$ into the completion ${\cal K}_k$ of ${\cal N}$ with
respect to $|\,\cdot \,|_k^{\prime }$ is a mapping bounded by $C$. We denote
by $i$ also its linear extension from ${\cal H}_{p,{\,}\Ckl }$ into ${\cal K}%
_{{\,}k,\Ckl }$. It follows that $i^{\otimes n}$ is bounded by $C^n$ from $%
{\cal H}_{{\,}p,\Ckl }^{\otimes n}$ into ${\cal K}_{{\,}k,\Ckl   }^{\otimes
n}$. Now we choose $q$ such that $2^{{\frac{q-l}2}}\geq C$. Then

\begin{eqnarray*}
\left\| \,\cdot \,\right\| _{k,l,\mu }^{\prime 2}&=&\sum_{n=0}^\infty %
(n!)^2\,2^{nl}\left| \,\cdot \,\right| _k^{\prime 2} %
\\&\leq & \sum_{n=0}^\infty (n!)^2\,2^{nl}C^{2n}\left| \,\cdot \,%
\right| _p^2 %
\\&\leq &\left\| \,\cdot \,\right\| _{p,q,\mu }^2\ , 
\end{eqnarray*}
which had to be proved.\TeXButton{End Proof}{\endproof}\bigskip\ 

\begin{lemma}
\label{L2NormPn}There exist $p,C,K>0$ such that for all $n$ 
\begin{equation}
\label{Pnx2norm}\int |P_n^\mu (x)|_{-p}^2\;{\rm d}\mu (x)\leq (n!)^2\,C^n\,K
\end{equation}
\end{lemma}

\TeXButton{Proof}{\proof} The estimate (\ref{Pnxnorm}) may be used for $\rho
\leq 2^{-q_0}$ and $\rho \leq 2\varepsilon _\mu $ ($\varepsilon _\mu $ from
Lemma \ref{equiLemma}).\\This gives 
$$
\int |P_n^\mu (x)|_{-p}^2\;{\rm d}\mu (x)\leq (n!)^2\left( \frac e\rho
\left\| i_{p,p_0}\right\| _{HS}\right) ^{2n}\int e^{2\rho |x|_{-p_0}}{\rm d}%
\mu (x) 
$$
which is finite because of Lemma \ref{equiLemma}.\TeXButton{End Proof}
{\endproof}

\begin{theorem}
There exist $p^{\prime },q^{\prime }>0$ such that for all $p\geq p^{\prime
},\ q\geq q^{\prime }$ the topological embedding $({\cal H}_p)_{q,\mu
}^1\subset L^2(\mu )$ holds.
\end{theorem}

\TeXButton{Proof}{\proof}Elements of the space $({\cal N})_\mu ^1$ are
defined as series convergent in the given topology. Now we need to study the
convergence of these series in $L^2(\mu )$. Choose $q^{\prime }$ such that $%
C>2^{q^{\prime }\text{ }}$ ($C$ from estimate (\ref{Pnx2norm})). Let us take
an arbitrary 
$$
\varphi =\sum_{n=0}^\infty \langle P_n^\mu ,\varphi ^{(n)}\rangle \in {\cal P%
}({\cal N}^{\prime }) 
$$
For $p>p^{\prime }$ ($p^{\prime }$ as in Lemma \ref{L2NormPn} ) and $%
q>q^{\prime }$ the following estimates hold%
\begin{eqnarray*}
\left\| \varphi \right\| _{L^2(\mu )}%
& \leq & \sum_{n=0}^\infty \left\| \langle%
P_n^\mu ,\varphi ^{(n)}\rangle \right\| _{L^2(\mu )}%
\\&\leq & \sum_{n=0}^\infty |\varphi ^{(n)}|_{-p}\left\| \,|P_n^\mu%
|_{-p}\right\| _{L^2(\mu )} %
\\&\leq & K\sum_{n=0}^\infty n!\,2^{nq/2}\left| \varphi ^{(n)}\right|%
_{-p}(C2^{-q})^{n/2} %
\\&\leq & K\left( \sum_{n=0}^\infty (C\,2^{-q})^n\right) ^{\frac 12}\left(%
\sum_{n=0}^\infty (n!)^2\,2^{qn}\left| \varphi ^{(n)}\right| _{-p}^2\right)%
^{\frac 12} %
\\&=& K\left( 1-C\,2^{-q}\right) ^{-1/2}\left\| \varphi \right\| _{p,q,\mu }%
\text{.}%
\end{eqnarray*}
Taking the closure the inequality extends to the whole space $({\cal H}%
_p)_q^1$.\TeXButton{End Proof}{\endproof}

\begin{corollary}
\label{N1inL2}$({\cal N})_\mu ^1$ is continuously and densely embedded in $%
L^2(\mu )$.
\end{corollary}

\example \newcounter{myexponent} \setcounter{myexponent}{\value{example}} 
{\it ($\mu $-exponentials as test functions)} \smallskip 
\\The $\mu $-exponential given in (\ref{Pgenerator}) has the following norm%
$$
||e_\mu (\theta ;\cdot )||_{p,q,\mu }^2=\sum_{n=0}^\infty 2^{nq}\,|\theta
|_p^{2n}\ ,\qquad \theta \in {\cal N}_{\Ckl } 
$$
This expression is finite if and only if $2^q|\theta |_p^2<1$. Thus we have $%
e_\mu (\theta ;\cdot )\notin ({\cal N})_\mu ^1$ if $\theta \neq 0$. But we
have that $e_\mu (\theta ;\cdot )$ is a test function of finite order i.e., $%
e_\mu (\theta ;\cdot )\in ({\cal H}_p)_q^1$ if $2^q|\theta |_p^2<1$. This is
in contrast to some useful spaces of test functions in Gaussian Analysis,
see e.g., \cite{BeKo88,HKPS93}.

The set of all $\mu $--exponentials $\{e_\mu (\theta ;\cdot )\;|\;2^q|\theta
|_p^2<1,\ \theta \in {\cal N}_{\Ckl }\}$ is a total set in $({\cal H}_p)_q^1$%
. This can been shown using the relation ${\rm d}^ne_\mu (0;\cdot )(\theta
_1,...,\theta _n)=\langle P_n^\mu ,\theta _1\hat \otimes \cdots \hat \otimes
\theta _n\rangle .$

\begin{proposition}
\label{N1inEmin}Any test function $\varphi $ in $({\cal N})_\mu ^1$ has a
uniquely defined extension to ${\cal N}_{\Ckl }^{\prime }$ as an element of $%
{\cal E}_{\min }^1\left( {\cal N}_{\Ckl }^{\prime }\right) $
\end{proposition}

\TeXButton{Proof}{\proof}Any element $\varphi $ in $({\cal N})_\mu ^1$ is
defined as a series of the following type 
$$
\varphi =\sum_{n=0}^\infty \langle P_n^\mu ,\varphi ^{(n)}\rangle \ ,\qquad
\varphi ^{(n)}\in {\cal N}_{\Ckl }^{\hat \otimes n} 
$$
such that%
$$
\left\| \varphi \right\| _{p,q,\mu }^2=\sum_{n=0}^\infty
(n!)^2\,2^{nq}\,|\varphi ^{(n)}|_p^2 
$$
is finite for each $p,q\in \N $ . In this proof we will show the convergence
of the series%
$$
\sum_{n=0}^\infty \langle P_n^\mu (z),\varphi ^{(n)}\rangle ,\quad z\in 
{\cal H}_{-p,\Ckl } 
$$
to an entire function in $z$.

Let $p>p_0$ such that the embedding $i_{p,p_0}:{\cal H}_p\hookrightarrow 
{\cal H}_{p_0}$ is Hilbert-Schmidt. Then for all $0<\varepsilon \leq
2^{-q_0}/e\left\| i_{p,p_0}\right\| _{HS}$ we can use (\ref{(P5)}) and
estimate as follows%
\begin{eqnarray*}
\sum_{n=0}^\infty |\langle P_n^\mu (z),\varphi ^{(n)}\rangle |%
& \leq & \sum_{n=0}^\infty |P_n^\mu (z)|_{-p}|\varphi ^{(n)}|_p %
\\& \leq & C_{p,\varepsilon }\,e^{\varepsilon |z|_{-p}}\sum_{n=0}^\infty%
n!\,|\varphi ^{(n)}|_p\,\varepsilon ^{-n} %
\\& \leq & C_{p,\varepsilon }\,\,e^{\varepsilon |z|_{-p}}\,\left(%
\sum_{n=0}^\infty (n!)^22^{nq}|\varphi ^{(n)}|_p^2\right) ^{1/2}\left(%
\sum_{n=0}^\infty 2^{-nq}\varepsilon ^{-2n}\right) ^{1/2} %
\\&=& C_{p,\varepsilon }\,\,\left( 1-2^{-q}\varepsilon ^{-2}\right)%
^{-1/2}\,\left\| \varphi \right\| _{p,q,\mu }\,\;e^{\varepsilon |z|_{-p}} %
\end{eqnarray*}
if $2^q>\varepsilon ^{-2}$. That means the series $\sum_{n=0}^\infty \langle
P_n^\mu (z),\varphi ^{(n)}\rangle $ converges uniformly and absolutely in
any neighborhood of zero of any space ${\cal H}_{-p,\Ckl }$ . Since each
term $\langle P_n^\mu (z),\varphi ^{(n)}\rangle $ is entire in $z$ the
uniform convergence implies that $z\mapsto \sum_{n=0}^\infty \langle P_n^\mu
(z),\varphi ^{(n)}\rangle $ is entire on each ${\cal H}_{-p,\Ckl }$ and
hence on ${\cal N}_{\Ckl }^{\prime }$. This completes the proof.%
\TeXButton{End Proof}{\endproof}\bigskip\ 

The following corollary is an immediate consequence of the above proof and
gives an explicit estimate on the growth of the test functions.

\begin{corollary}
\label{phi(z)Betrag}For all $p>p_0$ such that the norm $\left\|
i_{p,p_0}\right\| _{HS}$ of the embedding is finite and for all $%
0<\varepsilon \leq 2^{-q_0}/e\left\| i_{p,p_0}\right\| _{HS}$ we can choose $%
q\in \N $ such that $2^q>\varepsilon ^{-2}$ to obtain the following bound.%
$$
\left| \varphi (z)\right| \leq C\,\left\| \varphi \right\| _{p,q,\mu
}\,e^{\varepsilon |z|_{-p}}\ ,\qquad \varphi \in ({\cal N})_\mu ^1,\ z\in 
{\cal H}_{-p,\Ckl }\text{ ,} 
$$
where%
$$
C=C_{p,\varepsilon }\,\left( 1-2^{-q}\varepsilon ^{-2}\right) ^{-1/2}. 
$$
\end{corollary}

$\ $Let us look at Proposition \ref{N1inEmin} again. On one hand any
function $\varphi \in ({\cal N})_\mu ^1$ can be written in the form 
\begin{equation}
\label{phiPn}\varphi (z)=\sum_{n=0}^\infty \langle P_n^\mu (x),\varphi
^{(n)}\rangle \ ,\qquad \varphi ^{(n)}\in {\cal N}_{\Ckl }^{\hat \otimes n}\
, 
\end{equation}
on the other hand it is entire, i.e., it has the representation 
\begin{equation}
\label{phizn}\varphi (z)=\sum_{n=0}^\infty \langle z^{\otimes n},\tilde
\varphi ^{(n)}\rangle \ ,\qquad \tilde \varphi ^{(n)}\in {\cal N}_{\Ckl %
}^{\hat \otimes n}\ , 
\end{equation}
To proceed further we need the explicit correspondence $\left\{ \varphi
^{(n)},n\in \N \right\} \longleftrightarrow \left\{ \tilde \varphi
^{(n)},n\in \N \right\} $ which is given in the next lemma.

\begin{lemma}
\label{Reordering}{\bf (Reordering)} \smallskip
\\Equations (\ref{phiPn}) and (\ref{phizn}) hold iff%
$$
\tilde \varphi ^{(k)}=\sum_{n=0}^\infty \binom{n+k}k\left( P_n^\mu
(0),\varphi ^{(n+k)}\right) _{{\cal H}^{\hat \otimes n}} 
$$
or equivalently%
$$
\varphi ^{(k)}=\sum_{n=0}^\infty \binom{n+k}k\left( {\rm M}_n^\mu ,\tilde
\varphi ^{(n+k)}\right) _{{\cal H}^{\hat \otimes n}} 
$$
where $\left( P_n^\mu (0),\varphi ^{(n+k)}\right) _{{\cal H}^{\hat \otimes
n}}$ and $\left( {\rm M}_n^\mu ,\tilde \varphi ^{(n+k)}\right) _{{\cal H}%
^{\hat \otimes n}}$ denote contractions defined by (\ref{contraction}).
\end{lemma}

\noindent This is a consequence of (\ref{(P1)}) and (\ref{(P2)}). We omit
the simple proof.\bigskip\ \ 

Proposition \ref{N1inEmin} states%
$$
({\cal N})_\mu ^1\subseteq {\cal E}_{\min }^1({\cal N}^{\prime }) 
$$
as sets, where%
$$
{\cal E}_{\min }^1({\cal N}^{\prime })=\left\{ \varphi |_{{\cal N}^{\prime
}}\;\Big| \;\varphi \in {\cal E}_{\min }^1({\cal N}_{\Ckl }^{\prime
})\right\} \ . 
$$
Corollary \ref{phi(z)Betrag} then implies that the embedding is also
continuous. Now we are going to show that the converse also holds.

\begin{theorem}
\label{N1E1min}For all measures $\mu \in {\cal M}_a({\cal N}^{\prime })$ we
have the topological identity%
$$
({\cal N})_\mu ^1={\cal E}_{\min }^1({\cal N}^{\prime })\ . 
$$
\end{theorem}

\noindent To prove the missing topological inclusion it is convenient to use
the nuclear topology on ${\cal E}_{\min }^1({\cal N}_{\Ckl }^{\prime })$
(given by the norms $\lnorm \cdot \rnorm _{{p,q,1}}$) introduced in section 
\ref{Preliminaries}. Theorem \ref{Ekminprlim} ensures that this topology is
equivalent to the projective topology induced by the norms ${\rm n}_{p,l,k}$%
. Then the above theorem is an immediate consequence of the following norm
estimate.

\begin{proposition}
Let $p>p_\mu $ ($p_\mu $ as in Lemma \ref{equiLemma}) such that $\left\|
i_{p,p_\mu }\right\| _{HS}$ is finite and $q\in \N $ such that $2^{q/2}>K_p$
($K_p:=eC\left\| i_{p,p_\mu }\right\| _{HS}$ as in (\ref{MnmuNorm})). For
any $\varphi \in {\rm E}_{p,q}^1$ the restriction $\varphi |_{{\cal N}%
^{\prime }}$ is a function from $({\cal H}_p)_{q^{\prime },\mu }^1\ ,\
q^{\prime }<q$. Moreover the following estimate holds%
$$
||\varphi ||_{p,q^{\prime },\mu }\leq \lnorm  \varphi \rnorm
_{p,q,1}(1-2^{-q/2}K_p)^{-1}(1-2^{q^{\prime }-q})^{-1/2}\ . 
$$
\end{proposition}

\TeXButton{Proof}{\proof}Let $p,q\in \N $, $K_p$ be defined as above. A
function $\varphi \in {\rm E}_{p,q}^1$ has the representation (\ref{phizn}).
Using the Reordering lemma combined with (\ref{MnmuNorm}) and 
$$
\left| \tilde \varphi ^{(n)}\right| _p\leq \frac 1{n!}\,2^{-nq/2}\lnorm
\varphi \rnorm  _{p,q,1} 
$$
we obtain a representation of the form (\ref{phiPn}) where 
\begin{eqnarray*}
\left| \varphi ^{(n)}\right| _p & \leq & \sum_{k=0}^\infty \binom{n+k}k\left|%
{\rm M}_k^\mu \right| _{-p}\left| \tilde \varphi ^{(n+k)}\right| _p %
\\&\leq & \lnorm  \varphi \rnorm  _{p,q,1}\sum_{k=0}^\infty %
\binom{n+k}k\frac{k!}{(n+k)!}K_p^k\,2^{-(n+k)q/2} %
\\& \leq & \lnorm  \varphi \rnorm  _{p,q,1}%
\frac1{n!}2^{-nq/2}\sum_{k=0}^\infty (2^{-q/2}K_p)^k %
\\& \leq & \lnorm  \varphi \rnorm  _{p,q,1}%
\frac1{n!}2^{-nq/2}(1-2^{-q/2}K_p)^{-1} .%
\end{eqnarray*}
For $q^{\prime }<q$ this allows the following estimate%
\begin{eqnarray*}
||\varphi ||_{p,q^{\prime },\mu }^2%
&=&\sum_{n=0}^\infty (n!)^2\,2^{q^{\prime}n}\,|\varphi ^{(n)}|_p^2 %
\\& \leq & \lnorm  \varphi \rnorm ^2%
_{p,q,1}(1-2^{-q/2}K_p)^{-2}\sum_{k=0}^\infty 2^{n(q^{\prime }-q)}<\infty%
\end{eqnarray*}
This completes the proof.\TeXButton{End Proof}{\endproof}\bigskip\ 

Since we now have proved that the space of test functions $({\cal N})_\mu ^1$
is isomorphic to ${\cal E}_{\min }^1({\cal N}^{\prime })$ for all measures $%
\mu \in {\cal M}_a({\cal N}^{\prime })$, we will now drop the subscript $\mu 
$. The test function space $({\cal N})^1$ is the same for all measures $\mu
\in {\cal M}_a({\cal N}^{\prime })$.

\begin{corollary}
$({\cal N})^1$ is an algebra under pointwise multiplication.
\end{corollary}

\begin{corollary}
$({\cal N})^1$ admits `scaling' i.e., for $\lambda \in \C $ the scaling
operator $\sigma _\lambda :({\cal N})^1\rightarrow ({\cal N})^1$ defined by $%
\sigma _\lambda \varphi (x):=\varphi (\lambda x)$, $\varphi \in ({\cal N})^1$%
, $x\in {\cal N}^{\prime }$ is well--defined.
\end{corollary}

\begin{corollary}
For all $z\in {\cal N}_{\Ckl }^{\prime }$ the space $({\cal N})^1$ is
invariant under the shift operator $\tau _z:\varphi \mapsto \varphi (\cdot
+z)$.
\end{corollary}

\section{Distributions\label{Distributions}}

In this section we will introduce and study the space $({\cal N})_\mu ^{-1}$
of distributions corresponding to the space of test functions $({\cal N})^1$%
. Since ${\cal P}({\cal N}^{\prime })\subset ({\cal N})^1$ the space $({\cal %
N})_\mu ^{-1}$ can be viewed as a subspace of ${\cal P}_\mu ^{\prime }({\cal %
N}^{\prime })$%
$$
({\cal N})_\mu ^{-1}\subset {\cal P}_\mu ^{\prime }({\cal N}^{\prime }) 
$$
Let us now introduce the Hilbertian subspace $({\cal H}_{-p})_{-q,\mu }^{-1}$
of ${\cal P}_\mu ^{\prime }({\cal N}^{\prime })$ for which the norm 
$$
\left\| \Phi \right\| _{-p,-q,\mu }^2:=\sum_{n=0}^\infty 2^{-qn}\left| \Phi
^{(n)}\right| _{-p}^2\text{ } 
$$
is finite. Here we used the canonical representation%
$$
\Phi =\sum_{n=0}^\infty Q_n^\mu (\Phi ^{(n)})\in {\cal P}_\mu ^{\prime }(%
{\cal N}^{\prime })\text{ } 
$$
from Theorem \ref{PStrichRep}. The space $({\cal H}_{-p})_{-q,\mu }^{-1}$ is
the dual space of $({\cal H}_p)_q^1$ with respect to $L^2(\mu )$ (because of
the biorthogonality of $\p -$and $\Q -$systems). By general duality theory 
$$
({\cal N})_\mu ^{-1}:=\bigcup_{p,q\in \N }({\cal H}_{-p})_{-q,\mu }^{-1} 
$$
is the dual space of $({\cal N})^1$ with respect to $L^2(\mu )$. As we noted
in section \ref{Preliminaries} there exists a natural topology on co-nuclear
spaces (which coincides with the inductive limit topology). We will consider 
$({\cal N})_\mu ^{-1}$ as a topological vector space with this topology. So
we have the nuclear triple%
$$
({\cal N})^1\subset L^2(\mu )\subset ({\cal N})_\mu ^{-1}\ . 
$$
The action of $\Phi =\sum_{n=0}^\infty Q_n^\mu (\Phi ^{(n)})\in ({\cal N}%
)_\mu ^{-1}$ on a test function $\varphi =\sum_{n=0}^\infty \langle P_n^\mu
,\varphi ^{(n)}\rangle \in ({\cal N})^1$ is given by 
$$
\langle \!\langle \Phi ,\varphi \rangle \!\rangle _\mu =\sum_{n=0}^\infty
n!\langle \Phi ^{(n)},\varphi ^{(n)}\rangle \ . 
$$
\bigskip\ 

For a more detailed characterization of the singularity of distributions in $%
({\cal N})_\mu ^{-1}$ we will introduce some subspaces in this distribution
space. For $\beta \in [0,1]$ we define%
$$
({\cal H}_{-p})_{-q,\mu }^{-\beta }=\left\{ \Phi \in {\cal P}_\mu ^{\prime }(%
{\cal N}^{\prime })\;\bigg|\;\sum_{n=0}^\infty (n!)^{1-\beta }2^{-qn}\left|
\Phi ^{(n)}\right| _{-p}^2<\infty \text{ for }\Phi =\sum_{n=0}^\infty
Q_n^\mu (\Phi ^{(n)})\right\} 
$$
and 
$$
({\cal N})_\mu ^{-\beta }=\stackunder{p,q\in \N }{\bigcup }({\cal H}%
_{-p})_{-q,\mu }^{-\beta }\ , 
$$
It is clear that the singularity increases with increasing $\beta $:%
$$
({\cal N})^{-0}\subset ({\cal N})^{-\beta _1}\subset ({\cal N})^{-\beta
_2}\subset ({\cal N})^{-1} 
$$
if $\beta _1\leq \beta _2$.We will also consider $({\cal N})_\mu ^\beta $ as
equipped with the natural topology.\bigskip\ 

\example 
\newcounter{RadonNy} \setcounter{RadonNy}{\value{example}}{\it (Generalized
Radon--Nikodym derivative)} \smallskip 
\\We want to define a generalized function $\rho _\mu (z,\cdot )\in ({\cal N}%
)_\mu ^{-1}\ $, $z\in {\cal N}_{\Ckl }^{\prime }$ with the following property%
$$
\langle \!\langle \rho _\mu (z,\cdot ),\varphi \rangle \!\rangle _\mu =\int_{%
{\cal N}^{\prime }}\varphi (x-z)\;{\rm d}\mu (x)\ ,\qquad \varphi \in ({\cal %
N})^1\ . 
$$
That means we have to establish the continuity of $\rho _\mu (z,\cdot )$.
Let $z\in {\cal H}_{-p,\Ckl }$.\ If $p^{\prime }\geq p$ is sufficiently
large and $\varepsilon >0$ small enough, Corollary \ref{phi(z)Betrag}
applies i.e., $\exists q\in \N $ and $C>0$ such that%
\begin{eqnarray*}
\left| \int_{{\cal N}^{\prime }}\varphi (x-z){\rm d}\mu (x)\ \right| %
& \leq & C\left\| \varphi \right\| _{p^{\prime },q,\mu }%
\int_{{\cal N}^{\prime}}e^{\varepsilon |x-z|_{-p^{\prime }}}{\rm d}\mu (x) %
\\& \leq & C\left\| \varphi \right\| _{p^{\prime },q,\mu }e^{\varepsilon |z|_{-p^{\prime }}}\int_{{\cal N}^{\prime }}e^{\varepsilon |x|_{-p^{\prime }}}{\rm d}\mu (x) %
\end{eqnarray*}
If $\varepsilon $ is chosen sufficiently small the last integral exists.
Thus we have in fact $\rho (z,\cdot )\in ({\cal N})_\mu ^{-1}$. It is clear
that whenever the Radon--Nikodym derivative $\frac{{\rm d}\mu (x+\xi )}{{\rm %
d}\mu (x)}$ exists (e.g., $\xi \in {\cal N}$ in case $\mu $ is ${\cal N}$%
-quasi-invariant) it coincides with $\rho _\mu (\xi ,\cdot )$ defined above.
We will now show that in $({\cal N})_\mu ^{-1}$ we have the canonical
expansion%
$$
\rho _\mu (z,\cdot )=\sum_{n=0}^\infty \frac 1{n!}(-1)^nQ_n^\mu (z^{\otimes
n}). 
$$
It is easy to see that the r.h.s. defines an element in $({\cal N})_\mu
^{-1} $. Since both sides are in $({\cal N})_\mu ^{-1}$ it is sufficient to
compare their action on a total set from $({\cal N})^1$. For $\varphi
^{(n)}\in {\cal N}_{\Ckl }^{\hat \otimes n}$ we have 
\begin{eqnarray*}
\left\langle \!\!\left\langle \rho _\mu (z,\cdot ),\langle P_n^\mu ,\varphi%
^{(n)}\rangle \right\rangle \!\!\right\rangle _\mu %
&=& \int_{{\cal N}^{\prime }}\langle P_n^\mu (x-z),\varphi ^{(n)}\rangle%
 \;{\rm d}\mu (x) %
\\&=& \sum_{k=0}^\infty \binom nk(-1)^{n-k}\int_{{\cal N}^{\prime }}\langle %
P_k^\mu (x)\hat \otimes z^{\otimes n-k},\varphi ^{(n)}\rangle \;{\rm d}\mu %
(x) \\&=& (-1)^n\langle z^{\otimes n},\varphi ^{(n)}\rangle %
\\&=& \left\langle \!\!\left\langle \sum_{k=0}^\infty \frac 1{k!}(-1)^kQ_k^\mu %
(z^{\otimes k}),\langle P_n^\mu ,\varphi ^{(n)}\rangle \right\rangle%
\!\!\right\rangle _\mu \ ,%
\end{eqnarray*}
where we have used (\ref{(P3)}), (\ref{(P4)}) and the biorthogonality of $%
\p 
$- and $\Q $-systems. This had to be shown. In other words, we have proven
that $\rho _\mu (-z,\cdot )$ is the generating function of the $\Q $%
-functions 
\begin{equation}
\label{rhomyQn}\rho _\mu (-z,\cdot )=\sum_{n=0}^\infty \frac 1{n!}Q_n^\mu
(z^{\otimes n})\ . 
\end{equation}
Let use finally remark that the above expansion allows for more detailed
estimates. It is easy to see that $\rho _\mu \in ({\cal N})_\mu ^{-0}$.%
\bigskip\ 

\example {\it (Delta distribution)} \smallskip \\For $z\in {\cal N}_{\Ckl %
}^{\prime }$ we define a distribution by the following $\Q $-decomposition:%
$$
\delta _z=\sum_{n=0}^\infty \frac 1{n!}Q_n^\mu (P_n^\mu (z)) 
$$
If $p\in \N $ is large enough and $\varepsilon >0$ sufficiently small there
exists $C_{p,\varepsilon }>0$ according to (\ref{(P5)}) such that 
\begin{eqnarray*}
\left\| \delta _z\right\| _{-p,-q,\mu }^2 &=&\sum_{n=0}^\infty %
(n!)^{-2}2^{-nq}\left| P_n^\mu (z)\right| _{-p}^2 %
\\& \leq & C_{p,\varepsilon }^2\,e^{2\varepsilon |z|_{-p}}\sum_{n=0}^\infty%
2^{-nq}\varepsilon ^{-2n}\ ,\qquad z\in {\cal H}_{-p,\Ckl }\;, %
\end{eqnarray*}
which is finite for sufficiently large $q\in \N $. Thus $\delta _z\in ({\cal %
N})_\mu ^{-1}$.

For $\varphi =\sum_{n=0}^\infty \langle P_n^\mu ,\varphi ^{(n)}\rangle \in (%
{\cal N})^1$ the action of $\delta _z$ is given by 
$$
\langle \!\langle \delta _z,\varphi \rangle \!\rangle _\mu
=\sum_{n=0}^\infty \langle P_n^\mu (z),\varphi ^{(n)}\rangle =\varphi (z) 
$$
because of (\ref{QnPnPair}). This means that $\delta _z$ (in particular for $%
z$ real) plays the role of a ``$\delta $-function" (evaluation map) in the
calculus we discuss.

\section{Integral transformations}

\begin{sloppypar}
We will first introduce the Laplace transform of a function $\varphi \in
L^2(\mu )$. The global assumption $\mu \in {\cal M}_a({\cal N}^{\prime })$
guarantees the existence of $p_\mu ^{\prime }\in \N \ $, $\varepsilon _\mu
>0 $ such that $\int_{{\cal N}^{\prime }}\exp (\varepsilon _\mu |x|_{-p_\mu
^{\prime }})\,{\rm d}\mu (x)<\infty $ by Lemma \ref{equiLemma}. Thus $\exp
(\langle x,\theta \rangle )\in L^2(\mu )$ if $2|\theta |_{p_\mu ^{\prime
}}\leq \varepsilon _\mu \ ,\theta \in {\cal H}_{p_\mu ^{\prime },\Ckl }$.
Then by Cauchy--Schwarz inequality the Laplace transform defined by 
$$
L_\mu \varphi (\theta ):=\int_{{\cal N}^{\prime }}\varphi (x)\exp \langle
x,\theta \rangle \,{\rm d}\mu (x) 
$$
is well defined for $\varphi \in L^2(\mu )\ ,\theta \in {\cal H}_{p_\mu
^{\prime },\Ckl }$ with $2|\theta |_{p_\mu ^{\prime }}\leq \varepsilon _\mu $%
. Now we are interested to extend this integral transform from $L^2(\mu )$
to the space of distributions $({\cal N})_\mu ^{-1}$.
\end{sloppypar}

Since our construction of test function and distribution spaces is closely
related to $\p $- and $\Q $-systems it is useful to introduce the so called $%
S_\mu $-transform 
$$
S_\mu \varphi (\theta ):=\frac{L_\mu \varphi (\theta )}{l_\mu (\theta )}\ . 
$$
Since $e_\mu (\theta ;x)=e^{\langle x,\theta \rangle }/l_\mu (\theta )$ we
may also write%
$$
S_\mu \varphi (\theta )=\int_{{\cal N}^{\prime }}\varphi (x)\,e_\mu (\theta
;x)\,{\rm d}\mu (x)\ . 
$$
The $\mu $-exponential $e_\mu (\theta ,\cdot )$ is not a test function in $(%
{\cal N})^1$, see Example \arabic{myexponent} . So the definition of the $%
S_\mu $-transform of a distribution $\Phi \in ({\cal N})_\mu ^{-1}$ must be
more careful. Every such $\Phi $ is of finite order i.e., $\exists p,q\in 
\N 
$ such that $\Phi \in ({\cal H}_{-p})_{-q,\mu .}^{-1}$ As shown in Example 
\arabic{myexponent} $e_\mu (\theta ,\cdot )$ is in the corresponding dual
space $({\cal H}_p)_{q,\mu }^1$ if $\theta \in {\cal H}_{p,\Ckl }$ is such
that $2^q|\theta |_p^2<1$. Then we can define a consistent extension of $%
S_\mu $-transform.%
$$
S_\mu \Phi (\theta ):=\langle \!\langle \Phi ,\;e_\mu (\theta ,\cdot
)\rangle \!\rangle _\mu 
$$
if $\theta $ is chosen in the above way. The biorthogonality of $\p $- and $%
\Q $-system implies 
$$
S_\mu \Phi (\theta )=\sum_{n=0}^\infty \langle \Phi ^{(n)},\theta ^{\otimes
n}\rangle \ . 
$$
It is easy to see that the series converges uniformly and absolutely on any
closed ball $\left\{ \left. \theta \in {\cal H}_{p,\Ckl }\right| \;|\theta
|_p^2\leq r,\ r<2^{-q}\right\} $, see the proof of Theorem \ref{CharTh}.
Thus $S_\mu \Phi $ is holomorphic a neighborhood of zero, i.e., $S_\mu \Phi
\in {\rm Hol}_0({\cal N}_{\Ckl })$. In the next section we will discuss this
relation to the theory of holomorphic functions in more detail.

The third integral transform we are going to introduce is more appropriate
for the test function space $({\cal N})^1$. We introduce the convolution of
a function $\varphi \in ({\cal N})^1$ with the measure $\mu $ by%
$$
C_\mu \varphi (y):=\int_{{\cal N}^{\prime }}\varphi (x+y)\,{\rm d}\mu (x) ,
\quad y\in {\cal N}^{\prime } .%
$$
From Example \arabic{RadonNy} the existence of a generalized Radon--Nikodym
derivative $\rho _\mu (z,\cdot )$, $z\in {\cal N}_{\Ckl }^{\prime }$ in $(%
{\cal N})_\mu ^{-1}$ is guaranteed. So for any $\varphi \in ({\cal N})^1$, $%
z\in {\cal N}_{\Ckl }^{\prime }$ the convolution has the representation 
$$
C_\mu \varphi (z)=\langle \!\langle \rho _\mu (-z,\cdot ),\;\varphi \rangle
\!\rangle _\mu \;. 
$$
If $\varphi \in ({\cal N})^1$ has the canonical representation%
$$
\varphi =\sum_{n=0}^\infty \langle P_n^\mu ,\;\varphi ^{(n)}\rangle 
$$
we have by equation (\ref{rhomyQn}) 
$$
C_\mu \varphi (z)=\sum_{n=0}^\infty \langle z^{\otimes n},\varphi
^{(n)}\rangle \ . 
$$

In Gaussian Analysis $C_\mu $- and $S_\mu$-transform coincide. It is a
typical non-Gaussian effect that these two transformations differ from each
other.

\section{Characterization theorems \label{Characterization}}

Gaussian Analysis has shown that for applications it is very useful to
characterize test and distribution spaces by the integral transforms
introduced in the previous section. In the non-Gaussian setting first
results in this direction have been obtained by \cite{AKS93,ADKS94}.%
\bigskip 

We will start to characterize the space $({\cal N})^1$ in terms of the
convolution $C_\mu $.

\begin{theorem}
\label{CmuChar} The convolution $C_\mu $ is a topological isomorphism from $(%
{\cal N})^1$ on ${\cal E}_{\min }^1({\cal N}_{\Ckl }^{\prime })$.
\end{theorem}

\TeXButton{Remark }{\remark } Since we have identified $({\cal N})^1$ and $%
{\cal E}_{\min }^1({\cal N}^{\prime })$ by Theorem \ref{N1E1min} the above
assertion can be restated as follows. We have 
$$
C_\mu :{\cal E}_{\min }^1({\cal N}^{\prime })\rightarrow {\cal E}_{\min }^1(%
{\cal N}_{\Ckl }^{\prime }) 
$$
as a topological isomorphism.

\TeXButton{Proof}{\proof}The proof has been well prepared by Theorem \ref
{Ekminprlim}, because the nuclear topology on ${\cal E}_{\min }^1({\cal N}_{%
\Ckl }^{\prime })$ is the most natural one from the point of view of the
above theorem. Let $\varphi \in ({\cal N})^1$ with the representation%
$$
\varphi =\sum_{n=0}^\infty \langle P_n^\mu ,\varphi ^{(n)}\rangle \ . 
$$
From the previous section it follows%
$$
C_\mu \varphi (z)=\sum_{n=0}^\infty \langle z^{\otimes n},\varphi
^{(n)}\rangle \ 
$$
It is obvious from (\ref{3StrichNorm}) that%
$$
\lnorm  C_\mu \varphi \rnorm  _{p,q,1}=\left\| \varphi \right\| _{p,q,\mu }\ 
$$
for all $p,q\in \N _0$, which proves the continuity of 
$$
C_\mu :({\cal N})^1\rightarrow {\cal E}_{\min }^1({\cal N}_{\Ckl }^{\prime
})\ . 
$$

Conversely let $F\in {\cal E}_{\min }^1({\cal N}_{\Ckl }^{\prime })$. Then
Theorem \ref{Ekminprlim} ensures the existence of a sequence of generalized
kernels $\left\{ \varphi ^{(n)}\in {\cal N}_{\Ckl }^{\prime }\;|\;n\in \N %
_0\right\} $ such that 
$$
F(z)=\sum_{n=0}^\infty \langle z^{\otimes n},\varphi ^{(n)}\rangle \ . 
$$
Moreover for all $p,q\in \N _0$%
$$
\lnorm  F\rnorm  _{p,q,1}^2=\sum_{n=0}^\infty (n!)^2\,2^{nq}\left| \varphi
^{(n)}\right| _p^2 
$$
is finite. Choosing%
$$
\varphi =\sum_{n=0}^\infty \langle P_n^\mu ,\varphi ^{(n)}\rangle 
$$
we have $\left\| \varphi \right\| _{p,q,\mu }=\lnorm  F\rnorm
_{p,q,1}$. Thus $\varphi \in ({\cal N})^1$. Since $C_\mu \varphi =F$ we have
shown the existence and continuity of the inverse of $C_\mu $.%
\TeXButton{End Proof}{\endproof}\bigskip\ 

To illustrate the above theorem in terms of the natural topology on ${\cal E}%
_{\min }^1({\cal N}_{\Ckl }^{\prime })$ we will reformulate the above
theorem and add some useful estimates which relate growth in ${\cal E}_{\min
}^1({\cal N}_{\Ckl }^{\prime })$ to norms on $({\cal N})^1$.

\begin{corollary}
\hfill \\1) Let $\varphi \in ({\cal N})^1$ then for all $p,l\in \N _0$ and $%
z\in {\cal H}_{-p,\Ckl }$ the following estimate holds%
$$
\left| C_\mu \varphi (z)\right| \leq \left\| \varphi \right\| _{p,2l,\mu
}\exp (2^{-l}|z|_{-p}) 
$$
i.e., C$_\mu \varphi \in {\cal E}_{\min }^1({\cal N}_{\Ckl }^{\prime })$.%
\medskip\ \\2) Let $F\in {\cal E}_{\min }^1({\cal N}_{\Ckl }^{\prime })$.
Then there exists $\varphi \in ({\cal N})^1$ with $C_\mu \varphi =F$. The
estimate 
$$
\left| F(z)\right| \leq C\exp (2^{-l}|z|_{-p}) 
$$
for $C>0,\ p,q\in \N _0$ implies%
$$
\left\| \varphi \right\| _{p^{\prime },q,\mu }\leq C\left(
1-2^{q-2l}e^2\left\| i_{p^{\prime },p}\right\| _{HS}^2\right) ^{-1/2} 
$$
if the embedding $i_{p^{\prime },p}:{\cal H}_{p^{\prime }}\hookrightarrow 
{\cal H}_p$ is Hilbert-Schmidt and $2^{l-q/2}>e\left\| i_{p^{\prime
},p}\right\| _{HS}$.
\end{corollary}

\TeXButton{Proof}{\proof}The first statement follows from 
$$
\left| C_\mu \varphi (z)\right| \leq {\rm n}_{p,l,1}(C_\mu \varphi )\cdot
\exp (2^{-l}|z|_{-p}) 
$$
which follows from the definition of {\rm n}$_{p,l,1}$ and estimate (\ref
{nplk3StrichNorm}). The second statement is an immediate consequence of
Lemma \ref{3Strichnplk}. \TeXButton{End Proof}{\endproof}\bigskip\ 

The next theorem characterizes distributions from $({\cal N})_\mu ^{-1}$ in
terms of $S_\mu$-transform.

\begin{theorem}
\label{CharTh}The $S_\mu $-transform is a topological isomorphism from $(%
{\cal N})_\mu ^{-1}$ on ${\rm Hol}_0({\cal N}_{\Ckl })$.
\end{theorem}

\TeXButton{Remark }{\remark } The above theorem is closely related to the
second part of Theorem \ref{indlimEkmax}. Since we left the proof open we
will give a detailed proof here.

\TeXButton{Proof}{\proof}Let $\Phi \in ({\cal N})_\mu ^{-1}$ . Then there
exists $p,q\in \N $ such that 
$$
\left\| \Phi \right\| _{-p,-q,\mu }^2=\sum_{n=0}^\infty 2^{-nq}|\Phi
^{(n)}|_{-p}^2 
$$
is finite. From the previous section we have 
\begin{equation}
\label{SPhitheta}S_\mu \Phi (\theta )=\sum_{n=0}^\infty \langle \Phi
^{(n)},\theta ^{\otimes n}\rangle \;. 
\end{equation}
For $\theta \in {\cal N}_{\Ckl }$ such that $2^q|\theta |_p^2<1$ we have by
definition (Formula (\ref{3StrichNorm}))%
$$
\lnorm  S_\mu \Phi \rnorm  _{-p,-q,-1}=\left\| \Phi \right\| _{-p,-q,\mu \
}. 
$$
By Cauchy--Schwarz inequality 
\begin{eqnarray*}
\left| S_\mu \Phi (\theta )\right| & \leq & \sum_{n=0}^\infty |\Phi %
^{(n)}|_{-p}|\theta |_p^n %
\\&\leq & \left( \sum_{n=0}^\infty 2^{-nq}|\Phi ^{(n)}|_{-p}^2\right)%
^{1/2}\left( \sum_{n=0}^\infty 2^{nq}|\theta |_p^{2n}\right) ^{1/2} %
\\&=& \left\| \Phi \right\| _{-p,-q,\mu }\left( 1-2^q|\theta |_p^2\right)%
^{-1/2}\ . 
\end{eqnarray*}
Thus the series (\ref{SPhitheta}) converges uniformly on any closed ball $%
\left\{ \left. \theta \in {\cal H}_{p,\Ckl }\right| \;|\theta |_p^2\leq r,\
r<2^{-q}\right\} $. Hence $S_\mu \Phi \in {\rm Hol}_0({\cal N}_{\Ckl })$ and 
$$
{\rm n}_{p,l,\infty }(S_\mu \Phi )\leq \left\| \Phi \right\| _{-p,-q,\mu
}(1-2^{q-2l})^{-1/2} 
$$
if $2l>q$. This proves that $S_\mu $ is a continuous mapping from $({\cal N}%
)_\mu ^{-1}$ to ${\rm Hol}_0({\cal N}_{\Ckl })$. In the language of section 
\ref{Holomorphy} this reads%
$$
\stackunder{p,q\in \N }{\rm ind\ lim}\,{\rm E}_{-p,-q}^{-1}\subset {\rm Hol}%
_0({\cal N}_{\Ckl }) 
$$
topologically.\medskip\ 

Conversely, let $F\in {\rm Hol}_0({\cal N}_{\Ckl })$ be given, i.e., there
exist $p,l\in \N $ such that {\rm n}$_{p,l,\infty }(F)<\infty $. The first
step is to show that there exists $p^{\prime },q\in \N $ such that 
$$
\lnorm  F\rnorm  _{-p^{\prime },-q,-1}<{\rm n}_{p,l,\infty }(F)\cdot C\ , 
$$
for sufficiently large $C>0$. This implies immediately 
$$
{\rm Hol}_0({\cal N}_{\Ckl })\subset \ \stackunder{p,q\in \N }{\rm ind\ lim}%
\,{\rm E}_{-p,-q}^{-1} 
$$
topologically, which is the missing part in the proof of the second
statement in Theorem \ref{indlimEkmax}.

By assumption the Taylor expansion%
$$
F(\theta )=\sum_{n=0}^\infty \frac 1{n!}\widehat{{\rm d}^nF(0)}(\theta ) 
$$
converges uniformly on any closed ball $\left\{ \left. \theta \in {\cal H}%
_{p,\Ckl }\right| \;|\theta |_p^2\leq r,\ r<2^{-l}\right\} $ and 
$$
\left| F(\theta )\right| \leq {\rm n}_{p,l,\infty }(F)\ . 
$$
Proceeding analogously to Lemma \ref{nplk3Strich}, an application of
Cauchy's inequality gives%
\begin{eqnarray*}
\frac 1{n!}\widehat{{\rm d}^nF(0)}(\theta ) & \leq & 2^l|\theta |_p^n\sup %
_{|\theta |_p\leq 2^{-l}}|F(\theta )| %
\\& \leq & {\rm n}_{p,l,\infty }(F)\ \cdot 2^{nl}\cdot |\theta |_p^n 
\end{eqnarray*}
The polarization identity gives 
$$
\left| \frac 1{n!}{\rm d}^nF(0)(\theta _1,\ldots ,\theta _n)\right| \leq 
{\rm n}_{p,l,\infty }(F)\ \cdot e^n\cdot 2^{nl}\prod_{j=1}^n|\theta _j|_p 
$$
Then by kernel theorem (Theorem \ref{KernelTh}) there exist kernels $\Phi
^{(n)}\in {\cal H}_{-p^{\prime },\Ckl }^{\hat \otimes n}$ for $p^{\prime }>p$
with $\left\| i_{p^{\prime },p}\right\| _{HS}<\infty $ such that%
$$
F(\theta )=\sum_{n=0}^\infty \langle \Phi ^{(n)},\theta ^{\otimes n}\rangle
\ . 
$$
Moreover we have the following norm estimate 
$$
\left| \Phi ^{(n)}\right| _{-p^{\prime }}\leq {\rm n}_{p,l,\infty }(F)\
\left( 2^le\left\| i_{p^{\prime },p}\right\| _{HS}\right) ^n 
$$
Thus%
\begin{eqnarray*}
\lnorm  F\rnorm  _{-p^{\prime },-q,-1}^2%
&=& \sum_{n=0}^\infty%
2^{-nq}\left| \Phi ^{(n)}\right| _{-p^{\prime }}^2 %
\\& \leq & {\rm n}_{p,l,\infty }^2(F)\sum_{n=0}^\infty %
\left( 2^{2l-q}e^2\left\|%
i_{p^{\prime },p}\right\| _{HS}^2\right) ^n %
\\&=& {\rm n}_{p,l,\infty }^2(F)\left( 1-2^{2l-q}e^2%
\left\| i_{p^{\prime },p}\right\| _{HS}^2\right) ^{-1}\ 
\end{eqnarray*}
if $q\in \N $ is such that $\rho :=2^{2l-q}e^2\left\| i_{p^{\prime
},p}\right\| _{HS}^2\ <1$. So we have in fact 
$$
\lnorm  F\rnorm  _{-p^{\prime },-q,-1}\leq {\rm n}_{p,l,\infty }(F)(1-\rho
)^{-1/2}. 
$$
Now the rest is simple. Define $\Phi \in ({\cal N})_\mu ^{-1}$ by 
$$
\Phi =\sum_{n=0}^\infty Q_n^\mu (\Phi ^{(n)}) 
$$
then $S_\mu \Phi =F$ and 
$$
\left\| \Phi \right\| _{-p^{\prime },-q,\mu }=\lnorm  F\rnorm
_{-p^{\prime },-q,-1} 
$$
This proves the existence of a continuous inverse of the $S_\mu $%
--transform. Uniqueness of $\Phi $ follows from the fact that $\mu $%
-exponentials are total in any $({\cal H}_p)_q^1$. \TeXButton{End Proof}
{\endproof}\bigskip\ 

We can extract some useful estimates from the above proof which describe the
degree of singularity of a distribution.

\begin{corollary}
Let $F\in {\rm Hol}_0({\cal N}_{\Ckl })$ be holomorphic for all $\theta \in 
{\cal N}_{\Ckl }$ with $|\theta |_p\leq 2^{-l}$. If $p^{\prime }>p$ with $%
\left\| i_{p^{\prime },p}\right\| _{HS}<\infty $ and $q\in \N $ is such that 
$\rho :=2^{2l-q}e^2\left\| i_{p^{\prime },p}\right\| _{HS}^2<1$. Then $\Phi
\in ({\cal H}_{-p^{\prime }})_{-q}^{-1}$ and 
$$
\left\| \Phi \right\| _{-p^{\prime },-q,\mu }\leq {\rm n}_{p,l,\infty
}(F)\cdot (1-\rho )^{-1/2}. 
$$
\end{corollary}

For a more detailed discussion of the degree of singularity the spaces $(%
{\cal N})^{-\beta },\ \beta \in [0,1)$ are useful. In the following theorem
we will characterize these spaces by means of $S_\mu$-transform.

\begin{theorem}
\sloppy The $S_\mu $-transform is a topological isomorphism from $({\cal N}%
)_\mu ^{-\beta }$, $\beta \in [0,1)$ on ${\cal E}_{\max }^{2/(1-\beta )}(%
{\cal N}_{\Ckl })$.
\end{theorem}

\fussy \TeXButton{Remark }{\remark } The proof will also complete the proof
of Theorem \ref{indlimEkmax}.

\TeXButton{Proof}{\proof}Let $\Phi \in ({\cal H}_{-p})_{-q,\mu }^{-\beta }$
with the canonical representation $\Phi =\sum_{n=0}^\infty Q_n^\mu (\Phi
^{(n)})$ be given. The $S_\mu $-transform of $\Phi $ is given by 
$$
S_\mu \Phi (\theta ) =\sum_{n=0}^\infty \langle \Phi
^{(n)},\theta ^{\otimes n}\rangle . 
$$
Hence 
$$
\lnorm  S_\mu \Phi \rnorm  _{-p,-q,-\beta }^2=\sum_{n=0}^\infty
(n!)^{1-\beta }\,2^{-nq}|\Phi ^{(n)}|_{-p}^2 
$$
is finite. We will show that there exist $l\in \N $ and $C<0$ such that%
$$
{\rm n}_{-p,-l,2/(1-\beta )}(S_\mu \Phi )\leq C\,\lnorm  S_\mu \Phi \rnorm  %
_{-p,-q,-\beta }\ . 
$$

We can estimate as follows%
\begin{eqnarray*}
|S_\mu \Phi (\theta )| & \leq & \sum_{n=0}^\infty \left| \Phi ^{(n)}\right| %
_{-p}\left| \theta \right| _p^n %
\\& \leq & \left( \sum_{n=0}^\infty (n!)^{1-\beta }2^{-nq}|\Phi %
^{(n)}|_{-p}^2\right) ^{1/2}\left( \sum_{n=0}^\infty %
\frac 1{(n!)^{1-\beta }}2^{nq}\left| \theta \right| _p^{2n}\right) ^{1/2} %
\\&=& \lnorm  S_\mu \Phi \rnorm  _{-p,-q,-\beta }\left( %
\sum_{n=0}^\infty \rho ^{n\beta }\cdot \frac 1{(n!)^{1-\beta }}2^{nq}\rho %
^{-n\beta }\left| \theta \right| _p^{2n}\cdot \right) ^{1/2}, 
\end{eqnarray*}
where we have introduced a parameter $\rho \in (0,1)$. An application of
H\"older's inequality for the conjugate indices $\frac 1\beta $ and $\frac
1{1-\beta }$ gives%
\begin{eqnarray*}
\left| S_\mu \Phi (\theta )\right| & \leq & \lnorm  S_\mu \Phi \rnorm%
_{-p,-q,-\beta }\left( \sum_{n=0}^\infty \rho ^n\right) %
^{\beta /2}\cdot \left( \sum_{n=0}^\infty \frac 1{n!}\left( 2^q\rho %
^{-\beta }|\theta |_p^2\right) ^{\frac n{1-\beta }}\right) %
^{\frac{1-\beta }2} %
\\&=& \lnorm  S_\mu \Phi \rnorm  %
_{-p,-q,-\beta }\left( 1-\rho %
\right) ^{-\beta /2}\exp \left( \tfrac{1-\beta }2 \,2^{\frac q{1-\beta }}\,%
\rho ^{-\frac \beta {1-\beta }}\, |\theta |_p^{\frac 2{1-\beta }}\right) %
\end{eqnarray*}
If $l\in \N $ is such that%
$$
2^{l-\frac q{1-\beta }}>\tfrac{1-\beta }2\rho ^{-\frac \beta {1-\beta }} 
$$
we have%
\begin{eqnarray*}
{\rm n}_{-p,-l,2/(1-\beta )}(S_\mu \Phi )%
&=& \sup _{\theta \in {\cal H}_{p,\Ckkl }}\left| S_\mu \Phi (\theta )%
\right| \, \exp \left( -2^l|\theta %
|_p^{2/(1-\beta )}\right) %
\\& \leq & \left( 1-\rho \right) ^{-\beta /2}\lnorm  S_\mu \Phi \rnorm %
_{-p,-q,-\beta }%
\end{eqnarray*}
This shows that $S_\mu $ is continuous from $({\cal N})_\mu ^{-\beta }$ to $%
{\cal E}_{\max }^{2/(1-\beta )}({\cal N}_{\Ckl }).$ Or in the language of
Theorem \ref{indlimEkmax}%
$$
\stackunder{p,q\in \N }{\rm ind\ lim}\,{\rm E}_{-p,-q}^{-\beta }\subset 
{\cal E}_{\max }^{2/(1-\beta )}({\cal N}_{\Ckl }) 
$$
topologically.\medskip\ 

The proof of the inverse direction is closely related to the proof of Lemma 
\ref{3Strichnplk}. So we will be more sketchy in the following.

\noindent Let $F\in {\cal E}_{\max }^k({\cal N}_{\Ckl }),\ k=\frac 2{1-\beta
}$. Hence there exist $p,l\in \N _0$ such that%
$$
\left| F(\theta )\right| \leq {\rm n}_{-p,-l,k}(F)\exp (2^l|\theta |_p^k)\
,\qquad \theta \in {\cal N}_{\Ckl } 
$$
From this we have completely analogous to the proof of Lemma \ref
{3Strichnplk} by Cauchy inequality and kernel theorem the representation%
$$
F(\theta )=\sum_{n=0}^\infty \langle \Phi ^{(n)},\theta ^{\otimes n}\rangle 
$$
and the bound%
$$
\left| \Phi ^{(n)}\right| _{-p^{\prime }}\leq {\rm n}_{-p,-l,k}(F)%
\;(n!)^{-1/k}\left\{ (k2^l)^{1/k}e\left\| i_{p^{\prime },p}\right\|
_{HS}\right\} ^n\ , 
$$
where $p^{\prime }>p$ is such that $i_{p^{\prime },p}:{\cal H}_{p^{\prime
}}\hookrightarrow {\cal H}_p$ is Hilbert--Schmidt. Using this we have 
\begin{eqnarray*}
\lnorm  F\rnorm  _{-p^{\prime },-q,-\beta}^2 %
&=&  \sum_{n=0}^\infty (n!)^{1-\beta }2^{-qn}\left| \Phi ^{(n)}\right| %
_{-p^{\prime }}^2 %
\\& \leq & {\rm n}_{-p,-l,k}^2(F)\sum_{n=0}^\infty (n!)^{1-\beta -2/k}2^{-qn}%
\left\{ (k2^l)^{1/k}e\left\| i_{p^{\prime },p}\right\| %
_{HS}\right\} ^{2n} %
\\& \leq & {\rm n}_{-p,-l,k}^2(F)\sum_{n=0}^\infty \rho ^n 
\end{eqnarray*}
where we have set $\rho :=2^{-q+2l/k}k^{2/k}e^2\left\| i_{p^{\prime
},p}\right\| _{HS}^2$ . If $q\in \N $ is chosen large enough such that $\rho
<1$ the sum on the right hand side is convergent and we have 
\begin{equation}
\label{F3Strich}\lnorm  F\rnorm  _{-p^{\prime },-q,-\beta }\leq {\rm n}%
_{-p,-l,2/(1-\beta )}(F)\cdot (1-\rho )^{-1/2}\ . 
\end{equation}
That means%
$$
{\cal E}_{\max }^{2/(1-\beta )}({\cal N}_{\Ckl })\subset \ \stackunder{%
p,q\in \N _0}{\rm ind\ lim}\,{\rm E}_{-p,-q}^{-\beta } 
$$
topologically.

If we set 
$$
\Phi :=\sum_{n=0}^\infty Q_n^\mu (\Phi ^{(n)}) 
$$
then $S_\mu\Phi =F$ and $\Phi \in ({\cal H}_{-p^{\prime }})_{-q}^{-\beta }$
since%
$$
\sum_{n=0}^\infty (n!)^{1-\beta }2^{-qn}|\Phi ^{(n)}|_{-p^{\prime }}^2 
$$
is finite. Hence%
$$
S_\mu:({\cal N})_\mu ^{-\beta }\rightarrow {\cal E}_{\max }^{2/(1-\beta )}(%
{\cal N}_{\Ckl }) 
$$
is one to one. The continuity of the inverse mapping follows from the norm
estimate (\ref{F3Strich}).\TeXButton{End Proof}{\endproof}
\LaTeXparent{nga4.tex}

\section{The Wick product \label{Wick}}

In Gaussian Analysis it has been shown that $({\cal N})_{\gamma _{{\cal H}%
}}^{-1}$ (and other distribution spaces) is closed under so called Wick
multiplication (see \cite{KLS94} and \cite{BeS95,Ok94,Va95} for
applications). This concept has a natural generalization to the present
setting.

\begin{definition}
\ Let $\Phi ,\Psi \in $ $({\cal N})_\mu ^{-1}$. Then we define the Wick
product $\Phi \diamond \Psi $by 
$$
S_\mu (\Phi \diamond \Psi )=S_\mu \Phi \cdot S_\mu \Psi \ . 
$$
\end{definition}

This is well defined because ${\rm Hol}_0({\cal N}_{\Ckl })$ is an algebra
and thus by the characterization Theorem \ref{CharTh} there exists an
element $\Phi \diamond \Psi \in ({\cal N})_\mu ^{-1}$ such that $S_\mu (\Phi
\diamond \Psi )=S_\mu \Phi \cdot S_\mu \Psi $.

By definition we have 
$$
Q_n^\mu (\Phi ^{(n)})\diamond Q_m^\mu (\Psi ^{(m)})=Q_{n+m}^\mu (\Phi
^{(n)}\hat \otimes \Psi ^{(m)})\text{ ,} 
$$
$\Phi ^{(n)}\in ({\cal N}_{\Ckl }^{\hat \otimes n})^{\prime }$ and $\Psi
^{(m)}\in ({\cal N}_{\Ckl }^{\hat \otimes m})^{\prime }$. So in terms of $%
\Q 
$--decompositions $\Phi =\sum_{n=0}^\infty Q_n^\mu (\Phi ^{(n)})$ and $\Psi
=\sum_{n=0}^\infty Q_n^\mu (\Psi ^{(n)})$ the Wick product is given by 
$$
\Phi \diamond \Psi =\sum_{n=0}^\infty Q_n^\mu (\Xi ^{(n)}) 
$$
where%
$$
\Xi ^{(n)}=\sum_{k=0}^n\Phi ^{(k)}\hat \otimes \Psi ^{(n-k)} 
$$
This allows for concrete norm estimates.

\begin{proposition}
The Wick product is continuous on $({\cal N})_\mu ^{-1}$. In particular the
following estimate holds for $\Phi \in ({\cal H}_{-p_1})_{-q_1,\mu }^{-1}\
,\ \Psi \in ({\cal H}_{-q_2})_{-q_2}^{-1}$ and $p=\max (p_1,p_2),\
q=q_1+q_2+1$ 
$$
\left\| \Phi \diamond \Psi \right\| _{-p,-q,\mu }=\left\| \Phi \right\|
_{-p_1,-q_1,\mu }\left\| \Psi \right\| _{-p_2,-q_2,\mu }\text{ .} 
$$
\end{proposition}

\TeXButton{Proof}{\proof}We can estimate as follows%
\begin{eqnarray*}
\left\| \Phi \diamond \Psi \right\| _{-p,-q,\mu }^2 %
&=& \sum_{n=0}^\infty 2^{-nq}\left| \Xi ^{(n)}\right| _{-p}^2 %
\\&=& \sum_{n=0}^\infty 2^{-nq}\left( \sum_{k=0}^n\left| \Phi ^{(k)}\right| %
_{-p}\left| \Psi ^{(n-k)}\right| _{-p}\right) ^2 %
\\& \leq & \sum_{n=0}^\infty 2^{-nq}\,(n+1)\sum_{k=0}^n\left| \Phi ^{(k)} %
\right| _{-p}^2\left| \Psi ^{(n-k)}\right| _{-p}^2 %
\\& \leq & \sum_{n=0}^\infty \sum_{k=0}^n2^{-nq_1}\left| \Phi ^{(n)}\right| %
_{-p}^22^{-nq_2}\left| \Psi ^{(n-k)}\right| _{-p}^2 %
\\& \leq & \left( \sum_{n=0}^\infty 2^{-nq_1}\left| \Phi ^{(k)}\right| %
_{-p_1}^2\right) \left( \sum_{n=0}^\infty 2^{-nq_2}\left| \Psi ^{(n)}\right| %
_{-p_2}^2\right) %
\\&=& \left\| \Phi \right\| _{-p_1,-q_1,\mu }^2\left\| \Psi \right\|%
_{-p_2,-q_2,\mu }^2\text{ .} 
\end{eqnarray*}
\TeXButton{End Proof}{\endproof}\bigskip\ 

Similar to the Gaussian case the special properties of the space $({\cal N}%
)_\mu ^{-1}$ allow the definition of {\it Wick analytic functions }under
very general assumptions. This has proven to be of some relevance to solve
equations e.g., of the type $\Phi \diamond X=\Psi $ for $X\in ({\cal N})_\mu
^{-1}$ . See \cite{KLS94} for the Gaussian case.

\begin{theorem}
Let $F:\C \rightarrow \C $ be analytic in a neighborhood of the point $z_0=%
\E _\mu (\Phi )\ ,\ \Phi \in ({\cal N})_\mu ^{-1}$. Then $F^{\diamond }(\Phi
)$ defined by $S_\mu (F^{\diamond }(\Phi ))=F(S_\mu \Phi )$ exists in $(%
{\cal N})^{-1}$ .
\end{theorem}

\TeXButton{Proof}{\proof}By the characterization Theorem \ref{CharTh} $S_\mu
\Phi \in {\rm Hol}_0({\cal N}_{\Ckl })$. Then $F(S_\mu \Phi )\in {\rm Hol}_0(%
{\cal N}_{\Ckl })$ since the composition of two analytic functions is also
analytic. Again by characterization Theorem we find $F^{\diamond }(\Phi )\in
({\cal N})_\mu ^{-1}.$\TeXButton{End Proof}{\endproof}\bigskip\ 

\TeXButton{Remark }{\remark } If $F(z)=\sum_{n=0}^\infty a_k(z-z_0)^n$ then
the {\it Wick series} $\sum_{n=0}^\infty a_k(\Phi -z_0)^{\diamond n}$ (where 
$\Psi ^{\diamond n}=\Psi \diamond \ldots \diamond \Psi $ n-times converges
in $({\cal N})^{-1}$ and $F^{\diamond }(\Phi )=\sum_{n=0}^\infty a_k(\Phi
-z_0)^{\diamond n}$ holds.\bigskip\ 

\example  The above mentioned equation $\Phi \diamond X=\Psi $ can be solved
if $\E _\mu (\Phi )=S_\mu \Phi (0)\neq 0$. That implies $(S_\mu \Phi
)^{-1}\in {\rm Hol}_0({\cal N}_{\Ckl })$. Thus $\Phi ^{\diamond (-1)}=S_\mu
^{-1}\left( (S_\mu \Phi )^{-1}\right) \in ({\cal N})_\mu ^{-1}$. Then $%
X=\Phi ^{\diamond (-1)}\diamond \Psi $ is the solution in $({\cal N})_\mu
^{-1}$. For more instructive examples we refer the reader to \cite{KLS94}.
\LaTeXparent{dis3.tex}

\section{Positive distributions \label{PosDist}}

In this section we will characterize the positive distributions in $({\cal N}%
)_\mu ^{-1}$. We will prove that the positive distributions can be
represented by measures in ${\cal M}_a ({\cal N}^{\prime })$. In the case of
the Gaussian Hida distribution space $({\cal S})^{\prime }$ similar
statements can be found in works of Kondratiev \cite[b]{Ko80a} and Yokoi 
\cite{Yok90,Yok93}, see also \cite{Po87} and \cite{Lee91}. In the Gaussian
setting also the positive distributions in $({\cal N})^{-1}$ have been
discussed, see \cite{KoSW94}.

Since $({\cal N})^1={\cal E}_{\min }^1({\cal N}^{\prime })$ we say that $%
\varphi \in ({\cal N})^1\,$is positive ($\varphi \geq 0$) if and only if $%
\varphi (x)\geq 0$ for all $x\in {\cal N}^{\prime }$.

\begin{definition}
An element $\Phi \in ({\cal N})_\mu ^{-1}$ is positive if for any positive $%
\varphi \in ({\cal N})^{1\text{ }}$we have $\left\langle \!\left\langle \Phi
,\varphi \right\rangle \!\right\rangle _\mu \geq 0$ . The cone of positive
elements in $({\cal N})_\mu ^{-1}$ is denoted by $({\cal N})_{\mu ,+}^{-1}$.
\end{definition}

\begin{theorem}
\label{N1Posi} Let $\Phi \in ({\cal N})_{\mu ,+}^{-1}$ . Then there exists a
unique measure $\nu \in {\cal M}_a({\cal N}^{\prime })$ such that $\forall
\varphi \in ({\cal N})^1$ 
\begin{equation}
\label{Phiny}\left\langle \!\left\langle \Phi ,\varphi \right\rangle
\!\right\rangle _\mu =\int_{{\cal N}^{\prime }}\varphi (x)\ {\rm d}\nu (x)\ .
\end{equation}
Vice versa, any (positive) measure $\nu \in {\cal M}_a({\cal N}^{\prime })$
defines a positive distribution $\Phi \in ({\cal N})_{\mu ,+}^{-1}$ by (\ref
{Phiny}).
\end{theorem}

\TeXButton{Remarks}{\remarks } \smallskip
\\1. For a given measure $\nu $ the distribution $\Phi $ may be viewed as
the generalized Radon-Nikodym derivative $\frac{{\rm d}\nu }{{\rm d}\mu }$
of $\nu $ with respect to $\mu $. In fact if $\nu $ is absolutely continuous
with respect to $\mu $ then the usual Radon-Nikodym derivative coincides
with $\Phi .$ \smallskip
\\2. Note that the cone of positive distributions generates the same set of
measures ${\cal M}_a({\cal N}^{\prime })$ for all initial measures $\mu \in $
${\cal M}_a({\cal N}^{\prime })$. \medskip\ 

\TeXButton{Proof}{\proof}To prove the first part we define moments of a
distribution $\Phi $ and give bounds on their growth. Using this we
construct a measure $\nu $ which is uniquely defined by given moments%
\TeXButton{TeX}{\renewcommand{\thefootnote}{\fnsymbol{footnote}}}\footnote{%
Since the algebra of exponential functions is not contained in $({\cal N}%
)_\mu ^1$ we cannot use Minlos' theorem to construct the measure. This was
the method used in Yokoi's work \cite{Yok90}.}. The next step is to show
that any test functional $\varphi \in {\cal (N)}^1$ is integrable with
respect to $\nu $.

Since ${\cal P}({\cal N}^{\prime })\subset {\cal (N)}^1$ we may define
moments of a positive distribution $\Phi \in ({\cal N})_\mu ^{-1}$ by 
$$
{\rm M}_n(\xi _1,...,\xi _n)=\left\langle \!\!\!\left\langle \Phi ,\
\prod\limits_{j=1}^n\left\langle \cdot ,\xi _j\right\rangle \right\rangle
\!\!\!\right\rangle _\mu \ ,\quad \ n\in {\N},\quad \xi _j\in {\cal N,\ }%
1\leq j\leq n 
$$
$$
{\rm M}_0=\left\langle \!\left\langle \Phi ,\ \1 \right\rangle
\!\right\rangle \ \text{.} 
$$
We want to get estimates on the moments. Since $\Phi \in ({\cal H}%
_{-p})_{-q,\mu }^{-1}$ for some $p,q>0$ we may estimate as follows 
$$
\bigg|\left\langle \!\!\!\left\langle \Phi ,\left\langle x^{\otimes
n},\tbigotimes_{j=1}^n\xi _j\right\rangle \right\rangle \!\!\!\right\rangle
_\mu \bigg|\leq \left\| \Phi \right\| _{-p,-q,\mu }\left\| \left\langle
x^{\otimes n},\tbigotimes_{j=1}^n\xi _j\right\rangle \right\| _{p,q,\mu }%
\text{ .} 
$$
To proceed we use the property (\ref{(P2)}) and the estimate (\ref{MnmuNorm}%
) to obtain 
\begin{eqnarray*}
\left\| \left\langle x^{\otimes n},\tbigotimes_{j=1}^n\xi _j\right\rangle %
\right\| _{p,q,\mu }^2 %
&=& \sum_{k=0}^n\binom nk^2\left\| \left\langle P_k^\mu %
\hat \otimes {\rm M}_{n-k}^\mu ,\tbigotimes_{j=1}^n\xi _j\right\rangle %
\right\| _{p,q,\mu }^2 %
\\& \leq & \sum_{k=0}^n\binom nk^2(k!)^2 \, 2^{kq}\,|{\rm M}_{n-k}^\mu %
|_{-p}^2\prod_{j=1}^n|\xi _j|_p^2 %
\\&=& \tprod_{j=1}^n|\xi _j|_p^2\sum_{k=0}^n\binom nk^2(k!)^2\left( (n-k)! %
\right) ^2K^{2(n-k)}2^{kq} %
\\& \leq & \tprod_{j=1}^n|\xi %
_j|_p^2\;(n!)^2 \, 2^{nq}\sum_{k=0}^n2^{-(n-k)q}K^{2(n-k)} %
\\& \leq & \tprod_{j=1}^n|\xi _j|_p^2\;(n!)^2 \, 2^{nq}%
\sum_{k=0}^\infty 2^{-kq}K^{2k} 
\end{eqnarray*}
which is finite for $p,q$ large enough. Here $K$ is determined by equation (%
\ref{MnmuNorm}).

\noindent Then we arrive at 
\begin{equation}
\label{mombound}\Big|{\rm M}_n(\xi _1,...\xi _n)\Big|\leq K\ C^n\
n!\prod\limits_{j=1}^n\left| \xi _j\right| _p 
\end{equation}
for some $K,C>0$.

Due to the kernel theorem \ref{KernelTh} we then have the representation%
$$
{\rm M}_n(\xi _1,...\xi _n)=\left\langle {\rm M}^{(n)},\xi _1\otimes
...\otimes \xi _n\right\rangle \text{ ,} 
$$
where ${\rm M}^{(n)}\in ({\cal N}^{\hat \otimes n})^{\prime }$. The sequence 
$\left\{ {\rm M}^{(n)},\ n\in {\N}_0\right\} $ has the following property of
positivity: for any finite sequence of smooth kernels $\left\{ g^{(n)},n\in {%
\N}\right\} $ (i.e.,\ $g^{(n)}\in {\cal N}^{\hat \otimes n}$ and $g^{(n)}=0$%
\ $\forall \;\,n\geq n_0$ for some $n_0\in {\N}$) the following inequality
is valid 
\begin{equation}
\label{posmon}\sum_{k,j}^{n_0}\left\langle {\rm M}^{(k+j)}\;,g^{(k)}\otimes 
\overline{g^{(j)}}\right\rangle \geq 0\text{ .} 
\end{equation}
This follows from the fact that the left hand side can be written as $%
\left\langle \!\left\langle \Phi ,|\varphi |^2\right\rangle \!\right\rangle $
with%
$$
\varphi (x)=\sum_{n=0}^{n_0}\left\langle x^{\otimes n},g^{(n)}\right\rangle
,\quad x \in {\cal N}^{\prime }\text{ ,} 
$$
which is a smooth polynomial. Following \cite{BS71,BeKo88} inequalities (\ref
{mombound}) and (\ref{posmon}) are sufficient to ensure the existence of a
uniquely defined measure $\nu $ on $({\cal N}^{\prime },{\cal C}_\sigma (%
{\cal N}^{\prime }))$, such that for any $\varphi \in {\cal P}({\cal N}%
^{\prime })$ we have 
$$
\left\langle \!\left\langle \Phi ,\varphi \right\rangle \!\right\rangle _\mu
=\int_{{\cal N}^{\prime }}\varphi (x)\ {\rm d}\nu (x)\text{ .} 
$$

From estimate (\ref{mombound}) we know that $\nu \in {\cal M}_a({\cal N}%
^{\prime })$. Then Lemma \ref{equiLemma} shows that there exists $%
\varepsilon >0\,,\,p\in \N $ such that $\exp (\varepsilon |x|_{-p})$ is $\nu
-$integrable. Corollary \ref{phi(z)Betrag} then implies that each $\varphi
\in ({\cal N})^1$ is $\nu $-integrable.\bigskip\ 

Conversely let $\nu \in {\cal M}_a({\cal N}^{\prime })$ be given. Then the
same argument shows that each $\varphi \in ({\cal N})^1$ is $\nu $%
-integrable and from Corollary \ref{phi(z)Betrag} we know that%
$$
\left| \int_{{\cal N}^{\prime }}\varphi (x){\rm d}\nu (x)\right| \leq
\,C\,\left\| \varphi \right\| _{p,q,\mu }\int_{{\cal N}^{\prime }}\exp
(\varepsilon |x|_{-p})\,{\rm d}\nu (x) 
$$
for some $p,q\in \N \,,\,C>0$. Thus the continuity of $\varphi \mapsto \int_{%
{\cal N}^{\prime }}\varphi \;{\rm d}\nu $ is established, showing that $\Phi 
$ defined by equation (\ref{Phiny}) is in $({\cal N})_{\mu ,+}^{-1}$.%
\TeXButton{End Proof}{\endproof}
\LaTeXparent{nga4.tex}

\section{Change of measure}

Suppose we are given two measures $\mu ,\hat \mu \in {\cal M}_a({\cal N}%
^{\prime })$ both satisfying Assumption 2. Let a distribution $\hat \Phi \in
({\cal N})_{\hat \mu }^{-1}$ be given. Since the test function space $({\cal %
N})^1$ is invariant under changes of measures in view of Theorem \ref
{N1E1min}, the continuous mapping 
$$
\varphi \mapsto \langle \!\langle \hat \Phi ,\varphi \rangle \!\rangle
_{\hat \mu }\ ,\qquad \varphi \in ({\cal N})^1 
$$
can also be represented as a distribution $\Phi \in ({\cal N})_\mu ^{-1}$.
So we have the implicit relation $\Phi \in ({\cal N})_\mu
^{-1}\leftrightarrow \hat \Phi \in ({\cal N})_{\hat \mu }^{-1}$ defined by 
$$
\langle \!\langle \hat \Phi ,\varphi \rangle \!\rangle _{\hat \mu }=\langle
\!\langle \Phi ,\varphi \rangle \!\rangle _\mu \ . 
$$
This section will provide formulae which make this relation more explicit in
terms of re-decomposition of the $\Q $-series. First we need an explicit
relation of the corresponding $\p $-systems.

\begin{lemma}
Let $\mu ,\hat \mu \in {\cal M}_a({\cal N}^{\prime })$ then 
$$
P_n^\mu (x)=\sum_{k+l+m=n}\frac{n!}{k!\,l!\,m!}P_k^{\hat \mu }(x)\hat
\otimes P_l^\mu (0)\hat \otimes {\rm M}_m^\mu \ . 
$$
\end{lemma}

\TeXButton{Proof}{\proof}Expanding each factor in the formula%
$$
e_\mu (\theta ,x)=e_{\hat \mu }(\theta ,x)l_\mu ^{-1}(\theta )l_{\hat \mu
}(\theta )\ , 
$$
we obtain%
$$
\sum_{n=0}^\infty \frac 1{n!}\langle P_n^\mu (x),\theta ^{\otimes n}\rangle
=\sum_{k,l,m=0}^\infty \frac 1{k!\,l!\,m!}\langle P_k^\mu (x)\otimes
P_l^{\hat \mu }(0)\otimes {\rm M}_m^\mu ,\theta ^{\otimes (k+l+m)}\rangle \
. 
$$
A comparison of coefficients gives the above result.\TeXButton{End Proof}
{\endproof}\bigskip\ 

An immediate consequence is the next reordering lemma.

\begin{lemma}
Let $\varphi \in ({\cal N})^1$ be given. Then $\varphi $ has representations
in $\p  ^\mu $-series as well as $\p  ^{\hat \mu }$-series:%
$$
\varphi =\sum_{n=0}^\infty \langle P_n^\mu ,\varphi ^{(n)}\rangle
=\sum_{n=0}^\infty \langle P_n^{\hat \mu },\hat \varphi ^{(n)}\rangle  
$$
where $\varphi ^{(n)},\hat \varphi ^{(n)}$ $\in {\cal N}_{\Ckl }^{\hat
\otimes n}$ for all $n\in \N _0$, and the following formula holds: 
\begin{equation}
\label{phiHut}\hat \varphi ^{(n)}=\sum_{l,m=0}^\infty \frac{(l+m+n)!}{%
l!\,m!\,n!}\left( P_l^\mu (0)\hat \otimes {\rm M}_m^{\hat \mu },\varphi
^{(l+m+n)}\right) _{{\cal H}^{\otimes (l+m)}}\ .
\end{equation}
\end{lemma}

\noindent Now we may prove the announced theorem.

\begin{theorem}
Let $\hat \Phi =\sum_{n=0}^\infty \langle Q_n^{\hat \mu },\hat \Phi
^{(n)}\rangle \in ({\cal N)}_{\hat \mu }^{-1}$. Then $\Phi
=\sum_{n=0}^\infty \langle Q_n^\mu ,\Phi ^{(n)}\rangle $ defined by 
$$
\langle \!\langle \Phi ,\varphi \rangle \!\rangle _\mu =\langle \!\langle
\hat \Phi ,\varphi \rangle \!\rangle _{\hat \mu }\ ,\qquad \varphi \in (%
{\cal N})^1 
$$
is in $({\cal N})_\mu ^{-1}$ and the following relation holds%
$$
\Phi ^{(n)}=\sum_{k+l+m=n}\frac 1{l!\,m!}\hat \Phi ^{(k)}\hat \otimes
P_l^\mu (0)\hat \otimes {\rm M}_m^{\hat \mu } 
$$
\end{theorem}

\TeXButton{Proof}{\proof}We can insert formula (\ref{phiHut}) in the formula 
$$
\sum_{n=0}^\infty n!\, \langle \Phi ^{(n)},\varphi ^{(n)}\rangle
=\sum_{n=0}^\infty n!\, \langle \hat \Phi ^{(n)},\hat \varphi ^{(n)}\rangle 
$$
and compare coefficients again.\TeXButton{End Proof}{\endproof}

\LaTeXparent{dis3.tex}

\chapter{Gaussian analysis}

The primordial object of Gaussian analysis (e.g., \cite
{BeKo88,HKPS93,Ko80a,KT80}, ) is a real separable Hilbert space ${\cal H}$.
One then considers a rigging of ${\cal H}$, ${\cal N}\subset {\cal H}\subset 
{\cal N}^{\prime }$, where ${\cal N}$ is a real nuclear space (see below and 
\cite{GV68}), densely and continuously embedded into ${\cal H}$, and ${\cal N%
}^{\prime }$ is its dual (${\cal H}$ being identified with its dual). A
typical example (which appears for instance in white noise analysis) is the
rigging ${{\cal S}({I\!\!R})}\subset L^2({I\!\!R})\subset {{\cal S}^{\prime
}({I\!\!R})}$ of $L^2({I\!\!R})$ (with Lebesgue measure) by the Schwartz
spaces of test functions and tempered distributions. Via Minlos' theorem the
canonical Gaussian measure $\mu $ on ${\cal N}^{\prime }$ is introduced by
giving its characteristic function 
$$
C(\xi )=\int_{{\cal N}^{\prime }}e^{i\left\langle \omega ,\xi \right\rangle
}\,{\rm d}\mu (\omega )=e^{-{\frac 12}\left| \xi \right| _{{\cal H}%
}^2},\quad \xi \in {\cal N}. 
$$
Of course the basic setting has already been introduced in the previous
chapter, see e.g., Example \arabic{GaussA} on page \pageref{GaussAP}. For
traditional reasons we have some changes in the notation at this point. The
Gaussian measure connected with the Hilbert space ${\cal H}$ previously
denoted by $\gamma _{{\cal H}}$ is called $\mu $ from now on. Since it is
fixed throughout the rest of the work, we will drop some subscripts $\mu $.
Since $S_\mu $-transform and convolution $C_{\mu \text{ }}$coincide both
will be denoted by $S$. The letter $C$ will be reserved for the
characteristic function of the Gaussian measure, which will be used more
frequently than its Laplace transform. The basic variable of integration, in
the previous chapter called $x\in $ ${\cal N}^{\prime }$ will now be called $%
\omega \in {\cal N}^{\prime }$.

The space $L^2({\cal N}^{\prime },\mu )\equiv L^2(\mu )$ of (equivalence
classes of) complex valued functions on ${\cal N}^{\prime }$ which are
square-integrable with respect to $\mu $ has the well-known
Wiener--It\^o--Segal chaos decomposition \cite{Ne73,Si74,Se56}, and one has
the familiar Segal isomorphism ${\cal I}$ between $L^2(\mu )$ and the
complex Fock space $\Gamma ({\cal H)}$ over the complexification ${\cal H}_{%
\Ckl }$ of ${\cal H}$. Spaces of smooth functions on ${\cal N}^{\prime }$
can be constructed by mapping appropriate subspaces of $\Gamma ({\cal H)}$
into $L^2(\mu )$ via the unitary mapping ${\cal I}^{-1}:\Gamma ({\cal %
H)\rightarrow }L^2(\mu )$, see, e.g., the construction using second
quantized operators in \cite{BeKo88,HKPS93}. So $\varphi \in L^2(\mu )$ has
a representation%
$$
\varphi =\sum_{n=0}^\infty \langle :\omega ^{\otimes n}:,\varphi
^{(n)}\rangle \ ,\qquad \varphi ^{(n)}\in {\cal H}_{\Ckl }^{\hat \otimes n} 
$$
with norm%
$$
\left\| \varphi \right\| _{L^2(\mu )}^2=\sum_{n=0}^\infty n!\left| \varphi
^{(n)}\right| ^2\qquad . 
$$

\section{The Hida spaces 
\texorpdfstring{$({\cal N})$ and $({\cal N})^{\prime}$ }{(N) and (N)\textacute }}

\subsection{Construction and properties}

\noindent In the Gaussian setting it is of course also possible to study the
triple 
$$
\left( {\cal N}\right) ^1\subset L^2(\mu )\subset \left( {\cal N}\right)
^{-1}\qquad . 
$$
This will be done in the next section. But for Gaussian measures one also
has the very important possibility to construct the space of Hida test
functionals $\left( {\cal N}\right) \supset \left( {\cal N}\right) ^1$ which
is much bigger and was historically considered first. We will only sketch
the well known construction of $\left( {\cal N}\right) $.

Consider the space of smooth polynomials ${\cal P}\left( {\cal N}^{\prime
}\right) .$ For $\varphi =\sum_{n=0}^\mu \langle :\omega ^{\otimes
n}:,\varphi ^{(n)}\rangle \ \in {\cal P}\left( {\cal N}^{\prime }\right) \
,\ \varphi ^{(n)}\in {\cal N}_{\Ckl }^{\hat \otimes n}$ we introduce the
Hilbertian norm, $p,q,\in \N _0$%
\begin{equation}
\label{pqNorm}\left\| \varphi \right\| _{p,q}^2=\sum_{n=0}^\infty
n!\,2^{nq}\left| \varphi ^{(n)}\right| _p^2\;.
\end{equation}
\medskip\ 

\noindent {\bf Note}. Of course the notation is now in some sense
misleading, since $\left\| \cdot \right\| _{p,q}$ is different from $\left\|
\cdot \right\| _{p,q,\mu }$ from the previous chapter. Despite of this the
notation (\ref{pqNorm}) will be used to stay consistent with the literature.
The norm $\left\| \cdot \right\| _{p,q,\mu }$ will be called $\left\| \cdot
\right\| _{p,q,1}$ in the Gaussian setting. \bigskip\ 

\noindent Denote the closure of ${\cal P}\left( {\cal N}^{\prime }\right) $
with respect to $\left\| \cdot \right\| _{p,q}$ by $({\cal H}_p)_q$. Finally
we set%
$$
\left( {\cal N}\right) =\stackunder{p,q\in \N }{\ {\rm pr\ lim}}({\cal H}%
_p)_q\ . 
$$
\medskip\ 

\noindent {\bf Remark.} Evidently substitution of the value 2 in equation (%
\ref{pqNorm}) by any other number strictly larger than 1 produces the same
space $\left( {\cal N}\right) $. \bigskip\ 

It is worthwhile to note that $\left( {\cal N}\right) $ is continuously
embedded in $L^2(\mu )\,$in the Gaussian case. This is due to the fact that
our $\p $-system used here coincides with the orthonormal basis of Hermite
functions.

\begin{lemma}
\ $({\cal N})$ is nuclear.
\end{lemma}

\TeXButton{Proof}{\proof}\quad Nuclearity of $\left( {\cal N}\right) $
follows essentially from that of ${\cal N}.$ For fixed $p,q$ consider the
embedding 
$$
I:\left( {\cal H}_{p^{\prime }}\right) _{q^{\prime }}\rightarrow \left( 
{\cal H}_p\right) _{q^{\prime }} 
$$
where $p^{\prime }$ is chosen such that the embedding%
$$
i_{p^{\prime },p}:\,{\cal H}_{p^{\prime }}\hookrightarrow {\cal H}_p 
$$
is Hilbert--Schmidt. Then $I$ is given by 
$$
{\cal I}I{\cal I}^{-1}=\bigoplus_ni{}_{p^{\prime },p}^{\otimes n}. 
$$
where ${\cal I}$ is the Segal isomorphism. Its Hilbert--Schmidt norm is
easily estimated by using an orthonormal basis (cf., e.g., 
\cite[Appendix A. 2]{HKPS93}) of $({\cal H}_{p^{\prime }})_{q^{\prime }}$.
The result is the bound 
$$
\left\| I\right\| _{HS}^2\le \sum_{n=0}^\infty 2^{n(q-q^{\prime })}\left\|
i_{p^{\prime },p}\right\| _{HS}^{2n} 
$$
which is finite for suitably chosen $q^{\prime }$. \TeXButton{End Proof}
{\endproof}\medskip

\begin{theorem}
The topology on $\left( {\cal N}\right) $ is uniquely determined by the
topology on ${\cal N}$.
\end{theorem}

\TeXButton{Proof}{\proof}\quad Let us assume that we are given two different
systems of Hilbertian norms $\left| \,\cdot \,\right| _p$ and $\left|
\,\cdot \,\right| _k^{\prime }$ , such that they induce the same topology on 
${\cal N}$ . For fixed $k$ and $l$ we have to estimate $\left\| \,\cdot
\,\right\| _{k,l}^{\prime }$ by $\left\| \,\cdot \,\right\| _{p,q}$ for some 
$p,q$ (and vice versa which is completely analogous). Since $\left| \,\cdot
\,\right| _k^{\prime }$ has to be continuous with respect to the projective
limit topology on ${\cal N}$, there exists $p$ and a constant $C$ such that $%
\left| f\right| _k^{\prime }\leq C\left| f\right| _p$, for all $f\in {\cal N}
$, i.e., the injection $\iota $ from ${\cal H}_p$ into the completion ${\cal %
K}_k$ of ${\cal N}$ with respect to $|\,\cdot \,|_k^{\prime }$ is a mapping
bounded by $C$. We denote by $\iota $ also its linear extension from ${\cal H%
}_{{\,}\Ckl ,p}$ into ${\cal K}_{{\,}\Ckl ,k}$. It follows from a
straightforward modification of the proof of the Proposition on p.\ 299 in 
\cite{ReSi72}, that $\iota ^{\otimes n}$ is bounded by $C^n$ from ${\cal H}_{%
{\,}\Ckl ,p}^{\otimes n}$ into ${\cal K}_{{\,}\Ckl  ,k}^{\otimes n}$. Now we
choose $q$ such that $2^{{\frac{q-l}2}}\geq C$. Then

$$
\left\| \,\cdot \,\right\| _{k,l}^{\prime 2}=\sum_{n=0}^\infty
n!\,2^{nl}\left| \,\cdot \,\right| _k^{\prime 2}\leq \sum_{n=0}^\infty
n!\,2^{nl}C^{2n}\left| \,\cdot \,\right| _p^2\leq \left\| \,\cdot \,\right\|
_{p,q}^2, 
$$
which had to be proved. \TeXButton{End Proof}{\endproof}

\medskip
\noindent From general duality theory on nuclear spaces we know that the
dual of $\left( {\cal N}\right) $ is given by%
$$
\left( {\cal N}\right) ^{\prime }=\stackunder{p,q\in \N }{\ {\rm ind\ lim}}%
\left( {\cal H}_{-p}\right) _{-q}\text{ ,} 
$$
where%
$$
\left( {\cal H}_{-p}\right) _{-q}=({\cal H}_p)_q^{\prime }\; . 
$$
We shall denote the bilinear dual pairing on $({\cal N})^{\prime }\times (%
{\cal N})$ by $\left\langle \!\left\langle \cdot ,\cdot \right\rangle
\!\right\rangle :$%
$$
\left\langle \!\left\langle \Phi ,\varphi \right\rangle \!\right\rangle
=\sum_{n=0}^\infty n!\,\langle \Phi ^{(n)},\varphi ^{(n)}\rangle , 
$$
where $\Phi \in ({\cal H}_{-p})_{-q}$ corresponds to the sequence $(\Phi
^{(n)},\,n\in \N )$ with $\Phi ^{(0)}\in \C$, and $\Phi ^{(n)}\in {\cal H}_{%
\Ckl ,-p}^{\widehat{\otimes }n},\,n\in \N $. \medskip

\noindent {\bf Remark.} Consider the particular choice ${\cal N}={{\cal S}({%
I\!\!R})}$. Then $({\cal N})^{(\prime )}$ coincide with the well-known
spaces $({\cal S})^{(\prime )}$ of white noise functionals, see, e.g., \cite
{HKPS93,PS91a}. For the norms $\Vert \varphi \Vert _p\equiv \Vert \Gamma
(A^p)\varphi \Vert _0$ introduced there, we have $\Vert \,\cdot \,\Vert
_p=\Vert \,\cdot \,\Vert _{p,0}$, and $\Vert \,\cdot \,\Vert _{p,q}\le \Vert
\,\cdot \,\Vert _{p+{\frac q2}}$. More generally, if the norms on ${\cal N}$
satisfy the additional assumption that for all $p\ge 0$ and all $\varepsilon
>0$ there exists $p^{\prime }\ge 0$ such that $|\,\cdot \,|_p\le \varepsilon
|\,\cdot \,|_{p^{\prime }}$, then the construction of Kubo and Takenaka \cite
{KT80} (and other authors) leads to the same space $({\cal N})$. The
construction presented here has the advantage of being manifestly
independent of the choice of any concrete system of Hilbertian norms
topologizing ${\cal N}$. \medskip\ 

\noindent For Wick exponentials%
$$
:\exp \langle \cdot ,\xi \rangle :=e^{\langle \cdot ,\xi \rangle -\frac
12\langle \xi ,\xi \rangle }=\sum_{n=0}^\infty {\langle :\omega }^{\otimes
n}:,\frac 1{n!}\xi ^{\otimes n}\rangle 
$$
one calculates the norms%
$$
\left\| :\exp \langle \cdot ,\xi \rangle :\right\| _{p,q}^2=e^{2^q\left| \xi
\right| _p^2}, 
$$
and hence for all $\xi \in {\cal N}$ they are in $\left( {\cal N}\right) $.
This then allows for the following

\begin{definition}
\label{Strafo}Let $\Phi \in ({\cal N})^{\prime }$. The {\bf {S}--transform
of $\Phi $} is the mapping from ${\cal N}$ into ${\C}$ given by 
$$
S\Phi (\xi ):=\langle \!\langle \Phi ,:\exp \langle \cdot ,\xi \rangle
:\rangle \!\rangle ,\quad \xi \in {\cal N}. 
$$
\end{definition}

We note that the exponential vectors $\{:\exp \langle \cdot ,\xi \rangle
:,\,\xi \in {\cal N}\}$, are a total set ${\cal E}$ in $({\cal N})$, and
hence elements of $\left( {\cal N}\right) ^{\prime }$ are characterized by
their $S$--transforms. Furthermore, it is obvious that the $S$--transform of 
$\Phi \in ({\cal N})^{\prime }$ extends to ${\cal N}_{\Ckl }$: for $\theta
\in {\cal N}_{\Ckl }$ set $S\Phi (\theta )=\langle \!\langle \Phi ,:\exp
\langle \cdot ,\theta \rangle :\rangle \!\rangle $, where $:\exp \langle
\cdot ,\theta \rangle :$ $\in ({\cal N})$ has complex kernels.

\subsection{U--functionals and the characterization theorems}

\noindent
We begin with a definition.

\begin{definition}
Let $F:{\cal N}\rightarrow {\C}$ be such that \medskip
\newline C.1 for all $\xi ,\,\eta \in {\cal N}$, the mapping $l\longmapsto
F(\eta +l\xi )$ from ${\R}$ into ${\C}$ has an entire extension to $l\in {\C}
$, \newline C.2 for some continuous quadratic form $B$ on ${\cal N}$ there
exists constants $C,\,K>0$ such that for all $f\in {\cal N},\,z\in {\C}$, 
$$
|F(z\xi )|\le C\,\exp (K\,|z|^2|B(\xi )|). 
$$
Then F is called a {\bf U--functional}.
\end{definition}

\noindent {\bf Remark.}\quad Condition C.2 is actually equivalent to the
more conventional \medskip\newline {C.2$\,^{\prime }$} there exists
constants $C,\,K>0$ and $p\in {I\!\!N}_0$, so that for all $\xi \in {\cal N}%
,\,z\in \C $,

\begin{equation}
\label{Ubound}|F(z\xi )|\le C\,\exp (K\,|z|^2|\xi |_p^2). 
\end{equation}
\bigskip\ 

To proceed we need a result which is related to the celebrated ``cross
theorem" of Bernstein. For a review of such results we refer the interested
reader also to \cite{AR73}. The following is a special case of a result by
Siciak: if we make use of the fact that any segment of the real line in the
complex plane has strictly positive transfinite diameter, then Corollary 7.3
in \cite{Si69} implies

\begin{proposition}
\label{Siciak}Let $n\in {\N},\,n\ge 2$, and $f$ be a complex valued function
on ${\R}^n$. Assume that for all $k=1,2,\ldots ,n$, and $(x_1,\ldots
,x_{k-1},x_{k+1},\ldots ,x_n)\in {\R}^{n-1}$, the mapping 
$$
x_k\longmapsto f(x_1,\ldots ,x_{k-1},x_k,x_{k+1},\ldots ,x_n), 
$$
from ${\R}$ into ${\C}$ has an entire extension. Then $f$ has an entire
extension to ${\C}^n$.
\end{proposition}

\begin{lemma}
\label{Ufunctional}Every $U$--functional $F$ has a unique extension to an
entire function on ${\cal N}_{\Ckl }$. Moreover, if the bound on $F$ holds
in the form (\ref{Ubound}) then for all $\rho \in (0,1)$, 
$$
|F(\theta )|\le C^{\prime }\exp (K^{\prime }\,|\theta |_p^2),\quad \theta
\in {\cal N}_{\Ckl }, 
$$
with $C^{\prime }=C(1-\rho )^{-{\frac 12}},\,K^{\prime }=2\rho ^{-1}e^2K$.
\end{lemma}

\TeXButton{Proof}{\proof}\quad First we show that a $U$--functional $F$ has
a G--entire extension. The extension of $F$ (denoted by the same symbol) is
given by $F(\theta )=F(\xi _0+z\xi _1),\,\theta =\xi _0+z\xi _1\in {\cal N}_{%
\Ckl },\,\xi _0,\xi _1\in {\cal N},\,z\in \C $. Let $\theta \in {\cal N}_{%
\Ckl }$ be of the form $\theta =\xi _2+i\xi _3,\xi _2,\xi _3\in {\cal N}$.
Consider the mapping 
$$
(\lambda _1,\lambda _2,\lambda _3)\,\longmapsto \,F(\xi _0+\lambda _1\xi
_1+\lambda _2\xi _2+\lambda _3\xi _3), 
$$
from ${I\!\!R}^3$ into $\C $. Condition C.1 and Proposition \ref{Siciak}
imply that this function has an entire extension to $\C ^3$. In particular, $%
F$ is G--entire on ${\cal N}_{\Ckl }$. Let $\theta \in {\cal N}_{\Ckl }$,
and consider the Taylor expansion of $F(\theta )$ at the origin : 
\begin{equation}
\label{FTaylor}F(\theta )=\sum_{n=0}^\infty {\frac 1{n!}}\,\widehat{{\rm d}%
^nF(0)}(\theta ).
\end{equation}
For all $\xi \in {\cal N},\,n\in \N ,\,R>0$, we obtain from C.2$^{\prime }$
and Cauchy's inequality the estimate 
$$
|\widehat{{\rm d}^nF(0)}(\xi )|\le C\,n!\,R^{-n}e^{R^2K|\xi |_p^2}. 
$$
We choose $R=({\frac n{2K}})^{\frac 12}$, and get for $\xi \in {\cal N}$
with $|\xi |_p=1$ the inequality 
$$
|\widehat{{\rm d}^nF(0)}(\xi )|\le C\,n!\,\Big({\frac{2eK}n}\Big)^{n/2}. 
$$
A standard polarization argument (see, e.g., \cite[sec.3]{Na69}) and
homogeneity of $\widehat{{\rm d}^nF(0)}$ yield the following bound for the $n
$--linear form ${\rm d}^nF(0)$: 
\begin{equation}
\label{5}|{\rm d}^nF(0)(\xi _1,\ldots ,\xi _n)|\le C(n!\,(2e^2K)^n)^{{\frac
12}}\prod_{k=1}^n|\xi _k|_p,
\end{equation}
where $\xi _1,\ldots ,\xi _n\in {\cal N}$ (and we used ${\frac{n^n}{n!}}\le
e^n$). Since ${\rm d}^nF(0)$ is $n$--linear on ${\cal N}_{\Ckl }$, the last
inequality gives the estimate 
\begin{equation}
\label{6}|{\rm d}^nF(0)(\theta _1,\ldots ,\theta _n)|\le
C\,(n!\,(4e^2K)^n)^{\frac 12}\prod_{k=1}^n|\theta _k|_p,
\end{equation}
for $\theta _1,\ldots ,\theta _n\in {\cal N}_{\Ckl }$. In particular, the
Taylor coefficients in (\ref{FTaylor}) have absolute value bounded by 
$$
C\,\Big({\frac{(4e^2K|\theta |_p^2)^n}{n!}}\Big)^{\frac 12}, 
$$
and we get (by Schwarz' inequality) the following estimate for all $\rho \in
(0,1)$, 
$$
|F(\theta )|\le C(1-\rho )^{-{\frac 12}}e^{2\rho ^{-1}e^2K|\theta
|_p^2},\quad \theta \in {\cal N}_{\Ckl }. 
$$
Hence $F$ is locally bounded on ${\cal N}_{\Ckl }$, and therefore
Proposition \ref{GHolLocB} implies that $F$ is entire.\hfill\medskip\ 
\TeXButton{End Proof}{\endproof}

{Now we are ready to prove the following generalization of the main result
in \cite{PS91a} which characterizes the space $({\cal N})^{\prime }$ in
terms of its $S$--transform. }

\begin{theorem}
\label{CharacTh}A mapping $F:{\cal N}\rightarrow {\C}$ is the $S$--transform
of an element in $({\cal N})^{\prime }$ if and only if it is a U--functional.
\end{theorem}

\TeXButton{Proof}{\proof}\quad Let $\Phi \in ({\cal N})^{\prime }$. Then $%
\Phi \in ({\cal H}_{-p})_{-q}$ for some $p,\,q\in \N _0$. As we have
remarked after Definition \ref{Strafo}, the $S$--transform of $\Phi $
extends to ${\cal N}_{\Ckl }$, and therefore it makes sense to consider the
mapping $\theta \mapsto S\Phi (\theta )$ from ${\cal N}_{\Ckl }$ into ${\C}$%
. We shall show that this mapping is entire. We have 
$$
S\Phi (\theta )=\sum_{n=0}^\infty \langle \Phi ^{(n)},\theta ^{\otimes
n}\rangle ,\quad \theta \in {\cal N}_{\Ckl }. 
$$
We estimate as follows: 
\begin{eqnarray*}
|S\Phi (\theta )| & \le & \sum_{n=0}^\infty %
|\Phi ^{(n)}|_{-p}|\theta |_p^n %
\\ & \le & \Big(\sum_{n=0}^\infty n! %
\,2^{-qn}|\Phi ^{(n)}|_{-q}^2\Big)^{1/2} %
\Big( \sum_{n=0}^\infty {\frac 1{n!}}2^{qn}|\theta |_p^{2n}\Big)^{1/2} %
\\&=& \Vert \Phi \Vert _{-p,-q}\,e^{2^{q-1}|\theta |_p^2}. 
\end{eqnarray*}
The last estimation shows that the power series for $S\Phi $ on ${\cal N}_{%
\Ckl }$ converges uniformly on every bounded neighborhood of zero in ${\cal N%
}_{\Ckl }$, and therefore it defines an entire function on this space \cite
{Di81}. In particular, C.1 holds for $S\Phi $. Moreover, the choice $\theta
=z\xi ,\,z\in \C ,\,\xi \in {\cal N}$, shows that also C.2$^{\prime }$ is
fulfilled. Hence $S\Phi $ is a $U$--functional. \medskip 

Conversely let $F$ be a $U$--functional. We may assume the bound in the form
(\ref{Ubound}). Consider the $n$--linear form ${\rm d}^nF(0)$ on ${\cal N}_{%
\Ckl }$ constructed in the proof of Lemma \ref{Ufunctional}. The estimate (%
\ref{6}) shows that ${\rm d}^nF(0)$ is separately continuous on ${\cal N}_{%
\Ckl }$ in its $n$ variables. Hence by the Kernel Theorem \ref{KernelTh}
there exists $\Phi ^{(n)}\in ({\cal N}_{\Ckl }^{\prime })^{{\widehat{\otimes 
}}n}$ so that 
$$
\langle \Phi ^{(n)},\theta _1{\widehat{\otimes }}\cdots {\widehat{\otimes }}%
\theta _n\rangle ={\frac 1{n!}}\,{\rm d}^nF(0)(\theta _1,\ldots ,\theta
_n),\quad \theta _1,\ldots ,\theta _n\in {\cal N}_{\Ckl } 
$$
and from (\ref{5}) we have the norm estimate 
\begin{equation}
\label{7}\left| \Phi ^{(n)}\right| _{-p^{\prime }}\leq C(n!)^{\frac
12}\left( 2e^2K\left\| i_{p^{\prime },p}\right\| _{HS}^2\right) ^{n/2} 
\end{equation}
if $p^{\prime }>p$ is such that the embedding $i_{p^{\prime },p}$ $:{\cal H}%
_{p^{\prime }}\hookrightarrow {\cal H}_p$ is Hilbert Schmidt. For $\Phi $
given by the sequence $(\Phi ^{(n)},\,n\in \N _0)$ ($\Phi ^{(0)}\equiv F(0)$%
) we have%
\begin{eqnarray*}
\|\Phi\|^2_{-p \prime,-q}&=&\sum_{n=0}^\infty %
n!\,2^{-nq}\left| \Phi ^{(n)}\right| _{-p^{\prime }}^2 %
\\&\le & C^2\sum_{n=0}^\infty(2^{1-q}e^2K\,\|i_{p^{\prime },p}\|_{\rm HS}^2)^n %
        \\&=& C^2(1-2^{1-q}e^2K\,\|i_{p^{\prime },p}\|_{\rm HS}^2)^{-1}%
        \\&<&+\infty,
\end{eqnarray*}
if we choose $q$ large enough so that $2^{1-q}e^2K\,\Vert i_{p^{\prime
},p}\Vert _{HS}^2<1$. In particular, $\Phi \in ({\cal N})^{\prime }$, and
for $f\in {\cal N}$ we have by (\ref{FTaylor}),%
\begin{eqnarray*}
S\Phi(\xi)&=&\sum_{n=0}^\infty\langle\Phi^{(n)},\xi^{\otimes n}\rangle %
\\&=&\sum_{n=0}^\infty{1\over n!}\,\widehat{{\rm d}^nF(0)}(\xi) %
\\&=&F(\xi).
\end{eqnarray*}

Uniqueness of $\Phi =S^{-1}F$ follows from the fact that the exponential
vectors are total in $({\cal N})$.\TeXButton{End Proof}{\endproof}

\medskip
\noindent As a by-product of the above proof we obtain the following
localization result for generalized functionals.

\begin{corollary}
Given a U--functional $F$ satisfying C.2$\,^{\prime }$. Let $p\prime >p$ be
such that the embedding $i_{p^{\prime },p}:\,{\cal H}_{p\prime }\rightarrow 
{\cal H}_p$ is Hilbert--Schmidt, and $q\in {\N}_0$ so that $\rho
:=2^{1-q}e^2K\,\Vert i_{p^{\prime },p}\Vert _{HS}^2<1$. Then $\Phi
:=S^{-1}F\in ({\cal H}_{-p\prime })_{-q}$, and 
\begin{equation}
\label{8}\left\| \Phi \right\| _{-p\prime ,-q}\leq C(1-\rho )^{-1/2}.
\end{equation}
\end{corollary}

For analogous results in white noise analysis see, e.g., \cite
{KoS92,Ob91,Yan90}. \bigskip\ 

We close this section by the corresponding characterization theorem for $%
\left( {\cal N}\right) $. This result is independently due to \cite
{Ko80a,KPS91,Lee89}, and has been generalized and modified in various ways,
e.g., \cite{Ob91,Yan90,Zh92}.

\begin{theorem}
\label{Charac(N)}A mapping $F:{\cal N\rightarrow }{\C}$ is the $S$%
--transform of an element in $({\cal N)}$ if and only if it admits C.1 and
the following condition \medskip
\newline C.3 there exists a system of norms $(\left| \,\cdot \,\right|
_{-p},\,p\in {\N}_0)$, which yields the inductive limit topology on ${\cal N}%
^{\prime }$, and such that for all $p\geq 0$ and $\varepsilon >0$ there
exists $C_{p,\varepsilon }>0$ so that 
\begin{equation}
\label{9}\left| F(z\xi )\right| \leq C_{p,\varepsilon }\exp \left(
\varepsilon |z|^2\left| \xi \right| _{-p}^2\right) ,\quad \xi \in {\cal N}%
,\,z\in {\C}.
\end{equation}
\end{theorem}

If for $F$ conditions C.1 and C.3 are satisfied we say that $F$ is of order
2 and minimal type, i.e., $F\in {\cal E}_{\min }^2({\cal N}^{\prime })$.

\TeXButton{Proof}{\proof}If $\varphi \in ({\cal N)}$ then condition C.1 is
satisfied as a consequence of Theorem \ref{CharacTh}. For any $p,q\geq 0$ we
estimate as follows%
\begin{eqnarray*}
|S\varphi(zf)|&=&|\sum_{n=0}^\infty\langle\varphi^{(n)}, %
(z\xi)^{\otimes n}\rangle| %
\\&\le & \sum_{n=0}^\infty |z|^n|\varphi^{(n)}|_p\,|\xi|_{-p}^n %
\\&\le & (\sum_{n=0}^\infty n!\,2^{nq}\,|\varphi^{(n)}|_p^2)^{1/2} %
(\sum_{n=0}^\infty {1\over n!}\,(2^{-q}|z|^2|\xi|^2_{-p})^n)^{1/2} %
\\&= & \|\varphi\|_{p,q}\,\exp(2^{1-q}|z|^2|\xi|_{-p}^2).               
\end{eqnarray*}
Hence condition C.3, too, is necessary. \medskip

Conversely, let $F$ be a $U$--functional of order 2 and minimal type. From $%
F $, construct a sequence $\varphi =(\varphi ^{(n)},\,n\in {I\!\!N}_0)$ of
continuous linear forms $\varphi ^{(n)}$ on ${\cal N}^{{\widehat{\otimes }}%
n} $ as in the proof of Lemma \ref{Ufunctional}. We have to show that $%
\varphi $ belongs to $({\cal H}_r)_q$ for all $r,q\in {I\!\!N}_0$. Let $%
r,q\in {I\!\!N}_0$ be given. Choose $p>r$ such that the injection $i_{p,r}:\,%
{\cal H}_p\rightarrow {\cal H}_r$ is Hilbert--Schmidt. Then so is the
injection $i_{p,r}^{*}:\,{\cal H}_{-r}\rightarrow {\cal H}_{-p}$. $%
\varepsilon >0$ in (\ref{9}) is chosen so that $\rho :=\varepsilon
2^{1+q}e^2\,\Vert i_{p,r}^{*}\Vert _{HS}^2<1$. Then the analogue of (\ref{7}%
) reads%
\begin{eqnarray*}
\|\varphi^{(n)}\|_{r,q}^2 %
        &\le & C_{p,\varepsilon}^2 (2^{q+1}e^2\varepsilon\,\| i_{p,r}^*\|^2_                 {\rm HS})^n %
        \\&= & C^2_{p,\varepsilon}\rho^n,
\end{eqnarray*}
and we get%
\begin{eqnarray*}
\|\varphi\|_{r,q}&=& (\sum_{n=0}^\infty \|\varphi^{(n)}\|_{r,q}^2)^{{1/2}} %
        \\&\le & C_{p,\varepsilon}(1-\rho)^{-{1/2}}.
\end{eqnarray*}
Thus $\varphi \in ({\cal N})$, and the proof is complete. 
\TeXButton{End Proof}{\endproof}\bigskip\ 

Within the framework established here one can treat the following and
numerous other examples in a unified way. \medskip

\example
We choose the triplet%
$$
{\cal S}(\R ^n)\subset L^2(\R ^n)\subset {\cal S}^{\prime }(\R ^n), 
$$
and equip ${\cal S}^{\prime }(\R ^n)$ with the Gaussian measure with
characteristic functional 
$$
C(\xi )=e^{-{\frac 12}\int \xi ^2(t)\,{\rm d}^nt},\quad \xi \in {\cal S}%
\left( {I\!\!R}^n\right) . 
$$
Then the framework allows to discuss functionals of white noise with $n$%
--dimensional time parameter \cite{SW93}.\bigskip\ 

\example \newcounter{SdTriple} \setcounter{SdTriple}{\value{example}} \label
{SdTripleP}{\em (Vector valued white noise) }\smallskip\ 

The starting point is the real separable Hilbert space $L_d^2:=L^2(\R%
)\otimes \R^d,\ \ d\in \N $ which is isomorphic to a direct sum of $d$
identical copies of $L^2(\R).$ In this space we choose a densely imbedded
nuclear space. Here we fix this space to be ${\cal S}_d:={\cal S}(\R)\otimes 
\R^d$. A typical element $f\in {\cal S}_d$ is a $d$-dimensional vector where
each component $f_j\quad 1\leq j\leq d$ is a Schwartz test function. The
topology on ${\cal S}_d$ may be represented by a system of Hilbertian norms%
$$
\left| \vec f\right| _p^2=\sum_{j=1}^d\left| f_j\right| _{p\qquad }^2,\quad
p\geq 0,\ \vec f\in {\cal S}_d 
$$
where $\left| \cdot \right| _p$ on the r.h.s. is an increasing system of
Hilbertian norms topologizing ${\cal S}(\R)$. For notational simplicity we
identify $\left| \cdot \right| _0$ with the norm on $L_d^2$. Together with
the dual space ${\cal S}_d^{\prime }\equiv {\cal S}^{\prime }(\R)\otimes \R%
^d $ of ${\cal S}_d$ we obtain the basic nuclear triple 
$$
{\cal S}_d\subset L_d^2\subset {\cal S}_d^{\prime }\;. 
$$
On ${\cal S}_d^{\prime }$ the canonical Gaussian measure is introduced by
the characteristic function%
$$
C(\vec \xi )=e^{-{\frac 12}\langle \vec \xi ,\vec \xi \rangle },\quad \vec
\xi \in {\cal S}_d. 
$$
If we introduce the vector valued random variable%
$$
\vec B(t,\vec \omega ):=\langle \vec \omega ,\1 _{[0,t)}\rangle =\left(
\langle \omega _j,\1 _{[0,t)}\rangle ,\ j=1..d\right) ,\quad \vec \omega \in 
{\cal S}_d^{\prime } 
$$
a representation of an $d$--dimensional Brownian motion is obtained. In this
setting the above theorem gives the characterization of the space of Hida
distributions of the noise of an $d$--dimensional Brownian motion \cite{SW93}%
. \bigskip\ 

\example \newcounter{JzDef} \setcounter{JzDef}{\value{example}} \label
{JzDefP} For later use we are interested in the formal expression%
$$
\Phi =\exp \left( \frac 12(1-z^{-2})\langle \omega ,\omega \rangle \right) \
,\qquad z\in \C /\{0\}\ . 
$$
Using finite dimensional approximations to calculate its $S$--transform, we
see that the sequence factorizes in a convergent sequence of U-functionals
and a divergent pre-factor. So instead of constructing the ill defined
expression $\Phi $, we consider its multiplicative renormalization (see \cite
{HKPS93} for more details) ${\rm J}_z=\Phi /\E (\Phi )$ . So the divergent
pre-factor cancels in each step of approximation. For ${\rm J}_z$ we also
use the suggestive notation of {\it normalized exponential}%
$$
{\rm J}_z={\rm Nexp}\left( \frac 12(1-z^{-2})\langle \omega ,\omega \rangle
\right) 
$$
The resulting $S$--transform is given by 
$$
S{\rm J}_z(\theta )={\rm exp}\left( -\frac 12(1-z^2)\langle \theta ,\theta
\rangle \right) \ ,\qquad \theta \in {\cal H}_{\Ckl }\ . 
$$
The right hand side is obviously a U-functional and thus by characterization 
${\rm J}_z\in ({\cal N})^{\prime }$.

Let us now choose $z=\lambda \in \R _{+}$ and consider the $T$-transform:%
$$
T{\rm J}_\lambda (\theta )=C(\theta )\cdot S{\rm J}_\lambda (i\theta )=\exp
\left( -\frac{\lambda ^2}2\langle \theta ,\theta \rangle \right) =C_{\lambda
^2}(\theta ) 
$$
which is the characteristic function of the Gaussian measure $\mu _{\lambda
^2}$ with variance $\lambda ^2$. This implies 
$$
{\rm J}_\lambda =\frac{{\rm d}\mu _{\lambda ^2}}{{\rm d}\mu } 
$$
where the right hand side is the generalized Radon Nikodym derivative (see
Example \arabic{RadonNy} for this concept). The fact that ${\rm J}_\lambda
\notin L^2(\mu )$ for $\lambda \neq 1$ is in agreement with the fact that $%
\mu _{\lambda ^2}$ and $\mu $ are singular measures if $\lambda \neq 1$.%
\bigskip\ 

\example {\em (A simple second quantized operator)}\smallskip\ 

Let $z\in \C $ and $\Phi \in ({\cal N})^{\prime }$. Then $S\Phi $ has an
entire analytic extension and we may consider the function 
$$
\theta \mapsto S\Phi (z\theta )\ ,\qquad \theta \in {\cal N}_{\Ckl }^{\prime
}\text{ .} 
$$
This function is also an element of ${\cal E}_{\max }^2({\cal N}_{\Ckl %
}^{\prime })$. Thus we may define $\Gamma _z\Phi $ by 
$$
S\left( \Gamma _z\Phi \right) (\theta )=S\Phi (z\theta )\ . 
$$
Moreover $\Gamma _z$ is continuous from $({\cal N})^{\prime }$ into $({\cal N%
})^{\prime }$. $\Gamma _z$ is an extension of $\Gamma (z\1 )$ where $\Gamma $
is the usual second quantization, see e.g., \cite{Si74}.\bigskip\ 

\example {\em (Wick product)}\smallskip\ 

The characterization theorems give simple arguments why the spaces $({\cal N}%
)^{\prime }$ and $({\cal N})$ are closed under the so called Wick product
(already discussed in the non-Gaussian setting). Besides the defining
equation%
$$
S(\Phi \diamond \Psi )=S\Phi \cdot S\Psi \ ,\qquad \Phi ,\Psi \in ({\cal N}%
)^{(\prime )} 
$$
we only need to mention that ${\cal E}_{\max }^2({\cal N}_{\Ckl })$ and $%
{\cal E}_{\min }^2({\cal N}_{\Ckl }^{\prime })$ are both algebras under
pointwise multiplication.

\subsection{Corollaries \label{GaussCoro}}

\noindent
One useful application of Theorem \ref{CharacTh} is the discussion of
convergence of a sequence of generalized functionals. A first version of
this theorem is worked out in \cite{PS91a}. Here we use our more general
setting to state

\begin{theorem}
\label{conv}Let $(F_n,\,n\in {\N})$ denote a sequence of $U$--functionals
such that \medskip
\newline 1. $(F_n(\xi ),\,n\in {\N})$ is a Cauchy sequence for all $\xi \in 
{\cal N}$, \newline 2. there exists a continuous norm $\left| \,\cdot
\,\right| $ on ${\cal N}$ and $C,\,K>0$ such that $\left| F_n(z\xi )\right|
\leq Ce^{K\left| z|^2|\xi \right| ^2}$ for all $\xi \in {\cal N},\,z\in {\C}$%
, and for almost all $n\in {\N}$. \medskip \noindent 
Then $(S^{-1}F_n,\,n\in {\N})$ converges strongly in $({\cal N})^{\prime }$.
\end{theorem}

\TeXButton{Proof}{\proof}\quad The assumptions and inequality (\ref{8})
imply that there exist $p,q\geq 0$ and $\rho \in (0,1)$ such that for all $%
n\in {I\!\!N}$, 
$$
\left\| \Phi _n\right\| _{-p,-q}\leq C(1-\rho )^{-{\frac 12}} 
$$
where $\Phi _n=S^{-1}F_n$. Since ${\cal E}$ is total in $({\cal H}%
_{-p})_{-q} $, assumption 1 implies that $(\langle \!\langle \Phi _n,\varphi
\rangle \!\rangle $,$\,n\in \N )$ is a Cauchy sequence for all $\varphi \in (%
{\cal N})$. Since $\left( {\cal N}\right) ^{\prime }$ is the dual of the
countable Hilbert space $({\cal N})$, which is in particular Fr\'echet, it
follows from the Banach--Steinhaus theorem that $({\cal N})^{\prime }$ is
weakly sequentially complete. Thus there exists $\Phi \in \left( {\cal N}%
\right) ^{\prime }\,$ such that $\Phi $ is the weak limit of $(\Phi
_n,\,n\in \N )$. The proof is concluded by the remark that weak and strong
convergence of sequences coincide in the duals of nuclear spaces (e.g., \cite
{GV68}). \TeXButton{End Proof}{\endproof}

\medskip
\noindent As a second application we consider a theorem which concerns the
integration of a family of generalized functionals.

\begin{theorem}
\label{Bochner}Let $\left( \Lambda ,{\cal A},\nu \right) $ be a measure
space, and $\lambda \mapsto \Phi _\lambda $ a mapping from $\Lambda $ to $%
\left( {\cal N}\right) ^{\prime }$. We assume that the $S$--transform $%
F_\lambda =S\Phi _\lambda $ satisfies the following conditions: \medskip
\\1. for every $\xi \in {\cal N}$ the mapping $\lambda \mapsto F_\lambda
\left( \xi \right) $ is measurable,\smallskip\ \\2. there exists a
continuous norm $|\,\cdot \,|$ on ${\cal N}$ so that for all $l\in \Lambda
,\,F_\lambda $ satisfies the bound $\left| F_\lambda (z\xi )\right| \leq
C_\lambda e^{K_\lambda |z|^2|\xi |^2}$ , and such that $\lambda \mapsto
K_\lambda $ is bounded $\nu $--a.e., and $\lambda \mapsto C_\lambda $ is
integrable with respect to $\nu $. \medskip \\ \noindent
Then there are $q,p\geq 0$ such that $\Phi _{\cdot }$ is Bochner integrable
on $\left( {\cal H}_{-p}\right) _{-q}$. Thus in particular, 
$$
\int_\Lambda \Phi _\lambda \,{\rm d}\nu (\lambda )\in \left( {\cal N}\right)
^{\prime }, 
$$
and%
$$
S\left( \int_\Lambda \Phi _\lambda \,{\rm d}\nu (\lambda )\right) (\xi
)=\int_\Lambda S\Phi _\lambda (\xi )\,{\rm d}\nu (\lambda ),\quad \xi \in 
{\cal N}. 
$$
\end{theorem}

\TeXButton{Proof}{\proof}\quad In inequality (\ref{Ubound}) for $F_l(z\xi )$
we can replace $K_l$ by its bound. With this modified estimate and Corollary
12 we can find $p,q\geq 0$ and $\rho \in (0,1)$ such that for all $\lambda
\in \Lambda $, 
\begin{equation}
\label{10}\left\| \Phi _\lambda \right\| _{-p,-q}\leq C_\lambda (1-\rho )^{-{%
\frac 12}}. 
\end{equation}
Since the right hand side of (\ref{10}) is integrable with respect to $\nu $%
, we only need to show the weak measurability of $\lambda \mapsto \Phi
_\lambda $ (see \cite{Yo80}). But this is obvious because $\lambda \mapsto
\left\langle \!\left\langle \Phi _\lambda ,\varphi \right\rangle
\!\right\rangle $ is measurable for all $\varphi \in {\cal E}$ which is
total in $({\cal H}_p)_q.$ \TeXButton{End Proof}{\endproof}\bigskip\ 

\example Let us look at Donsker's delta function (see section \ref{DeltaSec}
for the definition) 
\begin{equation}
\label{deltaInt}\delta (\langle \omega ,\eta \rangle -a)=\frac 1{2\pi }\int_{%
{\R}}e^{i\lambda (\langle \omega ,\eta \rangle -a)}{\rm d}\lambda \ ,\qquad
\eta \in {\cal N},\ a\in \R  
\end{equation}
in the sense of Bochner integration (see \cite{HKPS93} and compare Theorem 
\ref{seqDelta}).\bigskip

\noindent {\bf Remark.}

\noindent For later use we have to define pointwise products of a Hida
distribution $\Phi $ with a Donsker delta function%
$$
\delta \left( \langle \omega ,\eta \rangle -a\right) \ ,\qquad \eta \in L^2(%
\R )\ ,\ a\in \R \ . 
$$
If $T\Phi $ has an extension to $L_{\Ckl }^2(\R )$ and the mapping $\lambda
\longmapsto T\Phi (\theta +\lambda \eta )$, $\theta \in {\cal N}_{\Ckl }$ is
integrable on ${\R}$ the following formula may be used to define the product 
$\Phi \cdot \delta $%
\begin{equation}
\label{phidelta}T\left( \Phi \cdot \delta (\langle \omega ,\eta \rangle
-a)\right) (\theta )=\frac 1{2\pi }\int_{{\R}}e^{-i\lambda a}\ T\Phi \left(
\theta +\lambda \eta \right) \ {\rm d}\lambda \text{,}
\end{equation}
in case the right hand integral is indeed a {\it U}-functional. 

This definition extends the usual definition of pointwise multiplication
where one factor is a test function. This is easily seen by use of (\ref
{deltaInt}) in the following calculation, $\varphi \in ({\cal N})$%
\begin{eqnarray*}
T\left( \varphi \cdot \delta (\langle \omega ,\eta \rangle -a)\right) %
(\theta )&=&\left\langle \!\left\langle \delta (\langle \omega ,\eta \rangle%
-a),\;\varphi \cdot e^{i\left\langle \omega ,\theta \right\rangle %
}\right\rangle \!\right\rangle  %
\\ &=& %
\frac 1{2\pi }\int_{{\R}}e^{-i\lambda a}\left\langle \!\left\langle %
e^{i\lambda \left\langle \omega ,\eta \right\rangle },\;\varphi \cdot %
e^{i\left\langle \omega ,\theta \right\rangle }\right\rangle \!\right\rangle %
\ {\rm d}\lambda  %
\\ &=& %
\frac 1{2\pi }\int_{{\R}}e^{-i\lambda a}\ T\varphi \left( \theta +\lambda %
\eta \right) \ {\rm d}\lambda \ . %
\end{eqnarray*}
\LaTeXparent{dis3.tex}

\section{The nuclear triple 
\texorpdfstring{$({\cal N})^1\subset L^2(\mu )\subset ({\cal N})^{-1}$}{of 
Kondratiew}}

\subsection{Construction}

Consider the space ${\cal P}({\cal N^{\prime }})$ of continuous polynomials
on ${\cal N^{\prime }}$, i.e.,\ any $\varphi \in {\cal P}({\cal N^{\prime }})
$ has the form $\varphi (\omega )=\sum_{n=0}^N\left\langle \omega ^{\otimes
n},\tilde \varphi ^{(n)}\right\rangle $ , $\omega \in {\cal N}^{\prime }$ , $%
N\in {\N}$ for kernels $\tilde \varphi ^{(n)}\in {\cal N}^{\hat \otimes n}$.
It is well-known that any $\varphi \in {\cal P}({\cal N^{\prime }})$ can be
written as a Wick polynomial i.e.,\ $\varphi (\omega
)=\sum_{n=0}^N\left\langle :\omega ^{\otimes n}:,\varphi ^{(n)}\right\rangle 
$, $\varphi ^{(n)}\in {\cal N}^{\hat \otimes n}$, $N\in {\N}$ (see e.g.,\
equations (\ref{(P1)}) and (\ref{(P2)})). To construct test functions we
define for $p,q\in {\N}$, $\beta \in \left[ 0,1\right] $ the following
Hilbertian norm on ${\cal P}({\cal N^{\prime }})$%
\begin{equation}
\label{norms}\left\| \varphi \right\| _{p,q,\beta }^2=\sum_{n=0}^\infty
(n!)^{(1+\beta )}2^{nq}\left| \varphi ^{(n)}\right| _p^2\quad ,\quad \varphi
\in {\cal P}({\cal N^{\prime }})\text{ .} 
\end{equation}
Then we define $({\cal H}_p)_q^\beta $ to be the completion of ${\cal P}(%
{\cal N^{\prime }})$ with respect to $\left\| \cdot \right\| _{p,q,\beta }$.
Or equivalently 
$$
({\cal H}_p)_q^\beta =\left\{ \varphi \in L^2(\mu )\biggm|\left\| \varphi
\right\| _{p,q,\beta }<\infty \right\} \text{ .} 
$$
Finally, the space of test functions $\left( {\cal N}\right) ^\beta $ is
defined to be the projective limit of the spaces $({\cal H}_p)_q^\beta $:%
$$
\left( {\cal N}\right) ^\beta =\stackunder{p,q\in \N }{\ {\rm pr\ lim}}\ (%
{\cal H}_p)_q^\beta \text{ . } 
$$
For $0\leq \beta <1$ the corresponding spaces have been studied in \cite
{KoS92} and in the special case of Gaussian product measures all the spaces
for $0\leq \beta \leq 1$ were introduced in \cite{Ko78}. For $\beta =0$ and $%
{\cal N}={\cal S}({\R})$ the well-known space $({\cal S})=\left( {\cal S}%
\right) ^0$ of Hida test functions is obtained (e.g.,\ \cite
{KoSa78,Ko80a,Ko80b,KT80,HKPS93,BeKo88,KLPSW94}), while in this section we
concentrate on the smallest space $\left( {\cal N}\right) ^1$.

Let $({\cal H}_{-p})_{-q}^{-1}$ be the dual with respect to $L^2(\mu )$ of $(%
{\cal H}_p)_q^1$ and let $\left( {\cal N}\right) ^{-1}$ be the dual with
respect to $L^2(\mu )$ of $\left( {\cal N}\right) ^1$. We denote by $%
\left\langle \!\left\langle \ .\ ,\ .\ \right\rangle \!\right\rangle $ the
corresponding bilinear dual pairing which is given by the extension of the
scalar product on $L^2(\mu )$. We know from general duality theory that%
$$
\left( {\cal N}\right) ^{-1}=\stackunder{p,q\in \N }{\ {\rm ind\ lim}}\ (%
{\cal H}_{-p})_{-q}^{-1}\quad . 
$$
In particular, we know that every distribution is of finite order i.e.,\ for
any $\Phi \in \left( {\cal N}\right) ^{-1}$ there exist $p,q\in {\N}$ such
that $\Phi \in ({\cal H}_{-p})_{-q}^{-1}$. The chaos decomposition
introduces the following natural decomposition of $\Phi \in \left( {\cal N}%
\right) ^{-1}$. Let $\Phi ^{(n)}\in ({\cal N}_{{\Ckl}}^{\prime })^{\widehat{%
\otimes }n}$ be given. Then there is a distribution $\left\langle :\omega
^{\otimes n}:,\Phi ^{(n)}\right\rangle $ in $\left( {\cal N}\right) ^{-1}$
acting on $\varphi \in \left( {\cal N}\right) ^1$ as 
$$
\left\langle \!\!\left\langle \left\langle :\omega ^{\otimes n}:,\Phi
^{(n)}\right\rangle ,\varphi \right\rangle \!\!\right\rangle =n!\left\langle
\Phi ^{(n)},\varphi ^{(n)}\right\rangle . 
$$
Any $\Phi \in \left( {\cal N}\right) ^{-1}$ then has a unique decomposition 
$$
\Phi =\sum_{n=0}^\infty \left\langle :\omega ^{\otimes n}:,\Phi
^{(n)}\right\rangle \quad , 
$$
where the sum converges in $\left( {\cal N}\right) ^{-1}$ and we have 
$$
\left\langle \!\left\langle \Phi ,\varphi \right\rangle \!\right\rangle
=\sum\limits_{n=0}^\infty n!\left\langle \Phi ^{(n)},\varphi
^{(n)}\right\rangle \quad ,\ \varphi \in \left( {\cal N}\right) ^1\quad . 
$$
From the definition it is not hard to see that $({\cal H}_{-p})_{-q}^{-1}$
is a Hilbert space with norm%
$$
\left\| \Phi \right\| _{-p,-q,-1}^2=\sum\limits_{n=0}^\infty 2^{-nq}\left|
\Phi ^{(n)}\right| _{-p}^2\quad . 
$$

\noindent {\bf Remark.} Considering also the above mentioned spaces $\left( 
{\cal N}\right) ^\beta $ and their duals $\left( {\cal N}\right) ^{-\beta }$
we have the following chain of spaces 
$$
\left( {\cal N}\right) ^1\subset ...\subset \left( {\cal N}\right) ^\beta
\subset ...\subset \left( {\cal N}\right) =\left( {\cal N}\right) ^0\subset
L^2(\mu )\subset \left( {\cal N}\right) ^{\prime }\subset ...\subset \left( 
{\cal N}\right) ^{-\beta }\subset ...\subset \left( {\cal N}\right)
^{-1}\quad . 
$$

\subsection{Description of test functions by infinite dimensional holomorphy}

We state a theorem proven in \cite{KLS94} which shows that functions from $%
\left( {\cal N}\right) ^1$ have a pointwise meaning on ${\cal N}^{\prime }$
and are even (real) analytic on this space. Since the space $\left( {\cal N}%
\right) ^1$ is discussion in great detail in the previous chapter, we can
refer to Theorem \ref{N1E1min}. But we will also give an independent proof
using different methods.

\begin{corollary}
\label{Description}Any test function in $\left( {\cal N}\right) ^1$ has a
pointwise defined version which has an analytic continuation onto the space $%
{\cal N}_{{\Ckl}}^{\prime }$ as an element of ${\cal E}_{\min }^1({\cal N}_{{%
\Ckl}}^{\prime })$. Vice versa the restriction of any function in ${\cal E}%
_{\min }^1({\cal N}_{{\Ckl}}^{\prime })$ to ${\cal N}^{\prime }$ is in $%
\left( {\cal N}\right) ^1$.
\end{corollary}

\noindent In the rest of the paper we identify any $\varphi \in \left( {\cal %
N}\right) ^1$ with its version in ${\cal E}_{\min }^1({\cal N}_{{\Ckl}%
}^{\prime })$. In this sense we may write 
$$
\left( {\cal N}\right) ^1={\cal E}_{\min }^1({\cal N}^{\prime })=\left\{ u|_{%
{\cal N}^{\prime }} \Bigm|u\in {\cal E}_{\min }^1({\cal N}_{{\Ckl}}^{\prime
})\right\} \text{ .} 
$$
We will give an independent and short proof of Corollary \ref{phi(z)Betrag}.

\begin{corollary}
\label{Corexptype} For all $\varphi \in \left( {\cal N}\right) ^1$ and $%
q\geq 0$ we have the following pointwise bound 
\begin{equation}
\label{exptype}\left| \varphi (\omega )\right| \leq C_{p,\varepsilon \
}\left\| \varphi \right\| _{p,q,1}e^{\varepsilon \left| \omega \right| _{-p}}%
\text{, }\omega \in {\cal H}_{-p}\text{ ,} 
\end{equation}
where $\varepsilon =2^{-\frac q2}$ and%
$$
C_{p,\varepsilon \ }=\int_{{\cal N}^{\prime }}e^{\varepsilon \left| \omega
\right| _{-p}}\ {\rm d}\mu (\omega )\text{ .} 
$$
Here $p>0$ is taken such that the embedding $i_{p,0}:{\cal H}_p{\cal %
\hookrightarrow H}_0$ is of Hilbert-Schmidt type.
\end{corollary}

\TeXButton{Proof}{\proof}Let us introduce the following function%
$$
w(z)=\sum_{n=0}^\infty (-i)^n\left\langle z^{\otimes n},\varphi
^{(n)}\right\rangle \text{ , }z\in {\cal N}_{{\Ckl}}^{\prime } 
$$
using the chaos decomposition $\varphi (\omega )$ $=\sum_{n=0}^\infty
\left\langle :\omega ^{\otimes n}:,\varphi ^{(n)}\right\rangle $ of $\varphi 
$. Using the inequality 
$$
\left| \varphi ^{(n)}\right| _p\leq \frac 1{n!}\varepsilon ^n\left\| \varphi
\right\| _{p,q,1}\text{ , }\varepsilon =2^{-\frac q2} 
$$
we may estimate $\left| w(z)\right| $ for $z\in {\cal H}_{-p}$ as follows%
\begin{eqnarray*}
\left| w(z)\right| &\leq & \sum_{n=0}^\infty \left| \varphi ^{(n)}\right| %
_p\left| z\right| _{-p}^n %
\\&\leq & \left\| \varphi \right\| _{p,q,1}\sum_{n=0}^\infty \frac 1{n!}\varepsilon ^n\left| z\right| _{-p}^n %
\\&=&\left\| \varphi \right\| _{p,q,1}\exp \left( \varepsilon \left| %
z\right|_{-p}\right) \text{ .} 
\end{eqnarray*}
To achieve a bound of the type (\ref{exptype}) we use the relation \cite
{KLS94,BeKo88} 
$$
\varphi (\omega )=\int_{{\cal N}^{\prime }}w(y+i\omega )\ {\rm d}\mu (y)%
\text{ , }\omega \in {\cal N}^{\prime }\text{ .} 
$$
This allows to estimate%
\begin{eqnarray*}
\left| \varphi (\omega)\right| &\leq &\left\| \varphi \right\| _{p,q,1} %
\int_{{\cal N}^{\prime }}\exp \left( \varepsilon \left| y+i\omega %
\right| _{-p}\right) \ {\rm d}\mu (y) %
\\&\leq &\left\| \varphi \right\| _{p,q,1}e^{\varepsilon \left| \omega \right| _{-p}}\int_{{\cal N}^{\prime }}e^{\varepsilon \left| y \right| _{-p}}\ {\rm d} \mu (y)\text{ } 
\end{eqnarray*}
We conclude the proof with the inequality 
$$
C_{p,\varepsilon \ }=\int_{{\cal N}^{\prime }}e^{\varepsilon \left| \omega
\right| _{-p}}\ {\rm d}\mu (\omega )\text{ }\leq e^{\frac{\varepsilon ^2}{%
4\alpha }}\int_{{\cal N}^{\prime }}e^{\alpha \left| \omega \right| _{-p}^2}\ 
{\rm d}\mu (\omega )\text{ } 
$$
for $\alpha >0$ . If $p>0$ is such that the embedding $i_{p,0}$ is of
Hilbert-Schmidt type and $\alpha $ is chosen sufficiently small the right
hand integral is finite, see e.g.,\ \cite[Fernique's theorem]{Kuo75}. 
\TeXButton{End Proof}{\endproof}
\LaTeXparent{dis3.tex}

\section{The spaces 
\texorpdfstring{${\cal G}$ and ${\cal M}$}{G and M}}

\subsection{Definitions and examples}

For applications it is often useful to have distribution spaces with kernels 
$\Phi ^{(n)}$ but not more singular than $\Phi ^{(n)}\in {\cal H}_{\Ckl %
}^{\hat \otimes n}$. To this end Potthoff and Timpel \cite{PT94} introduced
a triple 
$$
{\cal G}\subset L^2(\mu )\subset {\cal G}^{\prime }\,. 
$$
We will introduce a second triple which is embedded in the above chain 
\begin{equation}
\label{GinM}{\cal G}\subset {\cal M}\subset L^2(\mu )\subset {\cal M}%
^{\prime }\subset {\cal G}^{\prime }\,. 
\end{equation}
We compare the properties of the two triples and we will discuss some
interesting interplay between these spaces.\medskip\ 

Let use first note that there is no need in the definition of $({\cal H}%
)_q:=({\cal H}_0)_q$ to choose $q\in \N $. We will denote the new real
parameter replacing $q$ by $\alpha \in \R _{+}$

\begin{definition}
We can define 
$$
{\cal G}:=\stackunder{\alpha >0}{\ {\rm pr\ lim}}({\cal H})_\alpha  
$$
and 
$$
{\cal M}:=\stackunder{\alpha >0}{\ {\rm ind\ lim}}({\cal H})_\alpha  
$$
\end{definition}

Both spaces are Fr\'echet spaces continuously embedded in $L^2(\mu )$. By
general duality theory \cite{Sch71} we have the representations%
$$
{\cal G}^{\prime }=\stackunder{\alpha >0}{\ {\rm ind\ lim}}({\cal H}%
)_{-\alpha } 
$$
$$
{\cal M}^{\prime }=\stackunder{\alpha >0}{\ {\rm pr\ lim}}({\cal H}%
)_{-\alpha }\text{ .} 
$$
Obviously (\ref{GinM}) holds.\bigskip\ 

\example  \newcounter{phiGauss} \setcounter{phiGauss}{\value{example}} \label
{phiGaussP}Let $\eta \in {\cal H}\,,\,c\in \R \,,\,2c|\eta |^2<1$ and%
$$
\varphi =e^{c\langle \cdot ,\eta \rangle ^2}\,. 
$$
The $S$-transform is easy to calculate, $\xi \in {\cal N}$%
$$
S\varphi (\xi )=(1-2c|\eta |^2)^{-1/2}\exp \left( \frac c{1-2c|\eta
|^2}\langle \xi ,\eta \rangle ^2\right) \,, 
$$
expanding the exponential we obtain the kernels%
$$
\varphi ^{(2n)}=\frac 1{n!}\left( \frac c{1-2c|\eta |^2}\right) ^n\eta
^{\otimes 2n}\,\,,\quad \varphi ^{(n)}=0\text{ if }n\text{ is odd.} 
$$
Then%
$$
\left\| \varphi \right\| _{0,\alpha }^2=\sum_{n=0}^\infty n!\,2^{\alpha
n}|\varphi ^{(n)}|^2 \hspace{18mm} 
$$
$$
\hspace*{6mm}\leq \sum_{n=0}^\infty 2^{2\alpha n}\left( \frac{2c|\eta |^2}{%
1-2c|\eta |^2}\right) ^{2n} 
$$
which is finite if $2^\alpha \frac{2c|\eta |^2}{1-2c|\eta |^2}<1$.\smallskip%
\ 

\noindent From this it follows \smallskip \\ \hspace*{2mm} 
\begin{tabular}{lclllcl}
     $\varphi \notin {\cal G}$ & if & $c\neq 0$ & %
      but & $\varphi \in {\cal M}$ & if & $4c|\eta |^2<1$ %
\\   $\varphi\notin L^2(\mu )$ & if & $4c|\eta |^2=1$ & 
      but then & $\varphi \in {\cal M}^{\prime }$ & &
\\   $\varphi \notin {\cal M}^{\prime }$ & if & $4c|\eta |^2>1$ &
      but & $\varphi \in {\cal G}^{\prime }$ & if & $2c|\eta |^2<1$.
\end{tabular}
\bigskip

In section \ref{DeltaSec} we will prove that we also can define Donsker's
delta $\delta (\langle \cdot ,\eta \rangle -a)\in {\cal M}^{\prime }$ for $%
a\in \C \,,\,\eta \in {\cal H}_{\Ckl }\,,\,\arg \langle \eta ,\eta \rangle
\neq \pi $. This was in fact one of the main motivations to introduce ${\cal %
M}^{\prime }$. We wanted to study pointwise multiplication of $\delta $ with
other functions.\bigskip\ 

The next proposition will produce whole classes of examples.

\begin{proposition}
\label{LpGStrich} \hfill \\ 1) Let $\varphi \in L^p(\mu )$ for some $p>1$
then $\varphi \in {\cal G}^{\prime }$, i.e.,%
$$
\bigcup_{p>1}L^p(\mu )\subset {\cal G}^{\prime }\;. 
$$
2) Let $\varphi \in L^p(\mu )$ for all $1<p<2$ then $\varphi \in {\cal M}%
^{\prime }$, i.e.,%
$$
\bigcap_{p<2}L^p(\mu )\subset {\cal M}^{\prime }\;. 
$$
\end{proposition}

\TeXButton{Proof}{\proof}The argument is based on Nelson's
Hypercontractivity Theorem \cite{Ne73}. In particular we obtain $2^{-\alpha
N/2}:L^p(\mu )\rightarrow L^2(\mu )$ (here $N$ denotes the well known number
operator) is a contraction if $2^{-\alpha }\leq p-1$. Otherwise $2^{-\alpha
N/2}$ is unbounded. Hence%
$$
\left\| \varphi \right\| _{0,-\alpha }=\left\| 2^{-\alpha N/2}\varphi
\right\| _{L^2(\mu )}\leq \left\| \varphi \right\| _{L^p(\mu )}\text{ \quad
if \ }2^{-\alpha }\leq p-1 
$$
If $\varphi \in L^p(\mu )$ for some $p>1$ , there exists a $\alpha >0$ such
that the above inequality holds. Now we prove the second assertion. For any $%
\alpha >0$ we may choose $p\in (1,2)$ such that the above estimate holds,
hence $\varphi \in {\cal M}^{\prime }$.\TeXButton{End Proof}{\endproof}%
\bigskip\ 

\noindent {\bf Notes}.\\1. The first assertion is already proved in \cite
{PT94}. \smallskip\ \\2. From Example \arabic{phiGauss} we know that for all 
$p\in (1,2)$, there exists $\varphi \in {\cal M}^{\prime }$ such that $%
\varphi \notin L^p(\mu )$, i.e., $\forall p\in (1,2)$ 
$$
L^p(\mu )\not \subset {\cal M}^{\prime }\;. 
$$
Moreover $L^1(\mu )\not \subset {\cal G}^{\prime }$.\smallskip\ \\3. Let us
also mention the trivial consequence that any $L^p(\mu )$-function, $p>1$,
has a chaos expansion with all kernels contained in ${\cal H}_{\Ckl }^{\hat
\otimes n}$.\bigskip\ 

Now we state the `dual result' which may be proved along the same lines.

\begin{proposition}
\label{MGLp} \hfill \\ 1.%
$$
\varphi \in {\cal G}\Rightarrow \varphi \in \bigcap_{p>1}L^p(\mu )\;. 
$$
2.%
$$
\varphi \in {\cal M}\Rightarrow \exists p>2:\varphi \in L^p(\mu )\text{ .} 
$$
\end{proposition}

\noindent {\bf Note.} Assertion 1 is related to the inclusion ${\cal G}%
\subset {\cal D}$ where ${\cal D}$ is the so called Meyer-Watanabe space,
see \cite{HKPS93} for a definition and \cite{PT94} for a proof of the
inclusion. \\We can again refer to Example \arabic{phiGauss}. For all $p>2\ $%
there exists $\varphi \in {\cal M}$ such that $\varphi \notin L^p(\mu )$,
i.e., $\forall p>2$ 
$$
{\cal M}\not \subset L^p(\mu )\;, 
$$
in particular 
$$
{\cal M}\not \subset {\cal D\,\,.} 
$$


\subsection{The pointwise product}

It is well known that $({\cal N})$ and $({\cal N})^1$ are algebras under
pointwise multiplication (sometimes called Wiener Product). In \cite{PT94}
it has been shown that also ${\cal G}$ has this property. On the other hand
it is obvious that ${\cal M}$ can not be an algebra. To see this consider $%
\varphi =e^{\langle \cdot ,\eta \rangle ^2}$, $\left| \eta \right| ^2=1/8$
then $\varphi \in {\cal M}$ but $\varphi ^2=e^{2\langle \cdot ,\eta \rangle
^2}\notin L^2(\mu )$, see Example \arabic{phiGauss} on page \pageref
{phiGaussP}. We will show that the pointwise product can also be defined if
one factor is in ${\cal M}$ and the other in ${\cal G}$. To prove this we
found it useful to have a detailed discussion of pointwise products in \cite
{Ob94} which we were able to modify to the present setting.

The first question is, how does the pointwise product look like in terms of
chaos expansion?

\begin{lemma}
Let $\varphi ,\psi \in {\cal G}$ be given by 
$$
\varphi =\sum_{n=0}^\infty \langle :\omega ^{\otimes n}:,\varphi
^{(n)}\rangle \,,\quad \psi =\sum_{n=0}^\infty \langle :\omega ^{\otimes
n}:,\psi ^{(n)}\rangle \ . 
$$
Then the chaos expansion of 
$$
\varphi \psi =\sum_{l=0}^\infty \langle :\omega ^{\otimes l}:,f^{(l)}\rangle
$$
is given by 
\begin{equation}
\label{WienerKern}f^{(l)}=\sum_{m+n=l}\sum_{k=0}^\infty k!\,\binom{m+k}k%
\binom{n+k}k\varphi ^{(m+k)}\stackunder{k}{\hat \otimes }\psi ^{(n+k)}
\end{equation}
where the {\it contraction }$\varphi ^{(m+k)}\stackunder{k}{\hat \otimes }%
\psi ^{(n+k)}$ of the kernels $\varphi ^{(m+k)}$ and $\psi ^{(n+k)}$ is the
symmetrization of the partial scalar product $\left( \varphi ^{(m+k)},\psi
^{(n+k)}\right) _{{\cal H}^{\otimes k}}\in {\cal H}_{\Ckl }^{\otimes (m+n)}$.
\end{lemma}

Note that 
$$
\left| \varphi ^{(m+k)}\stackunder{k}{\hat \otimes }\psi ^{(n+k)}\right|
\leq \left| \varphi ^{(m+k)}\right| \left| \psi ^{(n+k)}\right| 
$$
Now we are giving a variant of Lemma 3.5.4 in \cite{Ob94}. The only
qualitative change is that we do not need smoother kernels in the estimate.

\begin{lemma}
Let $\alpha ,\beta \geq 0$, then $f^{(l)\text{ }}$defined by (\ref
{WienerKern}) can be estimated%
$$
l!\left| f^{(l)}\right| ^2\leq (l+1)(2^{-\alpha }+2^{-\beta })^l\;\left\|
\varphi \right\| _{0,\beta }^2\,\left\| \psi \right\| _{0,\beta
}^2\sum_{k=0}^\infty \binom{l+2k}{2k}2^{-k(\alpha +\beta )}\ . 
$$
\end{lemma}

Following the lines of the proof in \cite{Ob94} a little further we obtain.

\begin{theorem}
Let $\alpha ,\beta ,\gamma \geq 0$ satisfy $2^{-(\alpha +\beta
)/2}+2^{\gamma -\alpha }+2^{\gamma -\beta }<1$ then 
$$
\left\| \varphi \psi \right\| _{0,\gamma }\leq \frac{\sqrt{1-2^{-(\alpha
+\beta )/2}}}{1-2^{-(\alpha +\beta )/2}-2^{\gamma -\alpha }-2^{\gamma -\beta
}}\left\| \varphi \right\| _{0,\alpha }\left\| \psi \right\| _{0,\beta }\,. 
$$
\end{theorem}

\begin{corollary}
${\cal G}$ is closed under pointwise multiplication and multiplication is a
separately continuous bilinear map from ${\cal G}\times {\cal G}$ into $G$.
\end{corollary}

This result is also shown in \cite{PT94}.

\begin{corollary}
\label{GMtoM} The pointwise multiplication is a separately continuous
bilinear map from ${\cal G}\times {\cal M}$ into ${\cal M}$.
\end{corollary}

\TeXButton{Proof}{\proof}First fix the factor $\varphi \in {\cal G}$. To
prove that $\psi \mapsto \varphi \cdot \psi $ from ${\cal M}$ into itself is
continuous we have to show that for all $\beta >0$ there exists a $\gamma >0$
such that $\psi \mapsto \varphi \cdot \psi $ is continuous from $({\cal H}%
)_\beta $ into $({\cal H})_\gamma $. But this follows from the above
theorem, since we may choose $\gamma <\beta $ and $\alpha $ large enough. \\%
For fixed $\psi \in {\cal M}$ we have to show that there exist $\alpha >0$
and $\gamma >0$ such that $\varphi \mapsto \varphi \cdot \psi $ is
continuous from $({\cal H})_\alpha $ into $({\cal H})_\gamma $. Also this is
clear from the above theorem.\TeXButton{End Proof}{\endproof}\bigskip\ 

Now we can extend the concept of pointwise multiplication to products where
one factor is a distribution.\smallskip\ 

Let $\Phi \in {\cal M}^{\prime }\,,\,\varphi \in {\cal G}$ then $\Phi \cdot
\varphi \in {\cal M}^{\prime }$ defined by 
$$
\langle \!\langle \Phi \cdot \varphi ,\psi \rangle \!\rangle :=\langle
\!\langle \Phi ,\varphi \psi \rangle \!\rangle \,,\,\quad \psi \in {\cal M} 
$$
is well defined because of the previous corollary. More useful is the
following:\smallskip\ 

Let $\Phi \in {\cal M}^{\prime }\,,\,\psi \in {\cal M}$ then $\Phi \cdot
\psi \in {\cal G}^{\prime }$ given by 
\begin{equation}
\label{MStrichProd}\langle \!\langle \Phi \cdot \psi ,\varphi \rangle
\!\rangle :=\langle \!\langle \Phi ,\varphi \psi \rangle \!\rangle \,,\quad
\,\varphi \in {\cal G} 
\end{equation}
is well defined.

\subsection{Integrating out Donsker's delta}

Let $a\in \C \,,\,\eta \in {\cal N}$ such that $|\eta |=1$ and $\varphi \in (%
{\cal N})$ than a simple calculation yields 
\begin{equation}
\label{DeltaFormal}\langle \!\langle \delta (\langle \cdot ,\eta \rangle
-a),\varphi \rangle \!\rangle =\frac 1{\sqrt{2\pi }}e^{-\frac 12a^2}\E %
\left( \varphi (\cdot +(a-\langle \cdot ,\eta \rangle )\eta \right) \, . 
\end{equation}
Since $\varphi $ has a pointwise well defined version which has some
analytic continuation, we understand what $\varphi (\omega +a\eta )$ for $%
a\in \C $ means. So everything is well defined. In this section we want to
extend the above formula to $\varphi \in {\cal M}$ and $\eta \in {\cal H}$.
This raises at least the following questions\medskip\ \\1. What does $%
\varphi \left( \omega +(a-\langle \omega ,\eta \rangle )\eta \right) $ mean?
In particular in what sense do we have an analytic continuation?\smallskip\ 
\\2. Is the expectation value at the end of the procedure well defined?

\subsubsection{Analyticity of shifts}

In this section we want to define an operator%
$$
\tau _\eta :{\cal M}\rightarrow {\cal M}\,,\ \,\varphi \mapsto \varphi
(\cdot +\eta )\ \text{ for }\eta \in {\cal H}_{\Ckl } \, . 
$$
(Note that this operation surely has no sense pointwisely. Consider e.g., $%
\varphi =\langle :\omega ^{\otimes 2}:,\varphi ^{(2)}\rangle $ then $\tau
_\eta \varphi (0)=\langle :\eta ^{\otimes 2}:,\varphi ^{(2)}\rangle =\langle
\eta ^{\otimes 2},\varphi ^{(2)}\rangle -\langle {\rm Tr},\varphi
^{(2)}\rangle $ which is ill defined if $\varphi ^{(2)}\in {\cal H}_{\Ckl %
}^{\hat \otimes 2}$ allows no trace.)

To give a meaningful definition we use 
$$
:(\omega +\eta )^{\otimes n}:\ =\sum_{k=0}^n\binom nk:\omega ^{\otimes
(n-k)}:\otimes \eta ^{\otimes k} 
$$
and define $\forall \eta \in {\cal H}_{\Ckl }$%
$$
\tau _\eta \varphi =\varphi (\cdot +\eta ):=\sum_{k=0}^\infty
\sum_{l=0}^\infty \binom{k+l}k\langle :\omega ^{\otimes l}:,\left( \eta
^{\otimes k},\varphi ^{(k+l)}\right) _{{\cal H}^{\otimes k}}\rangle 
$$
whenever the series converges.

\begin{theorem}
\label{tauetaMG} \smallskip \hfill \\ 1. Let $\varphi \in {\cal M}\,,\,\eta
\in {\cal H}_{\Ckl }$ then $\tau _\eta \varphi \in {\cal M}$. Moreover the
mapping $\eta \mapsto \tau _\eta \varphi \in {\cal E}_{\max }^2({\cal H}_{%
\Ckl },{\cal M})$.\smallskip\ \\2. Let $\varphi \in {\cal G}\,,\,\eta \in 
{\cal H}_{\Ckl }$ then $\tau _\eta \varphi \in {\cal G}$ and moreover the
mapping $\eta \mapsto \tau _\eta \varphi \in {\cal E}_{\min }^2({\cal H}_{%
\Ckl },{\cal G})$.
\end{theorem}

\TeXButton{Proof}{\proof}Define the $k$-homogeneous polynomials on ${\cal H}%
_{\Ckl }$%
$$
\frac 1{k!}\widehat{{\rm d}^k{\rm \varphi }(\eta )}:=\sum_{l=0}^\infty 
\binom{k+l}k\langle :\omega ^{\otimes l}:,\left( \eta ^{\otimes k},\varphi
^{(k+l)}\right) _{{\cal H}^{\otimes k}}\rangle 
$$
with values in ${\cal M}$ (or ${\cal G}$ respectively). To show that this is
well defined let $\varphi \in ({\cal H})_{\alpha \,},\,\alpha >0.$ Such that 
$\left| \varphi ^{(k)}\right| ^2\leq \left\| \varphi \right\| _{0,\alpha
}^2\frac 1{k!}2^{-\alpha k}$ and choose $\gamma \in (0,\alpha )$ to estimate%
\begin{eqnarray*}
\left\| \frac 1{k!}\widehat{{\rm d}^k\varphi (\eta )}\right\| _{0,\gamma }^2 %
& \leq & \sum_{l=0}^\infty l!\,2^{\gamma l}\binom{k+l}k^2|\eta |^{2k}| %
\varphi ^{(k+l)}| %
\\ & \leq & \left\| \varphi \right\| _{0,\alpha }^2|\eta |^{2k}%
\sum_{l=0}^\infty  \binom{k+l}k^2\frac{l!}{(k+l)!}2^{l\gamma }%
2^{-\alpha (k+l)} %
\\ & \leq & \left\| \varphi \right\| _{0,\alpha }^2\frac 1{k!} %
2^{-\alpha k}|\eta |^{2k}\sum_{l=0}^\infty \binom{k+l}k %
2^{l(\gamma -\alpha )} %
\\ &=& \left\| \varphi \right\| _{0,\alpha }^2\frac 1{k!}2^{-\alpha k} %
|\eta |^{2k}(1-2^{\gamma -\alpha })^{-(k+1)} %
\\ &=& (1-2^{\gamma -\alpha })^{-1}\left\| \varphi \right\| _{0,\alpha }^2 %
\frac 1{k!}|\eta |^{2k}(2^\alpha -2^\gamma )^{-k}\,. %
\end{eqnarray*}
This shows that $\frac 1{k!}\widehat{{\rm d}^k{\rm \varphi }(\eta )}$ is in
fact a $k$-homogeneous continuous polynomial. If $\varphi \in {\cal G}$ and
if $\varphi \in {\cal M}$ then $\frac 1{k!}\widehat{{\rm d}^k\varphi (\eta )}%
\in {\cal M}$.

The next step is to show that $\sum_{k=0}^\infty \frac 1{k!}\widehat{{\rm d}%
^k\varphi (\eta )}$ converges uniformly on any ball in ${\cal H}_{\Ckl }$ in
the topology of ${\cal M}$ (or ${\cal G}$ respectively). So we estimate 
\begin{eqnarray*}
& & \hspace*{-1cm}\sum_{k=0}^\infty \left\| \frac 1{k!}\widehat{{\rm d}^k\varphi (\eta )}%
\right\| _{0,\gamma }\leq \ \left\| \varphi \right\| _{0,\alpha }(1-2^{\gamma -\alpha })^{-1/2}\sum_{k=0}^\infty \frac 1{\sqrt{k!}}(2^\alpha -2^\gamma %
)^{-k/2}\;|\eta |^k %
\\ & \leq & \left\| \varphi \right\| _{0,\alpha }(1-2^{\gamma -\alpha })^{-1/2}\left( \sum_{k=0}^\infty \frac 1{k!}2^{k(\alpha -\gamma )}(2^\alpha %
-2^\gamma )^{-k}|\eta |^{2k}\right) ^{1/2}\left( \sum_{k=0}^\infty %
2^{k(\gamma -\alpha )}\right) ^{1/2} %
\\ &=& \left\| \varphi \right\| _{0,\alpha }(1-2^{\gamma -\alpha })^{-1}\exp %
\left( \frac{2^{\alpha -\gamma }}{2(2^\alpha -2^\gamma )}|\eta |^2\right) %
\,\,. %
\end{eqnarray*}
showing the uniform and absolute convergence of the series. \smallskip\ \\If 
$\varphi \in {\cal M}$ we have shown that $\eta \mapsto \tau _\eta \varphi
\in {\cal E}_{\max }^2({\cal H}_{\Ckl },{\cal M})$. \smallskip\ \\If $%
\varphi \in {\cal G}\,$ choose e.g$.$, $\alpha =2\gamma $, then the type of
growth is bounded by $(2^{\gamma +1}-2)^{-1}$ which converges to zero for
growing $\gamma $. Thus $\eta \mapsto \tau _\eta \varphi \in {\cal E}_{\min
}^2({\cal H}_{\Ckl },{\cal G})$. \TeXButton{End Proof}{\endproof}\bigskip\ 

The second term in the Taylor series coincides with the G\^ateaux
derivative. So we have the following corollary.

\begin{corollary}
Let $\eta \in {\cal H}_{\Ckl }$ then the G\^ateaux derivative defined by 
$$
D_\eta \varphi :=\sum_{l=0}^\infty l\;\langle :\omega ^{\otimes l}:,\left(
\eta ,\varphi ^{(l+1)}\right) _{{\cal H}}\rangle \,,\qquad \varphi \in {\cal %
M} 
$$
is a well defined operator from ${\cal M}$ into itself and from ${\cal G}$
into itself.
\end{corollary}

\subsubsection{Composition with projection operators \label{ProjSec}}

Let $\eta \in {\cal N}$ with $|\eta |=1$. In view of the aim of this section
we also want to understand how to define $\varphi (\omega -\langle \omega
,\eta \rangle \eta )$ (if $\varphi \notin ({\cal N})$ this is non trivial)$.$
Note that 
$$
P_{\perp }:{\cal N}^{\prime }\rightarrow {\cal N}^{\prime }\,,\quad \omega
\mapsto P_{\perp }\omega =\omega -\langle \omega ,\eta \rangle \eta 
$$
is the projection on the orthogonal complement of the subspace spanned by $%
\eta \in {\cal H}_{\Ckl }$. For $\varphi \in {\cal G}$ (also for $\varphi
\in {\cal M}$) we want to define $P\varphi =\varphi \circ P_{\perp }$.

First we have to understand what happens in terms of chaos expansion.

\begin{lemma}
For $\eta \in {\cal N}$, $|\eta |=1$ and $\omega \in {\cal N}^{\prime }$ we
have $:(P_{\perp }\omega )^{\otimes n}:\ \in ({\cal N}^{\hat \otimes
n})^{\prime }$ and the following relation holds 
$$
:(P_{\perp }\omega )^{\otimes n}:=\sum_{k=0}^{[\frac n2]}\frac{n!}{%
k!\,(n-2k)!}(-\frac 12)^k\left( :\omega ^{\otimes (n-2k)}:\circ P_{\perp
}^{\otimes (n-2k)}\right) \hat \otimes \eta ^{\otimes 2k} 
$$
where $:\omega ^{\otimes l}:\circ P_{\perp }^{\otimes l}\in ({\cal N}^{\hat
\otimes l})^{\prime }$ is defined by 
$$
\langle :\omega ^{\otimes l}:\circ P_{\perp }^{\otimes l},\,\varphi
^{(l)}\rangle =\langle :\omega ^{\otimes l}:,\,P_{\perp }^{\otimes l}\varphi
^{(l)}\rangle \,. 
$$
\end{lemma}

\TeXButton{Proof}{\proof}Choose $\varphi =:\exp \langle \cdot ,\xi \rangle :$
then 
\begin{eqnarray*}
P\varphi &=& \exp \left( \langle P_{\perp }\omega ,\xi \rangle %
-\frac 12|\xi |^2\right)%
\\ &=& \ :\exp \langle \omega ,P_{\perp }\xi \rangle :\ \exp %
    \left( -\frac 12(|\xi |^2-|P_{\perp }\xi |^2)\right) %
\\ &=& \ :\exp \langle \omega ,P_{\perp }\xi \rangle :\ \exp \left( -\frac 12%
     \langle \eta ,\xi \rangle ^2\right) %
\end{eqnarray*}
Expanding both sides of this equation we obtain 
$$
\sum_{n=0}^\infty \frac 1{n!}\langle :(P_{\perp }\omega )^{\otimes n}:,\xi
^{\otimes n}\rangle = \sum_{l=0}^\infty \frac 1{l!}\langle :\omega ^{\otimes
l}:,(P_{\perp }\xi )^{\otimes l}\rangle \sum_{k=0}^\infty \frac 1{k!}(-\frac
12)^k\langle \eta ^{\otimes 2k},\xi ^{\otimes 2k}\rangle 
$$
$$
\hspace*{2cm} = \sum_{n=0}^\infty \sum_{k=0}^{[\frac n2]}\frac
1{k!\,(n-2k)!} (-\frac 12)^k\left\langle \left( :\omega ^{\otimes
(n-2k)}:\circ P_ {\perp }^{\otimes (n-2k)}\right) \otimes \eta ^{\otimes
2k},\xi ^ {\otimes n}\right\rangle \,. 
$$
A comparison of coefficients proves the lemma.\TeXButton{End Proof}
{\endproof}\bigskip\ 

An immediate consequence is the following lemma

\begin{lemma}
Let $\eta \in {\cal N}$, $|\eta |=1$ and $\varphi =\sum_{n=0}^N\langle
:\omega ^{\otimes n}:,\varphi ^{(n)}\rangle \in {\cal G}$ be a finite linear
combination. Then 
\begin{equation}
\label{Pphi}P\varphi =\sum_{n=0}^\infty \langle :\omega ^{\otimes n}:,\tilde
\varphi ^{(n)}\rangle 
\end{equation}
where 
\begin{equation}
\label{phinTilde}\tilde \varphi ^{(n)}=\sum_{k=0}^\infty \frac{(n+2k)!}{%
k!\,n!}\left( -\frac 12\right) ^kP_{\perp }^{\otimes n}\left( \eta ^{\otimes
2k},\varphi ^{(n+2k)}\right) _{{\cal H}^{\otimes 2k}}\,.
\end{equation}
(We have no problems of convergence since all sums are in fact finite.)
\end{lemma}

\TeXButton{Proof}{\proof}Follows from%
\begin{eqnarray*}
P\varphi &=& \sum_{n=0}^\infty \langle :(P_{\perp }\omega )^{\otimes n}:,\varphi ^{(n)}\rangle %
\\ &=& \sum_{n=0}^\infty \sum_{k=0}^{[\frac n2]}\frac{n!}{k!\,(n-2k)!} %
(-\frac 12)^k\left\langle \left( :\omega ^{\otimes (n-2k)}:\circ P_{\perp }^{\otimes (n-2k)}\right) %
 \otimes \eta ^{\otimes 2k},\;\varphi ^{(n)}\right\rangle %
\\ &=& \sum_{l=0}^\infty \sum_{k=0}^\infty \frac{(l+2k)!}{k!\,l!}\left( %
-\frac 12\right) ^k\left\langle :\omega ^{\otimes l}:,\;P_{\perp }^{\otimes l} %
\left( \eta ^{\otimes 2k},\varphi ^{(l+2k)}\right) _{{\cal H} ^{\otimes 2k}}\right \rangle \,. %
\end{eqnarray*}

\TeXButton{End Proof}{\endproof}\bigskip\ 

Now we observe that the kernels $\tilde \varphi ^{(n)}\in {\cal H}_{\Ckl %
}^{\hat \otimes n}$ in (\ref{phinTilde}) are well defined if we extend from $%
\eta \in {\cal N}$ to $\eta \in {\cal H}$. Also the definition of $P$ is not
restricted to finite linear combinations as the following theorem shows.

\begin{theorem}
\label{PMG} The linear mapping $P$ has the following well defined
extensions; for $\eta \in {\cal H}$, $|\eta |=1$%
$$
P:{\cal G}\rightarrow {\cal G} 
$$
$$
P:{\cal M}\rightarrow {\cal G}^{\prime } 
$$
$$
P:({\cal H})_\alpha \rightarrow {\cal M\,},\qquad \alpha >1, 
$$
more precisely%
$$
P:({\cal H})_\alpha \rightarrow ({\cal H})_\gamma {\cal \,}\qquad \text{if}%
\quad 2^\alpha -1>2^\gamma  
$$
\end{theorem}

\TeXButton{Proof}{\proof}We discuss the convergence of (\ref{Pphi}) with (%
\ref{phinTilde}). First note 
$$
\left| P_{\perp }^{\otimes l}\left( \eta ^{\otimes 2k},\varphi
^{(l+2k)}\right) _{{\cal H}^{\otimes 2k}}\right| \leq \left| \eta \right|
^{2k}\left| \varphi ^{(l+2k)}\right| \leq \left| \varphi ^{(l+2k)}\right| 
$$
since $\left| \eta \right| =1$. If we assume $\varphi \in ({\cal H})_\alpha $
then $n!\left| \varphi ^{(n)}\right| ^2\leq 2^{-\alpha n}\left\| \varphi
\right\| _{0,\alpha }^2$. For $\gamma <\alpha $ we have%
$$
\left\| \sum_{n=0}^\infty \frac{(n+2k)!}{k!\,n!}\left( -\frac 12\right)
^k\left\langle :\omega ^{\otimes n}:,P_{\perp }^{\otimes n}\left( \eta
^{\otimes 2k},\varphi ^{(n+2k)}\right) _{{\cal H}^{\otimes 2k}}\right\rangle
\right\| _{0,\gamma }^2\hspace{2cm} 
$$
\vspace{-6mm} 
\begin{eqnarray*}
&=& \sum_{n=0}^\infty n!\left( \frac{(n+2k)!}{k!\,n!}\right) ^2 %
2^{n\gamma }\left( \frac 12\right) ^{2k}\left| \varphi ^{(n+2k)}\right| ^2 %
\\ &=& \sum_{n=0}^\infty \binom{n+2k}{2k}\frac{(2k)!}{(k!)^22^{2k}} \, %
(n+2k)!\,\left| \varphi ^{(n+2k)}\right| ^22^{n\gamma } %
\\ & \leq & \left\| \varphi \right\| _{0,\alpha }^22^{-2k\alpha } %
\sum_{n=0}^\infty \binom{n+2k}{2k}2^{n(\gamma -\alpha )} %
\\ &=& \left\| \varphi \right\| _{0,\alpha }^22^{-2k\alpha }(1-2^{ \gamma -\alpha })^{-(2k+1)} %
\\ &=& (1-2^{\gamma -\alpha })^{-1}\left\| \varphi \right\| _{0,\alpha }^2(2^\alpha -2^\gamma )^{-2k} %
\end{eqnarray*}
Hence%
$$
\left\| P\varphi \right\| _{0,\gamma }\leq \sum_{k=0}^\infty \left\|
\sum_{n=0}^\infty \frac{(n+2k)!}{k!\,n!}\left( -\frac 12\right)
^k\left\langle :\omega ^{\otimes n}:,P_{\perp }^{\otimes n}\left( \eta
^{\otimes 2k},\varphi ^{(n+2k)}\right) _{{\cal H}^{\otimes 2k}}\right\rangle
\right\| _{0,\gamma } 
$$
$$
\leq (1-2^{\gamma -\alpha })^{-1/2}\left\| \varphi \right\| _{0,\alpha
}\sum_{k=0}^\infty (2^\alpha -2^\gamma )^{-k}\hspace*{30mm} 
$$
which is convergent if $2^\alpha -2^\gamma >1$. \\If $\varphi \in {\cal G}$
then for every $\gamma >0$ there exists an $\alpha >0$ (e.g., $\alpha
=\gamma +1$) such that $2^\alpha -2^\gamma >1$. Hence $P\varphi \in {\cal G}$%
.\smallskip\ \\If $\varphi \in {\cal M}$ there exists $\alpha >0$ such that $%
\left\| \varphi \right\| _{0,\alpha }<\infty $. Then we choose $\gamma \in 
\R $ such that $2^\alpha -1>2^\gamma $ ($\gamma $ possibly negative), to
obtain $\left\| P\varphi \right\| _{0,\gamma }<\infty $ i.e., $P\varphi \in 
{\cal G}^{\prime }$. \\If $\varphi \in ({\cal H})_\alpha \,,\,\alpha >1$
then we can find $\gamma >0$ such that $2^\alpha -1>2^\gamma \,$,\thinspace
i.e., $\varphi \in {\cal M}$.\TeXButton{End Proof}{\endproof}\bigskip\bigskip%
\ 

Now we are going back to our motivating example.

Let $\eta \in {\cal N}\,,\,|\eta |=1$ and $\varphi \in ({\cal N})$. Starting
from expression (\ref{DeltaFormal}) we calculate%
$$
\varphi (\omega +(a-\langle \omega ,\eta \rangle )\eta )=\varphi (P_{\perp
}\omega +a\eta )=P\varphi (\omega +a\eta )=P\tau _{a\eta }\varphi (\omega
)\,. 
$$
The last expression can be extended to $\eta \in {\cal H}$ , $\,|\eta |=1$, $%
\varphi \in {\cal M\ }$in view of Propositions \ref{tauetaMG} and \ref{PMG},
since 
$$
\varphi \in {\cal M}\Rightarrow \tau _{a\eta }\varphi \in {\cal M}%
\Rightarrow P\tau _{a\eta }\varphi \in {\cal G}^{\prime }. 
$$
Hence we can take expectation $\E (P\tau _{a\eta }\varphi )$ without
problems. So now we can formulate equation (\ref{DeltaFormal}) as a
proposition

\begin{proposition}
\label{DeltaMode} Let $\eta \in {\cal H}\,,\,|\eta |=1$ ,\thinspace $a\in 
\C 
$ and $\varphi \in {\cal M}$. Then 
$$
\langle \!\langle \delta (\langle \cdot ,\eta \rangle -a),\varphi \rangle
\!\rangle =\frac 1{\sqrt{2\pi }}e^{-\frac 12a^2}\E (P\tau _{a\eta }\varphi
)\,. 
$$
\end{proposition}

\TeXButton{Proof}{\proof}It is easy to show that both sides coincide if we
choose $\varphi =:\exp \langle \cdot ,\xi \rangle :\,,\,\xi \in {\cal H}$.
Then a continuity argument shows that they agree for all $\varphi \in {\cal M%
}$. \TeXButton{End Proof}{\endproof}
\LaTeXparent{dis3.tex}

\section{The Meyer\texorpdfstring{--}{-}Yan triple}

If the $S$-(or $T$-)transform of a distribution is well defined as an entire
function, but of infinite order of growth, then the distributions spaces $(%
{\cal N})^{-\beta }$ for $0\leq \beta <1$ are too small. Obviously we can
use $({\cal N})^{-1}$ but then we will only require that the $S$-transform
is analytic in a neighborhood of zero. Meyer and Yan \cite{MY90} introduced
a triple which helps to close this gap. In \cite{KoS93} this triple was
discussed in great detail. So we will only introduce the notation and quote
some useful results from the second work. The only new results are stated in
the two corollaries.

\begin{definition}
The Meyer--Yan space is defined by 
$$
{\cal Y}:=\ \stackunder{p\in \N }{\rm pr\ lim}\stackunder{q\in \N }{\ {\rm %
ind\ lim}}({\cal H}_p)_{-q}^1\,. 
$$
The dual space can be represented as 
$$
{\cal Y}^{\prime }=\ \stackunder{p\in \N }{\rm ind\ lim}\ \stackunder{q\in 
\N 
}{\rm pr\ lim}({\cal H}_{-p})_{+q}^{-1}\,. 
$$
\end{definition}

\noindent Then we have 
$$
({\cal N})^1\subset {\cal Y}\subset ({\cal N})^\beta \subset ({\cal N}%
)\subset L^2(\mu )\subset ({\cal N})^{\prime }\subset ({\cal N})^{-\beta
}\subset {\cal Y}^{\prime }\subset ({\cal N})^{-1} 
$$
for $\beta \in [0,1)$.\medskip\ 

We will only need a characterization of distributions.

\begin{theorem}
\label{MYChar} A mapping $F:{\cal N}_{\Ckl }\rightarrow \C $ is the $S$%
-transform of a distribution $\Phi \in {\cal Y}^{\prime }$ if and only if \\%
1. $F$ is entire on ${\cal N}_{\Ckl }$.\smallskip\ \\2. There exists $p\in 
\N _0$ such that for any $R>0$ exists $C>0:$%
\begin{equation}
\label{FBound}\left| F(\theta )\right| \leq C\,,\quad \,|\theta |_p\leq
R\,,\qquad \theta \in {\cal N}_{\Ckl }\,.
\end{equation}
In particular if inequality (\ref{FBound}) holds and $p^{\prime },q\in \N _0$
are such that $\left\| i_{p^{\prime },q}\right\| _{HS}<\infty $ and $%
2^q<(\left\| i_{p^{\prime },q}\right\| _{HS}R/e)^2$ then 
$$
\left\| \Phi \right\| _{-p^{\prime },q,-1}\leq \sqrt{2}C\,. 
$$
\end{theorem}

\noindent Of course the same is true for the $T$-transform.\bigskip\ 

As consequences of the above theorem we discuss the convergence of a
sequence of distributions as well as an integration theorem.

\begin{corollary}
\label{MYseq}Let $(F_n\,,\,n\in \N )$ denote a sequence of entire functions $%
F_n:{\cal N}_{\Ckl }\rightarrow \C $ such that\smallskip\\1. $(F_n\,(\theta
)\;,\,n\in \N )$ is a Cauchy sequence for all $\theta \in {\cal N}_{\Ckl }$.%
\smallskip\\2. There exists $p\in \N _0$ such that $\forall R>0\ \exists C>0:
$%
$$
|F_n(\theta )|\leq C\,,\quad \,|\theta |_p\leq R\,,\,\theta \in {\cal N}_{%
\Ckl } 
$$
uniformly in $n\in \N $. \smallskip\ \\Then the sequence $(\Phi
_n=S^{-1}F_n\,,\,n\in \N )$ converges weakly to a distribution $\Phi \in 
{\cal Y}^{\prime }$, i.e., 
\begin{equation}
\label{limPhin}\langle \!\langle \Phi ,\varphi \rangle \!\rangle =\lim
_{n\rightarrow \infty }\langle \!\langle \Phi _n,\varphi \rangle \!\rangle
\,,\qquad \varphi \in {\cal Y\,.}
\end{equation}
\end{corollary}

\TeXButton{Proof}{\proof}Let $\Phi _n:=S^{-1}F_n\in {\cal Y}^{\prime }$.
Since ${\cal E}$ is total in ${\cal Y}$ assumption 1 implies that $\left(
\langle \!\langle \Phi _n,\varphi \rangle \!\rangle \,\,,\,n\in \N \right) $
is a Cauchy sequence for all $\varphi \in {\cal Y}$. Theorem \ref{MYChar}
implies that there exists $p^{\prime }\in \N $ such that $\forall q\in \N \
\exists C>0:$%
$$
\left\| \Phi _n\right\| _{-p^{\prime },q,-1}\leq \sqrt{2}C\,. 
$$
Thus%
$$
\left| \lim _{n\rightarrow \infty }\langle \!\langle \Phi _n,\varphi \rangle
\!\rangle \right| \leq \sqrt{2}C\left\| \varphi \right\| _{p^{\prime
},-q,+1} 
$$
which proves the continuity of the linear functional $\Phi $ defined by (\ref
{limPhin}).\TeXButton{End Proof}{\endproof}\bigskip\ 

Now we are going to prove the analog of Theorem \ref{Bochner}. Since the
representation of ${\cal Y}^{\prime }$ involves a projective limit, it is
more convenient to use Pettis integration instead of Bochner integration.

\begin{corollary}
\label{PettisI}Let $(\Lambda ,{\cal A},\nu )$ be a measure space and $%
\lambda \mapsto \Phi _\lambda $ a mapping from $\Lambda $ to ${\cal Y}%
^{\prime }.$ We assume that the $S$-transform $F_\lambda =S\Phi _\lambda $
satisfies the following conditions:\smallskip\ \\1. for every $\theta \in 
{\cal N}_{\Ckl }$ the mapping $\lambda \mapsto F_\lambda (\theta )\,$is
measurable,\smallskip\ \\2. there exists $p\in \N $ such that $\forall R>0\
\exists C\in L^1(\nu )$:%
$$
|F_\lambda (\theta )|\leq C_\lambda \,,\quad \,|\theta |_p\leq R\,,\,\theta
\in {\cal N}_{\Ckl } 
$$
for almost all $\lambda \in \Lambda $. \smallskip\ \\Then $\lambda \mapsto
\Phi _\lambda $ is Pettis integrable i.e., $\exists \Phi \in {\cal Y}%
^{\prime }:$%
$$
\langle \!\langle \Phi ,\varphi \rangle \!\rangle =\int_\Lambda \langle
\!\langle \Phi _\lambda ,\varphi \rangle \!\rangle \;{\rm d}\nu (\lambda
)\,,\ \,\varphi \in {\cal Y}\,. 
$$
\end{corollary}

\TeXButton{Proof}{\proof}Let $\varphi \in {\cal E}$, then assumption 1
implies that $\lambda \mapsto \langle \!\langle \Phi _\lambda ,\varphi
\rangle \!\rangle $ is measurable. From Theorem \ref{MYChar} and assumption
2 we know that there exists $p^{\prime }\in \N $ such that $\forall q\in \N %
\;\exists C\in L^1(\nu ):\left\| \Phi _\lambda \right\| _{-p^{\prime
},q,-1}\leq \sqrt{2}C_\lambda $. Thus 
$$
\int_\Lambda \left| \langle \!\langle \Phi _\lambda ,\varphi \rangle
\!\rangle \right| {\rm d}\nu (\lambda )\,\leq \sqrt{2}\left\| \varphi
\right\| _{p^{\prime },-q,1}\int_\Lambda C_\lambda \,{\rm d}\nu (\lambda
)\,\,,\,\varphi \in {\cal E\,\,.} 
$$
This implies that $\Phi $ defined by 
$$
\langle \!\langle \Phi ,\varphi \rangle \!\rangle =\int_\Lambda \langle
\!\langle \Phi _\lambda ,\varphi \rangle \!\rangle \;{\rm d}\nu (\lambda ) 
$$
is well defined for all $\varphi \in {\cal E}$. The definition of $\Phi $
may now be extended by continuity to $\varphi \in {\cal Y}$.%
\TeXButton{End Proof}{\endproof}
\LaTeXparent{dis3.tex}

\section{The scaling operator}

In this section we collect some facts about the so called ``scaling
operator'', which has some interesting applications in the theory of Feynman
integrals. We first define this operator on a small domain, collect some
properties and afterwards extend the domain to include more interesting
examples. For the definition we follow \cite{HKPS93}.

Let $\varphi \in ({\cal N})$ be given. Without loss of generality we assume
that $\varphi $ coincides with its pointwisely defined, continuous version.
Let $z\in \C $ be given and define%
$$
(\sigma _z\varphi )(\omega ):=\varphi (z\omega )\qquad . 
$$

\begin{theorem}
\label{sigmaz}For all $z\in \C $ the mapping $\sigma _z:\varphi \mapsto
\sigma _z\varphi $ is continuous from $({\cal N})$ into itself.
\end{theorem}

We will give a proof later (which is related to the one in \cite{HKPS93}).
Let $\varphi \in ({\cal N})$ be given by its chaos expansion $\varphi
=\sum_{n=0}^\infty \langle :\omega ^{\otimes n}:,\varphi ^{(n)}\rangle $. It
is easy to calculate the expansion 
$$
\sigma _z\varphi =\sum_{n=0}^\infty \langle :\omega ^{\otimes n}:,\tilde
\varphi ^{(n)}\rangle \ , 
$$
\begin{equation}
\label{sigmaChaos}\ \tilde \varphi ^{(n)}=z^n\sum_{k=0}^\infty \frac{(n+2k)!%
}{k!\;n!}\left( \frac{z^2-1}2\right) ^k\,{\rm tr}^k\varphi ^{(n+2k)} 
\end{equation}
where ${\rm tr}^k\varphi ^{(n+2k)}$ is shorthand for the contraction (def.
eq. (\ref{contraction})) of iterated traces (def. eq. (\ref{Trace})) with $%
\varphi ^{(n+2k)}$:%
$$
{\rm tr}^k\varphi ^{(n+2k)}:=\left( {\rm Tr}^{\otimes k},\varphi
^{(n+2k)}\right) _{{\cal H}^{\otimes 2k}}\ \in {\cal N}_{\Ckl }^{\hat
\otimes n}\ . 
$$

\begin{lemma}
{\rm Tr}$\in {\cal H}_{-p}^{\hat \otimes 2}$ if and only if $i_{p,0}:{\cal H}%
_{p\text{ }}\rightarrow {\cal H}$ is of Hilbert--Schmidt type. Moreover%
$$
\left| {\rm Tr}\right| _{-p}=\left\| i_{p,0}\right\| _{HS}\ . 
$$
\end{lemma}

\TeXButton{Proof}{\proof}Let $\{e_j|\ j\in \N _0\}$ be an orthonormal basis
of ${\cal H}$. Then the expansion%
$$
{\rm Tr}=\sum_{j=0}^\infty e_j\otimes e_j 
$$
is valid, and we may calculate%
$$
\left| {\rm Tr}\right| _{-p}^2=\left| \sum_{j=0}^\infty e_j\otimes
e_j\right| _{-p}^2=\sum_{j=0}^\infty |e_j|_{-p}^2=\left\| i_{0,-p}\right\|
_{HS}^2=\left\| i_{p,0}\right\| _{HS}^2\ . 
$$
\TeXButton{End Proof}{\endproof}\bigskip\ 

For $p>0$ large enough, the estimate 
\begin{equation}
\label{trBound}\left| {\rm tr}^k\varphi ^{(n+2k)}\right| _p\leq \left| {\rm %
Tr}\right| _{-p}^k\left| \varphi ^{(n+2k)}\right| _p=\left\| i_{p,0}\right\|
_{HS}^k\cdot \left| \varphi ^{(n+2k)}\right| _p 
\end{equation}
shows that smooth kernels $\varphi ^{(n)}\in {\cal N}_{\Ckl }^{\hat \otimes
n}$ allow the action of iterated traces. Now we are ready to prove a
statement which is a little more general then Theorem \ref{sigmaz}.

\begin{theorem}
\label{sigmazHpq}Let $z\in \C $, and $p>0$ be such that $i_{p,0}$ is of
Hilbert--Schmidt type. If $q^{\prime }$ and $q^{\prime }-q$ are large
enough, $\sigma _z$ is continuous from $({\cal H}_p)_{q^{\prime }}$ into $(%
{\cal H}_p)_q$.
\end{theorem}

\TeXButton{Proof}{\proof}Note that $\frac{(2k)!}{(k!)^22^{2k}}\leq 1$ for
all $k\in \N _0$. Using this and the estimate (\ref{trBound}) we can
estimate as follows%
\begin{eqnarray*}
\left| \tilde \varphi ^{(n)}\right| _p & \leq & (n!)^{-1/2}|z|^n %
\sum_{k=0}^\infty {\binom{n+2k}{2k}}^{\frac12}\frac{\sqrt{(2k)!}}{k!\;2^k}|z^2-1|^k %
\sqrt{(n+2k)!}\left\| i_{p,0}\right\| _{HS}^k %
\left| \varphi ^{(n+2k)}\right| _p %
\\& \leq & (n!)^{-1/2}|z|^n\sum_{k=0}^\infty {\binom{n+k}k}^{\frac12}|z^2-1|^{k/2} %
\left\| i_{p,0}\right\| _{HS}^{k/2}\cdot \sqrt{(n+k)!}\left| %
\varphi ^{(n+k)}\right| _p %
\\& \leq & (n!)^{-1/2}2^{-nq^{\prime }/2}|z|^n\left( \sum_{k=0}^\infty %
\binom{n+k}k 2^{-q^{\prime }k}\left( |z^2-1|\left\| i_{p,0}\right\| _{HS} %
\right) ^{k}\right) ^{\frac 12} \cdot %
\\& & \hspace*{3cm} \cdot \left( \sum_{k=0}^\infty (n+k)! \,%
2^{q^{\prime }(n+k)}\left| \varphi ^{(n+k)}\right| ^2_p \right) ^{\frac 12} %
\\& \leq & \left\| \varphi \right\| _{p,q^{\prime }}(n!)^{-1/2} %
\, 2^{-nq^{ \prime }/2} |z|^n\left( 1-2^{-q^{\prime }} |z^2-1| %
\left\| i_{p,0} %
\right\| _{HS} \right) ^{-\frac{n+1}2} 
\end{eqnarray*}
if $q^{\prime }$ is such that $2^{q^{\prime }}>|z^2-1|\left\|
i_{p,0}\right\| _{HS}$. Then we get 
$$
\left\| \sigma _z\varphi \right\| _{p,q}^2\leq \left\| \varphi \right\|
_{p,q^{\prime }}^2\cdot \sum_{n=0}^\infty 2^{n(q-q^{\prime })}|z|^{2n}\left(
1-2^{-q^{\prime }}|z^2-1|\left\| i_{p,0}\right\| _{HS} \right) ^{-(n+1)}\ . 
$$
The sum on the right hand side converges if $q^{\prime }-q$ is large enough.%
\TeXButton{End Proof}{\endproof}\bigskip\ 

\noindent {\bf Note}. We can also give a completely different proof of
Theorem \ref{sigmaz} using the powerful theorem describing the space $({\cal %
N})$. Since $({\cal N})={\cal E}_{\min }^2({\cal N}^{\prime }),$ every test
function $\varphi \in ({\cal N})$ has a version which has an extension to a
function from ${\cal E}_{\min }^2({\cal N}_{\Ckl }^{\prime })$. The function%
$$
\omega \mapsto \sigma _z\varphi (\omega )=\varphi (z\omega ) 
$$
is also entire of the same growth. Since $\varphi $ is of minimal type also $%
\omega \mapsto \varphi (z\omega )$ is of minimal type. Thus $\sigma
_z\varphi \in ({\cal N})$. In fact $\sigma _z:{\cal E}_{\min }^2({\cal N}_{%
\Ckl }^{\prime })\rightarrow {\cal E}_{\min }^2({\cal N}_{\Ckl }^{\prime })$
is continuous. The same argument based on Theorem \ref{N1E1min} shows:

\begin{theorem}
For all $z\in \C $ the mapping $\sigma _z$ is continuous from $({\cal N})^1$
into $({\cal N})^1$.
\end{theorem}

\noindent This kind of argument also shows

\begin{theorem}
For $\varphi ,\psi \in ({\cal N})$ the following equation holds $\sigma
_z(\varphi \cdot \psi )=(\sigma _z\varphi )\cdot (\sigma _z\psi )$.
\end{theorem}

\noindent {\bf Note}. To prove these relations without referring to the
description of $({\cal N})$, only based on chaos expansions, requires much
more effort.\bigskip\ 

Since $\sigma _z$ is continuous from $({\cal N})$ into $({\cal N})$ it is
possible to define its adjoint operator $\sigma _z^{\dagger }:({\cal N}%
)^{\prime }\rightarrow ({\cal N})^{\prime }$ by 
\begin{equation}
\label{sigmazadj}\langle \!\langle \sigma _z^{\dagger }\Phi ,\psi \rangle
\!\rangle =\langle \!\langle \Phi ,\sigma _z\psi \rangle \!\rangle \ ,\quad
\psi \in ({\cal N})\ .
\end{equation}
Of course there also exists a well defined extension $\sigma _z^{\dagger }:(%
{\cal N})^{-1}\rightarrow ({\cal N})^{-1}$.\bigskip\ 

The next Lemma will be useful later.

\begin{lemma}
\label{scalJz} For $z\in \C \ ,\ \Phi \in ({\cal N})^{-1}$ we have 
$$
\sigma _z^{\dagger }\Phi ={\rm J}_z\diamond \Gamma _z\Phi  
$$
in particular%
$$
\sigma _z^{\dagger }\1 ={\rm J}_z 
$$
where ${\rm J}_z$ is defined in Example \arabic{JzDef} .
\end{lemma}

\TeXButton{Proof}{\proof}The following calculation is valid%
$$
S(\sigma _z^{\dagger }\Phi )(\xi )=\langle \!\langle \Phi ,:\exp \langle
\cdot ,z\xi \rangle :\rangle \!\rangle e^{-\frac 12(1-z^2)\langle \xi ,\xi
\rangle } 
$$
$$
=S\Phi (z\xi )\cdot S{\rm J}_z(\xi ) \hspace{3mm} 
$$
$$
=S\left( \Gamma _z\Phi \diamond {\rm J}_z\right) (\xi )\ . \hspace{2mm} 
$$
\TeXButton{End Proof}{\endproof}\bigskip\ 

We can also derive some useful formulae concerning the pointwise product of 
{\rm J}$_z\in ({\cal N})^{\prime }$ with a test functional.

\begin{lemma}
\label{JzphiLem} Let $\varphi \in ({\cal N})$ then 
\begin{equation}
\label{Jzphi}{\rm J}_z\varphi =\sigma _z^{\dagger }(\sigma _z\varphi )
\end{equation}
or if we prefer to rewrite the r.h.s. as a Wick product 
\begin{equation}
\label{JzWick}{\rm J}_z\varphi ={\rm J}_z\diamond \Gamma _z(\sigma _z\varphi
)\ .
\end{equation}
\end{lemma}

\TeXButton{Proof}{\proof}Let $\varphi ,\psi \in ({\cal N})$%
$$
\langle \!\langle {\rm J}_z\varphi ,\psi \rangle \!\rangle =\langle
\!\langle \sigma _z^{\dagger }\1 ,\varphi \cdot \psi \rangle \!\rangle
=\langle \!\langle \sigma _z\varphi ,\sigma _z\psi \rangle \!\rangle
=\langle \!\langle \sigma _z^{\dagger }(\sigma _z\varphi ),\psi \rangle
\!\rangle \ , 
$$
hence (\ref{Jzphi}) follows.\TeXButton{End Proof}{\endproof}\bigskip\ 

\example\newcounter{JzWick} 
\setcounter{JzWick}{\value{example}}
\label{JzWickP}
 Let us discuss the above formula for the concrete choice $\varphi
=\langle \cdot ,\eta \rangle ^{n\,}\,,\ \eta \in {\cal N}$. Then 
\begin{equation}
\label{JzMoments}\langle \cdot ,\eta \rangle ^{n\,}{\rm J}%
_z=z^{2n}\sum_{k=0}^{[n/2]}\frac{n!}{k!\,(n-2k)!}\left( \frac 1{2z^2}|\eta
|^2\right) ^k\langle \cdot ,\eta \rangle ^{\diamond (n-2k)\,}\diamond {\rm J}%
_z
\end{equation}
by use of formula (\ref{JzWick}) and the expansion 
$$
\langle \cdot ,\eta \rangle ^{n\,}=\sum_{k=0}^{[n/2]}\frac{n!}{k!\,(n-2k)!}%
\left( \frac 12|\eta |^2\right) ^k:\langle \cdot ,\eta \rangle ^{n-2k\,}:\,. 
$$
Formula (\ref{JzMoments}) allows to express pointwise products by Wick
products which are well defined in more general situations. The right hand
side immediately extends to $\eta \in {\cal H}_{\Ckl }$. Then formula (\ref
{JzMoments}) may serve as a definition of the pointwise product on the left
hand side. By polarization this is also possible for mixed products.
Examples:\\1)%
$$
\langle \cdot ,\eta \rangle {\rm J}_z=z^2\langle \omega ,\eta \rangle
^{\,}\diamond {\rm J}_z. 
$$
Note that pointwise multiplication has a non-trivial translation to Wick
multiptication.\\2) 
$$
\langle \cdot ,\eta \rangle \langle \cdot ,\xi \rangle {\rm J}_z=z^4\langle
\cdot ,\eta \rangle \diamond \langle \cdot ,\xi \rangle \diamond {\rm J}%
_z+z^2(\eta ,\xi ){\rm J}_z. 
$$
This formula allows to read off the ``covariance'' of ${\rm J}_z$%
$$
\E \left( \langle \cdot ,\eta \rangle \langle \cdot ,\xi \rangle {\rm J}%
_z\right) =z^2(\eta ,\xi )\,. 
$$

\bigskip\ 

For the applications we have in mind the domain of $\sigma _z$ given in
Theorem \ref{sigmazHpq} is too small. We want to apply $\sigma _z$ to
Donsker's delta and the interaction term in Feynman integrals. Both have
kernels in ${\cal H}_{\Ckl }^{\hat \otimes n}$ but obviously not in ${\cal H}%
_{p,\Ckl }^{\hat \otimes n}$ for $p>0$. Thus we need to study extensions of $%
\sigma _z.$ Of course this is not trivial, since we may construct elements
in ${\cal H}_{\Ckl }^{\hat \otimes n}$ where a contraction with iterated
traces is {\it not} well defined. On the other hand kernels consisting of
tensor products raise no problems in this context. Let $\xi _j\in {\cal H}_{%
\Ckl },\ 1\leq j\leq n+2k$ then 
$$
{\rm tr}^k(\xi _1\hat \otimes \cdots \hat \otimes \xi _{n+2k})=\frac
1{(n+2k)!}\sum_\pi \left( \xi _{\pi _1},\xi _{\pi _2}\right) \cdots \left(
\xi _{\pi _{2k-1}},\xi _{\pi _{2k}}\right) \;\xi _{\pi _{2k+1}}\hat \otimes
\cdots \hat \otimes \xi _{\pi _{n+2k}}\text{ ,} 
$$
where the sum extends over the symmetric group of order $n+2k$. Obviously
also finite sums of tensor products are allowed. The next step is to discuss
infinite sums of tensor products. We give a sufficient condition discussed
in \cite{JK93}.

\begin{proposition}
\label{JoKal}Let $\varphi ^{(n)}\in {\cal H}_{\Ckl }^{\hat \otimes n}$.
Suppose there exists a complete orthonormal system $\{e_j\,|\,\,j\in \N _0\}$
of ${\cal H}$ such that the expansion 
$$
\varphi ^{(n)}=\sum_{j_1,\cdots ,j_n=1}^\infty a_{j_1,\cdots
,j_n}\,e_{j_1}\otimes \cdots \otimes e_{j_n} 
$$
holds. If the coefficients $(a_{j_1,\cdots ,j_n})$ are in $l_1$, i.e., 
\begin{equation}
\label{CnDef}C_n:=\sum_{j_1,\cdots ,j_n=0}^\infty \left| a_{j_1,\cdots
,j_n}\right| 
\end{equation}
is finite, then for every $k,\ 0\leq k\leq [\frac n2]\qquad {\rm tr}%
^k\varphi ^{(n)}$ exists in ${\cal H}_{\Ckl }^{\hat \otimes (n-2k)}$ and is
given by 
\begin{equation}
\label{CnBound}{\rm tr}^k\varphi ^{(n)}=\sum_{j_{2k+1},\cdots ,j_n=0}^\infty
\left( \sum_{j_1,\cdots ,j_k=0}^\infty a_{\underbrace{j_1,j_1},\cdots ,%
\underbrace{j_k,j_k},\underbrace{j_{2k+1},\cdots ,j_n}}\right)
e_{j_{2k+1}}\otimes \cdots \otimes e_{j_n}\ .
\end{equation}
Moreover%
$$
\left| {\rm tr}^k\varphi ^{(n)}\right| _{{\cal H}_{\Ckkl }^{\hat \otimes
(n-2k)}}\leq C_n\ . 
$$
\end{proposition}

Now we are going to use Proposition \ref{JoKal} to extend $\sigma _z$. We
will take the chaos expansion of $\sigma _z\varphi $ given by equation (\ref
{sigmaChaos}) as the fundamental definition, whenever this is well defined.
The question arises to find sufficient conditions to ensure that $\sigma
_z\varphi \in {\cal M}$ (or $L^2(\mu )$ or ${\cal G}^{\prime }$) or at least
the existence of the expectation value%
$$
\E (\sigma _z\varphi )=\tilde \varphi ^{(0)}\ . 
$$
We will formulate this type of conditions in terms of the $C_n$ appearing in
Proposition \ref{JoKal}.

\begin{lemma}
Let $\alpha \in \R $ such that $2^\alpha >|z^2-1|$ and assume that $K_\alpha 
$ defined by 
$$
K_\alpha ^2=\sum_{n=0}^\infty n!\,2^{\alpha n}\,C_n^2 
$$
is finite. Then%
$$
|\tilde \varphi ^{(n)}|^2\leq K_\alpha ^2\frac 1{n!}|z|^{2n}2^{-\alpha
n}\left( 1-2^{-\alpha /2}|z^2-1|\right) ^{-(n+1)} 
$$
\end{lemma}

\TeXButton{Proof}{\proof}The proof is very similar to the proof of Theorem 
\ref{sigmazHpq}. We can use the bound (\ref{CnBound}) and $\frac 1{k!\,2^k}%
\sqrt{(2k)!}\leq 1$ in the following estimate:%
\begin{eqnarray*}
& & \hspace*{-1cm} |\tilde \varphi ^{(n)}|\leq |z|^n\frac 1{\sqrt{n!}}\sum_{k=0}^\infty %
\sqrt{ \frac{(n+2k)!}{n!\,(2k)!}}\frac{\sqrt{(2k)!}}{k!\,2^k}\,|z^2-1|^k %
\sqrt{(n+2k)}C_{n+2k} %
\\ & \leq & |z|^n(n!\,2^{\alpha n})^{-1/2}\left( \sum_{k=0}^\infty %
\binom{n+2k}{2k}%
|z^2-1|^{2k} 2^{-2\alpha k}\right) ^{1/2}\left( \sum_{k=0}^\infty %
(n+2k)!\,2^{\alpha (n+2k)}C_{n+2k}^2\right) ^{1/2} %
\\ & \leq & K_\alpha (n!\,2^{\alpha n})^{-1/2}|z|^n\left( %
\sum_{k=0}^\infty \binom{n+k}k|z^2-1|^{k}2^{-\alpha k}\right) ^{1/2} %
\\ &=& K_\alpha (n!\,2^{\alpha n})^{-1/2}|z|^n(1-2^\alpha |z^2-1|%
 )^{-\frac{n+1}2} %
\end{eqnarray*}
if $2^\alpha >|z^2-1|$.\TeXButton{End Proof}{\endproof}\medskip\ 

\begin{proposition}
\label{sigmaExt} Assume all definitions as before. \smallskip\ \\If $\alpha
\in \R $ is such that $2^\alpha >|z^2-1|$, then $K_\alpha <\infty $ implies $%
\sigma _z\varphi \in {\cal G}^{\prime }$. \smallskip\ \\If $\alpha \in \R $
is such that $2^\alpha >|z|^2+|z^2-1|$, then $K_\alpha <\infty $ implies $%
\sigma _z\varphi \in {\cal M}$.
\end{proposition}

\noindent {\bf Note.} In the case $z=\sqrt{i}$ we obtain 
$$
K_\alpha <\infty \ ,\ \alpha > 0.5 \ \Rightarrow \ \sigma _z\varphi \in 
{\cal G}^{\prime } 
$$
$$
K_\alpha <\infty \ ,\ \alpha \,\QTATOP{>}{\sim }\,1.27\ \Rightarrow \ \sigma
_z\varphi \in {\cal M}\text{ .} 
$$

\TeXButton{Proof}{\proof}For $\alpha ,\beta \in \R $ with $2^\alpha >|z^2-1|$
we can estimate 
$$
\left\| \sigma _z\varphi \right\| _{0,\beta }^2=\sum_{n=0}^\infty
n!\,2^{\beta n}|\tilde \varphi ^{(n)}|^2\hspace*{3.2cm} 
$$
$$
\hspace*{3.2cm} \leq K_\alpha ^2\sum_{n=0}^\infty |z|^{2n}2^{n(\beta -\alpha
)}(1-|z^2-1| \, 2^{-\alpha })^{-(n+1)}. 
$$
1) If $\beta <0$ is chosen small enough the series on the right hand side is
convergent, such that $\sigma _z\varphi \in {\cal G}^{\prime }$.\smallskip\ 
\\2) If $\alpha \in \R $ is such that $2^\alpha >|z|^2+|z^2-1|$ then $|z|^2
\, 2^{-\alpha }(1-|z^2-1| \, 2^{-\alpha })^{-1} <1$. Hence there exists $%
\beta >0$ such that $\left\| \sigma _z \varphi \right\| _{0,\beta }<\infty $%
, which proves $\sigma _z\varphi \in {\cal M}$.\TeXButton{End Proof}
{\endproof}\bigskip\ 

Now let us check if we get less restrictive conditions if we only define $%
\tilde \varphi ^{(0)}$ (and interpret this as $\E (\sigma _z\varphi )$). We
have%
$$
|\tilde \varphi ^{(0)}|\leq \sum_{k=0}^\infty \frac{(2k)!}{k! \, 2^k}
\,|z^2-1|^kC_{2k}\hspace*{8mm} 
$$
$$
\hspace*{8mm} \leq \sum_{k=0}^\infty \sqrt{(2k)!} \, |z^2-1|^k\,C_{2k}\;. 
$$
This series is convergent if $\sum_{k=0}^\infty (2k)!\,2^{2k\alpha }C_{2k}^2$
is finite for $\alpha \in \R $ such that $2^\alpha >|z^2-1|$, i.e., we get
the same type of growth, but conditions are only put on the kernels $\varphi
^{(2k)}$ of even order. \bigskip\ 

One important application of the scaling operator is the following theorem
from \cite{S93} (see also \cite{HKPS93}) which gives an explicit relation to
pointwise multiplication with ${\rm J}_z$:\medskip\ 

\begin{theorem}
\label{Jsigmaz}\ Let $\varphi _n$ be a sequence of test functionals in $%
\left( {\cal N}\right) $. Then the following statements are equivalent:%
\smallskip\ \\\ \ $\left( i\right) $ The sequence ${\rm J}_z\varphi
_n\rightarrow \Psi $ converges in $\left( {\cal N}\right) ^{\prime }$.%
\smallskip\ \\\ \ \ $\left( ii\right) $ The sequence $\sigma _z\varphi _n$
converges in $\left( {\cal N}\right) ^{\prime }$.\smallskip\ \\\ \ \ $\left(
iii\right) $ The sequence $\E \left( \psi \,\cdot \sigma _z\varphi _n\right) 
$ converges for all $\psi \in \left( {\cal N}\right) $.\smallskip\\The
action of $\Psi $ is given by 
$$
\left\langle \!\left\langle \Psi ,\psi \right\rangle \!\right\rangle =%
\stackunder{n\rightarrow \infty }{\lim }{\E}\left( \sigma _z\left( \varphi
_n\psi \right) \right) , 
$$
if one of the conditions $\left( i\right) $\ to $\left( iii\right) $\ holds.
\end{theorem}

\noindent The proof is an immediate consequence of Lemma \ref{JzphiLem}.%
\medskip\ 

\noindent One may be tempted to extend $\sigma _z$ by continuity arguments.
This has to be done with great care as the following illustrative example
shows.\medskip

\example  \newcounter{Warnung} \setcounter{Warnung}{\value{example}} \label
{WarnungP} We will construct a sequence $\{\varphi _n|\,n\in \N \}\subset (%
{\cal N})$ converging to zero in the topology of $L^2(\mu )$. But for $z\neq
1\,$the sequence $\{\sigma _z\varphi _n|\,n\in \N \}$ converges to a
constant different from zero in $L^2(\mu )$.

Let $\{\varphi _n^{(2)}\in {\cal N}_{\Ckl }^{\hat \otimes 2}|\ n\in \N \}$
denote a sequence converging to $\varphi ^{(2)}\in {\cal H}_{\Ckl }^{\hat
\otimes 2}$, such that the sequence $\{c_n:={\rm tr\,}\varphi _n^{(2)}|\
n\in \N \}\subset \C $ is divergent. We consider the sequence of test
functions%
$$
\varphi _n=\frac 1{c_n}\langle \omega ^{\otimes 2},\varphi _n^{(2)}\rangle 
$$
which by construction converges to zero in $L^2(\mu )$.

\noindent On the other hand%
$$
\sigma _z\varphi _n=\frac 1{c_n}\langle \omega ^{\otimes 2},z^2\varphi
_n^{(2)}\rangle +\frac 1{c_n}(z^2-1){\rm tr\,}\varphi _n^{(2)} 
$$
$$
=\frac 1{c_n}\langle \omega ^{\otimes 2},z^2\varphi _n^{(2)}\rangle +(z^2-1) 
$$
such that 
$$
\lim _{n\rightarrow \infty }\sigma _z\varphi _n=(z^2-1) 
$$
which is different from zero for $z\neq 1$, concluding the example.
\LaTeXparent{dis3.tex}

\section{Donsker`s delta
\texorpdfstring{``}{"}function" \label{DeltaSec}}

\subsection{Complex scaling of Donsker's delta \label{Deltascaling}}

\noindent Consider again the $S$-transform of Donsker's delta function:%
$$
F(\theta )=\frac 1{\sqrt{2\pi \langle \eta ,\eta \rangle }}\exp \left(
-\frac 1{2\langle \eta ,\eta \rangle }\left( \langle \theta ,\eta \rangle
-a\right) ^{{2}}\right) ,\qquad \theta \in {\cal N}_{\Ckl }\ ,\ \eta \in 
{\cal H}_{+}\ \text{ .} 
$$
This is clearly analytic in the parameter $a\in {\R}$. We can thus extend to
complex $a$ and the resulting expression is still a U-functional. The same
argument holds if we extend to $\eta \in {\cal H}_{\Ckl }$. We only have to
be careful with regard to the square root. For our purpose it is convenient
to cut the complex plane along the negative axis. So we have to exclude $%
\eta \in {\cal H}_{\Ckl }$ with $\langle \eta ,\eta \rangle $ negative.
Hence by Theorem \ref{CharacTh} it is possible to define $\delta \left(
\langle \omega ,\eta \rangle -a\right) $ for this choice of parameters.
First of all we calculate the chaos expansion of Donsker's delta.\ 

\begin{lemma}
Let $a\in {\C\ ,\ }\eta \in {\cal H}_{\Ckl }\ ,\ \arg \langle \eta ,\eta
\rangle \neq \pi $ , then 
$$
\delta (\langle \cdot ,\eta \rangle -a)=\sum_{n=0}^\infty \langle :\omega
^{\otimes n}:,f^{(n)}\rangle  
$$
where 
\begin{equation}
\label{DeltaChaos}f^{(n)}=\frac{e^{-\frac{a^2}{2\langle \eta ,\eta \rangle }}%
}{\sqrt{2\pi \langle \eta ,\eta \rangle }}\,\frac 1{n!}\,H_n\left( \frac a{%
\sqrt{2\langle \eta ,\eta \rangle }}\right) (2\langle \eta ,\eta \rangle
)^{-n/2}\,\eta ^{\otimes n}
\end{equation}
is in ${\cal H}_{\Ckl }^{\hat \otimes }$. Here $H_n$ denotes the $n^{th}$
Hermite polynomial (in the normalization of {\rm \cite{HKPS93}}).
\end{lemma}

\TeXButton{Proof}{\proof}We can expand the $S$-transform of $\delta $%
$$
S\delta (\langle \cdot ,\eta \rangle -a)(\theta )=\frac 1{\sqrt{2\pi \langle
\eta ,\eta \rangle }}\exp \left( -\frac 1{2\langle \eta ,\eta \rangle
}(\langle \theta ,\eta \rangle -a)^2\right) \hspace*{3.1cm} 
$$
$$
\hspace*{3.1cm}=\frac{e^{-\frac{a^2}{2\langle \eta ,\eta \rangle }}}{\sqrt{%
2\pi \langle \eta ,\eta \rangle }}\sum_{n=0}^\infty \frac 1{n!}H_n\left(
\frac a{\sqrt{2\langle \eta ,\eta \rangle }}\right) (2\langle \eta ,\eta
\rangle )^{-n/2}\;\langle \theta ^{\otimes n},\eta ^{\otimes n}\rangle \ . 
$$
Then it is easy to read of the kernels $f^{(n)\text{ }}$given by equation (%
\ref{DeltaChaos}).\TeXButton{End Proof}{\endproof}\bigskip\ 

To discuss the convergence of the chaos expansion we need estimates on the
growth of the sequence $\{H_n(\lambda )\;|\,\;n\in \N _0\}$ at a fixed
complex point $\lambda $. This is a well known fact for $\lambda \in \R $,
but for complex $\lambda $ the sequence grows faster.

\begin{lemma}
{\rm \cite[Th. 8.22.7 and eq. (8.23.4)]{Sz39}}\\Let $\lambda \in \C $, then%
$$
\lim _{n\rightarrow \infty }(2n)^{-1/2}\log \left\{ \frac{\Gamma (n/2+1)}{%
\Gamma (n+1)}|H_n(\lambda )|\right\} =\left| {\rm Im}(\lambda )\right| \ . 
$$
\end{lemma}

\noindent {\bf Note}. That means for $n$ large we have the asymptotic
behavior%
$$
\left| H_n(\lambda )\right| \sim \frac{\Gamma (n+1)}{\Gamma (n/2+1)}\cdot e^{%
\sqrt{2n}|{\rm Im}(\lambda )|} 
$$
\begin{equation}
\label{HnGrowth}\hspace*{4mm}\lesssim \sqrt{n!\,2^n}e^{\sqrt{2n}%
\,|{\rm Im}(\lambda )|}\ . 
\end{equation}

\begin{theorem}
\label{DeltaInMSt} Let $a\in {\C\ ,\ }\eta \in {\cal H}_{\Ckl }\ ,\ \arg
\langle \eta ,\eta \rangle \neq \pi $ be given. Then $\delta (\langle \cdot
,\eta \rangle -a)$ defined by the chaos expansion (\ref{DeltaChaos}) (or
equivalently by its $S$-transform) is in ${\cal M}^{\prime }$, i.e., for all 
$\alpha >0,\quad \left\| \delta (\langle \cdot ,\eta \rangle -a)\right\|
_{0,-\alpha }$ is finite.
\end{theorem}

\TeXButton{Proof}{\proof}We have%
\begin{eqnarray*}
\left\| \delta \right\| _{0,-\alpha }^2 &=& \sum_{n=0}^\infty %
n!\,2^{-n\alpha }|f^{(n)}|^2 %
\\ & \leq & \left| \frac {e^{-\frac{a^2}{2\langle \eta ,\eta \rangle }}}  {\sqrt{2\pi \langle \eta ,\eta \rangle }} %
\right| \sum_{n=0}^\infty \frac 1{n!}2^{-n\alpha }\left| H_n\left( %
\frac a{\sqrt{2\langle \eta ,\eta \rangle }}\right) \right| ^2\left| %
 2\langle \eta ,\eta \rangle \right| ^{-n}|\eta |^{2n} %
\\ & \lesssim & \frac {e^{\frac{|a|^2}{2|\eta |^2}}} {\sqrt{2\pi }|\eta |} %
\sum_{n=0}^\infty 2^{-n\alpha }\exp \left( 2\sqrt{2n}\;| \, %
{\rm Im}(\frac a{2\langle \eta ,\eta \rangle })|\right)  
\end{eqnarray*}
in view of (\ref{HnGrowth}). The series is convergent for any $\alpha >0$.%
\TeXButton{End Proof}{\endproof}\bigskip\ 

\noindent Now we intend to study complex scaling of a sequence of test
functionals converging to $\delta .$ This is done in the spirit of Theorem 
\ref{Jsigmaz}. Let $\eta _n\in {\cal N}$ be a sequence of real Schwartz test
functions converging to $\eta \in {\cal H}$ .

\noindent Choose $\left| \alpha \right| <\frac \pi 4$ and $z\in {\bf S}%
_\alpha \equiv \left\{ z\in {\C}\mid \arg z\in \left( -\frac \pi 4+\alpha
,\frac \pi 4+\alpha \right) \right\} $ and define 
\begin{equation}
\label{fin}\varphi _{n,z}\left( \omega \right) =\frac 1{2\pi
}\int\limits_{-ne^{-i\alpha }}^{ne^{-i\alpha }}e^{i\lambda \left(
z\left\langle \omega ,\eta _n\right\rangle -a\right) }{\rm d}\lambda \text{ .%
} 
\end{equation}
To shorten notation we call the basic sequence $\varphi _{n,1}$ simply $%
\varphi _n$. Note that given any $z$ with $\left| \arg z\right| <\frac \pi 2$
one can choose $\alpha $ such that $z$ and $1$ are in ${\bf S}_\alpha $. In
this section we will establish the following results:\medskip 

\begin{theorem}
\label{seqDelta} {\rm \cite{LLSW94b}} \\ For all $z\in {\bf S}_\alpha $ we
have:\smallskip\ \\i)$\quad \varphi _{n,z}\in \left( {\cal N}\right) $.%
\smallskip\ \\ii)$\quad \sigma _z\varphi _n=\varphi _{n,z}$.\smallskip\ \\%
iii)$\quad \varphi _n\rightarrow \delta $ in $\left( {\cal N}\right)
^{\prime }$.\smallskip\ \\iv)$\quad \sigma _z\varphi _n$ converges in $%
\left( {\cal N}\right) ^{\prime }$.\\ \TeXButton{10mm}{\hspace*{10mm}}The
limit element is called $\sigma _z\delta $.
\end{theorem}

\noindent {\bf Remark.} The limit elements in (iii) and (iv) do not depend
on $\alpha .$\medskip\ 

\TeXButton{Proof}{\proof}{\bf i)} First of all we calculate the $S$%
-transform of the integrand of equation (\ref{fin}), $\theta \in {\cal N}_{%
\Ckl }$:%
$$
S\left( \exp \left( i\lambda \left( z\left\langle \omega ,\eta
_n\right\rangle -a\right) \right) \right) \left( \theta \right) =\exp \left(
-\frac 12z^2\lambda ^2\left| \eta _n\right| _0^2+i\lambda \left( z\langle
\theta ,\eta _n\rangle -a\right) \right) \text{ .} 
$$
This fulfills the requirements of Theorem \ref{Bochner}, thus the integral (%
\ref{fin}) is well-defined. Hence%
\begin{eqnarray*}
S\varphi _{n,z}\left( \theta \right) &=&\frac 1{2\pi }\int\limits_{-ne^{-i \alpha }}^{ne^{-i\alpha }}S\left( \exp \left( i\lambda \left( z\left \langle \omega ,\eta _n\right\rangle -a\right) \right) \right) \left( %
\theta \right) {\rm d}\lambda %
\\&=&\frac 1{2\pi }\int\limits_{-ne^{-i\alpha }}^{ne^{-i\alpha }}\exp %
 \left( -\frac 12z^2%
\lambda ^2\left| \eta _n\right| _0^2+i\lambda \left( z\langle \theta , %
\eta _n\rangle -a\right)%
 \right) {\rm d}\lambda \text{ .}
\end{eqnarray*} 
We substitute $\nu =e^{i\alpha }\lambda $ , this leads to 
\begin{equation}
\label{esfi}S\varphi _{n,z}\left( \theta \right) =\frac{e^{-i\alpha }}{2\pi }%
\int\limits_{-n}^n\exp \left( -\frac 12z^2e^{-2i\alpha }\nu ^2\left| \eta
_n\right| _0^2+ie^{-i\alpha }\nu \left( z\langle \theta ,\eta _n\rangle
-a\right) \right) {\rm d}\nu . 
\end{equation}
Now take the absolute value%
\begin{eqnarray*}
  \left| S\varphi _{n,z}\left( \theta \right) \right|  %
& \leq &\frac 1{2\pi }\int\limits_{-n}^n %
\exp \left( \frac 12\left| z\right| ^2\nu ^2\left| \eta _n\right| _0^2 %
+\left| \nu \right| \left| z\right| \left| \theta \right|_{-p} \left| %
 \eta _n \right|_p %
+\left| \nu \right| \left| a\right| \right) {\rm d}\nu %
\\& \leq  & \frac 1{2\pi }\int\limits_{-n}^n\exp \left( \frac 12\left| %
z\right| ^2\nu ^2 \left| \eta _n\right| _0^2+\frac 1{2s^2}\nu ^2\left| %
z\right| ^2\left| \eta _n\right|_p ^2 %
+\frac 12s^2\left| \theta \right|_{-p}^{2}+\left| \nu \right| %
 \left| a\right| \right) {\rm d}\nu %
\\& \leq  & \frac n\pi \exp \left( \frac{n^2}2\left| z\right| ^2\left( %
\left| \eta_n \right|_0^2+\frac 1{s^2} \left| \eta_n \right|_p^2\right) %
+n\left| a\right| \right) \exp\left(+\frac 12s^2\left| \theta %
\right|_{-p}^{2}\right)\text{ .}
\end{eqnarray*}
This estimate holds for all $s\in {\R}$ and $p\in {\N}.$ Thus it fulfills
the requirements of the characterization Theorem \ref{Charac(N)} and we
arrive at $\varphi _{n,z}\in \left( {\cal N}\right) .$\medskip\ \ 

\noindent {\bf ii)} Now we study the action of $\sigma _z$ on $\varphi _n$.
A direct computation yields%
\begin{eqnarray*}
& & \hspace*{-1cm} \varphi _n\left( \omega \right) =\frac 1{2\pi }\int\limits_{-ne^{-i\alpha }}^{ne^{-i\alpha }}e^{i\lambda \left( \left\langle \omega ,\eta _n\right\rangle -a\right) }{\rm d}\lambda  %
\\ &=& \frac 1{2\pi i\left( \left\langle \omega ,\eta _n\right \rangle -a\right) }\left( \exp \left( ine^{-i\alpha }\left( \left\langle \omega ,\eta %
_n\right\rangle -a\right) \right) -\exp \left( -ine^{-i\alpha }\left( %
\left\langle \omega ,\eta _n\right\rangle -a\right) \right) \right) %
\\ &=& \frac 1{\pi \left( \left\langle \omega ,\eta _n\right\rangle -a\right)}\sin \left( ne^{-i\alpha }\left( \left\langle \omega ,\eta _n\right\rangle%
-a\right) \right) \text{ .} %
\end {eqnarray*}
This is defined pointwise and continuous. Thus%
$$
\sigma _z\varphi _n\left( \omega \right) =\frac 1{\pi \left( z\left\langle
\omega ,\eta _n\right\rangle -a\right) }\sin \left( ne^{-i\alpha }\left(
z\left\langle \omega ,\eta _n\right\rangle -a\right) \right) \text{ .} 
$$
On the other hand we get%
$$
\frac 1{2\pi }\int\limits_{-ne^{-i\alpha }}^{ne^{-i\alpha }}e^{i\lambda
\left( z\left\langle \omega ,\eta _n\right\rangle -a\right) }{\rm d}\lambda 
\text{ }=\frac 1{\pi \left( z\left\langle \omega ,\eta _n\right\rangle
-a\right) }\sin \left( ne^{-i\alpha }\left( z\left\langle \omega ,\eta
_n\right\rangle -a\right) \right) =\varphi _{n,z}\left( \omega \right) \text{%
.} 
$$
Hence%
$$
\sigma _z\varphi _n\left( \omega \right) =\varphi _{n,z}\left( \omega
\right) \text{ , }z\in {\bf S}_\alpha \text{.} 
$$

\noindent {\bf iii,iv)} Let us look at the convergence of (\ref{esfi}). The
following estimate holds:%
$$
\left| S\sigma _z\varphi _n\left( \theta \right) \right| \leq \frac 1{2\pi
}\int\limits_{-\infty }^\infty \exp \left( -\frac 12{\rm Re}\left(
z^2e^{-2i\alpha }\right) \nu ^2\left| \eta _n\right| _0^2+\nu {\rm Re}\left(
ie^{-i\alpha }\left( z\langle \theta ,\eta _n\rangle -a\right) \right)
\right) {\rm d}\nu \text{ .} 
$$
The integral exists if ${\rm Re}\left( z^2e^{-2i\alpha }\right) >0$.

\noindent This condition is satisfied for $-\frac \pi 4+\alpha <\arg z<\frac
\pi 4+\alpha $ . We get%
$$
\left| S\sigma _z\varphi _n\left( \theta \right) \right| \leq \frac 1{2\pi }%
\sqrt{\frac{2\pi }{{\rm Re}\left( z^2e^{-2i\alpha }\right) \left| \eta
_n\right| _0^2}}\exp \left( \frac{\left[ {\rm Re}\left( ie^{-i\alpha }\left(
z\langle \theta ,\eta _n\rangle -a\right) \right) \right] ^2}{2{\rm Re}%
\left( z^2e^{-2i\alpha }\right) \left| \eta _n\right| _0^2}\right) \text{ } 
$$
$$
\hspace*{10mm}\leq \frac 1{\sqrt{2\pi {\rm Re}\left( z^2e^{-2i\alpha
}\right) \frac 12\left| \eta \right| _0^2}}\exp \left( \frac{\left( \left|
z\right| \left| \theta \right| _02\left| \eta \right| _0+\left| a\right|
\right) ^2}{2{\rm Re}\left( z^2e^{-2i\alpha }\right) \frac 12\left| \eta
\right| _0^2}\right) \text{ ,} 
$$
for $n$ large enough.

\noindent Now the convergence Theorem \ref{conv} applies:%
\begin{eqnarray} \label{limphinz}
\lim \limits_{n\rightarrow \infty }S\sigma _z\varphi _n\left( \theta \right) %
&=&\frac 1{2\pi }e^{-i\alpha }\int\limits_{-\infty }^\infty \exp \left( %
-\frac 12z^2e^{-2i\alpha }\nu ^2\left| \eta \right| _0^2+i\nu e^{-i\alpha } %
\left( z\langle \theta ,\eta \rangle -a\right) \right) {\rm d}\nu %
\nonumber \\&=&e^{-i\alpha }\frac 1{\sqrt{2\pi }ze^{-i\alpha }\left| \eta \right| _0}\exp \left( \frac{-e^{-2i\alpha }\left( z\langle \theta ,\eta  \rangle -a\right) }{2\left| \eta \right| _0^2z^2e^{-2i\alpha }}^2\right) %
\nonumber \\&=&\frac 1{\sqrt{2\pi }z\left| \eta \right| _0}\exp \left( %
\frac{-\left( z\langle \theta ,\eta \rangle -a\right) }{2\left| \eta \right| _0^2z^2}^2 %
\right) \text{ .}
\end{eqnarray}
Note that the limit does not depend on $\alpha $.\TeXButton{End Proof}
{\endproof}

\ 

\begin{proposition}
\label{DeltaHom} $\delta $ is homogeneous of degree $-1$\ in $z\in {\bf S}%
_\alpha $:%
$$
\sigma _z\delta \left( \left\langle \omega ,\eta \right\rangle -a\right)
=\frac 1z\delta \left( \left\langle \omega ,\eta \right\rangle -\frac
az\right) . 
$$
\end{proposition}

\TeXButton{Proof}{\proof}From formula (\ref{limphinz}) we have 
\begin{equation}
\label{sz}S\sigma _z\delta \left( \left\langle \omega ,\eta \right\rangle
-a\right) \left( \theta \right) =\frac 1{\sqrt{2\pi }z\left| \eta \right|
_0}\exp \left( -\frac 12\frac{\left( a-z\langle \theta ,\eta \rangle \right)
^2}{z^2\left| \eta \right| _0^2}\right) 
\end{equation}
$$
\hspace{35mm}=\frac 1z\frac 1{\sqrt{2\pi }\left| \eta \right| _0}\exp \left(
-\frac 12\frac{\left( \frac az-\langle \theta ,\eta \rangle \right) ^2}{%
\left| \eta \right| _0^2}\right) \text{\ } 
$$
$$
\hspace{21mm}=S\left( \frac 1z\delta \left( \left\langle \omega ,\eta
\right\rangle -\frac az\right) \right) \left( \theta \right) \text{ .} 
$$
\TeXButton{End Proof}{\endproof}

\subsection{Products of Donsker`s deltas}

To define products of scaled Donsker`s deltas, we use the following ansatz 
\begin{equation}
\label{(nDeltaInt)}\Phi =\tprod\limits_{j=1}^n\sigma _z\delta \left(
\left\langle \cdot ,\eta _j\right\rangle -a_j\right) =\frac 1{\left( 2\pi
\right) ^n}\tprod\limits_{j=1}^n\dint\limits_\gamma \exp \left( i\lambda
_j\left( z\left\langle \cdot ,\eta _j\right\rangle -a_j\right) \right) {\rm d%
}\lambda _j\ , 
\end{equation}
here $\gamma =\left\{ e^{-i\alpha }t\mid t\in {\R}\right\} $, $z\in {\bf S}%
_\alpha $, where $\alpha $ is chosen such that $\left| \alpha \right| <\frac
\pi 4$, $\eta _j$ are real, linear independent elements of ${\cal H}$ and $%
a_j\in {\C}$. We use the notation%
$$
\exp \left( \sum\limits_{j=1}^ni\lambda _j\left( \left\langle \cdot ,\eta
_j\right\rangle -a_j\right) \right) =\exp \left( i\vec \lambda \left(
\left\langle \cdot ,\vec \eta \right\rangle -\vec a\right) \right) . 
$$
To prove that $\Phi $ is well-defined, we calculate it`s $T$-transform:%
\begin{eqnarray*}
\ T\Phi \left( \theta \right)  & = & \frac{e^{-i\alpha n}}{\left( 2\pi \right) ^n}%
\dint \dint \exp \left( ize^{-i\alpha }\left\langle \omega ,\vec{\lambda }%
\vec{\eta }\right\rangle -ie^{-i\alpha }\vec{\lambda }%
\vec{a} +i\left\langle \omega ,\theta \right\rangle \right) {\rm d}\mu %
\,{\rm d}^n\lambda  %
\\& = & \frac 1{\left( 2\pi \right) ^n}e^{-i\alpha n}\dint \dint \exp %
\left( i\left\langle%
\omega ,ze^{-i\alpha }\vec{\lambda }\vec{\eta }+\theta \right\rangle %
 -ie^{-i\alpha }\vec{\lambda }\vec{a}\right) {\rm d}\mu %
\, {\rm d}^n\lambda  \\& = & \frac 1{\left( 2\pi \right) ^n} %
e^{-i\alpha n}\dint C\left( ze^{-i\alpha }%
\vec{\lambda }\vec{\eta }+\theta \right) \exp \left(-ie^{-i\alpha }%
\vec{\lambda }\vec{a} \right) {\rm d}^n\lambda  
 \text{ .}%
\end{eqnarray*}
To calculate $C(ze^{-i\alpha }\vec \lambda \vec \eta +\theta )$ consider now%
$$
\langle ze^{-i\alpha }\vec \lambda \vec \eta ,ze^{-i\alpha }\vec \lambda
\vec \eta \rangle =z^2e^{-i2\alpha }\dsum\limits_{k,l}\lambda _k\lambda
_l\left( \eta _k,\eta _l\right) =z^2e^{-i2\alpha }\vec \lambda M\vec \lambda
, 
$$
where $M\equiv \left( \eta _k,\eta _l\right) _{k,l}$. This is a Gram matrix
of linear independent vectors and thus positive definite.%
\begin{eqnarray*}
  T\Phi \left( \theta \right) \; %
& = &\frac 1{\left( 2\pi \right) ^n}\,e^{-i\alpha n}%
e^{-\frac 12\langle\theta,\theta\rangle } %
\\& &\times \dint \exp \left[ -\frac 12z^2e^{-i2\alpha }\vec{\lambda }%
M\vec{\lambda }-ze^{-i\alpha }\vec{\lambda }%
\langle\vec{\eta },\theta \rangle -ie^{-i\alpha }\vec{\lambda }%
\vec{a}\right] {\rm d}^n\lambda  \hspace{15mm} %
\\& = &\sqrt{\frac{\left( 2\pi \right) ^n}{\left( z^2e^{-i2\alpha } \right) ^n\det M}}\,%
\frac{e^{-i\alpha n}}{\left( 2\pi \right) ^n}\;e^{-\frac 12\langle\theta, \theta\rangle }%
\end{eqnarray*}
$$
\hspace*{15mm}\times \exp \left[ \frac 12\left( ze^{-i\alpha }\langle \vec
\eta ,\theta \rangle +ie^{-i\alpha }\vec a\right) \left( z^2e^{-i2\alpha
}M\right) ^{-1}\left( ze^{-i\alpha }\langle \vec \eta ,\theta \rangle
+ie^{-i\alpha }\vec a\right) \right] , 
$$
this Gaussian integral exists if ${\rm Re}\left( z^2e^{-i2\alpha }\right) >0$%
, which is equivalent to $z\in {\bf S}_\alpha $. The last expression is a 
U-functional, so we get:

\begin{theorem}
\label{ProdDelta} {\rm \cite{LLSW94b}} \\ Let $a_j\in {\C}$, $\eta _j\in L^2(%
{\R)}$ linear independent and $M$ $=\left( \eta _k,\eta _l\right) _{k,l}$
the corresponding Gram matrix.\\Then for all $z\in {\bf S}_\alpha $ $\Phi
=\tprod\limits_{j=1}^n\sigma _z\delta \left( \left\langle \cdot ,\eta
_j\right\rangle -a_j\right) $ is a Hida distribution with $S$-transform 
\begin{equation}
\label{(ProdDelta)}S\Phi \left( \theta \right) =\frac 1{\sqrt{\left( 2\pi
z^2\right) ^n\det M}}\exp \left[ -\frac 12\left( \langle \vec \eta ,\theta
\rangle -\frac 1z\vec a\right) M^{-1}\left( \langle \vec \eta ,\theta
\rangle -\frac 1z\vec a\right) \right] \text{ .}\ 
\end{equation}
\end{theorem}

\subsection{Complex scaling of finite dimensional Hida distributions}

We can use Theorem \ref{ProdDelta} to extend the scaling operator. Let $\eta
_j\in {\cal H}\,,\,1\leq j\leq n$ be linear independent and $G:\R %
^n\rightarrow \C $ in $L_{\Ckl }^p\left( \R ^n,\exp (-\frac 12\vec x\cdot
M^{-1}\vec x)\,{\rm d}^nx\right) $ for some $p>1$ where $M=(\eta _k,\eta
_l)_{k,l}$ is positive definite. These assumptions allow to define 
$$
\varphi :=G\left( \langle \cdot ,\eta _1\rangle ,\ldots ,\langle \cdot ,\eta
_n\rangle \right) 
$$
such that $\varphi \in L_{\Ckl }^p(\mu )$. In view of Proposition \ref{MGLp}%
.1 $\varphi \in {\cal G}^{\prime }$. Since $\varphi $ depends only on a
finite number of ``coordinates'' $\langle \cdot ,\eta _j\rangle $ we call it
a finite dimensional Hida distribution (similar to \cite{KK92} where this
notion was restricted to smooth $\eta _j\in {\cal N}\,,\,1\leq j\leq \eta )$.

\begin{lemma}
In the case of the above assumptions the following representation holds 
$$
\varphi =\int_{\R ^n}G(\vec x)\,\delta ^n(\langle \cdot ,\vec \eta \rangle
-\vec x)\,{\rm d}^nx 
$$
where the integral in $({\cal N})^{\prime }$ is in the sense of Bochner and 
$$
\delta ^n(\langle \cdot ,\vec \eta \rangle -\vec x):=\prod_{j=1}^n\delta
(\langle \cdot ,\eta _j\rangle -x_j) 
$$
is defined in Theorem \ref{ProdDelta}.
\end{lemma}

The proof is postponed because the existence of the Bochner integral will
follow from the more general discussion in the next theorem. Then the
equality follows from a comparison on the dense set of exponential functions.%
\bigskip\ 

Now it is natural to try the following extension of $\sigma _z:$%
$$
\sigma _z\varphi =\int_{\R ^n}G(\vec x)\,\sigma _z\delta ^n(\langle \cdot
,\vec \eta \rangle -\vec x)\,{\rm d}^nx 
$$
whenever the right hand side is a well defined Bochner integral in $({\cal N}%
)^{\prime }$. To do this, stronger assumptions on $G$ are needed. In the
next theorem we will give a sufficient condition.

\begin{theorem}
\label{finiteScal}Let $z\in {\bf S}_0\quad $(i.e., ${\rm Re}\frac 1{z^2}>0$)
and let 
$$
G\in L_{\Ckl }^p\left( \R ^n,\exp \left( -\frac 12({\rm Re}\frac
1{z^2})\,\vec x\cdot M^{-1}\vec x\right) {\rm d}^nx\right) \text{for some }%
p>1. 
$$
Then 
$$
\int_{\R ^n}G(\vec x)\,\sigma _z\delta ^n(\langle \cdot ,\vec \eta \rangle
-\vec x)\,{\rm d}^nx 
$$
is a well defined Bochner integral in $({\cal N})^{\prime }.$
\end{theorem}

\TeXButton{Proof}{\proof}From equation (\ref{(ProdDelta)}) we can estimate%
$$
\left| S\sigma _z\delta ^n(\langle \cdot ,\vec \eta \rangle -\vec x)(\theta
)\right| \leq \hspace{11.5cm} 
$$
$$
\leq \frac 1{\sqrt{(2\pi |z|^2)^n\det M}}\exp \left( -\frac 12{\rm Re}\frac
1{z^2}\,\vec x\cdot M^{-1}\vec x+\frac 1{|z|}\left| \langle \theta ,\vec
\eta \rangle \cdot M^{-1}\vec x\right| +\frac 12|\theta |^2|\vec \eta
||M^{-1}\vec \eta |\ \right) . 
$$
The term linear in $\vec x$ can now be estimated using a general estimate
for positive quadratic forms; for all $\varepsilon >0$%
$$
\frac 1{|z|}\left| \langle \theta ,\vec \eta \rangle \cdot M^{-1}\vec
x\right| \leq \varepsilon \vec x\cdot M^{-1}\vec x+\frac 1{4\varepsilon
|z|^2}\langle \theta ,\vec \eta \rangle \cdot M^{-1}\langle \theta ,\vec
\eta \rangle 
$$
$$
\hspace{24mm}\leq \varepsilon \vec x\cdot M^{-1}\vec x+\frac 1{4\varepsilon
|z|^2}|\theta |^2|\vec \eta ||M^{-1}\vec \eta |\ . 
$$
Thus%
$$
\left| S\sigma _z\delta ^n(\langle \cdot ,\vec \eta \rangle -\vec x)(\theta
)\right| \leq \hspace{11.5cm} 
$$
$$
\leq \frac 1{\sqrt{(2\pi |z|^2)^n\det M}}\exp \left( -\left( \frac 12{\rm Re}%
\frac 1{z^2}-\varepsilon \right) \vec x\cdot M^{-1}\vec x+\frac 12\left(
\frac 1{2\varepsilon |z|^2}+1\right) |\vec \eta ||M^{-1}\vec \eta ||\theta
|^2\right) \ . 
$$
Now we choose $q>1$ with $\frac 1p+\frac 1q=1$ and $\varepsilon >0$ such
that $q\varepsilon <\frac 12{\rm Re}\frac 1{z^2}$. Then 
$$
\int_{\R ^n}|G(\vec x)|\exp \left( -\left( \frac 12{\rm Re}\frac
1{z^2}-\varepsilon \right) \vec x\cdot M^{-1}\vec x\right) {\rm d}^nx\leq 
\hspace{6.5cm} 
$$
$$
\left( \int |G(\vec x)|^p\exp \left( -\frac 12{\rm Re}\frac 1{z^2}\vec
x\cdot M^{-1}\vec x\right) {\rm d}^nx\right) ^{1/p}\cdot \left( \int \exp
\left( -\left( \frac 12{\rm Re}\frac 1{z^2}-q\varepsilon \right) \vec x\cdot
M^{-1}\vec x\right) {\rm d}^nx\right) ^{1/q} 
$$
is finite because of our assumptions. Hence Theorem \ref{Bochner} applies
and proves the theorem.\TeXButton{End Proof}{\endproof}\bigskip\ 

\noindent {\bf Notes.}\\1. Instead of integration with respect to $G(\vec
x)\cdot {\rm d}^nx$ we can use complex measures $v$ on $\R ^n$ to define the
more general distribution%
$$
\int_{\R ^n}\sigma _z\delta ^n(\langle \cdot ,\vec \eta \rangle -\vec x)\,%
{\rm d}^nv(x) \ . 
$$

\noindent 2. If $z\notin {\bf S}_0$ (not negative) it is sill possible to
define $\sigma _z\delta ^n(\langle \cdot ,\vec \eta \rangle -\vec x)$ (if we
do not insist on the existence of an integral representation of type (\ref
{(nDeltaInt)})). Then equation (\ref{(ProdDelta)}) defines this object. In
this case ${\rm Re}\frac 1{z^2}$ may be negative and the analog of Theorem 
\ref{finiteScal} would require rapid decrease of $|G|$ at infinity, more
precisely there has to be an $\varepsilon >0$ such that 
$$
\int |G(\vec x)|\exp \left( \left( \varepsilon -\frac 12{\rm Re}\frac
1{z^2}\right) \vec x\cdot M^{-1}\vec x\right) {\rm d}^nx 
$$
is finite.

\subsection{Series of Donsker`s deltas \label{Donskerseries}}

We set 
$$
\Phi _N=\sum\limits_{n=-N}^N\sigma _z\,\delta \left( \langle \omega ,\eta
\rangle -a+n\right) ,\ \text{ }a\in {\C}\text{ .} 
$$
This is a well-defined Hida distribution and its $S$-transform is given by 
$$
S\Phi _N\left( \theta \right) =\frac 1{\sqrt{2\pi |\eta |^2}%
z}\sum\limits_{n=-N}^N\exp \left( -\frac 1{2|\eta |^2z^2}\left( a-n-z\langle
\eta ,\theta \rangle \right) ^2\right) \text{ .} 
$$
We now assume ${\rm Re}\frac 1{z^2}>0.$ To study the limit $N\rightarrow
\infty $ we calculate a uniform bound (in $N$) for 
\begin{eqnarray*}
& & \hspace*{-1cm} \left| S\Phi _N\left( \theta \right) \right| \leq \frac 1{\left| z\right| \sqrt{2\pi |\eta |^2}}\sum\limits_{n=-N}^N\exp \left( -\frac 1{2|\eta |^2}%
{\rm Re}\left( \frac 1{z^2}\left( a-n-z\langle \eta ,\theta \rangle \right) %
^2\right) \right) \text{ } %
\\ &\leq & \frac 1{\left| z\right| \sqrt{2\pi |\eta |^2}}\sum\limits_{n=-N}^N\exp %
\left( \frac 1{2|\eta |^2}\left( -n^2{\rm Re}\frac 1{z^2}+2\left| \frac nz\right| \left| \frac az-\langle \eta ,\theta \rangle \right| +\left| \frac az-\langle \eta ,\theta \rangle \right| ^2\right) \right)  %
\\ &\leq & \frac 1{\left| z\right| \sqrt{2\pi |\eta |^2}}\sum\limits_{n=-\infty }^\infty \exp \left( \frac 1{2|\eta |^2}\left( -n^2\frac 12{\rm Re}\frac 1{z^2}+\left( 1+\frac{2\left| z\right| ^2}{{\rm Re}\ z^2}\right) \left| %
\frac az-\langle \eta ,\theta \rangle \right| ^2\right) \right)  %
\\ &=& \frac 1{\left| z\right| \sqrt{2\pi |\eta |^2}}\exp \left( \frac 1{2|\eta |^2}\left( 1+\frac{2\left| z\right| ^2}{{\rm Re}\ z^2}\right) \left( \left| %
\frac az\right| -|\eta |_0\left| \theta \right| _0\right) ^2\right) %
\sum\limits_{n=-\infty }^\infty \exp \left( -\frac 1{4|\eta |^2}{\rm Re}%
\frac 1{z^2}n^2\right) \text{.} %
\end{eqnarray*}
The infinite sum converges if ${\rm Re}\frac 1{z^2}>0$, i.e., if $z\in {\bf S%
}_0$. The sum can also be expressed as $\vartheta \left( 0,\frac i{4\pi
|\eta |^2}{\rm Re}\frac 1{z^2}\right) $ using the theta function (see \cite
{Mu79})%
$$
\vartheta \left( \rho ,\tau \right) =\sum\limits_{n=-\infty }^\infty \exp
\left( \pi in^2\tau +2\pi in\rho \right) \text{ .} 
$$
Now Theorem \ref{conv} applies and we get:%
\begin{eqnarray*}
& &  \hspace*{-1cm} S\Phi \left( \theta \right) =\frac 1{z\sqrt{2\pi |\eta |^2}}\sum\limits_{n=-\infty }^\infty \exp \left( -\frac 1{2z^2|\eta |^2}\left( %
a-n-z\langle \eta ,\theta \rangle \right) ^2\right)  %
\\ &=& \frac 1{z\sqrt{2\pi |\eta |^2}}\exp \left( -\frac 1{2|\eta |^2}\left( %
\langle \eta ,\theta \rangle -\frac az\right) ^2\right)%
\sum\limits_{n=-\infty }^\infty \exp \left( -\frac{n^2}{2z^2|\eta |^2}-\frac n{|\eta |^2}\left( \langle \eta ,\theta \rangle -\frac az\right) \right)  %
\\ &=& \frac 1{z\sqrt{2\pi |\eta |^2}}\exp \left( -\frac 1{2|\eta |^2}\left( %
\langle \eta ,\theta \rangle -\frac az\right) ^2\right) \,\vartheta \left( %
\frac i{2\pi |\eta |^2}\left( \langle \eta ,\theta \rangle -\frac az\right) %
\,,\frac i{2\pi z^2|\eta |^2}\right) %
\\ &=& S\sigma _z\delta \left( \theta \right) \cdot \,\vartheta \left( \frac i{2\pi |\eta |^2}\left( \langle \eta ,\theta \rangle -\frac az\right) %
\,,\frac i{2\pi z^2|\eta |^2}\right) \text{ .} %
\end{eqnarray*}
Thus we have proved:

\begin{theorem}
\label{DonSeries} {\rm \cite{LLSW94b}} \\ For all $a\in {\C}$ and all $z\in 
{\bf S}_{0\text{ }}$the infinite sum 
$$
\Phi =\sum\limits_{n=-\infty }^\infty \sigma _z\,\delta \left( \langle
\omega ,\eta \rangle -a+n\right) 
$$
exists as a Hida distribution with $S$-transform%
$$
S\Phi \left( \theta \right) =S\sigma _z\,\delta \left( \theta \right)
\,\cdot \vartheta \left( \frac i{2\pi |\eta |^2}\left( \langle \theta ,\eta
\rangle -\frac az\right) \,,\frac i{2\pi z^2|\eta |^2}\right) \text{ .} 
$$
\end{theorem}

\subsection{Local Time}

In the next two sections we choose the nuclear triple%
$$
{\cal S}_d\subset L_d^2\subset {\cal S}_d^{\prime } 
$$
and $\eta =\1 _{[0,t)}$ the indicator function of a real interval. As is
well known Brownian motion may be represented in the framework of Gaussian
analysis as $\vec B(t)=\left\langle \vec \omega ,\1 _{\left[ 0,t\right]
}\right\rangle $. Let us consider the local time, which intuitively should
measure the mean time a Brownian particle spends at a given point.
Informally the local time is given by ``Tanaka's formula''%
$$
L(\tau ,\vec a)=\frac 1\tau \int_0^\tau \delta ^d(\vec B(t)-\vec a)\,{\rm d}%
t\quad ,\quad \vec a\in \R ^d,\tau \in \R_{+}\quad , 
$$
where $\delta ^d(\vec B(t)-\vec a)$ is Donsker delta function with
S-transform given by 
$$
S\delta ^d(\vec B(t)-\vec a)(\vec \xi )=\left( \frac 1{2\pi t}\right)
^{\frac d2}\exp \left( -\frac 1{2t}\left[ \int_0^t\vec \xi (s){\rm d}s-\vec
a\right] ^2\right) \quad . 
$$
For dimension $d\geq 2$ this expression is usually treated by
renormalization, i.e., the cancellation of divergent terms, see e.g., \cite
{SW93,FHSW94}. However, for $\vec a\neq 0$ --- Brownian motion starts in 0,
thus one expects a strong divergence for $\vec a=0$ --- the local time can
be rigorously defined in $\left( {\cal S}_d\right) ^{-1}$ using Bochner
integrals. In fact we can estimate as follows%
$$
\left| S\delta ^d(\vec B(t)-\vec a)(\vec \theta )\right| \leq \left( \frac
1{2\pi t}\right) ^{\frac d2}\exp \left( \frac 1{2t}\left| \1 _{\left[
0,t\right] }\right| ^2\left| \vec \theta \right| ^2+\frac{\left| \vec
a\right| }tt\sup _{0\leq s\leq t}\left| \vec \theta (s)\right| -\frac{\vec
a^2}{2t}\right)  
$$
$$
\hspace*{22mm}\leq \left( \frac 1{2\pi t}\right) ^{\frac d2}\exp \left( -%
\frac{\vec a^2}{2t}\right) \exp \left( \left| \vec \theta \right|
_1^2\right) \exp \left( \frac 12\left| \vec a\right| ^2\right) \quad . 
$$
Now%
$$
C(t):=\left( \frac 1{2\pi t}\right) ^{\frac d2}\exp \left( -\frac{\vec a^2}{%
2t}\right) \exp \left( \frac 12\left| \vec a\right| ^2\right)  
$$
is integrable on any interval $\left[ 0,\tau \right] $ with respect to
Lebesgue measure ${\rm d}t.$ Hence the conditions of Theorem \ref{Bochner}
are satisfied and we have 
$$
L(\tau ,\vec a)=\frac 1\tau \int_0^\tau \delta ^d(\vec B(t)-\vec a)\;{\rm d}%
t\in \left( {\cal S}_d\right) ^{\prime }\quad ,\quad 0\neq \vec a\in \R%
^d,\tau \in \R_{+}\quad . 
$$
In the case $d=1$ and $a\in \C $ the relevant estimate becomes%
$$
\left| S\delta \left( B\left( t\right) -a\right) \left( \theta \right)
\right| \leq \frac 1{\sqrt{2\pi t}}e^{\frac 12\left| a\right| ^2}e^{-\frac{%
{\rm Re}\left( a^2\right) }{2t}}\exp \left( \left| \theta \right|
_1^2\right) \text{ .} 
$$
This fulfills the conditions of Theorem \ref{Bochner} if ${\rm Re}\left(
a^2\right) \geq 0$. Thus we have an analytic extension of $L\left( \tau
,a\right) =\frac 1\tau \int\limits_0^\tau \delta \left( B\left( t\right)
-a\right) \,{\rm d}t$ to $a\in \C $ with ${\rm Re}\left( a^2\right) \geq 0$.
That means the point $a=0$ is allowed in this case (Note that $%
\int\limits_0^\tau t^{-d/2}{\rm d}t$ only exists for $d=1$.).
\LaTeXparent{dis3.tex}

\chapter{Concept of path integration in a white noise framework}

\section{The free Feynman integrand \label{free}}

The idea of realizing Feynman integrals within the white noise framework
goes back to \cite{HS83}. The ``average over all paths'' is performed with a
Hida distribution as the weight. The existence of such Hida distributions
corresponding to Feynman integrands has been established in \cite{FPS91}.
There the Feynman integrand for the free motion (in one space dimension)
reads: 
$$
{\rm I}_{0,old}={\rm Nexp}\left( \frac{i+1}2\int_{t_0}^t\omega ^2\left( \tau
\right) \text{ }{\rm d}\tau \right) \delta \left( x\left( t\right) -x\right) 
\text{.} 
$$
However the distribution 
$$
{\rm J}={\rm Nexp}\left( \frac{i+1}2\int_{\R }\omega ^2\left( \tau \right) 
\text{ }{\rm d}\tau \right) 
$$
has recently been seen to be particularly useful in this context because of
its relation to complex scaling (see Theorem \ref{Jsigmaz}). It turns out
that it is unnecessary to use the time interval $\left[ t_0,t\right] $ in
the kinetic energy factor; the delta function introduces the interval into
the resulting distribution ${\rm I}_0:={\rm J}\delta $. Indeed it will be
shown that ${\rm I}_0$ produces the correct physical results. As the choice
of ${\rm I}_0$ rather than ${\rm I}_{0,old}$ as a starting point produces
only minor modifications in calculations and formulae, all the pertinent
results in \cite{FPS91} can be established in a completely analogous manner.

Let us look at the construction of the free Feynman integrand (in more than
one space dimension) in more detail. We are going to use a variant of white
noise analysis which allows vector valued white noise and hence the
possibility to build up Brownian paths in $d$-dimensional space, see Example 
\arabic{SdTriple} on page \pageref{SdTripleP}.

We introduce the heuristic term%
$$
\exp \left( \frac{1+i}2\int_{\R }\vec \omega ^2(\tau )\ {\rm d}\tau \right) 
$$
where $\vec \omega \in {\cal S}_d^{\prime }$ is a $d$--tuple of independent
white noises. Formal speaking we expect this term to consist of one factor
representing Feynman's factor introduced for the kinetic part and one factor
compensating the Gaussian fall-off of the white noise measure which is used
instead of Feynman's ill defined flat measure on path--space. The above term
is not well defined because of its infinite expectation. Formally this can
be cured by dividing out this infinite constant. This leads to the
normalized exponential ${\rm J}:={\rm J}_{\sqrt{i}}={\rm Nexp}\left( \frac{%
1+i}2\int_{\R }\vec \omega ^2(\tau )\ {\rm d}\tau \right) $ in $({\cal S}%
_d)^{\prime }$ studied in Example \arabic{JzDef} on page \pageref{JzDefP}.
It can be defined rigorously by its $T$-transform%
$$
\exp \left( -\frac i2\int_{\R }\vec \xi ^2(\tau )\ {\rm d}\tau \right)
,\qquad \vec \xi \in {\cal S}_d 
$$
where $\vec \xi ^2=\sum\limits_{j=1}^d\xi _i^2$ denotes the Euclidean inner
product. The version of Brownian motion we are going to use starts in point $%
\vec x_0$ at time $t_0$: 
\begin{equation}
\label{path}\vec x(\tau )=\vec x_0+\left\langle \vec \omega ,{\1}_{[t_0,\tau
)}\right\rangle \quad ,\quad \vec \omega \in {\cal S}_d^{\prime } 
\end{equation}
here ${\1}_{[t_0,\tau )}$ denotes the indicator function of the interval $%
[t_0,\tau ).$ Since we will discuss propagators we also have to fix the
endpoint $\vec x$ of the paths at time $t$. To this end we introduce
Donsker's delta function which is the formal composition of a delta function
and a Brownian motion:%
$$
\delta ^d(\vec x(t)-\vec x)\;. 
$$
This is a well defined distribution in $({\cal S}_d)^{\prime }$ as can be
verified by calculating its $T$-transform. Here we give instead the slightly
more general $T$-transform where the scaling operator $\sigma _z:\varphi
(\vec \omega )\longrightarrow \varphi (z\vec \omega )$ has been applied to
Donsker's delta. This can be justified following the lines of section \ref
{Deltascaling}; 
$$
T(\sigma _z\delta )(\vec \xi )=\left( 2\pi z^2(t-t_0)\right) ^{-\frac
d2}\exp \left( -\frac 12\left| \vec \xi \right| ^2-\frac 1{2(t-t_0)}\left(
i\int_{t_0}^t\vec \xi (\tau )\ {\rm d}\tau -\frac{\vec x-\vec x_0}z\right) ^{%
\D 2}\right) \text{.} 
$$
Now we have to justify the pointwise multiplication of ${\rm J}$ and $\delta 
$ to get the so called free Feynman integrand ${\rm I}_0$. This was
initially been done in \cite{FPS91} for $d=1$ and in \cite{SW93} for higher
dimensions. So we use a short way to reproduce this result. Using the
relation of ${\rm Nexp}$ and complex scaling which is condensed in the
following formula (Lemma \ref{JzphiLem}) 
\begin{equation}
\label{TI0xi}T({\rm I}_0)(\vec \xi )=T({\rm J}\delta )(\vec \xi )=T(\sigma _{%
\sqrt{i}}\delta )(\sqrt{i}\vec \xi )\text{ ,} 
\end{equation}
we arrive at the $T$-transform of the free Feynman integrand 
\begin{equation}
\label{TI0delta}T({\rm I}_0)(\vec \xi )=\left( 2\pi i(t-t_0)\right) ^{-\frac
d2}\exp \left( -\frac i2\left| \vec \xi \right| ^2-\frac 1{2i(t-t_0)}\left(
\int_{t_0}^t\vec \xi (\tau )\ {\rm d}\tau +(\vec x-\vec x_0)\right) ^{\D %
2}\;\right) . 
\end{equation}
This is clearly a $U$-functional and can be used to define ${\rm I}_0$ in $(%
{\cal S}_d)^{\prime }.$ If it is necessary to be more precise we will also
use the notation ${\rm I}_0\left( \vec x,t|\vec x_0,t_0\right) $.\medskip\ 

Furthermore the Feynman integral $\E \left( {\rm I}_0\right) =T{\rm I}%
_0\left( 0\right) $ is indeed the (causal) free particle propagator $\left(
2\pi i(t-t_0)\right) ^{-\frac d2}$ $\exp \left[ \frac i{2\left| t-t_0\right|
}\text{ }\left( \vec x-\vec x_0\right) ^2\right] $. \\Not only the
expectation but also the $T$--transform has a physical meaning. By a formal
integration by parts\ 
$$
T{\rm I}_0\left( \vec \xi \right) =\E \left( {\rm I}_0\text{ }e\text{ }%
^{-i\int_{t_0}^t\vec x\left( \tau \right) \stackrel{.}{\vec \xi }\left( \tau
\right) \,{\rm d}\tau }\right) \text{ }e^{i\vec x\vec \xi \left( t\right)
-i\vec x_0\vec \xi \left( t_0\right) }\text{ }e\text{ }^{-\frac i2\left|
\vec \xi _{\left[ t_0,t\right] }c\right| ^2}. 
$$
($\vec \xi _{\left[ t_0,t\right] }c$ denotes the restriction of $\vec \xi $
to the complement of $\left[ t_0,t\right] $). The term $e$ $%
^{-i\int_{t_0}^t\vec x\left( \tau \right) \stackrel{.}{\vec \xi }\left( \tau
\right) \,{\rm d}\tau }$ would thus arise from a time-dependent potential $%
W\left( \vec x,\tau \right) =\stackrel{.}{\;\vec \xi }\!\!(\tau )\vec x$.
And indeed it is straightforward to verify that\ 
\begin{equation}
\label{KoDef}\Theta (t-t_0)\cdot T{\rm I}_0\left( \vec \xi \right)
=K_0^{\left( \stackrel{.}{\vec \xi }\right) {}}\left( \vec x,t|\vec
x_0,t_0\right) \text{ }e^{i\vec x\vec \xi \left( t\right) -i\vec x_0\vec \xi
\left( t_0\right) }\text{ }e\text{ }^{-\frac i2\left| \vec \xi _{\left[
t_0,t\right] }c\right| ^2}, 
\end{equation}
where%
$$
K_0^{\left( \stackrel{.}{\vec \xi }\right) }\left( \vec x,t|\vec
x_0,t_0\right) =\frac{\Theta \left( t-t_0\right) }{\sqrt{2\pi i\left|
t-t_0\right| }}\;\times \hspace{6cm} 
$$
\begin{equation}
\label{Ko}\exp \left( i\vec x_0\vec \xi (t_0)-i\vec x\vec \xi (t)-\frac
i2\left| \vec \xi _{\left[ t_0,t\right] }\right| ^2+\frac i{2\left|
t-t_0\right| }\left( \int_{t_0}^t\vec \xi (\tau )\,{\rm d}\tau +\vec x-\vec
x_0\right) ^2\right) 
\end{equation}
is the Green's function corresponding to the potential $W$, i.e., $%
K_0^{\left( \stackrel{.}{\vec \xi }\right) }$ obeys the Schr\"odinger
equation\ 
$$
\left( i\partial _t+\frac 12\bigtriangleup -\stackrel{.}{\vec \xi }\left(
t\right) x\right) K_0^{\left( \stackrel{.}{\vec \xi }\right) }\left( \vec
x,t|\vec x_0,t_0\right) =i\,\delta \left( t-t_0\right) \,\delta ^d\left(
\vec x-\vec x_0\right) . 
$$
\medskip\ 

\noindent More generally one calculates (e.g., using Theorem \ref{ProdDelta}%
)\ 
\begin{equation}
\label{Tnd}T\left( {\rm J}\stackrel{n+1}{\stackunder{j=1}{\Pi }}\delta
^d\left( \vec x\left( t_j\right) -\vec x_j\right) \right) \left( \vec \xi
\right) =e^{-\frac i2\left| \vec \xi _{\left[ t_0,t\right] }c\right|
^2}{}e^{i\vec x\vec \xi \left( t\right) -i\vec x_0\vec \xi \left( t_0\right)
}\stackrel{n+1}{\stackunder{j=1}{\Pi }}K_0^{\left( \stackrel{.}{\vec \xi }%
\right) }\left( \vec x_j,t_j|\vec x_{j-1},t_{j-1}\right) .
\end{equation}
$$
\hspace*{1.6cm}=e^{\frac{in}2\left| \vec \xi \right| ^2}{}\stackrel{n+1}{%
\stackunder{j=1}{\Pi }}T{\rm I}_0\left( \vec x_j,t_j|\vec
x_{j-1},t_{j-1}\right) (\vec \xi ) 
$$
Here $t_0<t_1<...<t_n<t_{n+1}\equiv t$ and $\vec x_{n+1}\equiv \vec x$ .\ 

\section{The unperturbed harmonic oscillator}

In this section we first review some results of \cite{FPS91}. Then we
prepare a proposition on which we base the perturbative method in section 
\ref{perHarmOsc}.

To define the Feynman integrand 
$$
{\rm I}_h={\rm I}_0\exp \left( -i\int_{t_0}^tU\left( x(\tau )\right) \ {\rm d%
}\tau \right) \text{, \ }U(x)=\frac 12k^2x^2 
$$
of the harmonic oscillator (for space dimension $d=1$), at least two things
have to be done.

First we have to justify the pointwise multiplication of ${\rm I}_0$ with
the interaction term and secondly it has to be shown that ${\E}({\rm I}_h)$
solves the Schr\"odinger equation for the harmonic oscillator. Both has been
done in \cite{FPS91}. There the $T$-transform of ${\rm I}_h$ has been
calculated and shown to be a $U$-functional. Thus ${\rm I}_h\in ({\cal S}%
)^{\prime }$. Later we will use the following modified version of their
result: 
$$
T{\rm I}_h\left( \xi \right) =\sqrt{\frac k{2\pi i\sin k\left| \Delta
\right| }}\exp \left( -\frac i2\left| \xi \right| ^2\right) \exp \bigg\{ 
\frac{ik}{2\sin k\left| \Delta \right| }\bigg[ \left( x_0^2+x^2\right) \cdot 
\hspace*{2cm} 
$$
$$
\cdot \cos k\left| \Delta \right| -2x_0x+2x\int_{t_0}^t{\rm d}t^{\prime
}\,\xi (t^{\prime })\cos k(t^{\prime }-t_0)-2x_0\int_{t_0}^t{\rm d}t^{\prime
}\,\xi (t^{\prime })\cos k(t-t^{\prime }) 
$$
\begin{equation}
\label{17}+2\int_{t_0}^t{\rm d}s_1\int_{t_0}^{s_1}{\rm d}s_2\,\xi (s_1)\xi
(s_2)\cos k(t-s_1)\cos k(s_2-t_0)\bigg] 
\bigg\} , 
\end{equation}
with $0<k\left| \Delta \right| <\dfrac \pi 2$ , $\Delta =[t_0,t]$ , $|\Delta
|=|t-t_0|$ . $T{\rm I}_h$ is easily seen to be a $U$-functional.

For our purposes it is convenient to introduce 
$$
K_h^{\left( \dot \xi \right) }(x,t\mid x_0,t_0)=\theta (t-t_0)\ T{\rm I}%
_h(\xi )\cdot \exp {\T\frac i2}\left| \xi _\Delta c\right| ^2\cdot \exp
\left( ix_0\xi (t_0)-ix\xi (t)\right) \text{ ,} 
$$
which is the propagator of a particle in a time dependent potential $\frac
12k^2x^2+x\dot \xi (t).$ (Again $\xi _\Delta c$ denotes the restriction of $%
\xi $ to the complement of $\Delta $.) This allows for an independent check
on the correctness of the above result. In advanced textbooks of quantum
mechanics such as \cite{Ho92} the propagator for a harmonic oscillator
coupled to a source $j$ (forced harmonic oscillator) is worked out. Upon
setting $j=\dot \xi $ their result is easily seen to coincide with the
formula given above.

As in equation (\ref{Tnd}) we also need a definition of the (pointwise)
product 
$$
{\rm I}_h\tprod\limits_{j=1}^n\delta \left( x(t_j)-x_j\right)  
$$
in $({\cal S})^{\prime }$. The expectation of this object can be interpreted
as the propagator of a particle in a harmonic potential, where the paths all
are ``pinned'' such that $x(t_j)=x_j$, $1\leq j\leq n$. Following the ideas
of the remark at the end of the section \ref{GaussCoro} we will have to
apply (\ref{phidelta}) repeatedly. But due to the form of $T{\rm I}_h(\xi ),$
which contains $\xi $ only in the exponent up to second order, all these
integrals are expected to be Gaussian.

Using this we arrive at the following proposition.

\begin{proposition}
\label{TIhProdDelta}For $x_0<x_j<x$, $1\leq j\leq n,\quad $ $%
t_0<t_j<t_{j+1}<t$, $1\leq j\leq n-1$,\\ $\quad {\rm I}_h\prod_{j=1}^n\delta
\left( x(t_j)-x_j\right) $ is a Hida distribution and its $T$-transform is
given by%
$$
T\Big({\rm I}_h\tprod\limits_{j=1}^n\delta \left( x(t_j)-x_j\right) \Big)%
(\xi )=e^{-\frac i2\left| \xi _{\Delta ^c}\right| ^2}e^{i\left( x\xi
(t)-x_0\xi (t_0)\right) }\tprod\limits_{j=1}^{n+1}K_h^{\left( \dot \xi
\right) }\left( x_j,t_j|x_{j-1},t_{j-1}\right) \text{ .} 
$$
\end{proposition}

\TeXButton{Proof}{\proof}For $n=1$ we may check the assertion by direct
computation using formula (\ref{phidelta}). To perform induction one needs
the following lemma.

\begin{lemma}
Let $[t_0,t]\subset [t_0^{^{\prime }},t^{^{\prime }}]$ then%
$$
K_h^{\left( \big(\xi +\lambda {\1}_{[t_0^{^{\prime }},t^{^{\prime }}]}\big)^{%
\D\cdot }\right) }\left( x,t|x_0,t_0\right) =K_h^{\left( \dot \xi \right)
}\left( x,t|x_0,t_0\right) ,\text{ }\forall \lambda \in {\R}\text{ .} 
$$
\end{lemma}

The lemma is also proven by a lengthy but straightforward computation. On a
formal level the assertion of the lemma is obvious as both sides of the
equation are solutions of the same Schr\"odinger equation if $[t_0,t]\subset
[t_0^{^{\prime }},t^{^{\prime }}]$ . \TeXButton{End Proof}{\endproof}\bigskip%
\ 

The proposition states what one intuitively expects, ordinary propagation
from one intermediate position to the next.

\section{An example: Quantum mechanics on a circle}

\noindent In this section we study a free quantum system whose one degree of
freedom is constrained to a unit circle. Constructing a path integral for
such a system, one has to take into account paths with different winding
numbers $n$. Thus the following ansatz for the Feynman integrand seems to be
natural:%
$$
{\rm I}\left( \varphi _1,t|\varphi _0,0\right) \equiv \sum\limits_{n=-\infty
}^\infty {\rm J}\,\delta \left( \varphi \left( t\right) -\varphi _1+2\pi
n\right) \text{ ,\quad {\rm J=J}}_{\sqrt{i}} 
$$
where $\varphi (t)=\varphi _0+B(t)$ is the angle of position modulo $2\pi $.
(Other quantizations would arise if we summed up the contributions from
different winding numbers with a phase factor $e^{i\theta n}$ \cite{Ri87}.)
However multiplication by ${\rm J}$ corresponds to complex scaling by $z=%
\sqrt{i}$ and we have seen in section \ref{Donskerseries} that the series
does not converge for this value of $z$. A formal calculation (e.g., using
Theorem \ref{DonSeries} and equation (\ref{TI0xi})) would lead to the
following $S$-transform:%
$$
S{\rm I}\left( \varphi _1,t|\varphi _0,0\right) \left( \theta \right) =S{\rm %
I}_0\left( \varphi _1,t|\varphi _0,0\right) \left( \theta \right) \cdot
\vartheta \left( \frac 1t\left( i\int\limits_0^t\theta \left( s\right) {\rm d%
}s-\left( \varphi _1-\varphi _0\right) \right) ,\frac{2\pi }t\right) \text{ .%
} 
$$
However the $\vartheta $-function does not converge for these arguments, see 
\cite{Mu79}. To stay within the ordinary white noise framework we thus
consider as final states smeared wave packets $F$ instead of strictly
localized states. So let 
$$
F\left( \varphi \right) =\sum\limits_{l=-\infty }^\infty a_l\,e^{il\varphi }%
\text{ ,} 
$$
where $\sum\limits_{l=-\infty }^\infty \left| a_l\right| \exp \left( \frac
12s^2l^2\right) <\infty $ for some $s>0$. This leads to:%
\begin{eqnarray*}
{\rm I}&=&{\rm J}F\left( B\left( t\right) +\varphi _0\right) %
\\&=&\sum\limits_{l=-\infty }^\infty a_l{\rm J}^{}\exp \left( il\left( %
\left\langle %
 \omega , \1 _{\left[ 0,t\right) }\right\rangle +\varphi _0\right) \right)  .
\end{eqnarray*}It is then easy to calculate%
\begin{eqnarray*}
T\,{\rm I}\left( \theta \right) %
& = & \sum\limits_{l=-\infty }^\infty a_{l\,}e^{il\varphi _0} %
T\left( {\rm J}\exp \left( il\left\langle \omega ,\1 _{\left[ 0,t\right) } %
\right\rangle \right) \right) \left( \theta \right) %
\\& = &\sum\limits_{l=-\infty }^\infty a_l\,e^{il\varphi _0}\exp %
\left( -\frac i2 %
\int \left( \theta +l \1 _{\left[ 0,t\right) }\right) ^2{\rm d}\tau \right) %
\\& = & e^{-\frac i2\int \theta ^2{\rm d}\tau }\sum\limits_{l=- \infty }^\infty a_{l\,}\exp \left( -\frac i2l^2t+il\left( -\int %
\limits_0^t\theta \left( s\right) {\rm d}s+\varphi _0\right) \right) 
\text{ .}
\end{eqnarray*}
To ensure convergence of the series we estimate:%
\begin{eqnarray*}
  \left| T\,{\rm I}\left( \theta \right) \right|  %
& \leq & \sum\limits_{l=-\infty }^\infty \left| a_l %
\right| \,e^{ \,\left| l\right| \left| \left(  \1 _{\left[ 0,t\right) }, \theta \right) \right| } %
\,e^{\frac 12\left| \theta \right| _0^2} %
\\& \leq & \sum\limits_{l=-\infty }^\infty \left| a_l\right| \, %
e^{\left| l\right| \sqrt{t}\left| \theta \right| _0}\,e^{\frac 12\left|  \theta \right| _0^2} %
\\& \leq & \left( \sum\limits_{l=-\infty }^\infty \left| a_l\right| %
 e^{\frac 12s^2l^2}\right) e^{\frac 12\left( 1+\frac t{s^2}\right) \left| \theta \right| _0^2}\text{ .} 
\end{eqnarray*}
This is a uniform bound, sufficient for the application of Theorem \ref{conv}%
. Thus we have proved ${\rm I}\in \left( {\cal S}\right) ^{\prime }$. It is
straightforward to check that the Feynman integral%
$$
\E ({\rm I})=\sum_{t=-\infty }^\infty a_l\exp \left( -\frac i2l^2t+il\varphi
_0\right) 
$$
solves the corresponding Schr\"odinger equation.

\LaTeXparent{dis3.tex}

\chapter{Feynman integrals and complex scaling}

\section{General remarks}

It has been shown in section \ref{free} that the kinetic energy term and the
factor compensating the Gaussian fall-off of the white noise measure combine
to give a well-defined Hida distribution%
$$
{\rm J:=J}_{\sqrt{i}}={\rm Nexp}\left( \tfrac{i+1}2|\omega |^2\right) 
$$
So the central question in realizing Feynman integrals in terms of white
noise distributions is the definition of ${\rm J}\cdot \varphi $ for most
general $\varphi $ (e.g., $\varphi =\delta (\vec B(t)-\vec x)$ in order to
construct the free particle propagator).

A very elegant and general way of defining products of ${\rm J}$ and other
distributions has been suggested in \cite{S93}, where the connection between 
${\rm J}$ and complex scaling was noted. One has ${\rm J}=\sigma _{\sqrt{i}%
}^{\dagger }\1 $ by Lemma \ref{scalJz}. In order to define products with $%
{\rm J}$ one approximates the other factor by test functionals and then
studies the convergence of the scaled sequence according to Theorem \ref
{Jsigmaz}.

\noindent Here we have to remind the reader of Example \arabic{Warnung} on
page \pageref{WarnungP}. If we want to define the action of $\sigma _z$ on $%
\Phi \in ({\cal N})^{\prime }$ by a limiting procedure, the result depends
on the choice of the approximating sequence $\varphi _n\rightarrow \Phi $.
In view of Theorem \ref{Jsigmaz} the same care is necessary if we want to
define the pointwise product ${\rm J}\cdot \Phi $.\medskip\ 

If we choose ${\cal N=S}_d(\R )$ and 
$$
\varphi =\delta (\vec B(t)-\vec x)\cdot \exp (-i\int_0^tV(\vec B(\tau )){\rm %
d}\tau ) 
$$
then $\E ({\rm J}\cdot \varphi )$ is a representation of a Feynman integral,
i.e., it should coincide with the particle propagator $K(\vec x,t|\vec 0,0)$.

In this section we follow the idea suggested by Lemma \ref{JzphiLem}. There we
proved%
$$
{\rm J}_z\cdot \varphi =\sigma _z^{\dagger }(\sigma _z\varphi )\,,\quad
\,\varphi \in ({\cal N})\;. 
$$
We will use the right hand side as a definition of the left side for a
larger class of $\varphi $ if this makes sense. Since the functionals $%
\varphi $ in question have kernels in ${\cal H}_{\Ckl }^{\hat \otimes n}$ we
discussed extensions of the scaling operator, which led to sufficient
conditions (in Proposition \ref{sigmaExt}) on $\varphi $ to ensure the
existence of $\sigma _z\varphi $ in ${\cal G}^{\prime }$ or more restrictive
ones for $\sigma _z\varphi \in $ ${\cal M}$. The last case is interesting if
we choose $\varphi =\exp (-i\int_0^tV(\vec B(\tau )){\rm d}\tau ).$ Then we
can justify the multiplication with $\sigma _z\delta $ afterwards (Theorem 
\ref{DeltaInMSt}).\medskip\ 

Now assume $\varphi \in {\cal G}^{\prime }$ such that $\sigma _z\varphi $ is
well defined in ${\cal G}^{\prime }$. Then we define%
$$
{\rm J}_z\varphi :=\sigma _z^{\dagger }(\sigma _z\varphi )\in ({\cal N}%
)^{\prime }\text{ ,} 
$$
such that%
$$
\E ({\rm J}_z\varphi )=\E (\sigma _z^{\dagger }(\sigma _z\varphi )\,)=\E %
(\sigma _z\varphi )\;. 
$$
Using equation (\ref{sigmaChaos}) we obtain%
$$
\E (\sigma _z\varphi )=\tilde \varphi ^{(0)}=\sum_{k=0}^\infty \frac{(2k)!}{%
k!\,2^k}(z^2-1)^k\,{\rm tr}^k\varphi ^{(2k)}\,. 
$$
Considerations of this type have been used by Hu and Meyer \cite{HM88}. They
defined the Feynman integral by 
$$
\E ({\rm J}_{\sqrt{i}}\varphi )=\sum_{k=0}^\infty \frac{(2k)!}{k!\,2^k}%
\,(i-1)^k\,{\rm tr}^k\varphi ^{(2k)}\,, 
$$
whenever the right hand side is well defined. (Note that they used different
normalization in the definition of chaos expansion.) One central question in
this approach is the existence of iterated traces if $\varphi ^{(2k)}\in 
{\cal H}_{\Ckl }^{\hat \otimes 2k}$. This was one important motivation in
the work \cite{JK93}, see Proposition \ref{JoKal} for a brief account. The
second problem in this approach is that it depends on the knowledge of the
chaos expansion of $\varphi $, which often cannot be calculated explicitly
enough. An alternative approach is suggested by the work of Doss \cite{D80}.

\section{Inspection of the Doss approach}

Let $V:\R ^d\rightarrow \R $ denote a potential on $d$ dimensional space. We
assume that $V$ has an extension to an analytic function (also denoted by $V 
$) defined on the following ``strip''%
$$
{\bf S}=\left\{ \left. \vec x+\sqrt{i}\vec y \; \right| \ \vec x\in {\bf D}%
\text{ and }\vec y\in \R ^d\right\} 
$$
where{\bf \ }${\bf D}\subset \R ^d$ is a connected open set.

Doss studies the expression%
$$
\psi (t,\vec x)=\E \left\{ f \left( \vec x+\sqrt{i} \vec B(t)\right) \exp
\left( -i\int_0^tV\left( \vec x+\sqrt{i}\vec B(\tau )\right) {\rm d}\tau
\right) \right\}  
$$
to obtain Feynman Kac type solutions of the time dependent Schr\"odinger
equation, where $f:{\bf S}\rightarrow \C $ plays the role of the initial
wave function $f(\vec x)=\psi (0,\vec x)$. He introduces conditions on $V$
to define $\exp \left( -i\int_0^tV(\vec x+\sqrt{i}\vec B(\tau ))\,{\rm d}%
\tau \right) $ as a well defined random variable. Nevertheless these
conditions are not very transparent, so we will restrict ourselves to
sub--classes of potentials where the meaning of the conditions becomes more
obvious. But before we need to give the underlying lemma. On the space $%
{\cal C}\left( [0,\infty ),\R ^d\right) $ we introduce the norms%
$$
\lnorm  \vec g\rnorm  _t=\sup _{j=1,\ldots ,d}\,\sup _{\tau \in [0,t)}\left|
\vec g_j(\tau )\right| \,,\qquad \vec g\in {\cal C}\left( [0,t),\R ^d\right)
. 
$$

\begin{lemma}
{\rm \cite{D80}}\label{DossLemma}\\Let $k:\R _{+}\rightarrow \R _{+}$ denote
a measurable function. Then%
$$
\E \left( k(\lnorm  \vec B\rnorm  _t)\right) \leq 2d\sqrt{\frac 2{\pi t}}%
\cdot \int_0^\infty k(u)\exp (-\frac{u^2}{2t})\;{\rm d}u\,. 
$$
\smallskip\ 
\end{lemma}

\noindent Now we state our central assumption

\begin{definition}
Let $z\in \C $. An analytic function $V:{\bf S}\rightarrow \C $ is said to
be in the Doss class (with parameters $z,a,b$) if there exist $a,b\geq 0$
such that $V$ obeys the following bound 
\begin{equation}
\label{DossClass}{\rm Im\,}V(\vec x+z\vec y)\leq a+b|\vec y|^2\,,\quad
\,\vec x\in {\bf D}\,,\ \,\vec y\in \R ^d\text{ .}
\end{equation}
\end{definition}

The above definition is interesting in view of the following proposition.

\begin{proposition}
\label{DossProp} Let $V$ be in the Doss class with parameters $z\in \C ,\
a,b\geq 0$.\smallskip\\1. Let $\vec x\in {\bf D}$ $,\ \,p\geq 1$ and $%
b<\frac 3{2pt^2}$ then 
$$
\exp \left( -i\int_0^tV\left( \vec x+z\vec B(\tau )\right) {\rm d}\tau
\right) \in L^p(\mu )\text{ .} 
$$
\\2. Let ${\bf D}$ be convex and $\vec x,\vec y\in {\bf D},\ p\geq 1$ and $%
b<\frac 3{14pt^2}$ then 
$$
\exp \left( -i\int_0^tV\left( \vec x+\frac \tau t(\vec y-\vec x)+z\left(
\vec B(\tau )-\frac \tau t\vec B(t)\right) \right) {\rm d}\tau \right) \in
L^p(\mu )\,\text{ .} 
$$
\end{proposition}

\TeXButton{Proof}{\proof}We prove assertion 2.~the proof of statement 1.~is
completely analogous. The following holds%
$$
\left| \exp -i\int_0^tV\left( \vec x+\frac \tau t(\vec y-\vec x)+z\left(
\vec B(\tau )-\frac \tau t\vec B(t)\right) \right) {\rm d}\tau \right| 
\hspace{3cm} 
$$
\begin{eqnarray*}
&=& \exp \int_0^t{\rm Im\,}V\left( \vec x+\frac \tau t(\vec y-\vec x)+z\left( %
\vec B(\tau )-\frac \tau t\vec B(t)\right) \right) {\rm d}\tau  %
\\ & \leq & \exp \left( ta+b\int_0^t\lnorm  \vec B(\tau ) %
-\frac \tau t\vec B(t) \rnorm  _t^2{\rm d}\tau \right)  %
\\ & \leq & \exp \left( ta+b\int_0^t\left( \lnorm  \vec B(t) \rnorm %
_t^2+2\frac \tau t\lnorm  \vec B(t)  \rnorm  _t^2+ %
\frac{\tau ^2}{t^2}\lnorm  \vec B(t) \rnorm  _t^2\right) %
{\rm d}\tau \right)  \\ &=& \exp \left( ta+\frac 73bt\lnorm  \vec B(t) %
 \rnorm  _t^2\right) \text{ .} 
\end{eqnarray*}
Lemma \ref{DossLemma} gives%
$$
\E \left( \left| \exp -i\int_0^tV\left( \vec x+\frac \tau t(\vec y-\vec
x)+z\left( \vec B(\tau )-\frac \tau t\vec B(t)\right) \right) {\rm d}\tau
\right| ^p\right) \hspace{1cm} 
$$
$$
\hspace*{19mm} \leq \E \left( \exp \left( pta+\frac 73ptb\lnorm  \vec B(t) 
\rnorm
_t^2\right) \right) \hspace*{19mm} 
$$
$$
\hspace*{19mm}\leq 2d\sqrt{\frac 2{\pi t}}\int_0^\infty \exp \left(
pta-\left( \frac 1{2t}-\frac 73ptb\right) u^2\right) {\rm d}u 
$$
which is finite if $b<\frac 3{14pt^2}$.\TeXButton{End Proof}{\endproof}%
\bigskip\ 

\example Let $d=1\,,\,z=\sqrt{i}\,$, and $\,{\bf D}\,$ bounded. Then
consider the polynomial potential%
$$
V(x)=g\sum_{k=0}^nc_kx^k\,,\quad \,g,c_k\in \R \qquad k\leq n\quad \text{and}%
\quad c_n=1\,. 
$$
First we assume the harmonic oscillator potential, i.e., $n=2$. If $g<b$
then there exist $a=a({\bf D})>0$ such that (\ref{DossClass}) is fulfilled.
Note in particular that negative values of the coupling constant $g$ are
allowed. For positive $g$ the restriction $g<b$ is consistent with the fact
that the propagator $K_h(x,t|0,0)$ of the harmonic oscillator is only
defined for small times $t$ (compare (\ref{17}) with Proposition \ref
{DossProp}.2, which is the relevant case for propagators as we will see).

Now let $n=2+8m\,,\ m\in \N _0$ and $g<0$ {\bf or} $n=6+8m\,,\,\ m\in \N _0$
and $g>0$, then the dominant behavior of the highest power shows that ${\rm %
Im}V(x+\sqrt{i}y)<0$ for $y$ large enough. So due to the smoothness of $V$ 
$$
{\rm Im\,}V(x+\sqrt{i}y)<a 
$$
for some $a=a({\bf D})\geq 0$. \\Note that we have included an interesting
class of repulsive potentials i.e., $g<0$. \\Let us also mention that this
example allows a comparison with results in the recent monograph \cite{Us94}%
. He obtained a nice behaviour for sextic oscillators i.e., for some
polynomial interactions with leading power $x^6$. If the coefficients $c_k$
satisfy an additional condition, the Schr\"odinger equation becomes
quasi--exact solvable, i.e., a finite number of energy eigenvalues $E_m$ and
eigenfunctions can be calculated explicitly. On the other hand the work of
Bender and Wu \cite{BeWu69} demonstrated that potentials with leading power $%
x^4$ produce very complicated non--perturbative effects (e.g., rapid growth
of $\left| E_m\right| $ $\sim m!\,A^m$  and the ``horn structure'' of the
singularities of the function $g\mapsto E(g):=\sum_{m=0}^\infty E_mg^m$ in a
neighborhood of zero).\\For more examples see \cite{D80}. \bigskip\ 

To shorten notation we define 
$$
\varphi :=\exp -i\int_0^tV(\vec x_0+\vec B(\tau ))\,{\rm d}\tau \,,\quad
x_0\in {\bf D\;.} 
$$
Of course $\varphi \in L^p(\mu )$ for any $p\geq 0$. If $V$ satisfies the
conditions of Proposition \ref{DossProp} for some $p\geq 0$ we will write%
$$
\sigma _z\varphi \stackrel{\rm def}{=}\exp -i\int_0^tV(\vec x_0+z\vec B(\tau
))\,{\rm d}\tau \in L^p(\mu ) 
$$
since the right hand side may be viewed as a well defined extension of the
scaling operator $\sigma _z$. (Any useful extension of the scaling operator
is expected to reflect the structure of the original definition $\sigma
_z\varphi (\vec \omega )=\varphi (z\vec \omega )$ for $\varphi \in ({\cal S}%
_d).$)\bigskip\ 

\noindent {\it Smooth final wave function.} We want to define ${\rm J}%
_z\cdot \varphi \cdot \psi $ by extension of formula (\ref{Jzphi}):%
$$
{\rm J}_z\cdot \varphi \cdot \psi =\sigma _z^{\dagger }(\sigma _z\varphi
\cdot \sigma _z\psi )\text{ .} 
$$
If we assume $V$ in the Doss class for some $a$ and $b<\frac 3{2t^2}$ then
there exists $p>1$ such that $\sigma _z\varphi \in L^p(\mu )$, i.e., by
Proposition \ref{LpGStrich} $\sigma _z\varphi \in {\cal G}^{\prime }$. Let $%
\psi \in {\cal G}$ such that $\sigma _z\psi \in {\cal G}$ (for example $\psi
\in ({\cal S}_d)$) then $\sigma _z\varphi \cdot \sigma _z\psi \in {\cal G}%
^{\prime }$ and $\sigma _z^{\dagger }(\sigma _z\varphi \cdot \sigma _z\psi
)\in ({\cal S}_d)^{\prime }$. For example we can choose $\psi $ to be an
approximation of $\delta ^d(\vec B(t)-(\vec x-\vec x_0))$ and $z=\sqrt{i}$
then ${\rm J}_{\sqrt{i}}\varphi \psi $ defined above is an approximation of
the Feynman integrand. Thus 
$$
\E ({\rm J}_{\sqrt{i}}\varphi \psi )=\E \left( \sigma _{\sqrt{i}}^{\dagger
}(\sigma _{\sqrt{i}}\varphi \cdot \sigma _{\sqrt{i}}\psi )\right) =\E %
(\sigma _{\sqrt{i}}\varphi \cdot \sigma _{\sqrt{i}}\psi ) 
$$
is an approximation of the propagator $K_V(\vec x,t|\vec x_0,0)$. \bigskip\ 

\noindent {\it Rewriting the propagator. }Now assume $V$ in the Doss class
for some $a$ and $b<\frac 3{14t^2}$ such that $\sigma _z\varphi \in {\cal M}$%
. This condition is not easy to check but nevertheless at the end of this
consideration it will be possible to extend the validity of the result to
more general $V$.

We define%
$$
{\rm J}_z\cdot \delta \cdot \varphi :=\sigma _z^{\dagger }(\sigma _z\delta
\cdot \sigma _z\varphi ) 
$$
here $\delta $ is shorthand for $\delta ^d\left( \vec B(t)-(\vec x-\vec
x_0)\right) $. This is well defined since $\sigma _z\delta \in {\cal M}%
^{\prime }$ (Theorem \ref{DeltaInMSt}) and the pointwise product $\sigma
_z\delta \cdot \sigma _z\varphi $ is in ${\cal G}^{\prime }$ in view of (\ref
{MStrichProd}).

\noindent The homogeneity property in Proposition \ref{DeltaHom} writes 
$$
\sigma _z\delta ^d\left( \vec B(t)-(\vec x-\vec x_0)\right) =\left( \frac 1{z%
\sqrt{t}}\right) ^d\delta ^d\left( \langle \vec \omega ,\frac{\1 _t}{\sqrt{t}%
}\rangle -\frac{\vec x-\vec x_0}{z\sqrt{t}}\right) \,, 
$$
since $\delta ^d$ is no more than the product of $d$ independent Donsker
deltas. Using this and Proposition \ref{DeltaMode} we can calculate:%
\begin{eqnarray*}
\E \left( \sigma _z^{\dagger }(\sigma _z\varphi \cdot \sigma _z\delta %
)\right) &=& \E (\sigma _z\delta \cdot \sigma _z\varphi ) %
\\&=& \left( \frac 1{z\sqrt{t}}\right) ^d\left\langle \!\!\!\left\langle %
\delta ^d\left( \langle \cdot ,\frac{\1 _t}{\sqrt{t}}\rangle - %
\frac{\vec x-\vec x_0 }{z\sqrt{t}}\right) ,\sigma _z\varphi %
\right\rangle \!\!\!\right\rangle  %
\\&=& \left( \frac 1{\sqrt{2\pi t}\ z}\right) ^d e^{-\frac{(\vec x- \vec x_0)^2} {2z^2t }}%
\E \left( P\tau _{\frac{\vec x-\vec x_0}{zt}\1 _t}(\sigma _z %
\varphi )\right)  
\end{eqnarray*}
where $P:{\cal M}\mapsto {\cal G}^{\prime }$ is defined as in section \ref
{ProjSec} with $\eta =\1 _t/\sqrt{t}$. Using the definitions we obtain 
$$
P\tau _{\frac{\vec x-\vec x_0}{zt}\1 _t}(\sigma _z\varphi )=\exp
-i\int_0^tV\left( \vec x_0+z\left\langle \vec \omega -\langle \vec \omega ,%
\1 _t\rangle \1 _t/t+\frac{\vec x-\vec x_0}{zt}\1 _t,\1 _\tau \right\rangle
\right) {\rm d}\tau  
$$
$$
\hspace*{2cm}=\exp -i\int_0^tV\left( \vec x_0+\frac \tau t(\vec x-\vec
x_0)+z\left( \vec B(\tau )-\frac \tau tB(t)\right) \right) {\rm d}\tau \;. 
$$
This term can now be substituted in the above formula. Furthermore it is
possible to change the representation of Brownian motion in the expectation.
If $\vec B(\tau )$ is a Wiener process then also $l\vec B(\frac \tau
{l^2})\,,\,l>0$ and they have the same covariance $\min (\tau ,\tau ^{\prime
})$. This property is called scaling invariance of Brownian motion.

\noindent Thus we have derived 
\begin{eqnarray*}
\E ({\rm J}_z\delta \varphi ) &=& \left( \frac 1{\sqrt{2\pi z^2t}}\right) ^d %
\exp \left( -\frac{(\vec x-\vec x_0)^2}{2z^2t}\right) \cdot %
\\& &\hspace*{2cm} \cdot \E \left( \exp -i\int_0^tV\left( \vec x_0+ %
\frac \tau t (\vec x-\vec x_0) +z\left( \vec B (\tau )-\frac \tau t %
\vec B(t)\right) \right) {\rm d}\tau \right)  %
\\&=& \left( \frac 1{\sqrt{2\pi z^2t}}\right) ^d\exp \left( -%
\frac{(\vec x -\vec x_0)^2}{2z^2t}\right) \cdot %
\\& &\hspace*{2cm} \cdot %
\E \left( \exp -it\int_0^1V\left( \vec x_0+s(\vec x-\vec x_0)+z\sqrt{t}%
\left( \vec B(s)- s \vec B(1)\right) \right) {\rm d}s\right) %
\end{eqnarray*}
Note that $\tau \mapsto \vec B(\tau )-\frac \tau t\vec B(t)$ is a
representation of the Brownian bridge from zero to $t$. In the physically
relevant case $z=\sqrt{i}$ the factor%
$$
\left( \frac 1{\sqrt{2\pi it}}\right) ^d\exp \left( -\frac{(\vec x-\vec
x_0)^2}{2it}\right) =K_0(\vec x,t|\vec x_0,0)\, 
$$
appearing in the above formulae is the free particle propagator. So we
obtained a well defined probabilistic expression. To write down the right
hand side of the above equations it is only necessary that $V$ satisfied the
Doss condition for some $a$ and $b<\frac 3{14t^2}$. Then the functional in
the expectation is in $L^p(\mu )$ for some $p>1$. Hence the physical
relevant quantity is well defined, also if $\sigma _z\varphi \in {\cal M}$
is not true. It remains to show that this is in fact the fundamental
solution of the corresponding Schr\"odinger equation. We will only give a
partial answer to this question. Expressions of the type discussed above
were also studied in the work of Yan \cite{Yan93}. We will state his result
in our setting

\begin{proposition}
{\rm \cite[Th's 3.9 and 5.2]{Yan93}} \\ Let $V$ be as above. The expression%
$$
q(\vec x,\lambda t|\vec x_0,0)\,=\frac 1{\sqrt{2\pi \lambda t}}\exp \left( -%
\frac{(\vec x-\vec x_0)^2}{2\lambda t}\right) \cdot \hspace{6cm} 
$$
$$
\hspace*{3cm}\cdot \E \left( \exp -it\int_0^1V\left( \vec x_0+s(\vec x-\vec
x_0)+z\sqrt{\lambda t}(\vec B(s)-s\vec B(1))\right) {\rm d}s\right)  
$$
has an analytic continuation to all $\lambda \in \C $ such that ${\rm Re\,}%
\lambda >0$. Moreover $q(\vec x,\lambda t|\vec x_0,0)\,$ is the fundamental
solution of 
$$
\frac{\partial \psi }{\partial t}=\lambda (\frac 12\Delta -V)\psi \,,\qquad
\lambda \in \C _{+}\,. 
$$
\end{proposition}

So the quantities constructed above solve the right (partial) differential
equation if $z^2\left( \sim \lambda \right) $ is such that ${\rm Re\,}z^2>0$%
. The open question remains if this is also true for $z^2=i$.
\LaTeXparent{dis3.tex}

\chapter{Quantum mechanical propagators in terms of white noise
distributions \label{Quantum}}

\section{An extension of the Khandekar Streit method \label{quatum}}

In order to pass from the free motion to more general situations, one has to
give a rigorous definition of the heuristic expression\ 
$$
{\rm I}={\rm I}_0\exp \left( -i\int_{t_0}^tV\left( \vec x\left( \tau \right)
\right) \text{ }{\rm d}\tau \right) . 
$$
In \cite{KaS92} Khandekar and Streit accomplished this by perturbative
methods in the case $d=1$ and $V$ is a finite signed Borel measure with
compact support. (Note that singular potentials are included in this class.)
We generalize the construction by allowing time-dependent potentials and a
Gaussian fall--off instead of a bounded support. In section \ref{quatum2}
potentials of exponential fall--off are considered, for the price that we
need to use a larger distribution space.

Let $\De \equiv \left[ {\sf T_0},{\sf T}\right] \supset \Delta =[t_0,t]$ and
let $v$ be a finite signed Borel measure on $\R\times \De $ . Let $v_x$
denote the marginal measure\ 
$$
v_x\left( A\right) \equiv v\left( A\times \De \right) \ ,\qquad A\in {\cal B}%
(\R ) 
$$
similarly\ 
$$
v_t\left( B\right) \equiv v\left( \R \times B\right) \ ,\qquad B\in {\cal B}(%
\De ). 
$$

\subsection{The Feynman integrand as a Hida distribution \label{quatum1}}

\noindent We assume that $|v|_x$ and $|v|_t$ satisfy:\smallskip\ 

i ) $\exists \,R>0$ $\forall \,r>R:\left| v\right| _x\left( \left\{ x:\text{ 
}\left| x\right| >r\right\} \right) <e^{-\beta r^2}$ for some $\beta >0$ ,%
\smallskip\ 

ii ) $\left| v\right| _t$ has a $L^\infty $density.\medskip\ \ 

\noindent The essential bound of this density is denoted by $C_v$.\\Let us
first describe heuristically the construction by treating $v$ as an ordinary
function $V$ before stating the rigorous result Theorem \ref{3.1}. The
starting point is a power series expansion of $\exp \left(
-i\int_{t_0}^tV(x(\tau ),\tau )\,{\rm d}\tau \right) $ using $V(x(\tau
),\tau )=\int {\rm d}x\,V(x,\tau )\,\delta \left( x(\tau )-x\right) :$

$$
\exp \left( -i\int_{t_0}^tV(x(\tau ),\tau ){\rm d}\tau \right)
=\dsum\limits_{n=0}^\infty \left( -i\right) ^n\int_{\Lambda _n}{\rm d}^nt%
\stackrel{n}{\stackunder{i=1}{\Pi }}\int {\rm d}x_i\,V(x_i,t_i)\delta
(x(t_i)-x_i) 
$$
where 
\begin{equation}
\label{Deltan}\Lambda _n=\left\{ (t_1,...,t_n)|\,t_0<t_1<...<t_n<t\right\} .
\end{equation}
If necessary we will also use the notation $\Lambda _n(t,t_0).$\bigskip\ 

More generally we can show: \ 

\begin{theorem}
\label{3.1} 
\begin{equation}
\label{IDyson}{\rm I}={\rm I}_0+\dsum\limits_{n=1}^\infty \left( -i\right)
^n\int_{\R^n}\int_{\Lambda _n}\stackrel{n}{\stackunder{i=1}{\Pi }}v\left( 
{\rm d}x_i,{\rm d}t_i\right) \,{\rm I}_0\stackrel{n}{\stackunder{j=1}{\Pi }}%
\delta \left( x\left( t_j\right) -x_j\right) 
\end{equation}
exists as a Hida distribution in case $V$ obeys i) and ii).
\end{theorem}

\proof {\bf 1)} ${\rm I}_n=\int_{\R^n}\int_{\Lambda _n}\stackrel{n}{%
\stackunder{i=1}{\Pi }}v\left( {\rm d}x_i,{\rm d}t_i\right) \,{\rm I}_0%
\stackrel{n}{\stackunder{j=1}{\Pi }}\delta \left( x\left( t_j\right)
-x_j\right) $ is a Hida distribution for $n\geq 1$. This is shown by
applying Theorem \ref{Bochner}. \medskip\ \ 

Hence we have to establish a bound of the required type for the T-transform
of the integrand. From formulae (\ref{Ko}) and (\ref{Tnd}) we find ($\theta
\in {\cal S}_{\Ckl }$)%
\begin{eqnarray*}
T\left( {\rm I}_0\prod_{j=1}^n\delta (x(t_j)-x_j)\right) (\theta ) &=& %
\exp (- \frac{i}2|\theta |_0^2)\cdot \prod_{j=1}^{n+1}\frac 1{\sqrt{2\pi i(t_j-t_{j-1})}} \cdot %
\\ & & \cdot \exp \left( \sum_{j=1}^{n+1}i\frac{(x_j-x_{j-1})^2} {2(t_j-t_{j-1})} \right) %
\\ & & \cdot \exp \left( \sum_{j=1}^{n+1}i\frac{x_j-x_{j-1}}{t_j-t_{j-1}}%
\int_{t_{j-1}}^{t_j}\theta (s){\rm d}s\right)%
\\ & & \cdot \exp \left( \sum_{j=1}^{n+1} \frac i{2(t_j-t_{j-1})}\left[ %
\int_{t_{j-1}}^{t_j}\theta (s){\rm d}s\right] ^2\right) \text{ .} %
\end{eqnarray*}
It is easy to estimate the last term%
$$
\left| \sum_{j=1}^{n+1}\frac i{2(t_j-t_{j-1})}\left[
\int_{t_{j-1}}^{t_j}\theta (s){\rm d}s\right] ^2\right| \leq
\sum_{j=1}^{n+1}\frac 1{2(t_j-t_{j-1})}\left[ \int_{t_{j-1}}^{t_j}|\theta
(s)|^2{\rm d}s\cdot \int_{t_{j-1}}^{t_j}\1 _{[t_{j-1},t_j]}^2(s){\rm d}%
s\right] 
$$
$$
\hspace{17mm}\leq \sum_{j=1}^{n+1}\frac 12\int_{t_{j-1}}^{t_j}|\theta (s)|^2%
{\rm d}s\leq \frac 12|\theta |_0^2\text{ .} 
$$
In order to estimate the term 
\begin{equation}
\label{Problem}\exp \left( \sum_{j=1}^{n+1}i\frac{x_j-x_{j-1}}{t_j-t_{j-1}}%
\int_{t_{j-1}}^{t_j}\theta (s){\rm d}s\right) \text{ ,} 
\end{equation}
we proceed as follows%
$$
\sum_{j=1}^{n+1}i\frac{x_j-x_{j-1}}{t_j-t_{j-1}}\int_{t_{j-1}}^{t_j}\theta
(s){\rm d}s=\frac x{t-t_n}\int_{t_n}^t\theta (s){\rm d}s-\frac{x_0}{t_1-t_0}%
\int_{t_0}^{t_1}\theta (s){\rm d}s\ + 
$$
$$
\hspace*{5cm}\sum_{j=1}^nx_j\left( \frac{\int_{t_{j-1}}^{t_j}\theta (s){\rm d%
}s}{t_j-t_{j-1}}-\frac{\int_{t_j}^{t_{j+1}}\theta (s){\rm d}s}{t_{j+1}-t_j}%
\right) \text{ .} 
$$
By the mean value theorem%
$$
\sum_{j=1}^nx_j\left( \frac{\int_{t_{j-1}}^{t_j}\theta (s){\rm d}s}{%
t_j-t_{j-1}}-\frac{\int_{t_j}^{t_{j+1}}\theta (s){\rm d}s}{t_{j+1}-t_j}%
\right) =\sum_{j=1}^nx_j(\theta (\tau _j)-\theta (\tau _{j+1}))\text{ ,} 
$$
where $\tau _k\in (t_{k-1},t_k)$. Then we can estimate%
\begin{eqnarray*}
\left| \sum_{j=1}^nx_j(\theta (\tau _j)-\theta (\tau _{j+1}))\right| & \leq & %
\sum_{j=1}^n|x_j||\theta (\tau _j)-\theta (\tau _{j+1})| %
\\ & \leq & \left( \sup %
_{1\leq j\leq n}|x_j|\right) \cdot \sum_{j=1}^n\int_{\tau _j}^{\tau _{j+1}}|\theta ^{\prime }(s)|{\rm d}s %
\\ & \leq & \left( \sup _{1\leq j\leq n}|x_j|\right) \cdot %
\sum_{j=1}^n\int_{t_j}^{t_{j+1}}|\theta ^{\prime }(s)|{\rm d}s %
\\ & \leq & \left( %
\sup _{1\leq j\leq n}|x_j|\right) \cdot \int_{t_0}^t|\theta ^{\prime }(s)|%
{\rm d}s\text{ .} 
\end{eqnarray*}
Therefore we have 
$$
\left| \exp \left( \sum_{j=1}^{n+1}i\frac{x_j-x_{j-1}}{t_j-t_{j-1}}%
\int_{t_{j-1}}^{t_j}\theta (s){\rm d}s\right) \right| \hspace{65mm} 
$$
$$
\leq \exp \left( |x|\sup _{t_n\leq s\leq t}|\theta (s)|+|x_0|\sup _{t_0\leq
s\leq t_1}|\theta (s)|+\left( \sup _{1\leq j\leq n}|x_j|\right) \cdot
\int_{t_0}^t|\theta ^{\prime }(s)|{\rm d}s\right) 
$$
$$
\leq \exp \left( \left( \sup _{0\leq j\leq n+1}|x_j|\right) \cdot \left[
\sup _{t_n\leq s\leq t}|\theta (s)|+\sup _{t_0\leq s\leq t_1}|\theta
(s)|+\int_{t_0}^t|\theta ^{\prime }(s)|{\rm d}s\right] \right) \text{ .} 
$$
Let us introduce the following norm on ${\cal S}_{\Ckl }(\R )$%
$$
\left\| \theta \right\| \equiv \int_{t_0}^t|\theta ^{\prime }(s)|{\rm d}%
s+\sup _{t_0\leq s\leq t}|\theta (s)|\text{ .} 
$$
Clearly this is a continuous norm on ${\cal S}_{\Ckl }(\R )$. From the last
estimate we obtain%
$$
\left| \exp \left( \sum_{j=1}^{n+1}i\frac{x_j-x_{j-1}}{t_j-t_{j-1}}%
\int_{t_{j-1}}^{t_j}\theta (s){\rm d}s\right) \right| \leq \exp \left(
\left( \sup _{0\leq j\leq n+1}|x_j|\right) \cdot \left\| \theta \right\|
\right) 
$$
$$
\hspace*{5cm}\leq \exp \left[ \gamma \left( \sup _{0\leq j\leq
n+1}|x_j|\right) ^2\right] \cdot \exp \left( \frac 1\gamma \left\| \theta
\right\| ^2\right) \text{ ,} 
$$
where $0<\gamma $ is to be chosen later. Now we can estimate as follows%
$$
\left| T\left( {\rm I}_0\prod_{j=1}^n\delta (x(t_j)-x_j)\right) (\theta
)\right| \leq \exp \left( \frac 12|\theta |_0^2\right) \cdot
\prod_{j=1}^{n+1}\frac 1{\sqrt{2\pi (t_j-t_{j-1})}}\hspace{2cm} 
$$
$$
\hspace*{3cm}\cdot \exp \left[ \gamma \left( \sup _{0\leq j\leq
n+1}|x_j|\right) ^2\right] \cdot \exp \left( \frac 1\gamma \left\| \theta
\right\| ^2\right) \cdot \exp \left( \frac 12|\theta |_0^2\right) \ . 
$$
If we introduce the norm%
$$
\lnorm\theta \rnorm\equiv \left\| \theta \right\| +|\theta |_0\ , 
$$
which is obviously also continuous on ${\cal S}_{\Ckl }(\R )$, we have the
bound%
$$
\left| T\left( {\rm I}_0\prod_{j=1}^n\delta (x(t_j)-x_j)\right) (\theta
)\right| \hspace{7cm} 
$$
$$
\hspace{2cm}\leq \prod_{j=1}^{n+1}\frac 1{\sqrt{2\pi (t_j-t_{j-1})}}\cdot
\exp \left[ \gamma \left( \sup _{0\leq j\leq n+1}|x_j|\right) ^2\right]
\cdot \exp \left( \frac{1+\gamma }\gamma \lnorm\theta \rnorm^2\right) \ . 
$$
In order to apply Theorem \ref{Bochner} we have to show the integrability of
the first two factors with respect to $v $. To this end we will use
H\"older's inequality. \bigskip\ 

Choose $q$ $>2$ and $0<\gamma <\beta /q$ and $p$ such that $\frac 1p+\frac
1q=1.$ The property i) of $v $ yields that $e^{\gamma x^2}\in L^q(\R\times 
\De ,\left| v \right| )$. Let $Q\equiv \left( \int_{\R }\left| v \right| _x(%
{\rm d}x)\,e^{\gamma qx^2}\right) ^{1/q}$, then%
$$
\left( \int_{\R^n}\int_{\Lambda _n}\stackrel{n}{\stackunder{i=1}{\Pi }}%
\left| v \right| ({\rm d}x_i,{\rm d}t_i)e^{\gamma q\left( \stackunder{0\leq
i\leq n+1}{\sup }\left| x_i\right| \right) ^2}\right) ^{1/q}\leq e^{\gamma
\left| x_0\right| ^2}e^{\gamma \left| x\right| ^2}Q^n. 
$$
Using the property ii) of $v $ and the formula%
$$
\int_{\Lambda _n}{\rm d}^nt\stackrel{n+1}{\stackunder{j=1}{\Pi }}\frac
1{(2\pi \left| t_j-t_{j-1}\right| )^\alpha }=\left( \frac{\Gamma (1-\alpha )%
}{(2\pi )^\alpha }\right) ^{n+1}\frac{\left| t-t_0\right| ^{n(1-\alpha
)-\alpha }}{\Gamma \left( (n+1)(1-\alpha )\right) }\;,\text{ }\alpha <1 
$$
we obtain the following estimate:%
$$
\left( \int_{\R^n}\int_{\Lambda _n}\stackrel{n}{\stackunder{i=1}{\Pi }}%
\left| v \right| \left( {\rm d}x_i,{\rm d}t_i\right) \stackrel{n+1}{%
\stackunder{j=1}{\Pi }}\left( \frac 1{\sqrt{2\pi \left| t_j-t_{j-1}\right| }%
}\right) ^p\right) ^{1/p}\leq \,C_v ^{\frac np}\frac{\Gamma (\frac{2-p}2)^{%
\frac{n+1}p}}{(2\pi )^{\frac{n+1}2}}\frac{|\Delta |^{\frac np-\frac 12(n+1)}%
}{\Gamma \left( (n+1)(\frac{2-p}2)\right) ^{1/p}} 
$$
Let%
$$
C_n(x,|\Delta |)\equiv e^{\gamma \left| x_0\right| ^2}e^{\gamma \left|
x\right| ^2}Q^nC_v ^{\frac np}\frac{\Gamma (\frac{2-p}2)^{\frac{n+1}p}}{%
(2\pi )^{\frac{n+1}2}}\frac{|\Delta |^{\frac np-\frac 12(n+1)}}{\Gamma
\left( (n+1)(\frac{2-p}2)\right) ^{1/p}}\text{ .} 
$$
\begin{samepage}
H\"older's inequality yields the following estimate:%
$$
\int_{\R^n}\int_{\Lambda _n}\stackrel{n}{\stackunder{i=1}{\Pi }}|v |\left( 
{\rm d}x_i,{\rm d}t_i\right) \left| T\left( {\rm I}_0\stackrel{n}{%
\stackunder{j=1}{\Pi }}\delta (x(t_j)-x_j)\right) \left( \theta \right)
\right| \hspace{2cm} 
$$ 
\begin{equation}
\label{Bound1}\hspace*{7cm}\leq C_n\exp \left( \frac{1+\gamma }\gamma \lnorm%
\theta \rnorm^2\right) 
\end{equation}
\end{samepage}
This establishes the bound required for the application of Theorem \ref
{Bochner} and hence ${\rm I}_n$ exists as a Bochner integral in $\left( 
{\cal S}\right) ^{\prime }$.\medskip

\noindent {\bf 2)}{\rm \ }${\rm I}=\dsum\limits_{n=0}^\infty {\rm I}_n$
exists in $\left( {\cal S}\right) ^{\prime }.$\smallskip\ \ 

As the $C_n$ are rapidly decreasing in $n$ the hypotheses of Theorem \ref
{conv} are fulfilled and hence the convergence in $\left( {\cal S}\right)
^{\prime }$ is established.\endproof \bigskip

\noindent {\bf Remark.} Conditions i) and ii) allow for some rather singular
potentials, e.g.,\ $\tilde v=\sum e^{-n^2}\delta _{_n}$. For a cut-off
interaction, i.e., compactly supported $v_x$, condition i) is of course
valid. Note also that $v$ is not supposed to be a product measure, hence the
time dependence can be more intricate than simple multiplication by a
function of time. For example we can take two bounded continuous functions $%
f $ and $g$ on $\Delta $. Use one to move the potential around and the other
one to vary its strength: $v(x,t)=f(t)\,\tilde v(x-g(t))$.

\subsection{The Feynman integrand in 
\texorpdfstring{$({\cal S})^{-1}$}{(S)\textminus \textonesuperior  }
 \label{quatum2}}

Instead of $({\cal S})^{\prime }$ we can also use $({\cal S})^{-1}$ to
discuss the convergence of the perturbative expansion (\ref{IDyson}). In
this case some of the technical difficulties in estimating the term (\ref
{Problem}) disappear. Furthermore we obtain a larger class of potentials
which allows some weaker decrease in the space direction.

\begin{theorem}
Let $v$ be a finite signed Borel measure on $\R \times \De $ such that the
(absolute) marginal measures satisfy\smallskip\ \\\qquad {\rm i')} $\exists
\,R>0$ $\forall \,r>R:\left| v\right| _x\left( \left\{ x:\text{ }\left|
x\right| >r\right\} \right) <e^{-\beta r}$ for some $\beta >0$ , \smallskip\\%
\qquad {\rm ii)} $\left| v\right| _t$ has a $L^\infty $density.\medskip\\%
Then ${\rm I}$\ \ defined by (\ref{IDyson}) exists in $({\cal S})^{-1}$.
\end{theorem}

\TeXButton{Proof}{\proof}The term (\ref{Problem}) may be estimated 
$$
\left| \exp i\sum_{j=1}^{n+1}\frac{x_j-x_{j-1}}{t_j-t_{j-1}}%
\int_{t_{i-1}}^{t_j}\theta (s){\rm d}s\right| \leq \exp \left( 2|\theta
|_\infty \sum_{j=1}^{n+1}|x_j|\right) \,\,. 
$$
Using this we obtain%
$$
\left| T\left( {\rm I}_0\prod_{j=1}^n\delta (x(t_j)-x_j)\right) (\theta
)\right| \leq e^{|\theta |_0^2}\cdot \prod_{j=1}^{n+1}\frac 1{\sqrt{2\pi
(t_j-t_{j-1})}}\exp \left( 2|\theta |_\infty \sum_{j=1}^{n+1}|x_j|\right) . 
$$
To prove that this bound is integrable w.r.t. the ($n$--fold) product
measure we refer to H\"older's inequality again. Choose $q>2$ and $p$ such
that $\frac 1p+\frac 1q=1$. In this case it is sufficient to do this for all 
$\theta \in {\cal S}_{\Ckl }(\R )$ in a neighborhood of zero. A possible
choice is 
$$
\theta \in {\cal U}:=\left\{ \theta \in {\cal S}_{\Ckl }(\R );|\theta
|_\infty <\frac \beta {2q}\right\} . 
$$
Then 
$$
\left( \int_{\R^n}\int_{\Lambda _n}\stackrel{n}{\stackunder{i=1}{\Pi }}%
\left| v\right| ({\rm d}x_i,{\rm d}t_i)\exp \left( 2q|\theta |_\infty
\sum_{j=1}^{n+1}|x_j|\right) \right) ^{1/q}\leq e^{\frac \beta
q(|x|+|x_0|)}\left( \int_{\R}\left| v\right| _x({\rm d}x)\;e^{\beta
|x|}\right) ^{\frac nq} 
$$
for all $\theta \in {\cal U}$. The rest of the proof is along the lines of
the proof of Theorem \ref{3.1}. The main difference is that the convergence
of the integrals and of the series here are controlled by the corresponding
theorems for $({\cal S})^{-1}$ (Theorems 5 and 6 in \cite{KLS94}) 
\TeXButton{End Proof}{\endproof}

\section{Verifying the Schr\"odinger equation \label{Verifying}}

In this section we prove that the expectation of the Feynman integrand
constructed in section \ref{quatum1}, i.e., the Feynman integral, does
indeed solve the usual integral equation for quantum mechanical propagators,
which corresponds to the Schr\"odinger equation. (In this section we will
always assume the situation of section \ref{quatum1} for simplicity.)\medskip%
\ 

As in the case of the free motion we expect 
\begin{equation}
\label{KDef}K^{\left( \dot \xi \right) }\left( x,t|x_0,t_0\right) \equiv 
\text{ }e^{+\frac i2\left| \xi _\Delta c\right| ^2}\text{ }e^{-ix\xi \left(
t\right) +ix_0\xi \left( t_0\right) }\,\Theta (t-t_0)T{\rm I}\left(
x,t|x_0,t_0\right) \left( \xi \right) 
\end{equation}
to be the propagator corresponding to the potential $W(x,t)=V\left(
x,t\right) +\dot \xi (t)x$. More precisely we have to use the measure $w(%
{\rm d}x,{\rm d}t)=v\left( {\rm d}x,{\rm d}t\right) +\dot \xi (t)x$ ${\rm d}%
x\ {\rm d}t$.\ We now proceed to show some properties of $K^{\left( \dot \xi
\right) }$. As the propagators $K_0^{\left( \dot \xi \right) }$ are
continuous on $\R^2\times \Lambda _2$ (see (\ref{Ko})), the product $%
\stackrel{n+1}{\stackunder{j=1}{\Pi }}K_0^{\left( \dot \xi \right) }\left(
x_j,t_j|x_{j-1},t_{j-1}\right) $ is continuous on $\R
^{n+1}\times \Lambda _{n+1}$.\ Set 
\begin{equation}
\label{Series}K^{\left( \dot \xi \right) }\left( x,t|x_0,t_0\right)
=\dsum\limits_{n=0}^\infty K_n^{\left( \dot \xi \right) }\left(
x,t|x_0,t_0\right) 
\end{equation}
where\ 
$$
K_n^{\left( \dot \xi \right) }\left( x,t\mid x_0,t_0\right) =\left(
-i\right) ^n\int_{\R^n}\int_{\Lambda _n}\stackrel{n}{\stackunder{i=1}{\Pi }}%
v\left( {\rm d}x_i,{\rm d}t_i\right) \stackrel{n+1}{\stackunder{j=1}{\Pi }}%
K_0^{\left( \dot \xi \right) }\left( x_j,t_j|x_{j-1},t_{j-1}\right) \text{.} 
$$
As the test functions $\xi $ are real the explicit formula (\ref{Ko}) yields 
\begin{equation}
\label{bound2}|K_0^{\left( \dot \xi \right) }\left( x,t|x_0,t_0\right) |=%
\frac{\Theta (t-t_0)}{\sqrt{2\pi |t-t_0|}}\equiv M_0 
\end{equation}
and for $n\geq 1$ the bounds\ 
\begin{equation}
\label{bound3}\left| K_n^{\left( \dot \xi \right) }\left( x,t|x_0,t_0\right)
\right| \leq \int_{\R^n}\int_{\Lambda _n}\stackrel{n}{\stackunder{i=1}{\Pi }}%
|v|\left( {\rm d}x_i,{\rm d}t_i\right) \stackrel{n+1}{\stackunder{j=1}{\Pi }}%
\frac 1{\sqrt{2\pi \left| t_j-t_{j-1}\right| }} 
\end{equation}
$$
\hspace*{24mm}\leq C_v^n\frac{|t-t_0|^{\frac{n-1}2}}{2^{\frac{n+1}2}\Gamma (%
\frac{n+1}2)}\leq C_v^n\frac{|\Delta |^{\frac{n-1}2}}{2^{\frac{n+1}2}\Gamma (%
\frac{n+1}2)}\equiv M_n\text{.} 
$$
Hence $\stackrel{n+1}{\stackunder{j=1}{\Pi }}K_0^{\left( \dot \xi \right)
}\left( x_j,t_j|x_{j-1},t_{j-1}\right) $ is integrable on ${\R}^n\times
\Lambda _n$ with respect to $v^n$. (This is also established in the course
of a detailed proof of Theorem \ref{3.1} and we have reproduced the argument
here for the convenience of the reader.) Thus we can apply Fubini's theorem
to change the order of integration in $K_n^{\left( \dot \xi \right) }$\ to
obtain%
$$
K_n^{\left( \dot \xi \right) }\left( x,t|x_0,t_0\right) =-i\iint v\left( 
{\rm d}x_n,{\rm d}t_n\right) K_0^{\left( \dot \xi \right) }\left(
x,t|x_n,t_n\right) \text{ }\times \hspace{25mm} 
$$
$$
\hspace*{25mm}\left( -i\right) ^{n-1}\int_{\R^{n-1}}\int_{\Lambda
_{n-1}^{^{\prime }}}\stackrel{n-1}{\stackunder{i=1}{\Pi }}v\left( {\rm d}x_i,%
{\rm d}t_i\right) \stackrel{n}{\stackunder{j=1}{\Pi }}K_0^{\left( \dot \xi
\right) }\left( x_j,t_j|x_{j-1},t_{j-1}\right) 
$$
$(\Lambda _{n-1}^{^{\prime }}=\left\{ (t_1,...,t_{n-1})\mid
\,t_0<t_1<...<t_{n-1}<t_n\right\} )$. This establishes the following
recursion relation for $K_n^{\left( \dot \xi \right) }$\ 
\begin{equation}
\label{Rec}K_n^{\left( \dot \xi \right) }\left( x,t|x_0,t_0\right) =\left(
-i\right) \iint v\left( {\rm d}y,{\rm d}\tau \right) K_0^{\left( \dot \xi
\right) }\left( x,t|y,\tau \right) K_{n-1}^{\left( \dot \xi \right) }\left(
y,\tau |x_0,t_0\right) \text{.} 
\end{equation}
We now claim that the series $K^{\left( \dot \xi \right) }\left( y,\tau
|x_0,t_0\right) =\dsum\limits_nK_n^{\left( \dot \xi \right) }\left( y,\tau
|x_0,t_0\right) $ converges uniformly in $y,\tau $ on $\R\times \left( t_0,%
{\sf T}\right) $. To see this recall the above estimate (\ref{bound3}) which
is uniform in $y,\tau .$ Because the $M_n$ are rapidly decreasing it follows
that\ 
$$
\stackrel{\infty }{\dsum\limits_{n=1}}\sup \left\{ \left| K_n^{\left( \dot
\xi \right) }\left( y,\tau |x_0,t_0\right) \right| \bigg|(y,\text{ }\tau
)\in \R \times \left( t_0,{\sf T}\right) \text{ }\right\} \leq \stackrel{%
\infty }{\dsum\limits_{n=1}}\text{ }M_n<\infty \text{ .} 
$$
Due to the uniform convergence we may interchange summation and integration
in the following expression\ 
$$
-i\iint v\left( {\rm d}y,{\rm d}\tau \right) K_0^{\left( \dot \xi \right)
}\left( x,t|y,\tau \right) K^{\left( \dot \xi \right) }\left( y,\tau
|x_0,t_0\right) \hspace*{3cm} 
$$
$$
\hspace*{3cm}=-i\iint v\left( {\rm d}y,{\rm d}\tau \right) K_0^{\left( \dot
\xi \right) }\left( x,t|y,\tau \right) \dsum\limits_nK_n^{\left( \dot \xi
\right) }\left( y,\tau |x_0,t_0\right) 
$$
$$
\hspace*{3cm}=\dsum\limits_n-i\iint v\left( {\rm d}y,{\rm d}\tau \right)
K_0^{\left( \dot \xi \right) }\left( x,t|y,\tau \right) K_n^{\left( \dot \xi
\right) }\left( y,\tau |x_0,t_0\right) . 
$$
By the above recursion relation (\ref{Rec}) for $K_n^{\left( \dot \xi
\right) }$ this equals\ 
$$
\dsum\limits_nK_{n+1}^{\left( \dot \xi \right) }\left( x,t|x_0,t_0\right)
=K^{\left( \dot \xi \right) }\left( x,t|x_0,t_0\right) -K_0^{\left( \dot \xi
\right) }\left( x,t|x_0,t_0\right) \text{.} 
$$
\smallskip\ Hence we obtain the following

\begin{theorem}
\label{LSGlg}$K^{\left( \dot \xi \right) }$ as defined in (\ref{KDef}) obeys
the following integral equation:\ 
$$
K^{\left( \dot \xi \right) }\left( x,t|x_0,t_0\right) =K_0^{\left( \dot \xi
\right) }\left( x,t|x_0,t_0\right) -i\iint v\left( {\rm d}y,{\rm d}\tau
\right) K_0^{\left( \dot \xi \right) }\left( x,t|y,\tau \right) K^{\left(
\dot \xi \right) }\left( y,\tau |x_0,t_0\right) . 
$$
In particular the Feynman integral $\E \left( {\rm I}\right) \equiv K$ obeys
the well-known propagator equation:\ 
$$
K\left( x,t|x_0,t_0\right) =K_0\left( x,t|x_0,t_0\right) -i\iint v\left( 
{\rm d}y,{\rm d}\tau \right) K_0\left( x,t|y,\tau \right) K\left( y,\tau
|x_0,t_0\right) \text{.} 
$$
\end{theorem}

\ We now proceed to show that this corresponds to the Schr\"odinger
equation. To prove this we first prepare the following

\begin{lemma}
\label{3.3}The mapping $\left( x,t\right) \mapsto K^{\left( \dot \xi \right)
}\left( x,t|x_0,t_0\right) $ is continuous on $\R\times \left( t_0,\text{%
{\sf T}}\right) $.
\end{lemma}

\TeXButton{Proof}{\proof}Because the series (\ref{Series}) converges
uniformly it is sufficient to show the continuity of $K_n^{\left( \dot \xi
\right) }$. For $n=0,1$ this is straightforward from the explicit formula (%
\ref{Ko}). For $n>1$ we use (\ref{Rec}) and the estimate (\ref{bound3}) to
obtain\ 
$$
\left| K_n^{\left( \dot \xi \right) }\left( x^{\prime },t^{\prime
}|x_0,t_0\right) -K_n^{\left( \dot \xi \right) }\left( x,t|x_0,t_0\right)
\right| \hspace{65mm} 
$$
$$
\hspace*{1cm}\leq M_{n-1}\int_{\R }\int_\Delta \left| v\right| \left( {\rm d}%
x_n,{\rm d}t_n\right) \left| K_0^{\left( \dot \xi \right) }\left( x^{\prime
},t^{\prime }|x_n,t_n\right) -K_0^{\left( \dot \xi \right) }\left(
x,t|x_n,t_n\right) \right| \ . 
$$
Using the explicit form (\ref{Ko}) of $K_0^{\left( \dot \xi \right) }$it is
now straightforward to check that%
$$
\int_{\R }\int_\Delta \left| v\right| \left( {\rm d}x_n,{\rm d}t_n\right)
\left| K_0^{\left( \dot \xi \right) }\left( x^{\prime },t^{\prime
}|x_n,t_n\right) -K_0^{\left( \dot \xi \right) }\left( x,t|x_n,t_n\right)
\right| \hspace*{3cm} 
$$
$$
\hspace*{6cm}\leq \text{ }\mid x-x^{\prime }{}\mid C(x,t)\text{ }+\text{ }%
\mid t-t^{\prime }\mid ^\alpha C_\alpha (x,t) 
$$
where $0<\alpha <\frac 12$ and $x>x^{\prime }$, $t>t^{\prime }$.\endproof%
\bigskip\ \medskip\ 

An application of Lemma \ref{3.3} combined with the estimate (\ref{bound2})
shows that \\$K^{\left( \dot \xi \right) }\left( .,.|x_0,t_0\right) $ is
locally integrable on $\R \times ({\sf T}_0,{\sf T})$ with respect to both $%
v $ and Lebesgue measure. We can thus regard $K^{\left( \dot \xi \right) }$
as a distribution on ${\cal D}\left( \Omega \right) \equiv {\cal D}\left( \R%
\times ({\sf T}_0,{\sf T})\right) $:

$$
\left\langle K^{\left( \dot \xi \right) },\varphi \right\rangle =\iint {\rm d%
}x\,{\rm d}t\,K^{\left( \dot \xi \right) }\left( x,t|x_0,t_0\right) \varphi
\left( x,t\right) ,\quad \text{ }\varphi \in {\cal D}\left( \Omega \right) . 
$$
And we can also define a distribution $vK^{\left( \dot \xi \right) }$ by
setting\ 

$$
\left\langle vK^{\left( \dot \xi \right) },\varphi \right\rangle =\iint v(%
{\rm d}x,\,{\rm d}t)\,K^{\left( \dot \xi \right) }\left( x,t|x_0,t_0\right)
\varphi \left( x,t\right) ,\quad \text{ }\varphi \in {\cal D}\left( \Omega
\right) . 
$$
\ ($K^{\left( \dot \xi \right) }$ is locally integrable with respect to $v$, 
$\varphi $ is bounded with compact support and $v$ is finite, hence $\varphi
K^{\left( \dot \xi \right) }$ is integrable with respect to $v$).

We now proceed to show that $K^{\left( \dot \xi \right) }$ solves the
Schr\"odinger equation as a distribution. To abbreviate we set $\hat
L=\left( i\partial _t+\frac 12\partial _x^2-\dot \xi \left( t\right)
x\right) $ and let $\hat L^{*}$ denote its adjoint. Let $\varphi \in {\cal D}%
\left( \Omega \right) $. By Theorem \ref{LSGlg} we have\ 

$$
\left\langle \hat LK^{\left( \dot \xi \right) },\varphi \right\rangle
=\left\langle K_0^{\left( \dot \xi \right) }\left( x,t|x_0,t_0\right)
-i\iint v\left( {\rm d}y,{\rm d}\tau \right) K_0^{\left( \dot \xi \right)
}\left( x,t|y,\tau \right) K^{\left( \dot \xi \right) }\left( y,\tau
|x_0,t_0\right) ,\hat L^{*}\varphi \right\rangle . 
$$
\ By Fubini's theorem this equals\ 
$$
\left\langle K_0^{\left( \dot \xi \right) },\hat L^{*}\varphi \right\rangle
-i\iint v\left( {\rm d}y,{\rm d}\tau \right) \left[ \iint {\rm d}x\,{\rm d}%
t\,K_0^{\left( \dot \xi \right) }\left( x,t|y,\tau \right) \hat L^{*}\varphi
\left( x,t\right) \right] K^{\left( \dot \xi \right) }\left( y,\tau
|x_0,t_0\right) . 
$$
\ As $K_0^{\left( \dot \xi \right) }$ is a Green's function of $\hat L$ we
obtain\ 
$$
i\varphi \left( x_0,t_0\right) +\iint v\left( {\rm d}y,{\rm d}\tau \right)
\varphi \left( y,\tau \right) K^{\left( \dot \xi \right) }\left( y,\tau
|x_0,t_0\right) =\left\langle i\delta _{x_0}\delta _{t_0},\varphi
\right\rangle +\left\langle vK^{\left( \dot \xi \right) },\varphi
\right\rangle . 
$$
Hence we have the following

\begin{theorem}
$K^{\left( \dot \xi \right) }$is a Green's function for the full
Schr\"odinger equation, i.e.,\ 
$$
\left( i\,\partial _t+\frac 12\partial _x^2-\dot \xi \left( t\right)
x-v\right) K^{\left( \dot \xi \right) }\left( x,t|x_0,t_0\right) =i\,\delta
_{x_0}\,\delta _{t_0}. 
$$
In particular the Feynman integral $\E \left( {\rm I}\right) =K$ solves the
Schr\"odinger equation\ 
$$
i\,\partial _t\text{ }K\left( x,t|x_0,t_0\right) =\left( -\frac 12\partial
_x^2+v\right) K\left( x,t|x_0,t_0\right) \text{, \quad for }t>t_0. 
$$
\end{theorem}

Hence the construction proposed by Khandekar and Streit yields a rigorously
defined Feynman integrand whose expectation is the correct quantum
mechanical propagator.
\LaTeXparent{dis3.tex}

\section{The Feynman integrand for the perturbed harmonic oscillator \label
{perHarmOsc}}

In this section we carry the ideas of section \ref{quatum} over to
perturbations of the harmonic oscillator. Hence instead of constructing a
Dyson series around the free particle Feynman integrand we expand around the
Feynman integrand of the harmonic oscillator. The external potentials to
which the oscillator is submitted correspond to the wide class of
time-dependent singular potentials treated in section \ref{quatum}.

In \cite[chap 5]{AHK76} the path integral of the unharmonic oscillator is
defined within the theory of Fresnel integrals. Compared to our ansatz this
procedure has the advantage of being manifestly independent of the space
dimension. Despite the lack of a generalization to higher dimensional
quantum systems our construction has some interesting features:

\begin{itemize}
\item  The admissible potentials may be very singular.

\item  We are not restricted to smooth initial wave functions and may thus
study the propagator directly.
\end{itemize}

In this section we construct the Feynman integrand for the harmonic
oscillator in an external potential $V(x,t)$. Thus we have to define 
$$
{\rm I}_V={\rm I}_h\cdot \exp \left( -i\int_{t_0}^tV(x(\tau ),\tau )\;{\rm d}%
\tau \right) \text{ .} 
$$
As for the free particle we introduce the perturbation $V$ via the series
expansion of the exponential. Hence we have to find conditions for $V$ such
that the following object exists in $\left( {\cal S}\right) ^{\prime }$ 
$$
{\rm I}_V={\rm I}_h+\sum_{n=1}^\infty \left( -i\right) ^n\int_{{\R^{\mit n}}}%
{\rm d}^nx\int_{\Lambda _n}{\rm d}^nt\prod_{j=1}^nV\left( x_j,t_j\right)
\delta \left( x\left( t_j\right) -x_j\right) {\rm I}_h\text{ .} 
$$
We are able to treat the same class of singular time-dependent potentials as
in section \ref{quatum1} i.e., we consider $\nu $ a finite signed Borel
measure on ${\R}\times \De $, where $\De =[{\sf T}_0,{\sf T}]\supset \Delta
=[t_0,t]$. The following theorem contains conditions under which the Feynman
integrand ${\rm I}_V$ exists as a Hida distribution.

\begin{theorem}
Let $v$ be a finite signed Borel measure on $\R \times \De $ satisfying {\rm %
i)} and{\rm \ ii) }of section \ref{quatum1}. Then 
\begin{equation}
\label{result}{\rm I}_V={\rm I}_h+\sum_{n=1}^\infty \left( -i\right) ^n\int_{%
{\R^{\mit n}}}\int_{\Lambda _n}\bigg(\tprod\limits_{j=1}^nv({\rm d}x_j,{\rm d%
}t_j)\bigg){\rm I}_h\prod_{j=1}^n\delta \left( x\left( t_j\right)
-x_j\right) 
\end{equation}
is a Hida distribution.
\end{theorem}

\noindent {\bf Proof.} \\{\bf 1.\ part.} In the first part of the proof we
have to perform some technicalities which are necessary to establish the
central estimate (\ref{Main}). We have to use a very careful procedure to
achieve that (\ref{Main}) survives $n$-fold integration and summation in the
second part of the proof.

Let $\theta \in {\cal S}_{\Ckl }(\R )$. From Proposition \ref{TIhProdDelta}
and the explicit formula (\ref{17}) we find ($\Delta _j=[t_{j-1},t_j]$) 
$$
\Big|T\Big({\rm I}_h\tprod\limits_{j=1}^n\delta \left( B(t_j)-x_j\right) 
\Big)\left( \theta \right) \Big|\leq e^{\frac 12\left| \theta \right|
_0^2}\left( {\T \prod\limits_{j=1}^{n+1}\sqrt{\frac 1{4\left| \Delta
_j\right| }}}\right) \exp \left( (\left| x_{n+1}\right| +\left| x_0\right|
)\frac \pi 2\sup _\Delta \left| \theta \right| \right) \cdot 
$$
$$
\cdot \bigg|\exp \bigg( \bigg\{ \sum_{j=1}^nikx_j\Big[ \frac 1{\sin k\left|
\Delta _j\right| }\int_{\Delta _j}{\rm d}t\,\theta \left( t\right) \cos
k\left( t-t_{j-1}\right) \TeXButton{5cm}{\hspace*{5cm}} 
$$
$$
\TeXButton{5cm}{\hspace*{5cm}}-\frac 1{\sin k\left| \Delta _{j+1}\right|
}\int_{\Delta _{j+1}}\!{\rm d}t\,\theta \left( t\right) \cos k\left(
t-t_{j+1}\right) \Big] 
\bigg\} \bigg) \bigg|
\cdot 
$$
$$
\cdot \exp \left\{ \sum_{j=1}^n\frac \pi {2\left| \Delta _j\right|
}\int_{\Delta _j}{\rm d}s_1\int_{\Delta _j}{\rm d}s_2\left| \theta \left(
s_1\right) \right| \left| \theta \left( s_2\right) \right| \right\} 
$$
We define 
$$
X=\sup _{0\leq j\leq n+1}\left| x_j\right| 
$$
and%
$$
\left\| \theta \right\| \equiv \sup _\Delta \left| \theta \right| +\sup
_\Delta \left| \theta ^{\prime }\right| +\left| \theta \right| _0\;. 
$$
With these 
$$
\Big|T\Big({\rm I}_h\tprod\limits_{j=1}^n\delta \left( B(t_j)-x_j\right) 
\Big)\left( \theta \right) \Big|\leq e^{\frac 12\left\| \theta \right\|
^2}\left( {\T \prod\limits_{j=1}^{n+1}\sqrt{\frac 1{4\left| \Delta _j\right|
}}}\right) \exp \left( X\pi \left\| \theta \right\| +\frac \pi 2\left|
\Delta \right| \left\| \theta \right\| ^2\right) \cdot 
$$
$$
\cdot \bigg| \exp \bigg( \bigg\{ \sum_{j=1}^nikx_j\Big[ \frac 1{\sin k\left|
\Delta _j\right| }\int_{\Delta _j}{\rm d}t\,\theta \left( t\right) \cos
k\left( t-t_{j-1}\right) \TeXButton{4cm}{\hspace*{4cm}} 
$$
$$
\TeXButton{4cm}{\hspace*{4cm}}-\frac 1{\sin k\left| \Delta _{j+1}\right|
}\int_{\Delta _{j+1}}\!{\rm d}t\,\theta \left( t\right) \cos k\left(
t-t_{j+1}\right) \Big] 
\bigg\} \bigg) \bigg| \text{ .} 
$$
To estimate the last factor we proceed as follows:%
$$
\left| \sum_{j=1}^nikx_j\left[ \frac 1{\sin k\left| \Delta _{j+1}\right|
}\int_{\Delta _{j+1}}\!{\rm d}t\,\theta \left( t\right) \cos k\left(
t-t_j\right) -\frac 1{\sin k\left| \Delta _{j+1}\right| }\int_{\Delta
_{j+1}}\!{\rm d}t\,\theta \left( t\right) \cos k\left( t-t_{j+1}\right)
\right] \right| 
$$
\vspace{-5mm}%
\begin{eqnarray*}
&\leq& \sum_{j=1}^nk X\frac 1{\sin k\left| \Delta _{j+1}\right| }\left| \int_{\Delta _{j+1}}{\rm d}t \ \theta\left( t\right) %
\int_{t_j}^{t_{j+1}}k\sin %
k\left( t-\tau \right) d\tau \right| %
\\&\leq& \sum_{j=1}^nk X\sup _\Delta \left| \theta\right| \frac \pi 2\left| \Delta _{j+1}\right|%
\\&\leq& \frac \pi 2 Xk\left\| \theta\right\| \left| \Delta \right|
\end{eqnarray*}
To obtain a bound for the remaining term%
$$
\left| \sum_{j=1}^nikx_j\left[ \frac 1{\sin k\left| \Delta _j\right|
}\int_{\Delta _j}{\rm d}t\,\theta \left( t\right) \cos k\left(
t-t_{j-1}\right) -\frac 1{\sin k\left| \Delta _{j+1}\right| }\int_{\Delta
_{j+1}}{\rm d}t\,\theta \left( t\right) \cos k\left( t-t_j\right) \right]
\right| 
$$
we expand $F(t_{j-1})=\int_{t_{j-1}}^{t_j}{\rm d}t\,\theta (t)\cos k\left(
t-t_{j-1}\right) $ and $G(t_{j+1})=\int_{t_j}^{t_{j+1}}{\rm d}t\,\theta
(t)\cos k\left( t-t_j\right) $ around $t_j.$ This yields with $\eta _j\in
\Delta _j$ and $\eta _{j+1}\in \Delta _{j+1}$ 
\begin{eqnarray*}
& \leq & kX\left| \sum_{j=1}^n\theta \left( t_j\right) \left[ \frac{\left| \Delta _j\right| }{\sin k\left| \Delta _j\right| }-\frac{\left| \Delta _{j+1}\right| }{\sin k\left| \Delta _{j+1}\right| }\right] \right| + %
\\ & & +kX\sum_{j=1}^n\left[ \frac{\left( t_{j-1}-t_j\right) ^2}{2\sin k\left|\Delta _j\right| }\left( -\theta ^{\prime }\left( \eta _j\right)%
-k^2\int_{\eta _j}^{t_j}{\rm d}t\,\theta \left( t\right) \cos k\left( t-\eta %
_j\right) \right) \right] - %
\\ & & -kX\sum_{j=1}^n\left[ -\frac{\left( t_{j+1}-t_j\right) ^2} {2\sin k\left| \Delta _{j+1}\right| }\left( \theta ^{\prime }\left( \eta _{j+1}\right) \cos %
k\left( \eta _{j+1}-t_j\right) -k\theta \left( \eta _{j+1}\right) \sin %
k\left( \eta _{j+1}-t_j\right) \right) \right]  %
\end{eqnarray*}
Since $\stackunder{0\leq x\leq \frac \pi 2}{\sup }\left( \frac x{\sin
x}\right) ^{\prime }=1$ then the first term above is bounded by\vspace{-3mm}%
$$
2k\left| \Delta \right| X\left\| \theta \right\| 
$$
For the second term we obtain the bound%
$$
X\frac{\left| \Delta \right| }4\sup _\Delta \left| \theta ^{\prime }\right| +%
\frac{k^2\left| \Delta \right| ^2}4\sup _\Delta \left| \theta \right| +\frac
\pi 4\left| \Delta \right| \sup _\Delta \left| \theta ^{\prime }\right| +%
\frac{\pi k}4\left| \Delta \right| \sup _\Delta \left| \xi \right| \leq 
$$
$$
\leq \frac{X\left| \Delta \right| \pi }4\left( \left\| \theta \right\|
\left( 2+k^2\left| \Delta \right| +k\right) \right) 
$$
Putting all of this together we finally arrive at 
$$
\Big|T\Big({\rm I}_h\tprod\limits_{j=1}^n\delta \left( B(t_j)-x_j\right) 
\Big)\left( \theta \right) \Big|\leq \left( {\T \prod\limits_{j=1}^{n+1}%
\sqrt{\frac 1{4\left| \Delta _j\right| }}}\right) \exp \left( LX\left\|
\theta \right\| +\left( \frac \pi 2\left| \Delta \right| +\frac 12\right)
\left\| \theta \right\| ^2\right) 
$$
where $L=\pi +\frac 34\pi k\left| \Delta \right| +2k\left| \Delta \right|
+\frac \pi 4\left| \Delta \right| \left( 2+k^2\left| \Delta \right| \right) $
is a constant.\medskip\ 

Hence for $\theta \in {\cal S}_{\Ckl }(\R )$ we have the following estimate 
\begin{equation}
\label{Main}\Big|T\Big({\rm I}_h\tprod\limits_{j=1}^n\delta \left(
B(t_j)-x_j\right) \Big)\left( z\xi \right) \Big|\leq \left( {\T%
\prod\limits_{j=1}^{n+1}\sqrt{\frac 1{4\left| \Delta _j\right| }}}\right)
\exp \left( X^2\gamma \right) \exp \left[ \left| z\right| ^2\left\| \xi
\right\| ^2\left( \frac 12+\frac \pi 2\left| \Delta \right| +\frac{L^2}{%
2\gamma }\right) \right] 
\end{equation}
where $\gamma >0$ is chosen later.\medskip

\noindent {\bf 2. part.} In this final step we use the method developed in
the proof of Theorem \ref{3.1} to control the convergence of (\ref{result}).
Although the slight modification to our case is easy we give the basic steps
for the convenience of the reader.

In order to apply Theorem \ref{Bochner} to perform the integration we need
to show that%
$$
\left( {\T\prod\limits_{j=1}^{n+1}\sqrt{\frac 1{4\left| \Delta _j\right| }}}%
\right) \exp \left( X^2\gamma \right) 
$$
is integrable with respect to $v .$ To this end we choose $q>2$ and $%
0<\gamma <\frac \beta q.$ With this choice of $\gamma $ the property i) of $%
v $ yields that $\exp \left( \gamma X^2\right) \in L^q\left( {\R}^n\times
\Lambda _n,\left| v \right| \right) $ and with%
$$
Q\equiv \left( \dint_{{\R}}\dint_\Delta \left| v \right| \left( {\rm d}x,%
{\rm d}t\right) \exp \left( \gamma qx^2\right) \right) ^{\frac 1q} 
$$
we have%
$$
\left( \dint_{{\R^{\mit n}}}\dint_{\Lambda _n}\tprod\limits_{j=1}^n\left| v
\right| \left( {\rm d}x_j,{\rm d}t_j\right) \exp \left( \gamma qX^2\right)
\right) ^{\frac 1q}\leq \exp \left( \gamma \left( x_0^2+x^2\right) \right)
Q^n<\infty . 
$$

Now we choose $p$ such that $\frac 1p+\frac 1q=1.$ Using the property ii) of 
$v$ and the formula%
$$
\int_{\Lambda _n}{\rm d}^nt\tprod\limits_{j=1}^{n+1}\left( \frac 1{4\left|
t_j-t_{j-1}\right| }\right) ^\alpha =\left( \frac{\Gamma \left( 1-\alpha
\right) }{4^\alpha }\right) ^{n+1}\frac{\left| \Delta \right| ^{n\left(
1-\alpha \right) -\alpha }}{\Gamma \left( \left( n+1\right) \left( 1-\alpha
\right) \right) },\text{ }\alpha <1 
$$
we obtain the following bound%
$$
\left[ \int_{{\R^{\mit n}}}\int_{\Lambda _n}\tprod\limits_{j=1}^n\left|
v\right| \left( {\rm d}x_j,{\rm d}t_j\right) \tprod\limits_{j=1}^{n+1}\left(
\frac 1{4\left| t_j-t_{j-1}\right| }\right) ^{\frac p2}\right] ^{\frac
1p}\leq  
$$
$$
\leq C_v^{n/p}\frac{\Gamma \left( {\T\frac{2-p}2}\right) ^{\frac{n+1}%
p}\left| \Delta \right| ^{\frac np-\frac 12\left( n+1\right) }}{4^{\frac{n+1}%
2}\Gamma \left( \left( n+1\right) {\T\frac{2-p}2}\right) ^{\frac 1p}}<\infty
\  
$$
(remember: $C_v$ is the essential supremum of the $L^\infty $-density of $%
|v|_t$ ).

Finally an application of H\"older's inequality gives%
$$
\left| \left( \tprod\limits_{j=1}^{n+1}\sqrt{\T\frac 1{4\left|
t_j-t_{j-1}\right| }}\right) \exp \left( \gamma X^2\right) \right|
_{L^1(|v|)}\leq \hspace{4cm} 
$$
$$
\leq \exp \left( \gamma x_0^2+\gamma x^2\right) Q^nC_v^{n/p}\frac{\Gamma
\left( {\T \frac{2-p}2}\right) ^{\frac{n+1}p}\left| \Delta \right| ^{\frac
np-\frac 12\left( n+1\right) }}{2^{n+1}\Gamma \left( \left( n+1\right) {\T%
\frac{2-p}2}\right) ^{\frac 1p}}\equiv C_n<\infty 
$$
Hence Theorem \ref{Bochner} yields%
$$
{\rm I}_n\equiv \int_{{\R^{\mit n}}}\int_{\Lambda
_n}\tprod\limits_{j=1}^nv\left( {\rm d}x_j,{\rm d}t_j\right) \bigg({\rm I}%
_h\prod_{j=1}^n\delta \left( B(t_j)-x_j\right) \bigg)\in \left( {\cal S}%
\right) ^{\prime }. 
$$

As the $C_n$ are rapidly decreasing in $n$ the hypotheses of Theorem \ref
{conv} are fulfilled and hence

$$
{\rm I}_V=\dsum\limits_{n=0}^\infty {\rm I}_n\in \left( {\cal S}\right)
^{\prime}\text{ .} 
$$
\TeXButton{End Proof}{\endproof}
\LaTeXparent{dis3.tex}

\chapter{The Feynman integrand for the Albeverio H\o egh-Krohn class \label
{AHKclass}}

\section{Introduction}

First we will introduce a further class of potentials for which we will
define a path integral representation of the Green's function for the
Schr\"odinger equation. Potentials of this type already appeared in earlier
(mathematically rigorous) works on Feynman integrals, see e.g., the works 
\cite{AHK76,Ga74,Ito66}. The most elegant construction of a path integral
for this class of potentials has been proposed by Albeverio and H\o
egh-Krohn \cite{AHK76}. They used the so-called Fresnel integral, an
extension of the Lebesgue integral. Thus I will call potentials of that kind
the Albeverio H\o egh-Krohn class.

Besides the fact that we use completely different methods than \cite
{AHK76,Ga74,Ito66} this approach differs from previous ones in two main
points.

\begin{enumerate}
\item  In the works \cite{AHK76,Ga74} smooth initial wave functions were
used and their propagation was handled by construction of a path integral.
In our white noise framework we are able to introduce delta like initial
wave functions. Thus we can go back to Feynman's original idea to treat
propagators by path integrals.

\item  We wish to give a meaning to the integrand itself. Its expectation
yields the desired propagator. We will see later that this seems to be the
reason why we will have to put one additional assumption on the class of
potentials we are able to handle.
\end{enumerate}

\begin{definition}
Let $\m $ denote a bounded complex measure on the Borel sets of $\R^d$, $%
d\geq 1$. A complex valued function $V$ on $\R^d$ is called Fresnel
integrable (following \cite{AHK76}) if 
\begin{equation}
\label{Fresnel}V(\vec x)=\int_{\R^d}e^{i\vec \alpha \vec x}\;{\rm d}^d\m %
(\alpha )\text{ }
\end{equation}
Since the bounded complex Borel measures form an algebra under convolution,
the Fresnel integrable functions are an algebra ${\cal F}(\R ^d)$ under
pointwise multiplication. ${\cal F}(\R ^d)$ is called the Albeverio H\o
egh-Krohn class.\\We will call $V\in {\cal F}(\R ^d)$ admissible, if 
\begin{equation}
\label{admV}\int_{\R ^d}e^{\varepsilon \left| \vec \alpha \right| }{\rm d}^d|%
\m |(\alpha )
\end{equation}
is finite for some $\varepsilon >0.$\\For later use we need also the
condition 
\begin{equation}
\label{adm3}\int_{\R ^d}\exp (\varepsilon |\vec \alpha |^{1+\delta })\,{\rm d%
}^d|\m |(\alpha )
\end{equation}
is finite for some $\varepsilon ,\delta >0$.
\end{definition}

\noindent {\bf Remark.} It is clear from the definition that our admissible
potentials are ${\cal C}^\infty (\R ^d)$ since all moments of the
corresponding measure $\m $ exist. In fact admissible potentials are
analytic in the open ball of radius $\varepsilon $. Since formula (\ref
{Fresnel}) now makes sense for all $\vec x\in \C ^d:\left| {\rm Im}(\vec
x\,)\right| <\varepsilon $, an admissible potential is regular in this strip
containing the real axis. A useful reference on (analytic) characteristic
functions has proven to be \cite{Lu70}. Condition (\ref{adm3}) implies that $%
V$ in fact is an entire function.

In their well known work \cite{AHK76} Albeverio and H\o egh-Krohn made
extensive use of the fact that a suitable choice of norm makes ${\cal F}(\R%
^d)$ to be a Banach algebra. Since the measures $\m $ have to be bounded, $%
{\cal F}(\R^d)$ contains only functions bounded on the real line. \bigskip

\example  One particular example is given by the 2 dimensional (periodic)
potential 
$$
V(\vec x)=\{\cos (x_1)\cdot \sin (x_2)\}^\beta 
$$
which is of special interest in the theory of {\it antidot (super) lattices,}
see \cite{FGKP95} for a recent review. The integer valued even parameter $%
\beta $ determines if the potential is ``soft or hard''. This potential
generates chaotic behavior in classical systems and the Hamiltonian has a
fractal spectrum of eigenvalues in a quantum mechanical treatment. Since the
measure $\m $ corresponding to $V$ is a linear combination of (products of)
delta measures and thus has compact support, the condition (\ref{adm3}) is
satisfied. Details will be presented in a forthcoming ``Diplomarbeit'' of M.
Grothaus.

\section{The Feynman integrand as a generalized white noise functional}

Now we proceed to introduce these interactions into the Feynman integral.
Mathematically speaking this means to give a rigorous definition to the
pointwise product%
$$
{\rm I}={\rm I}_0\cdot \exp \left[ -i\int_{t_0}^tV(\vec x(\tau ))\;{\rm d}%
\tau \right] \;. 
$$
In Chapter \ref{Quantum} we already saw a method which works in the one
dimensional case. There the potential was ``expanded in terms of delta
functions''. In a second step the expansion of the exponential led to a
convergent series of Hida distributions. Since in higher dimensions $d$ the
delta functions cause problems (in respect to the $t$-integration) we here
use a Fourier decomposition of the potential. Then we will proceed as in
Chapter \ref{Quantum} but we have to face the fact that the occurring
integrals in the perturbation series are only convergent in some larger
distribution space $({\cal S}_d)^{-1}$ (at least ${\cal Y}^{\prime }$ is
necessary).

\begin{theorem}
\label{Iin(Sd)-1}Let $V$ be an admissible potential in the Albeverio H\o
egh-Krohn class, i.e., there exists a bounded complex Borel measure $\m $
satisfying (\ref{admV}). Then 
\begin{equation}
\label{AlbDyson}{\rm I}={\rm I}_0+\sum_{n=1}^\infty (-i)^n\int_{\Lambda _n}%
{\rm d}^n\tau \int_{\R ^{dn}}\tprod\limits_{j=1}^n{\rm d}^d\m (\alpha _j)\ 
{\rm I}_0\cdot \tprod\limits_{j=1}^ne^{i\vec \alpha _j\vec x(\tau _j)}
\end{equation}
exists as a generalized white noise functional i.e., ${\rm I}\in ({\cal S}%
_d)^{-1}$. If also (\ref{adm3}) is satisfied then {\rm I} is a Meyer-Yan
distribution i.e., ${\rm I}\in {\cal Y}^{\prime }$. The integrals we are
using here are in sense of Theorems \ref{Bochner} and \ref{PettisI}
respectively. $\Lambda _n$ is defined by eq. (\ref{Deltan}).
\end{theorem}

\noindent {\bf Notes.} \\1.\ From physical reasons $V$ may supposed to be
real, but in mathematical respect this is irrelevant here.\smallskip\ \\2.\
The integral $\int_{\Lambda _n}$ can be replaced by $\frac
1{n!}\int_{[t_0,t]^n}=:\frac 1{n!}\int_{\Box _n}$ .\smallskip\ \\3.\ The
fact, that $\E ({\rm I})$ is physically reasonable i.e., it is the
elementary solution of the (time dependent) Schr\"odinger equation is well
known. It coincides with the series developed in \cite{AHK76}. In \cite{Ga74}
it is proved explicitly that this series is in fact the physical solution.

\TeXButton{Proof}{\proof} As a first step we have to justify the pointwise
product 
\begin{equation}
\label{PhinDef}\Phi _n={\rm I}_0\cdot \tprod\limits_{j=1}^ne^{i\vec \alpha
_j\vec x(\tau _j)}
\end{equation}
Since the explicit formula (\ref{TI0delta}) allows an extension of $T({\rm I}%
_0)(\vec \xi )$ to all $\vec \xi \in L_d^2$ we may use the following ansatz
to calculate%
\begin{eqnarray*}
T\Phi _n(\vec \xi ) & = & T(%
{\rm I}_0 )\left( \vec \xi +\sum\limits_{j=1}^n\vec \alpha _j{\1}%
_{[t_0,\tau _j)}\right) \cdot \exp \left( i\vec x_0\tsum\limits_{j=1}^n\vec \alpha%
_j\right)  %
\\  & = & (2\pi %
i(t-t_0))^{-\frac d2}\exp \left( i\vec x_0\tsum\limits_{j=1}^n\vec \alpha%
_j\right) \exp \left( -\frac i2\dint_{\R }\left[ \vec \xi (s)+\tsum\limits_{j=1}^n\vec \alpha _j{\1}%
_{[t_0,\tau _j)}(s)\right] ^{\D 2}{\rm d}s\right)  %
\\&  & \hspace*{1cm} \cdot \exp \left( -\frac1{2i(t-t_0)}\left[ \int_{t_0}^t\vec \xi (s){\rm d}s+\left\{%
\tsum\limits_{j=1}^n\vec \alpha _j(\tau _j-t_0)+\vec x-\vec x_0\right\}%
\right] ^{\D 2}\right) 
\end{eqnarray*}
Obviously this has an extension in $\vec \xi \in {\cal S}_d(\R )$ to all $%
\vec \theta \in {\cal S}_{d,\Ckl }(\R )$ and is locally bounded in a
neighborhood of zero (see estimate (\ref{est1}) below.) Thus $\Phi _n\in (%
{\cal S}_d)^{-1}$ (in fact $\Phi _n\in ({\cal S}_d)^{\prime }$). Now we want
to apply Theorem \ref{Bochner}. Since $T\Phi _n(\vec \theta )$ is a
measurable function of $(\tau _1,...,\tau _n;\vec \alpha _1,...,\vec \alpha
_n)$ for all $\vec \theta \in {\cal S}_{d,\Ckl }(\R )$, we only have to find
an integrable local bound%
\begin{eqnarray} \label{est1}
|T\Phi _n(\vec \theta )| & \leq  & (2\pi (t-t_0))^{-\frac d2}\exp %
\bigg(\frac 12\left| \vec \theta \right| _0^2+\sum\limits_{j=1}^n %
\left| (\vec \theta ,\vec \alpha _j\1_{[t_0,\tau_j)})\right| + %
\frac 1{2(t-t_0)}\bigg[ %
(t-t_0)\left| \vec \theta \right| _0^2 \nonumber %
\\ &  & \TeXButton{3cm}{\hspace*{3cm}} %
+2\left| \int_{t_0}^t\vec \theta (s){\rm d}s\cdot \left\{ %
\sum\limits_{j=1}^n\vec \alpha _j(\tau _j-t_0)+\vec x-\vec x_0\right\} %
\right| \bigg]\bigg) \nonumber %
\\ & \leq  & (2\pi (t-t_0))^{-\frac d2}\exp \bigg( %
\left| \vec \theta \right| _0^2+2 %
\sqrt{t-t_0}\left| \vec \theta \right| _0\sum\limits_{j=1}^n\left| %
\vec \alpha %
_j\right| +\frac{\left| \vec x-\vec x_0\right| }{\sqrt{t-t_0}}%
\left| \vec \theta \right| _0\bigg) \nonumber %
\\ & =: & {\rm C}_n(\vec \alpha _1,...,\vec \alpha _n,\vec \theta ) %
\end{eqnarray}
Since $\m $ has the property (\ref{admV}) we can find neighborhood of zero 
\begin{equation}
\label{U0r}{\cal U}_{0,r}=\left\{ \vec \theta \in {\cal S}_{d,\Ckl }\Big|%
\left| \vec \theta \right| _0<r\right\} \quad ,\ r=\frac \varepsilon {2\sqrt{%
t-t_0}}
\end{equation}
such that%
$$
\int_{\R ^d}{\rm d}^d|\m |(\alpha )\ e^{2\left| \vec \alpha \right| \sqrt{%
t-t_0}\left| \vec \theta \right| } 
$$
is finite for all $\vec \theta \in {\cal U}_{0,r}$. So we have 
\begin{equation}
\label{intbar}\int_{\Lambda _n}{\rm d}^n\tau \int_{\R ^{dn}}\tprod%
\limits_{j=1}^n{\rm d}^d|\m |(\alpha _j)\ {\rm C}_n(\vec \alpha _1,...,\vec
\alpha _n,\vec \theta )\leq \hspace{3cm}
\end{equation}
$$
\leq \frac 1{n!}(t-t_0)^n(2\pi (t-t_0))^{-\frac d2}\exp \left( \left| \vec
\theta \right| _0^2+\frac{\left| \vec x-\vec x_0\right| }{\sqrt{(t-t_0)}}%
\left| \vec \theta \right| _0\right) \left( \int_{\R ^d}{\rm d}^d|\m
|(\alpha )\ e^{2\left| \vec \alpha \right| \sqrt{t-t_0}\left| \vec \theta
\right| _0}\right) ^{\D n} 
$$
Thus we have proved the existence of an integrable bound. Also the
convergence of the series in $n$ is established because the right hand side
may be summed up. Thus we have proved ${\rm I}$ defined by (\ref{AlbDyson})
is in $({\cal S}_d)^{-1}$ and established the bound%
$$
\left| T{\rm I}(\vec \theta )\right| \leq \frac 1{(2\pi (t-t_0))^{\frac
d2}}\exp \left( \left| \vec \theta \right| _0+\frac{\left| \vec x-\vec
x_0\right| }{\sqrt{(t-t_0)}}\left| \vec \theta \right| _0+(t-t_0)\int_{\R ^d}%
{\rm d}^d|\m |(\alpha )\ e^{2\left| \vec \alpha \right| \sqrt{t-t_0}\left|
\vec \theta \right| _0}\right)  
$$
for all $\vec \theta \in {\cal U}.$\bigskip\ 

Now we assume that also (\ref{adm3}) is satisfied. Then it is useful to
estimate ${\rm C}_n$ in (\ref{est1}). We use the elementary estimate 
$$
\alpha \cdot \beta \leq \frac 1p\alpha ^p+\frac 1q\beta ^q\,,\qquad \alpha
,\beta >0,\,\frac 1p+\frac 1q=1 
$$
to show%
$$
2\sqrt{t-t_0}|\vec \theta |_0|\vec \alpha _j|\leq \varepsilon |\vec \alpha
_j|^{1+\delta }+\delta \varepsilon ^{-1/\delta }\left( \frac{2\sqrt{t-t_0}%
|\theta |_0}{1+\delta }\right) ^{\T 
\frac{1+\delta }\delta } 
$$
then 
$$
\int_{\Lambda _n}{\rm d}^n\tau \int_{\R ^{dn}}\tprod\limits_{j=1}^n{\rm d}^d|%
\m |(\alpha _j)\ {\rm C}_n(\vec \alpha _1,...,\vec \alpha _n,\vec \theta
)\leq \hspace{5cm} 
$$
$$
\leq \frac 1{n!}(t-t_0)^n(2\pi (t-t_0))^{-\frac d2}\exp \left( \left| \vec
\theta \right| _0^2+\frac{\left| \vec x-\vec x_0\right| }{\sqrt{(t-t_0)}}%
\left| \vec \theta \right| _0+n\delta \varepsilon ^{-1/\delta }\left( \frac{%
(2\sqrt{t-t_0}|\theta |_0}{1+\delta }\right) ^{\T \frac{1+\delta }\delta
}\right) \cdot 
$$
$$
\cdot \left( \int_{\R ^d}{\rm d}^d|\m |(\alpha )\ \exp (\varepsilon |\vec
\alpha |^{1+\delta })\right) ^n 
$$

This shows that the assumption of Theorem \ref{PettisI} are satisfied so
that the integrals in equation (\ref{AlbDyson}) are well defined Pettis
integrals in ${\cal Y}^{\prime }$. The series in $n$ now is no problem in
view of Corollary \ref{MYseq} since the right hand side of the above bound
can be summed up. Thus $I\in {\cal Y}^{\prime }$ (defined by (\ref{AlbDyson}%
)) and we have the bound 
$$
|T{\rm I}(\vec \theta )|\leq (2\pi (t-t_0))^{-d/2}\cdot \exp \left( |\vec
\theta |_0^2+\frac{|\vec x-\vec x_0|}{\sqrt{t-t_0}}|\vec \theta |_0\right)
\cdot \hspace{4cm} 
$$
$$
\cdot \exp \left( (t-t_0)\cdot \exp \left( \delta \varepsilon ^{-1/\delta
}\left( \frac{2\sqrt{t-t_0}|\theta |_0}{1+\delta }\right) ^{\T \frac{%
1+\delta }\delta }\right) \cdot \int_{\R ^d}{\rm d}|\m |(\alpha )\exp
(\varepsilon |\vec \alpha |^{1+\delta })\right) \ . 
$$
\TeXButton{End Proof}{\endproof}\bigskip\ 

The above construction may be generalized to include also explicitly time
dependent potentials.

\begin{theorem}
\label{TIin(Sd)-1} Let $\m $ denote a complex measure on $\R^d\times [t_0,t]$
, $d\geq 1$ such that 
\begin{equation}
\label{adm2}\int_{\R ^d}\int_{t_0}^te^{\varepsilon \left| \vec \alpha
\right| }\ |\m |({\rm d}^d\alpha ,{\rm d}\tau )<\infty 
\end{equation}
Then 
\begin{equation}
\label{dyson}{\rm I}={\rm I}_0+\sum_{n=1}^\infty (-i)^n\int_{\Lambda
_n}\int_{\R ^{dn}}\tprod\limits_{j=1}^n\m ({\rm d}^d\alpha _j,{\rm d}\tau
_j)\ {\rm I}_0\cdot \tprod\limits_{j=1}^ne^{i\vec \alpha _j\vec x(\tau _j)} 
\end{equation}
exists as in $({\cal S}_d)^{-1}$ i.e., as a generalized white noise
functional.
\end{theorem}

\TeXButton{Proof}{\proof}The proof can be done in the same way as in the
previous theorem. But inequality (\ref{intbar}) has to be modified%
$$
\int_{\Lambda _n}\int_{\R ^{dn}}\tprod\limits_{j=1}^n|\m |({\rm d}^d\alpha
_j,{\rm d}\tau _j)\ {\rm C}_n(\vec \alpha _1,...,\vec \alpha _n,\vec \theta )\leq 
\hspace{75mm} 
$$
$$
\leq (2\pi (t-t_0))^{-\frac d2}\exp \left( \left| \vec \theta \right| _0^2+%
\frac{\left| \vec x-\vec x_0\right| }{\sqrt{(t-t_0)}}\left| \vec \theta
\right| _0\right) \int_{\Lambda _n}\int_{\R ^{dn}}\tprod\limits_{j=1}^n|\m |(%
{\rm d}^d\alpha _j,{\rm d}\tau _j)\ e^{2\left| \vec \alpha _j\right| \sqrt{%
t-t_0}\left| \vec \theta \right| _0} 
$$
$$
\leq \frac 1{n!}(2\pi (t-t_0))^{-\frac d2}\exp \left( \left| \vec \theta
\right| _0^2+\frac{\left| \vec x-\vec x_0\right| }{\sqrt{(t-t_0)}}\left|
\vec \theta \right| _0\right) \left( \int_{t_0}^t\int_{\R ^d}|\m |({\rm d}%
^d\alpha ,{\rm d}\tau )\ e^{2\left| \vec \alpha \right| \sqrt{t-t_0}\left|
\vec \theta \right| _0}\right) ^{\displaystyle n}\text{.} 
$$
This shows that the integration and the summation in equation (\ref{dyson})
are well defined in $({\cal S}_d)^{-1}$.\TeXButton{End Proof}{\endproof}%
\bigskip\ 

\noindent {\bf Remark.} Note that the time dependence of our `potentials'
may be very singular here. Let us consider an admissible potential $V$ in $%
{\cal F}(\R ^d)$ represented by the measure $\tilde {\m }$ 
$$
V(\vec x)=\int_{\R ^d}e^{i\vec \alpha \vec x}{\rm d}^d\tilde {\m
}(\alpha )\text{ } 
$$
We wish to study a quantum mechanical system which is `kicked' a finite
number of times $s_j\in (t_0,t)$. This is done by multiplication with a
delta measure in time. Thus we have to introduce%
$$
\m ({\rm d}^d\alpha ,{\rm d}\tau ):=\sum_j\tilde {\m }({\rm d}^d\alpha
)\cdot \delta _{s_j}({\rm d}\tau )\text{,} 
$$
which clearly fulfills (\ref{adm2}).
\LaTeXparent{dis3.tex}

\chapter{A new look at Feynman Hibbs \label{look}}

\section{Transition amplitudes}

Since \cite{FH65} it is quite common to discuss so called transition
amplitudes which in our framework would read%
$$
\E (F\cdot {\rm I}) 
$$
where $F:\vec x(\cdot )\mapsto F(\vec x(\cdot ))$ is a function of the path.
Of course this is well defined (by writing $\E (F\cdot {\rm I})=\langle
\!\langle {\rm I},F\rangle \!\rangle $) whenever $F\in ({\cal S}_d)^1$.
Since this is too restrictive for relevant cases we shall discuss some
special extensions of this pairing before we start to discuss the physical
interpretation of the transition amplitudes we have defined. Throughout this
chapter we assume the setting of Theorem \ref{Iin(Sd)-1} for simplicity.

First let us introduce some convenient notations which will help to keep
formulas little shorter:%
$$
\vec \zeta :=\sum_{j=1}^n\vec \alpha _j{\1}_{\left[ t_0,\tau _j\right) }%
\text{,}\qquad \vec \chi :=\sum_{j=1}^n\vec \alpha _j(\tau _j-t_0) 
$$
and 
$$
\vec X:=\vec x-\vec x_0\text{,}\qquad \Delta =\left[ t_0,t\right] ,\qquad {\
|\Delta |}=t-t_{0\,}. 
$$

For later use we collect some useful formulas.

\begin{lemma}
Let $\vec \eta ,\vec \eta _1,\vec \eta _2\in {\cal S}_d(\R )\,,\,\vec \theta
\in {\cal S}_{d,\Ckl }(\R )\,$ then 
\begin{equation}
\label{1.Moment}T\big( \langle \cdot ,\vec \eta \rangle \Phi _n\big) (\vec
\theta )=-\left[ \langle \vec \eta ,\vec \theta +\vec \zeta \rangle -\tfrac
1{t-t_0}\tint_\Delta \vec \eta \left\{ \tint_\Delta \vec \theta +\vec \chi
+\vec X\right\} \right] T\Phi _n(\vec \theta )
\end{equation}
and%
$$
T\big( \langle \cdot ,\vec \eta _1\rangle \langle \cdot ,\vec \eta _2\rangle
\Phi _n\big) (\vec \theta )=\hspace{9cm} 
$$
$$
=T\Phi _n(\vec \theta )\cdot \bigg\{i(\vec \eta _1,\vec \eta _2)+\frac
i{|\Delta |}\tint_\Delta \vec \eta _1\tint_\Delta \vec \eta _2+\left[ (\vec
\eta _1,\vec \theta +\vec \zeta )-\tfrac 1{|\Delta |}\tint_\Delta \vec \eta
_1\left\{ \tint_\Delta \vec \theta +\vec \chi +\vec X\right\} \right] \cdot  
$$
\begin{equation}
\label{2.Moment}\hspace*{55mm}\cdot \left[ (\vec \eta _2,\vec \theta +\vec
\zeta )-\tfrac 1{|\Delta |}\tint_\Delta \vec \eta _2\left\{ \tint_\Delta
\vec \theta +\vec \chi +\vec X\right\} \right] \bigg\} 
\end{equation}
of course $\langle \cdot ,\vec \eta \rangle \cdot \Phi _n$ and $\langle
\cdot ,\eta _1\rangle \langle \cdot ,\eta _2\rangle \cdot \Phi _n\in ({\cal S%
}_d)^{\prime }$.
\end{lemma}

\TeXButton{Proof}{\proof} Use the formula 
$$
T(\langle \cdot ,\vec \eta \rangle ^k\Phi _n)(\vec \theta )=\left( \frac{%
{\rm d}}{i\,{\rm d}\lambda }\right) ^kT\Phi _n(\vec \theta +\lambda \vec
\eta )\bigg|_{\lambda =0} 
$$
and polarization identity.\TeXButton{End Proof}{\endproof}\bigskip\ 

Furthermore we need the possibility for an intermediate pinning of the
paths, which is prepared by the next proposition. This is a generalization
of formula (\ref{Tnd}).

\begin{proposition}
\label{Endpunktfix}The distribution%
$$
\Phi _n(\vec x,t|\vec x_0,t_0)\cdot \prod_{l=1}^m\delta ^d(\vec x(t_l)-\vec
x_l)\ ,\qquad t_0<t_l\leq t,\ \,\vec x_l\in \R ^d;\,1\leq l\leq m 
$$
is well defined in $({\cal S}_d)^{\prime }$ with $T$-transform%
$$
T\left( \Phi _n(\vec x,t|\vec x_0,t_0)\prod_{l=1}^m\delta ^d(\vec
x(t_l)-\vec x_l)\right) (\vec \theta )=e^{i\frac m2|\vec \theta
|^2}\prod_{l=1}^{m+1}T\Phi _{n_l}(\vec x_l,t_l|\vec x_{l-1},t_{l-1})(\vec
\theta ) 
$$
here $\vec x_{m+1}\equiv \vec x\,,\,t_{m+1}\equiv t\,,\ \,n_l=\#\left\{
j\;|\;t_{l-1}<\tau _j\leq t_l\right\} $ and $\Phi _{n_l}$ depends on the
parameters $\left\{ \alpha _j,\tau _j\;|\;t_{l-1}<\tau _j\leq t_l\right\} $.
\end{proposition}

\TeXButton{Proof}{\proof}To simplify the calculation we propose the use of 
$$
\prod_{l=1}^{m+1}\delta ^d(\vec x(t_l)-\vec x_l)\prod_{j=1}^n\exp \left(
i\vec \alpha _j(\vec x_0+\langle \cdot ,\1 _{[t_0,\tau _j)}\rangle )\right) =%
\hspace{7cm} 
$$
$$
\hspace*{41mm}=\prod_{l=1}^{m+1}\left\{ \delta ^d(\vec x(t_l)-\vec
x_l)\prod_{t_{l-1}<\tau _j\leq t_l}\exp \left( i\vec \alpha _j(\vec
x_{l-1}+\langle \cdot ,\1 _{[t_{l-1},\tau _j)}\rangle )\right) \right\} 
$$
which is simply checked by a comparison $T$-transforms. In view of formula (%
\ref{Tnd})\\ ${\rm J}\prod_{l=1}^{m+1}\delta ^d(\vec x(t_l)-\vec x_l)$ is
well defined due to its explicit $T$-transform. Then the starting point of
the calculation is 
$$
T\left( \Phi _n\prod_{l=1}^{m+1}\delta ^d(\vec x(t_l)-\vec x_l)\right) (\vec
\theta )=\hspace{9cm} 
$$
$$
T\left( {\rm J}\prod_{l=1}^{m+1}\delta ^d(\vec x(t_l)-\vec x_l)\right)
\left( \vec \theta +\sum_{l=1}^{m+1}\sum_{\{j|t_{l-1}<\tau _j\leq t_l\}}\vec
\alpha _j\1 _{[t_{l-1},\tau _j)}\right) \exp \left( \sum_{l=0}^{m+1}\vec
x_{l-1}\sum_{\{j|t_{l-1}<\tau _j\leq t_l\}}\vec \alpha _j\right) 
$$
which can be evaluated without problems.\TeXButton{End Proof}{\endproof}%
\bigskip\ 

Now we continue the discussion of Feynman integrands defined in Theorem \ref
{Iin(Sd)-1}. Pointwise products of ${\rm I}={\rm I}(\vec x,t|\vec x_0,t_0)$
with $\vec x(s)$, $\stackrel{\cdot }{\vec x}(s)$ and $\stackrel{\cdot \cdot 
}{\vec x}(s)$ for $t_0<s<t$ have natural interpretations in usual quantum
mechanics as we will see in the next section.

In the white noise framework $\vec x(s)$ is represented as $\vec x_0+\langle
\cdot ,\1 _{[t_0,s)}\rangle \,,\,\stackrel{\cdot }{\vec x}(s)$ as $\langle
\cdot ,\delta _s \rangle $ and by a formal partial integration we
obtain $\stackrel{\cdot \cdot }{\vec x}(s)=-\langle \cdot ,\delta _s^{\prime
}\rangle $. So we have to study products of the form $\langle \cdot ,\vec
T\rangle \cdot {\rm I}$ for suitable distributions $\vec T\in {\cal S}%
_d^{\prime }$. This can conveniently be done by approximating the first
factor by test functions.

\begin{definition}
\label{ProdDef}Let $\vec T\in {\cal S}_d^{\prime }$ and $\{\vec \eta _l\in 
{\cal S}_d\,,\,l\in \N \}$ a sequence of test functions such that $\lim
_{l\rightarrow \infty }\vec \eta _l=\vec T$. We define 
$$
\langle \cdot ,\vec T\rangle \cdot {\rm I}:=\lim _{l\rightarrow \infty
}\langle \cdot ,\vec \eta _l\rangle \cdot {\rm I} 
$$
if the limit exists and is independent of the sequence.
\end{definition}

\noindent That this is fulfilled in the three mentioned cases is shown in
the following proposition.

\begin{proposition}
\label{xsProd}In the sense of the above definition we have $x_k(s)\cdot {\rm %
I\,}$, $\dot x_k\,(s)\cdot {\rm I\,}$, $\ddot x_k(s)\cdot {\rm I}\in ({\cal S%
}_d)^{-1}$ (the index $1\leq k\leq d$ indicates the $k^{th}$-component) and 
\begin{equation}
\label{xksI}x_k\,(s)\cdot {\rm I}=\sum_{n=0}^\infty (-i)^n\int_{\Lambda _n}%
{\rm d}^n\tau \int_{\R ^{nd}}\tprod_{j=1}^n{\rm d}^d\m (\alpha
_j)\;x_k(s)\cdot \Phi _n
\end{equation}
$$
\dot x_k\,(s)\cdot {\rm I}=\sum_{n=0}^\infty (-i)^n\int_{\Lambda _n}{\rm d}%
^n\tau \int_{\R ^{nd}}\tprod_{j=1}^n{\rm d}^d\m (\alpha _j)\;\dot
x_k(s)\cdot \Phi _n\,\,. 
$$
\end{proposition}

\TeXButton{Proof}{\proof}Since $\langle \cdot ,\vec \eta _l\rangle $ is in $(%
{\cal S}_d)$ pointwise multiplication intertwines with Bochner integration
and the infinite sum: 
$$
\langle \cdot ,\vec \eta _l\rangle {\rm I}=\sum_{n=0}^\infty
(-i)^n\int_{\Lambda _n}{\rm d}^n\tau \int_{\R ^{nd}}\tprod_{j=1}^n{\rm d}^d%
\m (\alpha _j)\,\langle \cdot ,\vec \eta _l\rangle \cdot \Phi _n\ . 
$$
In the case $\vec \eta _l\rightarrow \1 _{[t_0,s)}\vec e_k$ $\ $($\{\vec
e_k,1\leq k\leq d\}$ denotes the canonical basis of $\R ^d$) equation (\ref
{1.Moment}) gives the estimate 
$$
\left| T\left( \langle \cdot ,\vec \eta _l\rangle \cdot \Phi _n\right) (\vec
\theta )\right| \leq 2\left| \vec \eta _l\right| \left( \left| \vec \theta
\right| +\sqrt{t-t_0}\sum_{j=1}^n\left| \alpha _{j,k}\right| +\frac{|\vec X|%
}{2\sqrt{t-t_0}}\right) \left| T\Phi _n(\vec \theta )\right|  
$$
$$
\hspace*{4.6cm}\leq 2\left| \vec \eta _l\right| \left( \left| \vec \theta
\right| +\sqrt{t-t_0}\sum_{j=1}^n\left| \alpha _{j,k}\right| +\frac{|\vec X|%
}{2\sqrt{t-t_0}}\right) {\rm C}_n(\vec \alpha _1,\ldots ,\vec \alpha _n,\vec
\theta ) 
$$
where $\alpha _{j,k}=\vec \alpha _j\cdot \vec e_k$. It is easy to see that
for $\vec \theta \in {\cal U}_{0,r}$ (defined in (\ref{U0r}) the right hand
side of the estimate is integrable on $\Lambda _n\times \R ^{nd}$ w.r.t. $%
{\rm d}^n\tau \prod_{j=1}^n{\rm d}^d|\m|(\alpha _j)$ such that 
$$
\left| T(\langle \cdot ,\vec \eta _l\rangle {\rm I})(\vec \theta )\right|
\leq \left| \vec \eta _l\right| \cdot K_n 
$$
on ${\cal U}_{0,r}$ for some rapidly decreasing sequence $K_n$. This is
sufficient to ensure the convergence of the sequence $l\rightarrow \langle
\cdot ,\vec \eta _l\rangle {\rm I}$ . The limit of the sequence $%
l\rightarrow (x_{0,k}+\langle \cdot ,\vec \eta _l\rangle ){\rm I}$ is given
by (\ref{xksI}) with 
$$
T(\vec x(s)\cdot \Phi _n)(\vec \theta )=\bigg\{ \frac{{t-s}}{{t-t_0}}\left(
\vec x_0-\int_{t_0}^s\vec \theta -\sum_{j=1}^k\vec \alpha _j(\tau
_j-t_0)\right)  
$$
$$
\hspace*{5cm}+\frac{s-t_0}{t-t_0}\left( \vec x+\int_s^t\vec \theta
-\sum_{j=k+1}^n\vec \alpha _j(t-\tau _j)\right) \bigg\} T\Phi _n(\vec \theta
)\;. 
$$

By a similar argument we discuss the product with $\dot x_k(s)$. Here the
basic formula is 
$$
T(\dot x_k(s)\Phi _n)(\vec \theta )=\bigg( -\theta _k(s)-\sum_{j=1}^k\alpha
_{j,k}\1 _{[t_0,\tau _j)}(s)+\hspace*{5.3cm} 
$$
$$
\hspace*{5cm}+\frac 1{t-t_0}\left( \int_{t_0}^t\theta _k+\sum_{j=1}^n\alpha
_{j,k}(\tau _j-t_0)+x_k-x_{0,k}\right) \bigg) T\Phi _n(\vec \theta )\;. 
$$
Again we can find a neighborhood of zero in ${\cal S}_{d\text{ }}$such that
the resulting estimate 
\begin{equation}
\label{xDotEst}\left| T(\dot x_k(s)\Phi _n)(\vec \theta )\right| \leq
2\left( \left| \vec \theta \right| _\infty +\sum_{j=1}^n\left| \alpha
_{j,k}\right| +\frac{|\vec X|}{2\cdot |t-t_0|}\right) {\rm C}_n(\vec \alpha
_1,\ldots \vec \alpha _n,\vec \theta ) 
\end{equation}
is integrable and can be summed up ($\left| \cdot \right| _\infty $denotes
the sup--norm). In particular this is sufficient to show that the
requirement of Definition \ref{ProdDef} is fulfilled.

In the third case we have 
$$
T(\ddot x_k(s)\Phi _n)(\vec \theta )=\left( -\dot \theta
_k(s)+\sum_{j=1}^n\alpha _{j,k}\cdot \delta _s(\tau _j)\right) T\Phi _n(\vec
\theta )\,. 
$$
The term $\dot \theta (s)\cdot T\Phi _n(\vec \theta )$ causes no problem. To
ensure integrability one integration (w.r.t. $\tau _j$) has to be regarded
as integration with respect to Dirac measure. This is possible because the
bound (\ref{est1}) is independent of $\tau _j$. The rest of the proof is as
before.\TeXButton{End Proof}{\endproof}\bigskip\ 

\noindent {\bf Note}. In terms of Wick products we may write%
$$
\langle \cdot ,\vec \eta \rangle \cdot {\rm I}=i\left( \langle \cdot ,\vec
\eta \rangle -\frac 1{t-t_0}\int_{t_0}^t\vec \eta \left( \langle \cdot ,\1 %
_{[t_0,t)}\rangle -\vec x+\vec x_0\right) \right) \diamond {\rm I}%
\hspace{5cm} 
$$
$$
\hspace*{2.5cm}+\sum_{n=0}^\infty (-i)^n\int_{\Lambda _n}{\rm d}^n\tau \int_{%
\R ^{dn}}\tprod_{j=1}^n{\rm d}^d\m (\alpha _j)\left( \sum_{j=1}^n\vec \alpha
_j\left( \frac{\tau _j-t_0}{t-t_0}\int_{t_0}^t\vec \eta -\int_{t_0}^{\tau
_j}\vec \eta \right) \right) \Phi _n\ . 
$$
\bigskip\ 

In fact the bound (\ref{xDotEst}) proves that $\dot x_k(s)\cdot \Phi _n$ is
integrable with respect to the product measure ${\rm d}^n\tau \prod_{j=1}^n%
{\rm d}^d|\m |(\alpha _j)\cdot {\rm d}s$ on the domain $\Lambda _n\times \R %
^{dn}\times [t_0,t]$. This implies that $\dot x_k(s)\cdot {\rm I}$ is
Bochner integrable w.r.t. ${\rm d}s$, i.e.,%
$$
\int_{t_0}^s\left( \dot x_k(s)\cdot {\rm I}\right) {\rm d}s=x_k(s)\cdot {\rm %
I}-x_{0,k}{\rm I\ ,} 
$$
in particular 
\begin{equation}
\label{ddsxs}\frac{{\rm d}}{{\rm d}s}\E (\vec x(s)\cdot {\rm I})=\E (%
\stackrel{.}{\vec x}(s)\cdot {\rm I})\ . 
\end{equation}
\medskip\ 

Furthermore we need pointwise products of the type $F(\vec x(s))\cdot {\rm I}
$ for fixed $s\in (t_0,t]$ and appropriate functions $F$. Since $F(\vec x(s))
$ is not in $({\cal S}_d)^1$ we have to give an extension of this pointwise
product. We will give two alternative definitions which have different
advantages such that later we can use the most convenient one.

\begin{definition}
Let $F\in {\cal F}(\R ^d)$ such that $F(\vec x)=\int_{\R ^d}e^{i\vec \alpha
\vec x}{\rm d}^d\m _F(\alpha )$ and $\Phi \in ({\cal S}_d)^{-1}$. If the
product $\Phi \cdot e^{i\vec \alpha \vec x(s)}$ is well defined in $({\cal S}%
_d)^{-1}$ and the Bochner integral with respect to ${\rm d}^d|\m _F|(\alpha )
$ exists we define%
$$
F(\vec x(s))\cdot \Phi :=\int_{\R ^d}\Phi \cdot e^{i\vec \alpha \vec x(s)}%
{\rm d}^d\m _F(\alpha )\qquad . 
$$
\end{definition}

\noindent {\bf Remark}. One can show that this definition extends the usual
definition of pointwise multiplication. Without loss of generality let $F\in 
{\cal F}(\R )$ and $\eta \in {\cal S}$. Further we assume that $|\m_F|$
satisfies the following integrability condition: $\forall K>0$ 
$$
\int e^{K\alpha ^2}{\rm d}|\m _F|(\alpha )<\infty \ . 
$$
Then 
$$
\left| \int e^{i\alpha \langle z,\eta \rangle }{\rm d}\m _F(\alpha )\right|
\leq \left( \int e^{K\alpha ^2}{\rm d}|\m _F|(\alpha )\right) \exp \left(
\frac 1{4K}|z|_{-p}^2|\eta |_p^2\right) 
$$
for all $p>0$. This shows that $F(\langle \cdot ,\eta \rangle )=\int
e^{i\alpha \langle \cdot ,\eta \rangle }{\rm d}\m _F(\alpha )$ is in ${\cal E%
}_{\min }^2({\cal S}^{\prime })$ which is the same as $({\cal S})$. For this
class of multipliers the coincidence of the two definitions can now easily
be seen. Let $\Phi \in ({\cal S})^{\prime }$ be arbitrary and $\varphi \in (%
{\cal S})$%
\begin{eqnarray*}
\left\langle \!\!\left\langle \int \Phi \cdot e^{i\alpha \langle \cdot ,\eta \rangle }{\rm d}\m _F(\alpha ),\varphi \right\rangle \!\!\right\rangle %
&=& %
\int \left\langle \!\!\left\langle \Phi ,\varphi \cdot e^{i\alpha \langle \cdot ,\eta \rangle }\right\rangle \!\!\right\rangle {\rm d}\m _F(\alpha ) %
\\ &=& %
\left\langle \!\!\left\langle \Phi ,\varphi \cdot \int e^{i\alpha \langle \cdot ,\eta \rangle }{\rm d}\m _F(\alpha )\right\rangle \!\!\right\rangle %
\\ &=& %
\left\langle \!\left\langle \Phi ,\varphi \cdot F(\langle \cdot ,\eta %
\rangle )\right\rangle \!\right\rangle \ . %
\end{eqnarray*}

\begin{lemma}
Let $F\in {\cal F}(\R ^d)$ be admissible i.e., there exists $\varepsilon >0$
such that\\ $\int_{\R ^d}e^{\varepsilon |\vec \alpha |}{\rm d}^d|\m %
_F|(\alpha )$ is finite. Then $F(\vec x(s))\cdot {\rm I}\in ({\cal S}_d)^{-1}
$ in the sense of the above definition.\\Moreover%
$$
F(\vec x(s))\cdot {\rm I}=\sum_{n=0}^\infty (-i)^n\int_{\Lambda _n}{\rm d}%
^n\tau \int_{\R ^{dn}}\tprod_{j=1}^n{\rm d}^d\m (\alpha _j)\int_{\R ^d}{\rm d%
}^d\m _F(\beta )\;\Phi _n\cdot e^{i\vec \beta \vec x(s)}\ . 
$$
\end{lemma}

\noindent The proof is a simple modification of the proof of Theorem \ref
{Iin(Sd)-1}. \bigskip\ 

A different class of multipliers is obtained by the following definition.

\begin{definition}
\label{ProdDefF}Let $F:\R ^d\rightarrow \C $ denote a Borel measurable
function and $\Phi \in ({\cal S}_d)^{-1}$. If the product $\Phi \cdot \delta
^d(\vec x(s)-\vec y)$ is well defined and Bochner integrable with respect to 
$|F(\vec y)|\;{\rm d}^dy$ then 
$$
F(\vec x(s))\cdot \Phi :=\int_{\R ^d}F(\vec y)\,\Phi \cdot \delta ^d(\vec
x(s)-\vec y)\,{\rm d}^dy\ . 
$$
\end{definition}

\begin{theorem}
\label{FxsI}Let $F:\R ^d\rightarrow \C $ such that 
$$
\int e^{\varepsilon |\vec y|}\,|F(\vec y)|\;{\rm d}^dy<\infty  
$$
for some $\varepsilon >0$. Then $F(\vec x(s))\cdot {\rm I}$ is defined in
the sense of the above definition. Moreover 
\begin{equation}
\label{FHT}T\left( F(\vec x(s))\cdot {\rm I}\right) (\vec \theta )=e^{\frac
i2|\vec \theta |^2}\int_{\R ^d}T{\rm I}(\vec x,t|\vec y,s)(\vec \theta
)\,F(\vec y)\,T{\rm I}(\vec y,s|\vec x_0,t_0)(\vec \theta )\;{\rm d}^dy\ .
\end{equation}
\end{theorem}

\noindent {\bf Remark.} If the last formula is evaluated at $\vec \theta
=\vec 0$ we obtain 
\begin{equation}
\label{keyFH}\E (F(\vec x(s))\cdot {\rm I})=\int K(\vec x,t|\vec
y,s)\,F(\vec y)\,K(\vec y,s|\vec x_0,t_0)\;{\rm d}^dy 
\end{equation}
which is one of the key formulas in Feynman Hibbs \cite{FH65}.\medskip\ 

\TeXButton{Proof}{\proof}The expression ${\rm I}\cdot \delta ^d(\vec
x(s)-\vec y)$ has a natural sense in view of Proposition \ref{Endpunktfix}.
More precisely we have 
\begin{equation}
\label{TPhinDelta}T\left( \Phi _n\cdot \delta (\vec x(s)-\vec y)\right)
(\vec \theta )=\sum_{k=0}^n\1 _{(\tau _k,\tau _{k+1}]}(s)\;e^{\frac i2|\vec
\theta |^2}T\Phi _{n-k}(\vec x,t|\vec y,s)(\vec \theta )\;T\Phi _k(\vec
y,s|\vec x_0,t_0)(\vec \theta )
\end{equation}
which can be estimated as follows (similar to (\ref{est1}))%
$$
T\left( \Phi _n\cdot \delta ^d(\vec x(s)-\vec y)\right) (\vec \theta )\leq
n\cdot \left( 4\pi ^2(t-s)(s-t_0)\right) ^{-d/2}\hspace{5.6cm} 
$$
$$
\hspace*{1.5cm}\cdot \exp \left( |\vec \theta |^2+2\sqrt{t-t_0}|\vec \theta
|\sum_{j=1}^n|\vec \alpha _j|+|\vec \theta ||\vec y|\left( \tfrac 1{\sqrt{t-s%
}}+\tfrac 1{\sqrt{s-t_0}}\right) +|\vec \theta |\left( \tfrac{|\vec x|}{%
\sqrt{t-s}}+\tfrac{|\vec y|}{\sqrt{s-t_0}}\right) \right) \;. 
$$
Analogous to the proof of Theorem \ref{Iin(Sd)-1} we can find a neighborhood
of zero in ${\cal S}_d$ such that this bound is integrable with respect to $%
\int_{\Lambda _n}{\rm d}^n\tau \int_{\R ^{dn}}\prod_{j=1}^n{\rm d}^d|\m %
|(\alpha _j)\int_{\R ^d}|F(\vec y)|{\rm d}^dy$. In particular%
$$
{\rm I}\cdot \delta ^d(\vec x(s)-\vec y)=\sum_{n=0}^\infty
(-i)^n\int_{\Lambda _n}{\rm d}^n\tau \int_{\R ^{nd}}\tprod_{j=1}^n{\rm d}^d%
\m (\alpha _j)\,\Phi _n\cdot \delta ^d(\vec x(s)-\vec y) 
$$
is well defined and integrable w.r.t. $\left| F(\vec y)\right| {\rm d}^dy$.

Now we deduce formula (\ref{FHT}). Note that each term in the sum in
equation (\ref{TPhinDelta}) factorizes in one factor depending on $\tau
_j,\vec \alpha _j\ ,$ $1\leq j\leq k$ and a second factor depending on the
remaining $\tau _j,\alpha _j\ ,$ $k+1\leq j\leq n$. So it is natural to use
the corresponding decomposition of the domain of integration%
$$
\Lambda _n(t,t_0)=\bigcup_{k=0}^n\Lambda _{n-k}^{\prime }(t,s)\times \Lambda
_k(s,t_0) 
$$
where the prime indicates that $\Lambda _{n-k}^{\prime }(t,s)$ is a set in
the $\tau _{k+1}\times \ldots \times \tau _n$--plane. Then%
$$
\int_{\Lambda _n}T\left( \Phi _n\delta ^d(\vec x(s)-\vec y)\right) (\vec
\theta )\,{\rm d}^n\tau =\hspace{8.8cm} 
$$
$$
\hspace*{3.5cm}=e^{\frac i2|\theta |^2}\sum_{k=0}^n\int_{\Lambda
_{n-k}^{\prime }(t,s)}T\Phi _{n-k}(\vec x,t|\vec y,s)(\vec \theta )\cdot
\int_{\Lambda _k(s,t_0)}T\Phi _k(\vec y,s|\vec x_0,t_0)(\vec \theta )\;. 
$$
This implies 
\begin{equation}
\label{TIDelta}T\left( {\rm I\cdot }\delta ^d(\vec x(s)-\vec y)\right) (\vec
\theta )=e^{\frac i2|\vec \theta |^2}T{\rm I}(\vec x,t|\vec y,s)(\vec \theta
)\cdot T{\rm I}(\vec y,s|\vec x_0,t_0)(\vec \theta )\ . 
\end{equation}
Integration with respect to $F(\vec y)\cdot {\rm d}^dy$ gives (\ref{FHT}).%
\TeXButton{End Proof}{\endproof}\medskip\ 

\noindent {\bf Remark.} Since the situation in Proposition \ref{Endpunktfix}
is more general, we can show the following generalization of Theorem \ref
{FxsI}. Let $t_0<s_1<s_2<\ldots <s_n<t$ and $F_j:\R ^d\rightarrow \C ,\quad
1\leq j\leq n$ as in Theorem \ref{FxsI}. Then $\prod_{j=1}^nF_j(\vec
x(s_j))\cdot {\rm I}${\rm \ }is in $({\cal S}_d)^{-1}$. Moreover (\ref{FHT})
generalizes to%
\begin{equation}
\label{FHTn}
T\left( \prod_{j=1}^nF_j(\vec x(s_j))\cdot {\rm I\ }\right) (\vec \theta )=%
\hspace{11cm} 
\end{equation}
$$
\hspace*{2.4cm}=e^{\frac{in}2|\vec \theta |^2}\int_{\R ^{dn}}\,T%
{\rm I}(\vec x,t|\vec y_n,s_n)(\vec \theta )\prod_{j=1}^nF_j(\vec y_j)\;T%
{\rm I}(\vec y_j,s_j|\vec y_{j-1},s_{j-1})(\vec \theta )\;{\rm d}^dy_j\ .
$$
\bigskip\ 

There are two relevant cases which are not covered by the previous theorem.
First of all we want to discuss (\ref{keyFH}) for the constant function $F=1$%
. In other words we want to apply the identity 
$$
\int_{\R ^d}\delta ^d(\vec x(s)-\vec y)\,{\rm d}^dy=1 
$$
to (\ref{TIDelta}). The second case is $F(\vec y)=y_k$. Here we want to
compare the two definitions \ref{ProdDef} and \ref{ProdDefF}. In both cases
the integral in (\ref{keyFH}) is not absolutely convergent (easily seen in
the free case) but has a sense as a Fresnel integral. Since the notion of a
Fresnel integral of a family white noise distributions is not yet developed
we will use a regularization procedure.

\begin{proposition}
Let $F_\varepsilon (\vec y)=e^{-\frac{\varepsilon ^2}2y^2}$ and $s\in (t_0,t)
$. Then the following two limits exist in $({\cal S}_d)^{-1}$ and are given
by\\1) 
$$
\lim _{\varepsilon \rightarrow 0}(F_\varepsilon (\vec x(s))\cdot {\rm I})=%
{\rm I} 
$$
2)%
$$
\lim _{\varepsilon \rightarrow 0}(x_k(s)F_\varepsilon (\vec x(s))\cdot {\rm I%
})=x_k(s)\cdot {\rm I} 
$$
\end{proposition}

\TeXButton{Proof}{\proof}Let $g(\vec y)$ be a real valued function which
either is identical $1$ or equal to $y_k$. By definition 
$$
T(g(\vec x(s))F_\varepsilon (\vec x(s))\cdot {\rm I})(\theta )=e^{\frac
i2|\vec \theta |^2}\int_{\R ^d}g(\vec y)F_\varepsilon (\vec y)\;T{\rm I}%
(\vec x,t|\vec y,s)(\vec \theta )\;T{\rm I}(\vec y,s|\vec x_0t_0)(\vec
\theta )\;{\rm d}^dy 
$$
$$
\hspace*{4.3cm}=\sum_{n=0}^\infty \sum_{k=0}^n(-i)^n\int_{\Lambda
_{n-k}^{\prime }(t,s)\times \Lambda _k(s,t_0)}{\rm d}^n\tau \int_{\R %
^{nd}}\tprod_{j=1}^n{\rm d}^d\m (\alpha _j)\;G_\varepsilon  
$$
with%
$$
G_\varepsilon =\int_{\R ^d}g(\vec y)F_\varepsilon (\vec y)\;T\Phi
_{n-k}(\vec x,t|\vec y,s)(\vec \theta )\;T\Phi _k(\vec y,s|\vec x_0t_0)(\vec
\theta )\;{\rm d}^dy\ . 
$$
The Gaussian integral $G_\varepsilon $ can be evaluated explicitly:%
$$
G_\varepsilon =\left( \frac{2\pi }{a_\varepsilon }\right) ^{d/2}g\left( 
\frac{i\vec b}{a_\varepsilon }\right) T\Phi _{n-k}(\vec x,t|\vec 0,s)(\vec
\theta )\;T\Phi _k(\vec 0,s|\vec x_0,t_0)(\vec \theta )\cdot \exp \left( -%
\frac{\vec b^2}{2a_\varepsilon }\right)  
$$
with 
$$
a_\varepsilon =\varepsilon ^2-\frac{i(t-t_0)}{(t-s)(s-t_0)} 
$$
$$
\vec b=\frac 1{s-t_0}\left( \int_{t_0}^s\vec \theta +\sum_{j=1}^k\vec \alpha
_j(\tau _j-t_0)-\vec x_0\right) -\frac 1{t-s}\left( \int_s^t\vec \theta
-\sum_{j=k+1}^n\vec \alpha _j(t-\tau _j)+\vec x\right) \ . 
$$
From this the following bound can be calculated%
$$
|G_\varepsilon |\leq (2\pi |\Delta |)^{d/2}{\rm C}_{n-k}(\vec x,t|\vec 0,s)\;%
{\rm C}_k(\vec 0,s|\vec x_0,t_0)\cdot \left| g\left( \frac{i\vec b}a\right)
\right| \hspace*{3cm} 
$$
$$
\hspace*{3cm}\cdot \exp \frac{|\Delta |}2\left( |\vec \theta |_\infty
^2+2|\vec \theta |_\infty \left( \sum_{j=1}^n|\vec \alpha _j|+\frac{|\vec x|%
}{|\Delta |}+\frac{|\vec x_0|}{|\Delta |}\right) \right)  
$$
where $\left| g\left( \frac{i\vec b}a\right) \right| $ is either equal to $1$
or 
$$
\left| g\left( \frac{i\vec b}a\right) \right| \leq \left( |\vec x_0|+|\vec
x|+|\Delta ||\vec \theta |_\infty +|\Delta |\sum_{j=1}^n|\vec \alpha
_j|\right) . 
$$
It is easy to see that in both cases this bound can be integrated for $\vec
\theta $ in some neighborhood of zero w.r.t. ${\rm d}^n\tau \prod_{j=1}^n%
{\rm d}^d|\m |(\alpha _j)$ on $\R ^{nd}\times \Lambda _{n-k}^{\prime
}(t,s)\times \Lambda _k(s,t_0)$ and stays finite after $\sum_{n=0}^\infty
\sum_{k=0}^n$. Thus the limit $\varepsilon \rightarrow 0$ exists in both
cases. The limit itself can be identified by an elementary calculation of $%
\lim _{\varepsilon \rightarrow 0}G_\varepsilon $.\TeXButton{End Proof}
{\endproof}\bigskip\ 

\noindent {\bf Consequences}. The above proposition together with (\ref
{keyFH}) gives%
$$
K(\vec x,t|\vec x_0,t_0)=\E ({\rm I}(\vec x,t|\vec x_0,t_0))\hspace*{8cm} 
$$
$$
=\lim _{\varepsilon \rightarrow 0}\int_{\R ^d}\E ({\rm I}(\vec x,t|\vec
y,s))\;F_\varepsilon (\vec y)\;\E ({\rm I}(\vec y,s|\vec x_0,t_0))\;{\rm d}%
^d\vec y 
$$
$$
=\lim _{\varepsilon \rightarrow 0}\int_{\R ^d}K(\vec x,t|\vec
y,s)\;F_\varepsilon (\vec y)\;K(\vec y,s|\vec x_0,t_0)\;{\rm d}^dy\qquad . 
$$
We will use this as a substitute of Feynman's%
$$
``\ K(\vec x,t|\vec x_0,t_0)=\int_{\R ^d}K(\vec x,t|\vec y,s)\;K(\vec
y,s|\vec x_0,t_0)\;{\rm d}^dy\ " 
$$
which is not absolutely convergent. The second consequence we want to
mention is 
\begin{equation}
\label{OrtTransAmp}\E (\vec x(s)\cdot {\rm I)}=\lim _{\varepsilon
\rightarrow 0}\int_{\R ^d}K(\vec x,t|\vec y,s)\;\vec y\,F_\varepsilon (\vec
y)\;K(\vec y,s|\vec x_0,t_0)\;{\rm d}^dy
\end{equation}
which allows to calculate the transition element directly from the
propagator.

\section{Relation to operator notation}

In usual quantum mechanics time evolution can be represented by an unitary
group $U(t,t_0)$ acting on a suitable Hilbert space. The infinitesimal
generator of $U(t,t_0)$ is assumed to be the Hamiltonian $H$. In the
Schr\"odinger representation the matrix element of $U(t,t_0)$ is given by
the propagator%
$$
\langle \vec x|U(t,t_0)|\vec x_0\rangle =K(\vec x,t|\vec x_0,t_0)\ . 
$$
(We have not proved this explicitly since this question is discussed in \cite
{Ga74} in great detail.) The above formula may be viewed as the standard
connection of path integral techniques to usual quantum mechanics.

For our discussion we will choose the Heisenberg picture where states are
time independent and observables evolve in time according to the time
evolution operator. Concretely the position operator is given by 
$$
\vec q(t)=U^{*}(t,t_0)\;\vec q\;U(t,t_0) 
$$
with%
$$
\vec q\;|\vec x\rangle =\vec x\;|\vec x\rangle  
$$
and the star denotes the adjoint operator. Now we are ready to connect the
transition amplitudes from the previous section to quantum mechanical
observables. From (\ref{OrtTransAmp}) it follows 
\begin{eqnarray*}
\E (\vec x(s)\cdot {\rm I)}
&=& %
\lim _{\varepsilon \rightarrow 0}\int_{\R ^d}\langle \vec x|U(t,s)|\vec y\rangle \;F_\varepsilon (\vec y)\,\vec y\;\langle \vec y|U(s,t_0)|\vec x_0\rangle \;{\rm d}^dy %
\\ &=& %
\lim _{\varepsilon \rightarrow 0}\int_{\R ^d}\langle \vec x|U(t,s)F_\varepsilon (\vec q)\,\vec q\,|\vec y\rangle \;\langle \vec y|U(s,t_0)|\vec x_0\rangle \;{\rm d}^dy %
\\&=& %
\lim _{\varepsilon \rightarrow 0}\langle \vec x|U(t,s)\,F_\varepsilon (\vec q)\,\vec q\,U(s,t_0)|\vec x_0\rangle %
\\ &=& %
\langle \vec x|U(t,t_0)\,U^{*}(s,t_0)\vec qU(s,t_0)|\vec x_0\rangle 
\\ &=& %
\langle \vec x|\,U(t,t_0)\,\,\vec q(s)\,|\vec x_0\rangle \ . 
\end{eqnarray*}

More generally we can show, based on (\ref{FHTn})%
$$
\E (\vec x(s_1)\ldots \vec x(s_n)\cdot {\rm I})=\langle \vec x|U(t,t_0)\;%
{\sf T}\vec q(s_1)\ldots \vec q(s_n)\,|\vec x_0\rangle \;, 
$$
where the usual time ordering of operators appears {\sf T}$\vec q(s_1)\ldots
\vec q(s_n)=\vec q(s_n)\ldots \vec q(s_1)$ if $s_n>s_{n-1}>\ldots >s_1$. 
\medskip\ 

Before $\E (\stackrel{\cdot }{\vec x}(s)\cdot {\rm I})$ is discussed we need
one more assumption. Assume the Hamiltonian $H$ not to be explicitly time
dependent. Then the Heisenberg equation of motion 
$$
\frac{{\rm d}}{{\rm d}s}\vec q(s)=\frac i\hbar [H,\vec q(s)] 
$$
holds where the square brackets denote the commutator. If furthermore the
Hamiltonian is of the form 
$$
H=\frac 1{2m}\vec p^2+V(\vec q) 
$$
we can use the relation 
$$
[p_l^n,q_k]=-ni\hbar p_l^{n-1}\delta _{k,l} 
$$
to find 
$$
[H,\vec q(s)]=-i\frac \hbar m\vec p(s)\ . 
$$
Hence we have 
$$
\frac{{\rm d}}{{\rm d}s}\vec q(s)=\frac 1m\vec p(s)\ . 
$$

The starting point of the following calculation is formula (\ref{ddsxs}):%
$$
\E (\stackrel{\cdot }{\vec x}(s)\cdot {\rm I})=\frac{{\rm d}}{{\rm d}s}\E %
(\vec x(s)\cdot {\rm I})=\frac{{\rm d}}{{\rm d}s}\langle \vec
x|U(t,t_0)\,\vec q(s)\,|\vec x_0\rangle =\frac 1m\langle \vec
x|U(t,t_0)\,\vec p(s)\,|\vec x_0\rangle \ . 
$$

To give more evidence on these relations between transition elements and
operator notation we will try to identify the canonical commutation
relations in the language of transition elements.

\section{A functional form of the canonical commutation relations}

A well-known fact from quantum mechanics is the non commutativity of
momentum and position operators at equal times. This seems to have no direct
translation in a path-integral formulation of quantum mechanics. But on a
heuristic level Feynman and Hibbs \cite{FH65} found an argument to show that 
$\E (\dot x(s+\varepsilon )x(s){\rm I})\neq \E (\dot x(s-\varepsilon )x(s)%
{\rm I})$ for infinitesimal small $\varepsilon $ and that the difference is
given by the commutator. In this section we will prove this fact rigorously
for the class of potentials introduced in the previous chapter.\bigskip\ 

First of all we use the convergence theorem to extend the validity of
formula (\ref{2.Moment}). We study the two limits $\vec \eta _1\rightarrow {%
\1}_{\left[ t_0,s\right) }\vec e_k$, $\vec \eta _2\rightarrow \delta _{s\pm
\varepsilon }\vec e_l$, $\ t_0<s<t$ . To avoid further terms in our formulas
we assume without loss of generality $\vec x_0=0$. Then we have

\begin{lemma}
$$
\dot x_l(s\pm \varepsilon )x_k(s)\Phi _n\in ({\cal S}_d)^{\prime } 
$$
with $T$-transform%
$$
T\big( \dot x_l(s\pm \varepsilon )x_k(s)\Phi _n\big) (\vec \theta )=\lim _{%
\T \QATOP{\vec \eta _1\rightarrow {\1}_{\left( t_0,s\right] }\vec e_k}{\vec
\eta _2\rightarrow \delta _{s\pm \varepsilon }\vec e_l}}T\big( \langle \cdot
,\vec \eta _1\rangle \langle \cdot ,\vec \eta _2\rangle \Phi _n\big) (\vec
\theta ) 
$$
$$
=T\Phi _n(\vec \theta )\bigg\{\big( \theta _l(s\pm \varepsilon )+\zeta
_l(s\pm \varepsilon )\big) \left[ \tint_{t_0}^s\theta _k+\tint_{t_0}^s\zeta
_k-\frac{s-t_0}{|\Delta |}\left( \tint_\Delta \theta _k+\chi _k+X_k\right)
\right]  
$$
$$
-\frac 1{|\Delta |}\left( \tint_\Delta \theta _l+\chi _l+X_l\right) \left[
\tint_{t_0}^s\theta _k+\tint_{t_0}^s\zeta _k-\tfrac{s-t_0}{|\Delta |}\left(
\tint_\Delta \theta _k+\chi _k+X_k\right) \right]  
$$
$$
+i\tfrac{s-t_0}{|\Delta |}\delta _{kl}+i{\1}_{\left[ t_0,s\right) }(s\pm
\varepsilon )\delta _{kl}\bigg\} 
$$
where the dependence $\tau _1,...,\tau _n\rightarrow \dot x_l(s\pm
\varepsilon )x_k(s)\cdot \Phi _n$ is now only defined in $L^2(\R ^n)$ sense.
(The value at the point $\tau _j=s\pm \varepsilon $ is not uniquely defined,
which causes no problems in respect to later integration.)
\end{lemma}

\TeXButton{Proof}{\proof}Let us look to the terms which may cause problems.

1) The sequence $\{\vec \eta _{2,m},m\in {\N}\}$ may be chosen such that the
support of each $\vec \eta _{2,m}$ does not contain the point $s$ where ${\1}%
_{\left[ t_0,s\right) }$ has its jump. Thus the convergence of $(\eta
_1,\eta _2)$ causes no problems.

2) We may find uniform bounds in $m$ for $(\vec \zeta ,\vec \eta _{1,m})$.
For simplicity suppose $\vec \eta _{1,m}\cdot \vec e_k\leq {\1}_{\left[
t_0,s\right) },$ $m\in {\N}$ then%
$$
\left| (\vec \zeta ,\vec \eta _{1,m})\right| \leq \left| \sum_{j=1}^n\alpha
_{j,k}\int_{t_0}^{\tau _j}{\1}_{\left[ t_0,s\right) }\right| \leq
(s-t_0)\sum_{j=1}^n\left| \vec \alpha _j\right| . 
$$

3) The limit $\vec \eta _{2,m}\rightarrow \delta _{s\pm \varepsilon }\vec
e_l $ in the term $(\vec \zeta ,\vec \eta _{2,m})$ is more subtle. The form $%
\vec \zeta =\sum_{j=1}^n\vec \alpha _j{\1}_{\left[ t_0,\tau _j\right) }$
urges us to study the action of a delta sequence on a step function. If we
write 
$$
\lim _{m\rightarrow \infty }(\vec \eta _{2,m},{\1}_{\left[ t_0,\tau
_j\right) })={\1}_{\left[ t_0,\tau _j\right) }(s)={\1}_{\left[ s,t\right)
}(\tau _j) 
$$
this formula is only valid (pointwise) for $s\neq \tau _j$ or in respect to $%
\tau _j$-dependence in $L^2(\R )$-sense. (In the point $s=\tau _j$ we may
find delta-sequences which produce any value between 0 and 1.)%
\TeXButton{End Proof}{\endproof}\bigskip\ 

\noindent {\bf Remark.} Compare to (\ref{DskPHI}) where a similar extension
procedure forces us to view $\tau _j$-dependence as a distribution. \medskip%
\ 

Now we are interested to study the difference%
$$
T\big(  \big(  \dot x_l(s+\varepsilon )x_k(s)-x_k(s)\dot x_l(s-\varepsilon )%
\big) \Phi _n\big) (\vec \theta )=\hspace{4cm} 
$$
$$
=iT\Phi _n(\vec \theta )\cdot \delta _{kl}+\big(  \theta _l(s+\varepsilon
)-\theta _l(s-\varepsilon )+\zeta _l(s+\varepsilon )-\zeta _l(s-\varepsilon )%
\big)  
$$
\begin{equation}
\label{TCCR}\cdot \left( \int_{t_0}^s\theta _k+\int_{t_0}^s\zeta _k-\frac{%
s-t_0}{|\Delta |}\left( \tint_\Delta \theta _k+\chi _k+X_k\right) \right)
T\Phi _n(\vec \theta )\ . 
\end{equation}

\begin{lemma}
The series%
$$
\sum_{n=0}^\infty \frac{(-i)^n}{n!}\int_{\Box _n}{\rm d}^n\tau \int_{\R %
^{dn}}\tprod_{j=1}^n{\rm d}^d\m (\alpha _j)\ \big(  \dot x_l(s+\varepsilon
)x_k(s)-x_k(s)\dot x_l(s-\varepsilon )\big) \Phi _n 
$$
converges in $({\cal S}_d)^{-1}$.
\end{lemma}

\TeXButton{Proof}{\proof}The convergence of the above stated limits of the
first term on the r.h.s. of (\ref{TCCR}) is proven in Theorem \ref{Iin(Sd)-1}%
. The second term may be bounded by 
\begin{eqnarray}\label{TCCRbound}
2{ | \Delta | }\Big(  2\varepsilon |\vec \theta ^{\prime }|_\infty &+& \sum_{k=1}^n|\vec \alpha _k|{\1}_{[s-\varepsilon ,s+\varepsilon ]}(\tau _k)\Big)  \Big(  |\vec \theta |_\infty +\tsum\limits_{j=1}^n|\vec \alpha _j|+\tfrac{|\vec X|}{2{ | \Delta | }}%
\Big)  {\rm C}_n(\alpha _1,...,\alpha _n,\vec \theta ) \leq
\nonumber \\ & \leq & %
{ | \Delta | }\sum_{k=1}^n\sum_{j=1}^n{\1}_{[s-\varepsilon ,s+\varepsilon ]}(\tau %
_k)\ (|\vec \alpha _k|^2+|\vec \alpha _j|^2)\ {\rm C}_n(\vec \alpha %
_1,...,\vec \alpha _n,\vec \theta ) %
\nonumber \\ & & %
+2{ | \Delta | }\sum_{k=0}^n\left( 2\varepsilon |\vec \theta ^{\prime }|_\infty +\left( %
|\vec \theta |_\infty +\tfrac{|\vec X|}{2{ | \Delta | }}\right) {\1}_{[s-\varepsilon ,s+\varepsilon ]}(\tau _k)\right) |\alpha _k|\ {\rm C}_n(\vec \alpha %
_1,...,\vec \alpha _n,\vec \theta ) %
\nonumber \\ & & %
+4{ | \Delta | }\varepsilon |\vec \theta ^{\prime }|_\infty \ \left( %
|\vec \theta |_\infty +\tfrac{|\vec X|}{2{ | \Delta | }}\right) {\rm C}_n(\vec \alpha %
_1,...,\vec \alpha _n,\vec \theta ) \ .%
\end{eqnarray}
For all $\vec \theta \in {\cal S}_d(\R )$ such that $2|\vec \theta |_0\sqrt{%
t-t_0}<\varepsilon $ we know that%
$$
\int e^{\varepsilon |\vec \alpha |}\ |\m |({\rm d}\alpha ):=K_1,\quad \int
|\vec \alpha |\ e^{\varepsilon |\vec \alpha |}\ |\m |({\rm d}\alpha
):=K_2\quad \text{and}\quad \int |\vec \alpha |^2e^{\varepsilon |\vec \alpha
|}\ |\m |({\rm d}\alpha ):=K_3 
$$
all are finite. Thus the above bound is integrable w.r.t. $\prod_{j=1}^n|\m %
|({\rm d}\alpha _j)$ and w.r.t. $\tau _1...\tau _n.$ The convergence of the
sum causes no problems.\TeXButton{End Proof}{\endproof}\bigskip\ 

In the last step we want to prove that the last term of (\ref{TCCR})

$$
F_\varepsilon (\vec \theta ):=\left( \big(  \theta _l(s+\varepsilon )-\theta
_l(s-\varepsilon )+\zeta _l(s+\varepsilon )-\zeta _l(s-\varepsilon )\big)  %
\right) \hspace{3cm} 
$$
$$
\hspace*{3cm}\cdot \left( \int_{t_0}^s\theta _k+\int_{t_0}^s\zeta _k-\frac{%
s-t_0}{ | \Delta | }\left( \tint_\Delta \theta _k+\chi _k+X_k\right) \right)
T\Phi _n(\vec \theta ) 
$$
has the following property $\sum \frac 1{n!}\int_{\Box _n}{\rm d}^n\tau
\int_{\R ^{dn}}\tprod_{j=1}^n|\m |({\rm d}\alpha _j)\ |F_\varepsilon (\vec
\theta )|$ is of order $\varepsilon $, i.e., the part of equation (\ref{TCCR}%
) connected to $F_\varepsilon (\vec \theta )$ vanishes in the limit $%
\varepsilon \rightarrow 0$.

\TeXButton{Proof}{\proof} Let us consider (\ref{TCCRbound}) after $%
\prod_{j=1}^n|\m |({\rm d}\alpha _j)$-integration is performed%
\begin{eqnarray*}
\int_{\R ^{dn}}{}\tprod_{j=1}^n|\m |({\rm d}\alpha _j)\ |F_\varepsilon %
(\vec \theta )|& \leq & { | \Delta | }\sum_{k=1}^n\sum_{j=1}^n{\1}_{[s-\varepsilon ,s+\varepsilon ]}(\tau _k)\ 2K_3K_1^{n-1} %
\\ & & %
+2{ | \Delta | }\sum_{k=0}^n\left( 2\varepsilon |\vec \theta ^{\prime }|_\infty +\left( %
|\vec \theta |_\infty +\tfrac{|\vec X|}{2{ | \Delta | }}\right) {\1}_{[s-\varepsilon ,s+\varepsilon ]}(\tau _k)\right) K_2K_1^{n-1} %
\\ & & %
+4{ | \Delta | }\varepsilon |\vec \theta ^{\prime }|_\infty \left( |\vec \theta |_\infty +%
\tfrac{|\vec X|}{2{ | \Delta | }}\right) K_1^n 
\end{eqnarray*}
the whole estimate will now be integrated w.r.t. $\tau _1,...,\tau _n$. Here
each integration produces a factor of $2\varepsilon $ iff ${\1}%
_{[s-\varepsilon ,s+\varepsilon ]}(\tau _k)$ appears in the integrand. Thus 
\begin{eqnarray*}
& & \int_{\Box _n}{\rm d}^n\tau \int {}_{\R ^{dn}} \tprod_{j=1}^n  %
|\m |({\rm %
d}\alpha _j)\ |F_\varepsilon (\vec \theta )|\leq 2{ | \Delta | }K_3K_1^{n-1}\cdot %
{ | \Delta | }^{n-1}\cdot 2\varepsilon \cdot n^2 %
\\ & &  \hspace{3cm}%
+2{ | \Delta | }\left( 2{ | \Delta | }\varepsilon |\vec \theta ^{\prime }|_\infty +2\varepsilon \left( %
|\vec \theta |_\infty +\tfrac{|\vec X|}{2{ | \Delta | }}\right) \right) %
K_2K_1^{n-1}{ | \Delta | }^{n-1}\cdot n %
\\ & &  \hspace{3cm}%
+4{ | \Delta | }\varepsilon |\vec \theta ^{\prime }|_\infty \left( |\vec \theta |_\infty +%
\tfrac{|\vec X|}{2{ | \Delta | }}\right) K_1^n{ | \Delta | }^n %
\\ & &  %
=4\varepsilon { | \Delta | }^nK^{n-1}\left\{ K_3n^2+\left( |\vec \theta ^{\prime }|_\infty { | \Delta | }+|\vec \theta |_\infty +\tfrac{|\vec X|}{2{ | \Delta | }}\right) K_2\cdot %
n+K_1|\vec \theta ^{\prime }|_\infty { | \Delta | }\left( |\vec \theta |_\infty +\tfrac{|\vec X|}{2{ | \Delta | }}\right) \right\} \text{.} %
\end{eqnarray*}
The sum converges due to the rapidly decreasing factor $\frac 1{n!}$. The
additional quadratic polynomial (in braces) in $n$ does not prevent
convergence.\TeXButton{End Proof}{\endproof}\bigskip\ 

Thus for $\vec \theta \in {\cal S}_d(\R )$ with $2\sqrt{t-t_0}|\theta
|_0<\varepsilon $ we have

$$
\lim _{\varepsilon \rightarrow 0}T\big( \big( \dot x_k(s+\varepsilon
)x_l(s)-x_l(s)\dot x_k(s-\varepsilon )\big) \;{\rm I}\big) (\vec \theta )=-iT%
{\rm I}(\vec \theta )\cdot \delta _{kl}. 
$$
Using the characterization theorem we have the following result.

\begin{theorem}
In the space $({\cal S}_d)^{-1}$ we have the following identity 
$$
\lim _{\varepsilon \rightarrow 0}\big( \big( \dot x_k(s+\varepsilon
)x_l(s)-x_l(s)\dot x_k(s-\varepsilon )\big) {\rm I}\big) =-i\delta _{kl}{\rm %
I}. 
$$
In particular we have in terms of expectation values%
$$
\lim _{\varepsilon \rightarrow 0}{\E}\big( \big( \dot x_k(s+\varepsilon
)x_l(s)-x_l(s)\dot x_k(s-\varepsilon )\big) {\rm I}\big) =-i\delta _{kl}\E 
({\rm I}). 
$$
\end{theorem}

This reflects the well-known fact that the quantum mechanical observables
position and momentum do {\it not} commute. Thus we have proved a functional
integral form of the canonical commutation relations, which was derived by a
heuristic argument in \cite{FH65}. The above theorem also shows that the
important sample paths in the mean value can not have a continuous
derivative. Hence this form of the canonical commutation relations reflects
the lack of smoothness of the sample paths.

\section{Ehrenfest's theorem}

This section is intended to demonstrate that it is worthwhile to work in a
white noise framework for the discussion of path integrals. The underlying
ideas are simple. We exploit identities from general Gaussian analysis like%
$$
{\E}(D_{\vec T}^{*}{\rm I})=0\ ,\quad {\rm I}\in ({\cal S}_d)^{-1},\ \vec
T\in {\cal S}_d^{\prime }(\R )\ . 
$$
A good choice of $\vec T$ and a calculation of the derivative may lead to
interesting quantum mechanical relations if{\rm \ I }is chosen to be a
Feynman integrand. The above formula may be viewed as a partial integration
formula in functional integrals.\medskip\ 

We start with the Feynman integrand defined in Theorem \ref{Iin(Sd)-1} and
choose 
\begin{equation}
\label{Tee}\vec T:=\delta _s^{\prime }\cdot \vec e_k\text{ ,\qquad }t_0<s<t 
\end{equation}
where $\left\{ \vec e_k,\ 1\leq k\leq d\right\} $ is the canonical basis of $%
\R ^d$. For convenience of notation we will introduce the following
abbreviation $D_{s,k}:=D_{\delta _s^{\prime }\cdot \vec e_k}$. This
differential operator has the following interesting property, which is the
basic motivation for our choice (\ref{Tee}) 
\begin{equation}
\label{partial}D_{s,k}\vec x(t)=\delta (t-s)\;\vec e_k\ . 
\end{equation}
So this represents a kind of ``partial derivative sensitive to the paths at
given time''. In textbooks of theoretical physics (\ref{partial}) is often
used as a definition of the so called functional derivative.

Now we apply $D_{\vec T}^{*}$ to{\rm \ I }and interchange it with limit and
integration. This is allowed because 
\begin{eqnarray*}
T(D_{\vec T}^{*}{\rm I})(\vec \theta )&=& i\,\langle \vec T,\vec \theta %
\rangle \ T{\rm I}(\vec \theta ) %
\\ &=& %
i\sum (-i)^n\int_{\Lambda _n}{\rm d}^n\tau \int_{\R ^{{\rm d}n}}\tprod_{j=1}^n{\rm d}^d\m (\alpha _j)\;\langle \vec T,\vec \theta %
\rangle \;T\Phi _n(\vec \theta ) %
\\ &=& %
\sum (-i)^n\int_{\Lambda _n}{\rm d}^n\tau \int_{\R ^{{\rm d}n}}\tprod_{j=1}^n%
{\rm d}^d\m (\alpha _j)\;T(D_{\vec T}^{*}\Phi _n)(\vec \theta ). %
\end{eqnarray*}
In the next step we apply the (slightly generalized) product rule with $\vec
\eta \in {\cal S}_d(\R )$ to $\Phi _n$ from (\ref{PhinDef})%
$$
D_{\vec \eta }^{*}\Phi _n=D_{\vec \eta }^{*}({\rm I}_0)\prod_{j=1}^n\exp
(i\vec \alpha _j\vec x(\tau _j))-{\rm I}_0D_{\vec \eta }\exp \Big( %
i\sum_{j=1}^n\vec \alpha _j\vec x(\tau _j)\Big)  
$$
\begin{equation}
\label{DeltaPHI}=-i\left\langle \vec \omega ,\vec \eta \right\rangle \cdot
\Phi _n-i\Big( \sum_{j=1}^n\vec \alpha _j\langle {\1}_{\left[ t_0,\tau
_j\right) },\vec \eta \rangle \Big) \Phi _n\ .\hspace{17mm}
\end{equation}
Here we had to be careful because the product rule in this form requires
smooth directions $\vec \eta $ $\in {\cal S}_d(\R )$ of differentiation. To
ensure that the second equality holds, we used only such $\vec \eta $ for
which $\int_{t_0}^t\vec \eta (\tau )\,{\rm d}\tau =\vec 0$ in order to avoid
an additional term coming from the differentiation of Donsker's delta.%
\medskip\ 

In the next step a careful discussion of the limit $\vec \eta
\longrightarrow \delta _s^{\prime }\vec e_k$ is necessary. As we have seen
in Proposition \ref{xsProd} the first term in (\ref{DeltaPHI}) becomes $%
i\ddot x_k(s)\cdot \Phi _n$ Thus let us fix our intermediate result 
\begin{equation}
\label{DskPHI}D_{s,k}^{*}\Phi _n=i\ddot x_k(s)\Phi
_n-i\tsum\limits_{j=1}^n\alpha _{j,k}\delta _s(\tau _j)\Phi _n\ . 
\end{equation}

\noindent {\bf Remark.} In fact the whole argument holds in a more general
situation. Let $\vec T\in {\cal S}_d^{\prime }(\R )$ of order $1$ such that
there exist neighborhoods ${\cal U}_0$ and ${\cal U}$ of $0$ and $t$
respectively with 
$$
T=0\text{ on }{\cal U}_0\text{ and }{\cal U}\text{.} 
$$
Then there exists $\vec S\in {\cal S}_d^{\prime }(\R )$ of order $0$ (i.e.,
a Radon measure), such that 
$$
\frac{{\rm d}}{{\rm d}t}\vec S=\vec T\text{ and }\vec S=0\text{ on }{\cal U}%
_0 
$$
If we assume further $\vec S=0$ on ${\cal U}$ then we have 
$$
D_{\vec T}^{*}\Phi _n=-i\left\langle \vec \omega ,\vec T\right\rangle \Phi
_n-i\sum_{j=1}^n\vec \alpha _j\vec S(\tau _j)\cdot \Phi _n 
$$
(We restricted the order of the distributions $\vec T$ and $\vec S$ to have
a $\tau _j$-dependence in the last formula which allows $\tau _j$%
-integration later in this section).\medskip\ 

Let us now consider the second term in (\ref{DskPHI}) after integration and
summing up%
$$
\sum_{n=0}^\infty \frac{\left( -i\right) ^n}{n!}\sum_{l=1}^n\int_{\Box _n}%
{\rm d}^n\tau \int \Big( \tprod_{j=1}^n{\rm d}^d\m (\alpha _j)\Big) \alpha
_{l,k}\;\delta _s(\tau _l)\;\Phi _n=:(*)\ . 
$$
$\Phi _n$ contains a factor $e^{i\vec \alpha _l\vec x(\tau _l)}$ which is
continuous in $\tau _l$ $\mu $-a.e.. Thus $\tau _l$-- integration amounts
to substitution of $\tau _l$ by $s$. (A different argument can easily be
produced by considering the integration of the corresponding $T$-transform).
Then one has to do a renumbering of integration variables:%
\begin{eqnarray*}
(*) & = & %
\sum_{n=0}^\infty \frac{\left( -i\right) ^n}{n!}n\int_{\Box _{n-1}}{\rm d%
}^{n-1}\tau \int_{\R ^{d(n-1)}}\tprod\limits_{j=1}^{n-1}{\rm d^d}\m (\alpha %
_j)\;\Phi _{n-1}\int {\rm d^d}\m (\alpha )\;\alpha _{,k}\;e^{i\vec \alpha \vec x(s)} %
\\ &=& %
-\left( \int {\rm d^d}\m (\alpha )\;i\,\alpha _{,k}\;e^{i\vec \alpha \vec x(s)}
\right) \sum_{n=0}^\infty \frac{\left( -i\right) ^{n-1}}{(n-1)!}%
\int_{\Box _{n-1}}{\rm d}^{n-1}\tau \int_{\R ^{d(n-1)}}\tprod\limits_{j=1}^{n-1}%
{\rm d^d}\m (\alpha _j)\;\Phi _{n-1} %
\\ &=& %
-(\nabla _kV)(\vec x(s))\cdot {\rm I} 
\end{eqnarray*} 
in the sense of Definition \ref{ProdDefF}. Hence together with Proposition 
\ref{xsProd} we have derived 
$$
D_{s,k}^{*}{\rm I}=i\ddot x_k(s)\cdot {\rm I}+i(\nabla _kV)(\vec x(s))\cdot 
{\rm I}\text{ .} 
$$
Collecting all components and taking expectation we have 
$$
{\E}(\stackrel{..}{\vec x}(s)\cdot {\rm I})=-{\E}(\vec \nabla V(\vec
x(s))\cdot {\rm I})\ . 
$$
This is a variant of the well-known Ehrenfest theorem of quantum mechanics.
The mean-values of the quantum observables satisfy the classical law of
motion.


\small \frenchspacing

\begin{thebibliography}{HL\O UZ93a}
\addcontentsline{toc}{chapter}{Bibliography}
\addtolength{\itemsep}{-1mm}

\bibitem[AR73]{AR73}  Akhiezer, N.I. and Ronkin, L.I. (1973), {\it On
separately analytic functions of several variables and theorems on ``the
thin edge of the wedge''}. Russian Math.\ Surveys {\bf 28}, No.\ 3, 27--44

\bibitem[ADKS94]{ADKS94}  Albeverio, S., Daletzky, Y., Kondratiev, Yu. G. and
Streit, L. (1994), {\it Non-Gaussian infinite dimensional analysis. }%
Preprint, to appear in J. Funct. Anal..

\bibitem[AHPS89]{AHPS89}  Albeverio, S., Hida, T., Potthoff, J. and Streit,
L. (1989),{\it \ The vacuum of the Hoegh-Krohn model as a generalized white
noise functional.} Phys. Lett. {\bf B 217}, 511--514.

\bibitem[AHPRS90a]{AHPRS90a}  Albeverio, S., Hida, T., Potthoff, J.,
R\"ockner, M. and Streit, L. (1990){\it , Dirichlet forms in terms of white
noise analysis I: Construction and QFT examples.} Rev. Math. Phys. {\bf 1},
291--312

\bibitem[AHPRS90b]{AHPRS90b}  Albeverio, S., Hida, T., Potthoff, J.,
R\"ockner, M. and Streit, L. (1990),{\it \ Dirichlet forms in terms of white
noise analysis II: Closability and diffusion processes}. Rev. Math. Phys. 
{\bf 1}, 313--323.

\bibitem[AHK76]{AHK76}  Albeverio, S.A., H\NEG oegh-Krohn, R. (1976), {\it %
Mathematical Theory of Feynman Integrals}. LNM {\bf 523}, Springer Verlag.

\bibitem[AKS93]{AKS93}  Albeverio, S., Kondratiev, Yu.G. and Streit, L.
(1993), {\it How to generalize White Noise Analysis to Non-Gaussian Spaces}.
In: \cite{BSST93}, 120-130.

\bibitem[BeWu69]{BeWu69}  Bender, C.M. and Wu, T.T. (1969), {\it Anharmonic
Oscillator.} Phys. Rev. {\bf 184}, 1231--1260.


\bibitem[BeS95]{BeS95}  Benth, F. and Streit, L. (1995), {\it The Burgers
Equation with a Non-Gaussian Random Force. }UMa preprint.

\bibitem[BeKo88]{BeKo88}  Berezansky, Yu. M. and Kondratiev, Yu. G. (1988), 
{\it Spectral Methods in Infinite-Dimensional Analysis}, (in Russian),
Naukova Dumka, Kiev. English translation, 1995, Kluwer Academic Publishers,
Dordrecht.

\bibitem[BeLy93]{BeLy93}  Berezansky, Yu.M. and Lytvynov, E.V. (1993), {\it %
Generalized White Noise Analysis connected with perturbed field operators},
Dopovidy AN Ukrainy, No {\bf 10}.

\bibitem[BS71]{BS71}  Berezansky, Yu. M. and Shifrin, S.N. (1971), {\it The
generalized degree symmetric Moment Problem,} Ukrainian Math. J. {\bf 23}
N3, 247-258.

\bibitem[BSST93]{BSST93}  Blanchard, Ph., Sirugue--Collin, M., Streit, L.
and Testard, D. (Eds., 1993), {\it Dynamics of complex and Irregular Systems.}
World Scientific.

\bibitem[Bo76]{Bo76}  Bourbaki, N. (1976), {\it Elements of mathematics.
Functions of a real variable. }Addison-Wesley.

\bibitem[C60]{C60}  Cameron, R.H. (1960), {\it A Family of Integrals Serving
to Connect Wiener and Feynman Integrals.} J. Math. Phys. {\bf 39}, 126--140.

\bibitem[C62]{C62}  Cameron, R.H. (1962), {\it The Ilstow and Feynman
Integrals.} J. Anal. Math. {\bf 10}, 287.

\bibitem[CFPSS94]{CFPSS94}  Cardoso, A.I., de Faria, M., Potthoff, J.,
S\'en\'eor, R. and Streit, L. (Eds., 1994), {\it Stochastic Analysis and
Applications in Physics.} Kluwer, Dordrecht.

\bibitem[CLP93]{CLP93}  Cochran, G., Lee, J.--S. and Potthoff, J. (1993), 
{\it Stochastic Volterra equations with singular kernels}. Preprint.

\bibitem[Co82]{Co82}  Colombeau, J.-F. (1982), {\it Differential calculus
and holomorphy. }Mathematical Studies {\bf 64}, North--Holland, Amsterdam.

\bibitem[Co53]{Co53}  Cook, J. (1953), {\it The mathematics of second
quantization.} Trans. Amer. Math. Soc. {\bf 74}, 222--245.

\bibitem[CDLSW95]{CDLSW95}  Cunha, M., Drumond, C., Leukert, P., Silva, J.L.
and Westerkamp, W. (1995), {\it The Feynman integrand for the perturbed
harmonic oscillator as a Hida distribution},
Ann. Physik {\bf 4}, 53--67.

\bibitem[Da91]{Da91}  Daletsky, Yu.L. (1991),{\it \ A biorthogonal analogy
of the Hermite polynomials and the inversion of the Fourier transform with
respect to a non Gaussian measure}, Funct. Anal. Appl. {\bf 25}, 68-70.

\bibitem[FHSW94]{FHSW94}  de Faria, M., Hida, T., Streit, L. and Watanabe,
H. (1995), {\it Intersection local times as Generalized White Noise
Functionals}. BiBoS preprint 642.

\bibitem[FPS91]{FPS91}  de Faria, M., Potthoff, J. and Streit, L. (1991), 
{\it The Feynman integrand as a Hida distribution.} J. Math. Phys. {\bf 32},
2123-2127.

\bibitem[Di81]{Di81}  Dineen, S. (1981), {\it Complex Analysis in Locally
Convex Spaces.} Mathematical Studies {\bf 57}, North Holland, Amsterdam.

\bibitem[D80]{D80}  Doss, H. (1980), {\it Sur une r\'esolution stochastique
de l`\'equation de Schr\"odinger \`a coefficients analytiques. }Comm. math.
Phys. {\bf 73}, 247-264.

\bibitem[Ex85]{Ex85}  Exner, P. (1985), {\it Open Quantum Systems and
Feynman Integrals.} Reidel, Dordrecht.

\bibitem[FH65]{FH65}  Feynman, R.P. and Hibbs, A.R. (1965), {\it Quantum
Mechanics and Path Integrals}. McGraw-Hill, New York.

\bibitem[FGKP95]{FGKP95}  Fleischmann, R., Geisel, T., Ketzmerich, R. and
Petschel, G. (1995), {\it Chaos und fraktale Energiespektren in
Antidot-Gittern.} Physikalische Bl\"atter {\bf 51}, 177-181.

\bibitem[Ga74]{Ga74}  Gawedzki, K. (1974), {\it Construction of
Quantum-Mechanical Dynamics by Means of Path Integrals in Path Space. }Rep.
Math. Phys. {\bf 6} 327--342.

\bibitem[GV68]{GV68}  Gel'fand, I.M. and Vilenkin, N.Ya. (1968),{\it \
Generalized Functions}, Vol. IV. Academic Press, New York and London.

\bibitem[Hi75]{Hi75}  Hida, T. (1975), {\it Analysis of Brownian
Functionals, }Carleton Math. Lecture Notes No. 13, Carleton.

\bibitem[Hi80]{Hi80}  Hida, T. (1980), {\it Brownian Motion}. Springer, New
York.

\bibitem[Hi89]{Hi89}  Hida, T. (1989), {\it Infinite-dimensional rotation
group and unitary group.} LNM {\bf 1379}, Springer, 125--134.

\bibitem[HKPS90]{HKPS90}  Hida, T., Kuo, H.-H., Potthoff, J., and Streit, L.
(Eds.,1990), {\it White Noise - Mathematics and Applications}. World
Scientific, Singapore.

\bibitem[HKPS93]{HKPS93}  Hida, T., Kuo, H.H., Potthoff, J. and Streit, L.
(1993),{\it \ White Noise. An infinite dimensional calculus}. Kluwer,
Dordrecht.

\bibitem[HPS88]{HPS88}  Hida, T., Potthoff, J. and Streit, L.(1988), {\it %
Dirichlet Forms and white noise analysis.} Commun. Math. Phys. {\bf 116},
235--245.

\bibitem[HS83]{HS83}  Hida, T. and Streit, L. (1983), {\it Generalized
Brownian functionals and the Feynman integral}. Stoch. Proc. Appl. {\bf 16},
55--69.

\bibitem[HL\O UZ93a]{HLOUZ93a}  Holden, H., Lindstr\o m, T., \O ksendal, B.,
Ub\o e, J. and Zhang, T.--S. (1993), {\it Stochastic boundary value
problems: A white noise functional approach}; Probab. Th. Rel. Fields {\bf 95%
}, 391--419.

\bibitem[HL\O UZ93b]{HLOUZ93b}  Holden, H., Lindstr\o m, T., \O ksendal, B.,
Ub\o e, J. and Zhang, T.--S. (1993), {\it The pressure equation for fluid
flow in a stochastic medium,} Preprint.

\bibitem[Ho92]{Ho92}  Holstein, B.R. (1992), {\it Topics in advanced Quantum
mechanics.} Redwood City, Ca., Addison-Wesley.

\bibitem[HM88]{HM88}  Hu, Y.Z. and Meyer, P.A. (1988), {\it Chaos de Wiener
et integrale de Feynman. }S\'eminaire de Probabili\'es XXII, LNM {\bf 1321},
Springer, 51-71.


\bibitem[Ito66]{Ito66}  It\^o, K. (1966), {\it Generalized Uniform Complex
Measures in the Hilbertian Metric Space with their Application to the
Feynman Integral. }In: Proceedings of the 5th Berkley Symposium on Statistcs
and Probability. Vol. II, part 1, 145--161.

\bibitem[Ito88]{Ito88}  It\^o, Y. (1988), {\it Generalized Poisson
Functionals.} Prob. Th. Rel. Fields {\bf 77}, 1-28.

\bibitem[IK88]{IK88}  It\^o, Y. and Kubo, I. (1988), {\it Calculus on Gaussian and
Poisson White Noises.} Nagoya Math. J. {\bf 111,} 41-84.

\bibitem[JK93]{JK93}  Johnson, G.W. and Kallianpur G. (1993), {\it %
Homogeneous Chaos, p-Forms, Scaling and the Feynman Integral}, Trans. AMS. 
{\bf 340}, 503-548.

\bibitem[KaS92]{KaS92}  Khandekar, D.C. and Streit, L. (1992),{\it \
Constructing the Feynman integrand}. Ann. Physik {\bf 1}, 49--55.

\bibitem[Ko78]{Ko78}  Kondratiev, Yu.G. (1978), {\it Generalized functions
in problems of infinite dimensional analysis}. Ph.D. thesis, Kiev University.

\bibitem[Ko80a]{Ko80a}  Kondratiev, Yu.G. (1980), {\it Spaces of entire
functions of an infinite number of variables, connected with the rigging of
a Fock space. }In: ``Spectral Analysis of Differential Operators.'' Math.
Inst. Acad. Sci. Ukrainian SSR, p. 18-37. English translation: Selecta Math.
Sovietica {\bf 10} (1991), 165-180.

\bibitem[Ko80b]{Ko80b}  Kondratiev, Yu.G. (1980), {\it Nuclear spaces of
entire functions in problems of infinite dimensional analysis.} Soviet Math.
Dokl. {\bf 22}, 588-592.

\bibitem[KLPSW94]{KLPSW94}  Kondratiev, Yu.G., Leukert, P., Potthoff, J.,
Streit, L. and Westerkamp, W. (1994), {\it Generalized Functionals in
Gaussian Spaces - the Characterization Theorem Revisited. }Manuskripte
175/94, Uni Mannheim.

\bibitem[KLS94]{KLS94}  Kondratiev, Yu.G., Leukert, P. and Streit, L. (1994),%
{\it \ Wick Calculus in Gaussian Analysis,} BiBoS preprint 637, to appear in
Acta Applicandae Mathematicae.


\bibitem[KoSa78]{KoSa78}  Kondratiev, Yu.G. and Samoilenko, Yu.S. (1978), 
{\it Spaces of trial and generalized functions of an infinite number of
variables}, Rep. Math. Phys. {\bf 14}, No.~3, 325-350.

\bibitem[KoS92]{KoS92}  Kondratiev, Yu.G. and Streit, L. (1992),{\it \ A
remark about a norm estimate for White Noise distributions.} Ukrainian Math.
J. 1992 No.7.

\bibitem[KoS93]{KoS93}  Kondratiev, Yu.G. and Streit, L. (1993), {\it Spaces
of White Noise distributions: Constructions, Descriptions, Applications.} I.
Rep. Math. Phys. {\bf 33}, 341-366.

\bibitem[KoSW95]{KoSW94}  Kondratiev, Yu.G., Streit, L. and Westerkamp, W.
(1995), {\it A Note on Positive Distributions in Gaussian Analysis},
Ukrainian Math. J. {\bf 47} No. 5.

\bibitem[KSWY95]{KSWY95}  Kondratiev, Yu.G., Streit, L., Westerkamp, W. and
Yan, J.-A. (1995), {\it Generalized Functions in Infinite Dimensional
Analysis. }IIAS preprint.

\bibitem[KoTs91]{KoTs91}  Kondratiev, Yu.G. and Tsykalenko T.V. (1991), {\it %
Dirichlet Operators and Associated Differential Equations.} Selecta Math.
Sovietica {\bf 10}, 345-397.



\bibitem[KMP65]{KMP65}  Kristensen, P., Mejlbo, L., and Poulsen, E.T.
(1965), {\it Tempered Distributions in Infinitely Many Dimensions. I.
Canonical Field Operators. }Commun. math. Phys. {\bf 1}, 175--214.

\bibitem[KT80]{KT80}  Kubo, I. and Takenaka, S. (1980), {\it Calculus on
Gaussian white noise I, II.} Proc. Japan Acad. {\bf 56}, 376-380 and 411-416.

\bibitem[Ku83]{Ku83}  Kubo, I. (1983), {\it It\^o formula for generalized
Brownian functionals}. In: Theory and Application of Random Fields. Ed.: G.
Kallianpur. Springer, Berlin, Heidelberg, New York.

\bibitem[KK92]{KK92} Kubo, I. and Kuo, H.-H. (1992), {\it Finite Dimensional 
Hida Distributions.} Preprint.

\bibitem[KY89]{KY89}  Kubo, I. and Yokoi, Y. (1989), {\it A remark on the
space of testing random variables in the White Noise calculus}. Nagoya Math.
J. {\bf 115}, 139-149.

\bibitem[Kuo75]{Kuo75}  Kuo, H.-H. (1975), {\it Gaussian Measures in Banach
Spaces.} LNM {\bf 463}, Springer, New York.

\bibitem[Kuo92]{Kuo92}  Kuo, H.-H. (1992), {\it Lectures on white noise
analysis}. Soochow J. Math. {\bf 18}, 229-300.


\bibitem[KPS91]{KPS91}  Kuo, H.-H., Potthoff, J. and Streit, L. (1991), {\it %
A characterization of white noise test functionals.} Nagoya Math. J. {\bf 121%
},185--194.

\bibitem[LLSW94a]{LLSW94a}  Lascheck, A., Leukert, P., Streit, L. and
Westerkamp, W. (1993) {\it Quantum mechanical propagators in terms of Hida
distributions}. Rep.\ Math.\ Phys.\ {\bf 33}, 221--232

\bibitem[LLSW94b]{LLSW94b}  Lascheck, A., Leukert, P., Streit, L. and
Westerkamp, W. (1994), {\it More about Donsker's Delta Function}. Soochow J.
Math. {\bf 20,} 401-418.

\bibitem[Lee89]{Lee89}  Lee, Y.-J. (1989), {\it Generalized Functions of
Infinite Dimensional Spaces and its Application to White Noise Calculus}. J.
Funct. Anal. {\bf 82}, 429--464.

\bibitem[Lee91]{Lee91}  Lee, Y.-J. (1991), {\it Analytic Version of Test
Functionals, Fourier Transform and a Characterization of Measures in White
Noise Calculus. }J. Funct. Anal. {\bf 100}, 359-380.

\bibitem[Lu70]{Lu70}  Lukacs, E. (1970), {\it Characteristic Functions}, 2nd
edition, Griffin, London.

\bibitem[MY90]{MY90}  Meyer, P.A. and Yan, J.-A. (1990),{\it \ Les
``fonctions caract\'eristiques'' des distributions sur l'\'espace de Wiener.}
Seminaire de Probabilites XXV, Eds.: J. Azema, P.A. Meyer, M. Yor, Springer,
p. 61--78.

\bibitem[Mu79]{Mu79}  Mumford, D. (1979), {\it Tata Lectures on Theta I}.
Birkh\"auser, Boston, Basel, Stuttgart.

\bibitem[Na69]{Na69}  Nachbin, L. (1969), {\it Topology on spaces of
holomorphic mappings.} Springer, Berlin.

\bibitem[Ne73]{Ne73}  Nelson, E. (1973), {\it Probability theory and
Euclidean quantum field theory. }In: ``Constructive Quantum Field Theory.''
Eds.: Velo, G. and Wightman, A., Springer, Berlin, Heidelberg, New York.

\bibitem[Ob91]{Ob91}  Obata, N. (1991), {\it An analytic characterization of
symbols of operators on white noise functionals. }J. Math. Soc. Japan 
{\bf 45} No. 3, 421--445.

\bibitem[Ob94]{Ob94}  Obata, N. (1994), {\it White Noise Calculus and Fock
Space.} LNM {\bf 1577}. Springer, Berlin.


\bibitem[\O k94]{Ok94}  \O ksendal, B. (1994),{\it \ Stochastic Partial
Differential Equations and Applications to Hydrodynamics. }In: \cite{CFPSS94} .

\bibitem[Ou91]{Ou91}  Ouerdiane, H. (1991),{\it \ Application des m\'ethodes
d'holomorphie et de distributions en dimension quelconque \'a l'analyse sur
les espaces Gaussiens}. BiBoS preprint 491.

\bibitem[Pi69]{Pi69}  Pietsch, A (1969), {\it Nukleare Lokal Konvexe R\"aume}%
, Berlin, Akademie Verlag.

\bibitem[Po87]{Po87}  Potthoff, J. (1987), {\it On positive generalized
functionals. }J. Funct. Anal. {\bf 74}, 81-95.

\bibitem[Po91]{Po91}  Potthoff, J. (1991), {\it Introduction to white noise
analysis}. In: ``Control Theory, Stochastic Analysis and Applications.''
Eds.: S. Chen, J. Yong; Singapore, World Scientific.

\bibitem[Po92]{Po92}  Potthoff, J. (1992), {\it White noise methods for
stochastic partial differential equations.} In: ``Stochastic Partial
Differential Equations and Their Applications." Eds.: B.L. Rozovskii, R.B.
Sowers; Berlin, Heidelberg, New York, Springer.

\bibitem[Po94]{Po94}  Potthoff, J. (1994), {\it White noise approach to
parabolic stochastic differential equations. }In: \cite{CFPSS94}.

\bibitem[PS91]{PS91a}  Potthoff, J. and Streit, L. (1991), {\it A
characterization of Hida distributions.} J. Funct. Anal. {\bf 101}, 212-229.


\bibitem[PS93]{PS93}  Potthoff, J. and Streit, L. (1993), {\it Invariant
states on random and quantum fields: }$\phi -${\it bounds and white noise
analysis}. J. Funct. Anal. {\bf 101}, 295-311.

\bibitem[PT94]{PT94}  Potthoff, J. and Timpel, M. (1994), {\it On a Dual
Pair of Spaces of Smooth and Generalized Random Variables}. Preprint:
Manuskripte No. 168/93 Uni Mannheim.

\bibitem[ReSi72]{ReSi72}  Reed, M. and Simon, B. (1972), {\it Methods of
Modern Mathematical Physics I: Functional Analysis}. Academic Press, New
York and London.

\bibitem[Ri87]{Ri87}  Rivers, R. (1987), {\it Path integral methods in
quantum field theory.} Cambridge University Press, Cambridge, New York,
Sydney.

\bibitem[Sch71]{Sch71}  Schaefer, H.H. (1971),{\it \ Topological Vector
Spaces}. Springer, New York.

\bibitem[Se56]{Se56}  Segal, I. (1956), {\it Tensor algebras over Hilbert
spaces}. Trans. Amer. Math. Soc. {\bf 81}, 106-134.

\bibitem[Si69]{Si69}  Siciak, J. (1969), {\it Separately analytic functions
and envelopes of holomorphy of some lower dimensional subsets of }$\C ^n $,
Ann.\ Polonici Math.\ {\bf 22}, 145--171.

\bibitem[Si74]{Si74}  Simon, B. (1974), {\it The }$P(\Phi )_2${\it \
Euclidean (Quantum) Field Theory}. Princeton University Press, Princeton.

\bibitem[Sk74]{Sk74}  Skorohod, A.V. (1974), {\it Integration in Hilbert
Space}, Springer, Berlin.

\bibitem[S93]{S93}  Streit, L. (1993), {\it The Feynman Integral - Recent
Results. }In: \cite{BSST93}, 166-173.

\bibitem[S94]{S94}  Streit, L. (1994), {\it Introduction to White Noise
Analysis}. In: \cite{CFPSS94}, 415--440.

\bibitem[SW93]{SW93}  Streit, L. and Westerkamp, W. (1993),{\it \ A
generalization of the characterization theorem for generalized functionals
of White Noise}. In: \cite{BSST93}, 174-187.

\bibitem[Sz39]{Sz39}  Szeg\"o, G. (1939), {\it Orthogonal Polynomials.} 3rd 
edition, Am. Math. Soc., Providence, Rhode Island.

\bibitem[Ta75]{Ta75}  Tarski, J. (1975), {\it Definitions and selected
applications of Feynman-type integrals. }In: ``Functional integration and
its application.'' Ed.: A.M. Arturs, Oxford, 169--180.

\bibitem[Us94]{Us94}  Ushveridze, A.G. (1994), {\it Quasi--exactly solvable
Models in Quantum Mechanics.} Institute of Physics Publishing, Bristol and
Philadelphia.

\bibitem[Va95]{Va95}  V\aa ge, G. (1995), {\it Stochastic Differential
Equations and Kondratiev Spaces. }Ph.D. thesis, Trondheim University. 

\bibitem[VGG75]{VGG75}  Vershik, A.M., Gelfand, I.M. and Graev, M.I. (1975), 
{\it Representations of diffeomorphisms groups}. Russian Math. Surveys {\bf %
30}, No 6, 3-50.

\bibitem[Wa91]{Wa91}  Watanabe, H. (1991), {\it The local time of
self-intersections of Brownian Motions as generalized Brownian functionals},
Lett. Math. Phys. {\bf 23}, 1--9.

\bibitem[Wa93]{Wa93}  Watanabe, H. (1993), {\it Donsker`s delta function and
its application in the theory of White Noise Analysis.} Kallianpur
Festschrift Springer, 338.

\bibitem[W93]{W93}  Westerkamp, W. (1993), {\it A Primer in White Noise
Analysis}. In: \cite{BSST93}, 188-202.

\bibitem[Yan90]{Yan90}  Yan, J.-A. (1990), {\it A characterization of white
noise functionals.} Preprint.

\bibitem[Yan93]{Yan93}  Yan, J.-A. (1993),{\it \ From Feynman-Kac Formula to
Feynman Integrals via Analytic Continuation}. Preprint.

\bibitem[Yok90]{Yok90}  Yokoi, Y.(1990), {\it Positive generalized white
noise functionals}. Hiroshima Math. J. {\bf 20}, 137-157.

\bibitem[Yok93]{Yok93}  Yokoi, Y. (1993), {\it Simple setting for white
noise calculus using Bargmann space and Gauss transform.} Preprint.

\bibitem[Yo80]{Yo80}  Yosida, K. (1980), {\it Functional Analysis}.
Springer, Berlin.

\bibitem[Za76]{Za76}  Zaharjuta, V.P. (1976), {\it Separately analytic
functions, generalizations of Hartogs' theorem, and envelopes of holomorphy}%
. Math.\ USSR Sbornik {\bf 30}, 51--67.

\bibitem[Zh92]{Zh92}  Zhang, T.--S. (1992), {\it Characterization of white
noise test functions and Hida distributions.} Stochastics {\bf 41}, 71--87.

\end{thebibliography}
\end{document}